\documentclass[pdflatex,sn-apa,table]{sn-jnl}

\usepackage{mathpazo}
\usepackage{parskip}
\usepackage{graphicx}
\usepackage{multirow}
\usepackage{amsmath,amssymb,amsfonts}
\usepackage{xcolor}
\usepackage{textcomp}
\usepackage{booktabs}
\usepackage{listings}
\usepackage{array}
\usepackage{tabularx}
\usepackage{adjustbox}
\usepackage{makecell}
\usepackage{subcaption}
\usepackage{xspace}
\usepackage[utf8]{inputenc}
\usepackage{tcolorbox}
\usepackage{enumitem}
\usepackage{cleveref}
\usepackage{bibunits}
\usepackage{fancyhdr}

\crefname{paragraph}{paragraph}{paragraphs}
\Crefname{paragraph}{Paragraph}{Paragraphs}
\tcbuselibrary{breakable}

\definecolor{colorControversy}{HTML}{F4C3CB}
\definecolor{colorCertainty}{HTML}{8FB8ED}
\definecolor{colorCC}{HTML}{C2BBF0}
\definecolor{colorAccept}{HTML}{EAF6EF}
\definecolor{colorReject}{HTML}{FDF2E5}
\definecolor{condControl}{HTML}{BBBBBB}
\definecolor{condControversy}{HTML}{CC6699}
\definecolor{condCertainty}{HTML}{6699CC}
\definecolor{condCC}{HTML}{9999FF}
\definecolor{condNA}{HTML}{EEEEEE}

\newcommand{\emptybox}[1]{{\setlength{\fboxsep}{\dimexpr\fboxsep-\fboxrule\relax}\textcolor{#1}{\fbox{}}}}
\newcommand{\condition}[4]{%
  \ifthenelse{\equal{#1}{1}}{\colorbox{condControl}{}}{\emptybox{condControl}}\hspace{2pt}%
  \ifthenelse{\equal{#2}{1}}{\colorbox{condControversy}{}}{\emptybox{condControl}}\hspace{2pt}%
  \ifthenelse{\equal{#3}{1}}{\colorbox{condCertainty}{}}{\emptybox{condControl}}\hspace{2pt}%
  \ifthenelse{\equal{#4}{1}}{\colorbox{condCC}{}}{\emptybox{condControl}}%
}
\newcommand{\dataset}{{{\sc MoralMoments }\xspace}}

\begin{document}
\begin{bibunit}[sn-apacite]
\title[Controversy and Group Certainty in Everyday Moral Dilemmas]{
  \centering
  Controversy and Group Certainty Jointly Shape Everyday Moral Judgments\\
    }

\author[1]{\fnm{Ziyu} \sur{Chen}}
\author[1]{\fnm{Minjeong} \sur{Shin}}
\author[2]{\fnm{Tuan Dung} \sur{Nguyen}}
\author[1]{\fnm{Colin} \sur{Klein}}
\author[3]{\fnm{Chenhao} \sur{Tan}}
\author[4]{\fnm{Nick} \sur{Schuster}}
\author[1]{\fnm{Nicholas George} \sur{Carroll}}
\author[1]{\fnm{Alasdair} \sur{Tran}}
\author[1]{\fnm{Lexing} \sur{Xie}}

\affil[1]{The Australian National University, Australia}
\affil[2]{University of Pennsylvania, United States}
\affil[3]{University of Chicago, United States}
\affil[4]{University of Georgia, United States}

\abstract{
Everyday moral life rarely resembles a trolley problem. It involves disputes about families, relationships, work, money, and care, where people often encounter the judgments of others. We examined how judgments about nuanced interpersonal dilemmas respond to social information that conveys collective opinion without revealing the arguments behind it. Specifically, we studied two signals: \textit{controversy}, the extent to which community judgments are divided between two opposing verdicts; and \textit{group certainty}, the confidence expressed by each side.
We derived these signals from 54,827 judgments on 135 dilemmas posted to Reddit’s r/AmItheAsshole and presented them separately or together in a preregistered randomized experiment (N = 2{,}159).
Relative to the control condition, all three treatments increased both weakening, a changed verdict or reduced confidence, and strengthening, increased confidence without a verdict change.
Thus, aggregated social information can reshape moral judgments even without arguments.
Controversy alone tripled the weakening rate, from 5.3\% to 15.9\%.
Shown alone, certainty on either side was associated with movement toward that side. Shown alongside controversy, only like-minded certainty remained clearly associated with less weakening.
This asymmetry of influence is double-edged: the confidence of like-minded others may help minorities resist majority pressure, but it may also insulate mistaken judgments from correction.
}

\makeatletter
\renewcommand{\keywords}[1]{
  \ifx#1\empty\else
    \def\@keywords{\par\addvspace{2pt}{\keywordfont{\bfseries\keywordname:} #1\par}}
  \fi
}
\makeatother
\enlargethispage{2\baselineskip}
\maketitle
\addtocontents{toc}{\protect\setcounter{tocdepth}{-1}}
\clearpage
\newgeometry{left=3cm,right=3cm,top=3cm,bottom=3.5cm}

\newpage
\section*{Main}
\label{sec:Main}
\addcontentsline{toc}{section}{Main}

We do not make moral judgments in a social vacuum.
Our judgments on moral issues can change quietly, under subtle influences from the beliefs of our peers. How malleable are opinions about everyday moral dilemmas? How much can we be swayed by learning about the degree of controversy and the certainty of others?

Moral judgments are generally considered difficult to change because of their close ties to identity, emotion, and deeply held values.
Experimental research on moral judgment often uses canonical dilemmas, such as variants of the trolley problem~\citep{Awad2018TheMM}, or standardized vignettes designed to elicit particular moral concerns~\citep{clifford2015moral}.
The everyday interpersonal dilemmas examined here are often less stark and more context-dependent.
They range from relatively ordinary questions, such as choosing an appropriate gift or taking sides in a family dispute, to more consequential decisions about caring for a relative’s child or reporting suspected wrongdoing.
Complementary research on opinion change has examined political attitudes~\citep{Balietti2021ReducingOP}, conspiracy beliefs~\citep{costello2024durably}, and contentious social issues discussed with or facilitated by AI~\citep{salvi2025conversational,tessler2024ai}, typically through exposure to arguments or dialogue.
Less is known about whether judgments of nuanced interpersonal dilemmas change in response to thinner forms of social evidence that reveal collective opinion without providing the reasons behind it.

Collective opinion can convey at least two kinds of social information that are separable from the substance of the underlying debate.
The first is \emph{controversy}: the distribution of judgments across a population~\citep{oktar2025aggregated}. It indicates whether a person’s initial view aligns with a majority or minority. The second is group-level \emph{certainty}: how confidently people on each side express their judgments.
Processing uncertainty is a fundamental feature of human cognition~\citep{kahneman1982variants}, and communicating uncertainty has important consequences in domains such as health, medicine, and climate change~\citep{vanderbles2020effects,Budescu2012EffectiveCO,Johnson1995PresentingUI}.
Philosophical accounts have considered how disagreement and confidence ought to interact~\citep{christensen2007epistemology,vavova2014confidence}.
Empirical work shows that confidence shapes interpersonal influence \citep{Kappes2020}, consistent minorities can influence majority judgments \citep{Moscovici1969Minority},
and that presenting group size and confidence together can improve factual judgments~\citep{HeLienZheng2026}.
Yet it remains unknown whether displaying group certainty changes moral judgments, or how it interacts with controversy in nuanced interpersonal dilemmas.

To address this question, we conducted a preregistered randomized controlled experiment. We drew 135 varied interpersonal moral dilemmas from Reddit’s r/AmItheAsshole, spanning 32 topics including {\it money, communication, relationships, children, work, and food}. From 54,827 community judgments, we derived aggregate controversy and group-certainty signals for each dilemma.

Participants (N=2{,}159) were randomly assigned to receive information about controversy, group certainty, both signals, or an active Control task without social information. Each participant evaluated two dilemmas, reporting an initial verdict and confidence, viewing the assigned display or completing the Control task, and then reporting their verdict and confidence again, yielding 4,318 judgment observations. We classified each response as \emph{weakened} when the verdict changed or confidence decreased, \emph{maintained} when neither changed, or \emph{strengthened} when confidence increased without a verdict change. Throughout, “in-group” and “out-group” refer solely to whether a community verdict agrees or disagrees with a participant’s initial verdict; they do not refer to demographic or identity groups.

We report two main findings. First, random assignment to any of the three social-information conditions (Controversy, Certainty, and Controversy+Certainty) increased both weakening and strengthening relative to Control, showing that brief, argument-free summaries can shift moral judgments in either direction across diverse interpersonal dilemmas.
Second, within the treatment conditions, disagreement and side-specific certainty showed distinct associations with judgment change.
When the signals were displayed separately, greater disagreement was associated with more weakening and less strengthening; higher in-group certainty was associated with less weakening and more strengthening, while higher out-group certainty showed the reverse pattern.
When Controversy and Certainty were displayed together, disagreement retained its associations with both outcomes, but the certainty pattern became asymmetric: higher in-group certainty remained associated with less weakening, while the associations with out-group certainty were no longer clear.
Consistent with this pattern, minority judgment holders weakened less often when their own side was equally or more certain than the majority.
 Participants’ written explanations showed a corresponding pattern: controversy was mentioned most often, and in-group certainty more often than out-group certainty.
Together, these findings show that people do not respond to group certainty in isolation: its association with judgment change differs when they can also see how divided the group is and whether their initial view is in the majority or minority.

\section*{Experimental setup}
\label{sec:expsetup}
\addcontentsline{toc}{section}{Experimental setup}

We draw real-life moral dilemmas from the r/AmItheAsshole (AITA) subreddit, a Reddit community with 2.8M weekly visitors~\citep{reddit_aita_stats_2026}.
In this community, each post contains a title and narrative describing a personal conflict, followed by other users' verdicts and comments. We use \textbf{post} for this Reddit source material and \textbf{dilemma} for the moral situation presented to participants; each selected post supplies one dilemma in the experiment. For example, one post title reads ``\textit{AITA for or not telling my significant other that I am a millionaire?}'' AITA posts cover a wide range of topics such as communication, family, money, work, and relationships (see \Cref{subsec:aita} for details).
Because of its scale and diversity, AITA has become a source of everyday moral dilemmas and judgments \citep{Lourie2020ScruplesAC,Candia2022SocialNO,Nguyen2022MappingTI,sachdeva2025normative} and an alternative to expert-crafted moral vignettes \citep{clifford2015moral}. However, prior work has not studied whether people's moral judgments can change.

\begin{figure}[!htbp]
    \centering
    \includegraphics[width=\textwidth]{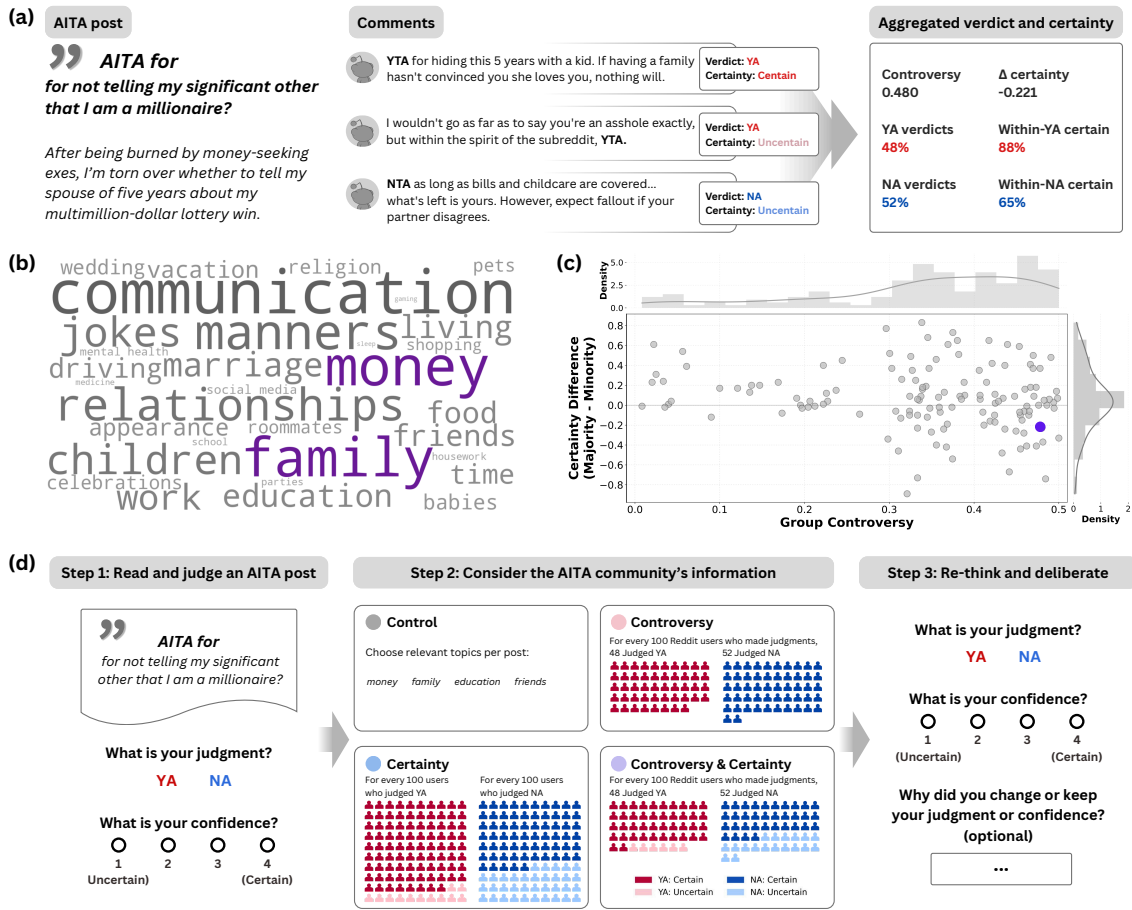}
    \caption{
    \textbf{Overview: source posts, dataset profile, and study design.} The \dataset{} dataset is constructed by aggregating verdict and certainty signals from AITA source posts and comments (a). It contains 135 dilemmas across 32 topics (b), each profiled by controversy and certainty difference (c), and supplies the dilemmas and social signals used in the randomized survey experiment (d).
    \textbf{(a) From an AITA post to aggregated social signals.} An AITA post presents a personal conflict followed by user comments. Each top-level comment contributes a verdict (YA: original poster at fault; NA: original poster not at fault) and is classified as certain or uncertain. Aggregating all comments yields verdict shares, certainty within each verdict group, group controversy, and certainty difference. The three comments shown illustrate the coding; the aggregate values are computed from all comments on the selected post.
    \textbf{(b) Topic distribution.} Word cloud of the 32 topics covered by the 135 dilemmas in the \dataset{} dataset. Text size is proportional to the number of dilemmas per topic; purple marks the topics assigned to the example in (a).
\textbf{(c) Group controversy versus certainty difference.} Scatterplot of the 135 dilemmas in the \dataset{} dataset, one point per dilemma. The x-axis shows group controversy: the proportion of verdicts in the minority verdict group, ranging from 0 (full consensus) to 0.5 (maximum division). The y-axis shows certainty difference: the proportion of certain comments in the majority verdict group minus that in the minority verdict group. Marginal histograms show the distribution of dilemmas across controversy (top) and certainty difference (right); the purple point marks the example in (a).
\textbf{(d) Study design of the moral judgment task.} For each dilemma, participants provide an initial verdict and confidence. They are then exposed to treatment information (Controversy and/or Certainty) or, in the Control condition, complete a topic-selection task. Each signal is presented as an infographic of 100 human icons scaled to the corresponding percentage, with a numeric label and brief description. Controversy is shown by coloring icons by verdict side (YA in red, NA in blue); Certainty is shown by shading icons (dark for certain, light for uncertain). Participants then resubmit their verdict and confidence, with an optional written explanation.
}
\label{fig:aggregated_signals}
                        \end{figure}

\textbf{Verdicts and judgments.} Top-level AITA comments usually contain a verdict label, such as \textit{YTA} (the original poster is at fault), \textit{ESH} (everyone involved is at fault), \textit{NTA} (the original poster is not at fault), or \textit{NAH} (no one is at fault). We combine these labels into two binary \textbf{verdicts}: \textit{YA} (the original poster is at fault), combining \textit{YTA} and \textit{ESH}, and \textit{NA} (the original poster is not at fault), combining \textit{NTA} and \textit{NAH}. For each post, the verdict with more responses on Reddit defines the \textit{majority verdict group}, and the other defines the \textit{minority verdict group}. For participants, a \textbf{judgment} refers to their verdict and confidence rating together, each recorded before and after treatment.

\textbf{The \dataset dataset.}
We curated the \dataset dataset\footnote{A website presenting an interactive overview of the \dataset dataset, study design, and findings is available at \url{https://moralmoments.vercel.app/}.}, comprising 135 real-life moral dilemmas drawn from AITA posts (\Cref{fig:aggregated_signals}b). Each post contains at least 50 verdicts, totaling 54{,}827. To ensure topic diversity, we selected 32 topics from an existing 47-topic taxonomy of moral dilemmas \citep{Nguyen2022MappingTI}, retaining those that are frequently discussed and highly relevant to everyday life (\Cref{fig:aggregated_signals}b).
 To ensure adequate coverage across the controversy distribution, we oversampled highly controversial posts.
We used a large language model (LLM) to extract a second key social signal for each post: \textit{Group Certainty}, from the comment accompanying each verdict (see validation in Methods and \Cref{subsec:uncertainty}). This represents the proportion of comments classified as certain within each verdict group.

\Cref{fig:aggregated_signals}c profiles the 135 dilemmas in the \dataset dataset using controversy and certainty difference derived from their source posts, with each point representing one dilemma.
Group controversy is the proportion of verdicts in the minority, ranging from 0 (full consensus) to 0.5 (an even split).
Certainty difference is defined as majority group certainty minus minority group certainty: positive values indicate higher majority certainty, negative values indicate higher minority certainty, and values near zero indicate similar certainty between the two groups. Overall, moral dilemmas show wide variation in both Controversy and Certainty. When Controversy is low (minority size below roughly 30\%), the majority is usually more certain than the minority. However, as Controversy approaches an even split, minority-group certainty can be either higher or lower than majority-group certainty.

\textbf{Study design.}
To test how participants respond to these signals, we used a randomized controlled design with the three-step moral judgment task illustrated in \Cref{fig:aggregated_signals}d. See \hyperref[sec:methods]{Methods} and \Cref{sec:surveyexperiment} for full design details.
In Step 1, participants read the dilemma and provided an initial moral verdict along with a confidence rating. In Step 2, participants were presented with information corresponding to their randomly assigned treatment condition (\Cref{fig:aggregated_signals}d).
The treatment information were presented as percentages, brief textual descriptions, and infographics; see \hyperref[sec:methods]{Methods} and \Cref{subsec:survey_flow}.
\begin{itemize}
    \item \textbf{Controversy}: The fraction of AITA users who agreed and disagreed with the participant's initial verdict for that dilemma.
        \item \textbf{Certainty}: The fractions of AITA users who were certain and uncertain within each verdict group, shown separately for the group who agreed with the participant's initial verdict and the group who disagreed.
        \item \textbf{Controversy+Certainty}: Both of the aforementioned signals presented together.
    \item \textbf{Control}: no information about others’ verdicts or certainty was shown; participants instead completed a topic-selection task for each dilemma.
    \end{itemize}

In Step 3, participants recorded their post-treatment verdict and confidence. They could optionally explain whether and why they changed or maintained their responses.
After the task, participants completed questionnaires measuring decision-making styles, personality traits, and intellectual humility.

\begin{table}[!tbp]
\centering
\footnotesize
\renewcommand{\arraystretch}{1.5}

\raisebox{2pt}{\colorbox{condControl}{}} Control \quad
\raisebox{2pt}{\colorbox{condControversy}{}} Controversy \quad
\raisebox{2pt}{\colorbox{condCertainty}{}} Certainty \quad
\raisebox{2pt}{\colorbox{condCC}{}} Controversy + Certainty \quad
\raisebox{2pt}{\emptybox{condControl}} Not applicable
\par\vspace{0.1cm}

\begin{tabular}{>{\raggedright\arraybackslash}p{1.3cm} >{\raggedright\arraybackslash}p{1.8cm}
@{\hspace{4pt}}>{\raggedright\arraybackslash}p{1.2cm}@{\hspace{4pt}}
>{\raggedright\arraybackslash}p{6cm} >{\raggedright\arraybackslash}p{3cm}}
\toprule
\textbf{Category} & \textbf{Variables} & \textbf{Treatment} & \textbf{Explanation} & \textbf{Data type} \\
\midrule
\makecell[tl]{Outcome}
    & Weakened vs. Maintained vs. Strengthened & \condition{1}{1}{1}{1} & \textit{Weakened}: verdict changed, or verdict did not change and confidence decreased; \textit{Maintained}: verdict and confidence stayed the same; \textit{Strengthened}: verdict did not change and confidence increased & Categorical \\
\midrule
\makecell[tl]{Pre-treatment}
    & Initial Confidence & \condition{1}{1}{1}{1} & Participants' initial confidence before the exposure to the treatment or control & Continuous, 0-1 (4-pt likert rescaled, 1 dimension) \\
\midrule
\makecell[tl]{Treatment\\variables}
    & Disagreement rate & \condition{0}{1}{0}{1} & Fraction of community verdicts opposing the participant's initial verdict & Continuous, 0-100\% \\
    & In-group certainty & \condition{0}{0}{1}{1} & Fraction of confident verdicts among users who agree with the participant & Continuous, 0-100\% \\
    & Out-group certainty & \condition{0}{0}{1}{1} & Fraction of confident verdicts among users who disagree with the participant & Continuous, 0-100\% \\
\midrule
\makecell[tl]{Participant's\\social\\position}
    & \makecell[tl]{\textit{In the Majority}\\vs.\\\textit{In the Minority}}  & \condition{0}{1}{0}{1} & \textit{In the Majority} if fewer than half of community verdicts oppose the participant (disagreement rate $<50\%$); \textit{In the Minority} otherwise & Binary \\
    & $In \ge Out$ vs. $In < Out$ & \condition{0}{0}{1}{1} & $In \ge Out$ if the in-group was equally or more certain than the out-group; $In < Out$ otherwise & Binary \\
\midrule
\makecell[tl]{Participant's\\profile}
    & Intellectual Humility & \condition{1}{1}{1}{1} & The degree to recognizing that a particular personal belief may be fallible & Continuous, 0-1 (5-pt likert rescaled, 1 dimension) \\
    & Decision Making styles & \condition{1}{1}{1}{1} & Five decision-making styles: \textit{Rational}, \textit{Dependent}, \textit{Intuitive}, \textit{Avoidant}, \textit{Spontaneous} & Continuous, 0-1 (5-pt likert rescaled, 5 dimensions) \\
    & Big Five Personality & \condition{1}{1}{1}{1} & Five personality traits: \textit{Extraversion}, \textit{Openness}, \textit{Conscientiousness}, \textit{Agreeableness}, \textit{Emotional Stability} & Continuous, 0-1 (7-pt likert rescaled, 5 dimensions) \\
        & Demographics & \condition{1}{1}{1}{1} & \textit{Age}, \textit{Sex}, \textit{Country} & Continuous; Categorical; Categorical \\
\bottomrule
\end{tabular}
\caption{\textbf{Variables used in the analysis.} The table details the applicable treatment conditions, definition, and data type for each outcome, pre-treatment measure, treatment variable, derived social position, and participant profile. Colored boxes in the \textit{Treatment} column indicate the conditions in which each variable applies; an outlined box indicates that it does not apply.}
\label{tab:variables}
\end{table}

\textbf{Experimental measures and derived variables.}
\Cref{tab:variables} summarizes the pre-treatment measure, treatment variables, judgment-change outcome, derived social-position variables, and participant-profile measures used in the analysis. Initial confidence and participant-profile measures were collected directly in the survey. All treatment variables and judgment-change outcomes were coded relative to each participant's initial verdict, so YA and NA did not enter the analysis as separate predictors. YA and NA refer to different moral content across dilemmas and are therefore not directly comparable, whereas majority and minority status is defined relative to each dilemma's own verdict distribution and provides a consistent construct across all 135 dilemmas.
The judgment-change outcome classifies each response into one of three types: a response is {\it maintained} when verdict and confidence both stay the same; {\it weakened} when the participant changes their verdict, or keeps the same verdict but reports lower confidence; and {\it strengthened} when the participant keeps the same verdict and reports higher confidence. We model verdict and confidence jointly as a single outcome rather than as two separate outcomes (first whether the verdict changed and then, if not, whether confidence changed), because both reflect one process of reconsidering a dilemma rather than two independent decisions.
See \Cref{app:overall} for detailed profiles and \Cref{app:j_c_change} for statistical analysis of verdict and confidence changes separately.

Experimental conditions were represented by three treatment variables. \emph{Disagreement rate}, shown in the Controversy and Controversy+Certainty conditions, was the fraction of AITA verdicts opposing the participant's initial verdict. \emph{In-group certainty} and \emph{out-group certainty}, shown in the Certainty and Controversy+Certainty conditions, were the proportions of comments classified as certain within the verdict groups that agreed and disagreed with the participant, respectively. Here, in-group and out-group refer only to verdict agreement rather than demographic groups, social identities, or group affiliations in other related work \citep{pryor2019even,feldmanhall2019resolving}.
To examine how majority status and relative group certainty shaped judgment change, we derived two social-position variables for each participant--dilemma observation. In Controversy and Controversy+Certainty conditions, an observation was classified as \textit{In the Majority} if more than 50\% of AITA verdicts agreed with the participant's initial verdict and as \textit{In the Minority} otherwise. In Certainty and Controversy+Certainty conditions, an observation was classified as \textit{In $\geq$ Out} when the in-group was equally or more certain than the out-group and as \textit{In $<$ Out} otherwise.

\section*{Results}
\label{sec:results}
\addcontentsline{toc}{section}{Results}

We first test whether brief summaries of controversy and group certainty change moral judgments overall. We then examine how participants’ social positions were associated with weakening and strengthening, before turning to exploratory analyses of written explanations, dilemma consequences, and participant feedback.

\subsection*{Overall treatment effect}\label{sec:overall}

The randomized experiment included $N=2{,}159$ participants, each of whom evaluated two dilemmas, yielding $n=4{,}318$ judgment observations.
We compared each participant’s verdict and confidence before and after the assigned display or Control task, classifying responses as maintained, weakened, or strengthened (\Cref{tab:variables}).

\begin{figure}[!htbp]
    \centering
    \includegraphics[width=\linewidth]{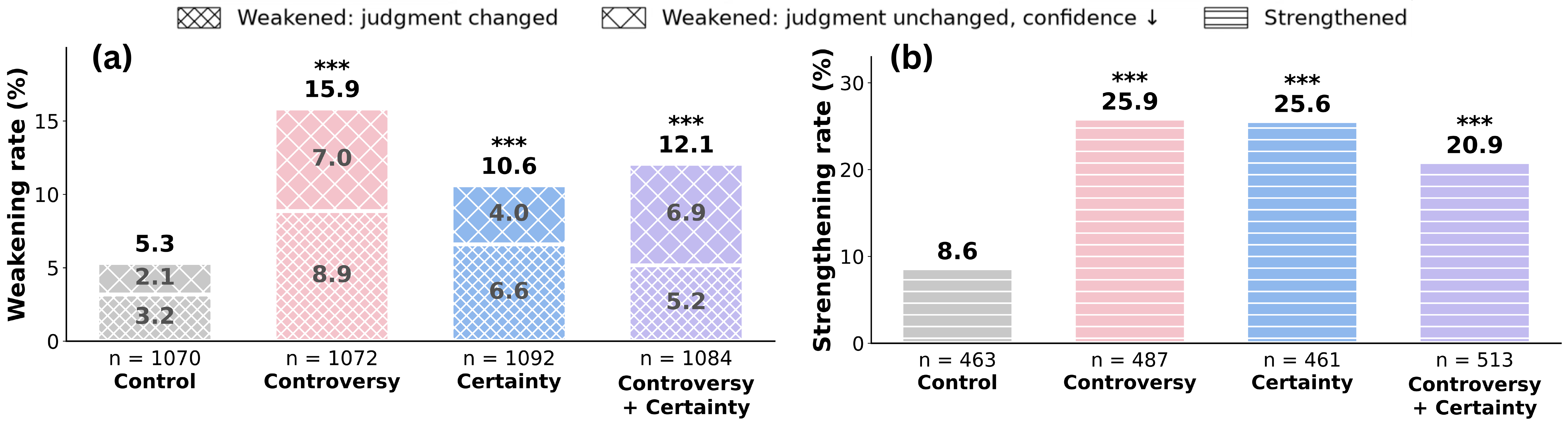}
    \caption{\textbf{Judgment weakening and strengthening across experimental conditions.}
    \textbf{(a)} Weakening among all judgment observations ($n = 4{,}318$). Stacked bars distinguish verdict changes (bottom, dense cross-hatching) from unchanged verdicts accompanied by decreased confidence (top, sparse cross-hatching).
    \textbf{(b)} Strengthening, defined as an unchanged verdict accompanied by increased confidence, among observations with initial confidence below the maximum and therefore eligible for strengthening ($n = 1{,}924$).
    Numbers above the bars report total percentages, numbers within the stacked bars report component percentages, and $n$ below each bar reports the number of observations in that condition. Asterisks indicate differences in total weakening in \textbf{(a)} or strengthening in \textbf{(b)} relative to Control. Values are from separate two-sided Barnard's exact tests for weakening and strengthening, with Benjamini--Hochberg false-discovery-rate correction ($\alpha = 0.05$). Significance after correction: $^{*}p < .05$, $^{**}p < .01$, and $^{***}p < .001$.}
    \label{fig:weakening_strengthening_rate}
\end{figure}

\Cref{fig:weakening_strengthening_rate}
shows that all three treatments increased both weakening and strengthening relative to Control. Weakening rose from 5.3\% in Control to 10.6\% under Certainty, 15.9\% under Controversy, and 12.1\% under Controversy+Certainty. Strengthening was assessed only among the 1,924 observations whose initial confidence could still increase; within this subset, it rose from 8.6\% in Control to 25.6\%, 25.9\%, and 20.9\%, respectively. These weakening and strengthening rates use different denominators and should therefore not be compared directly. Statistical details are reported in \Cref{app:weakening-strengthening-overall}.
The combined condition had numerically lower weakening and strengthening rates than Controversy alone. We next examine how participants’ majority--minority position and relative group certainty were related to this pattern.

\begin{figure}[!htbp]
    \centering
    \includegraphics[width=\linewidth]{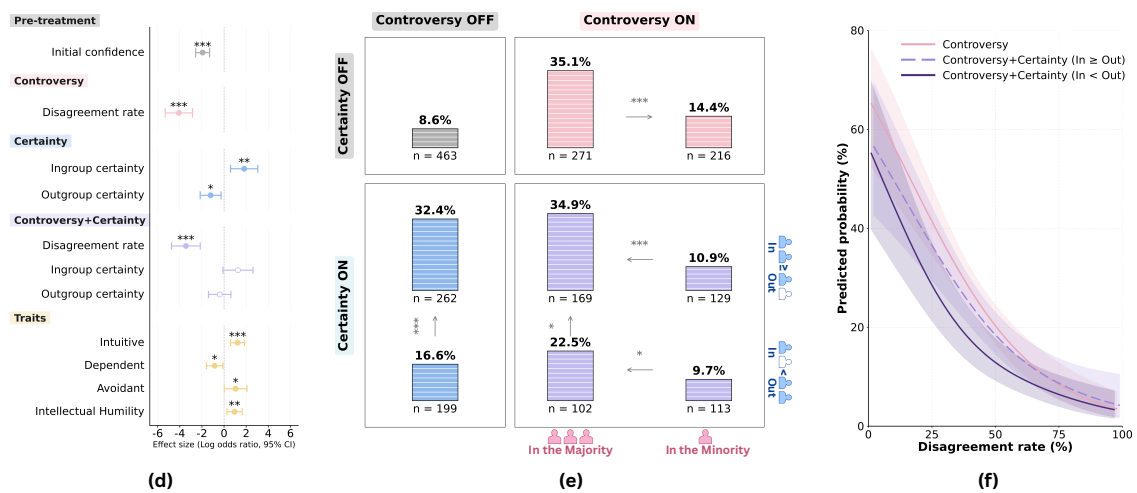}
    \caption{
        \textbf{Associations of controversy and group certainty with moral judgment change.}
        Panels \textbf{a}--\textbf{c} show judgment weakening across all observations ($n = 4{,}318$); panels \textbf{d}--\textbf{f} show judgment strengthening among observations with room to increase confidence ($n = 1{,}924$).
        \textbf{(a, d)} Log odds ratios (points) and 95\% confidence intervals from logistic regressions predicting weakening (a) and strengthening (d). Predictors include initial confidence, treatment variables, and participant traits.
        \textbf{(b, e)} Weakening (b) and strengthening (e) rates by treatment condition and social position. The outer $2 \times 2$ layout shows whether Controversy and Certainty were displayed. Within each condition, bars distinguish the social-position information made visible: \textit{In the Majority} versus \textit{In the Minority} when Controversy was displayed, and \textit{In $\geq$ Out} versus \textit{In $<$ Out} when Certainty was displayed. Bar heights show outcome rates; $n$ gives subgroup size. Arrows mark significant within-treatment contrasts from two-sided Barnard's exact tests with Benjamini--Hochberg correction ($\alpha = 0.05$).
        \textbf{(c, f)} Model-predicted probabilities of weakening (c) and strengthening (f) across disagreement rates, with other predictors set to their empirical values. Pink lines show the Controversy condition; purple lines show the Controversy+Certainty condition, separated by relative certainty (\textit{In $\geq$ Out}, dashed; \textit{In $<$ Out}, solid). Shaded bands show 95\% confidence intervals. Asterisks indicate statistical significance: * $p < .05$, ** $p < .01$, *** $p < .001$.}
    \label{fig:weakening_strengthening_results}
\end{figure}

\subsection*{When do judgments weaken or strengthen?}\label{sec:weakening}

To examine which features of the displayed information were associated with judgment change, we fitted separate logistic regressions for weakening and strengthening. The weakening model used all observations; the strengthening model was restricted to the 1,924 observations in which initial confidence could still increase. Full model specifications and coefficient estimates are reported in \hyperref[sec:methods]{Methods} and \Cref{app:weakening,app:strengthening}.

\Cref{fig:weakening_strengthening_results}(a) summarizes adjusted associations with judgment weakening. Participants with higher initial confidence were less likely to weaken: moving from the lowest to the highest confidence level was associated with 46\% lower odds of weakening. Within the Controversy condition, greater disagreement was associated with more weakening ($\beta = 4.52$, 95\% CI [3.51, 5.54]).
When Certainty was displayed alone, higher in-group certainty was associated with less weakening, whereas higher out-group certainty was associated with more weakening. When both signals were displayed, disagreement and in-group certainty retained these directions of association, but out-group certainty showed no clear independent association with weakening.
Full coefficient estimates including all participant characteristics are in Extended Fig. \ref{tab:full-weakening} and \Cref{app:weakening-regression}.

We next examine how these associations appear across the social positions made visible by each treatment (\Cref{fig:weakening_strengthening_results}b). Controversy alone revealed whether participants were in the majority or minority: minority participants weakened more often than majority participants (27.3\% versus 7.9\%). Certainty alone revealed the relative certainty of the two sides: participants weakened more often when their in-group was less certain than the out-group than when the in-group was equally or more certain than the out-group (18.5\% versus 5.3\%).
When both signals were displayed, minority participants again weakened more often than majority participants. Among majority participants, weakening was lower when the in-group was equally or more certain than the out-group than when it was less certain (4.0\% versus 11.7\%). Among minority participants, the corresponding difference was smaller and not statistically significant (18.1\% versus 20.3\%).
Thus, when controversy was visible, relative certainty differentiated weakening among majority participants but not among minority participants.

The model-predicted probabilities in \Cref{fig:weakening_strengthening_results}(c) clarify this pattern across the full range of disagreement rates. Predicted weakening increased as disagreement increased, particularly once participants were in the minority. However, the predicted probability was lower in the Controversy+Certainty condition than in the Controversy condition, and within the combined condition it was lower when the in-group was equally or more certain than the out-group. Thus, there was no evidence that adding Certainty increased controversy-related weakening; instead, its main association was with lower weakening when like-minded others appeared equally or more certain than opponents. Full comparisons are reported in \Cref{app:weakening-quadrant}.

\Cref{fig:weakening_strengthening_results}(d) shows adjusted associations with judgment strengthening among observations with room for confidence to increase. Because participants who began closer to the maximum had less remaining room to increase confidence, initial confidence was included as a covariate; higher initial confidence was associated with less strengthening. The social-information pattern was partly reciprocal to weakening: greater disagreement was associated with more weakening but less strengthening in both the Controversy and Controversy+Certainty conditions. When Certainty was displayed alone, higher in-group certainty was associated with more strengthening, whereas higher out-group certainty was associated with less strengthening (the reverse of their associations with weakening). In the combined condition, the certainty coefficients had the same directions, but neither independently predicted strengthening.
Full estimates are reported in Extended Fig. \ref{tab:full-strengthening} and \Cref{app:strengthening-regression}.

The social-position and predicted-probability results for strengthening are consistent with this reciprocal pattern (\Cref{fig:weakening_strengthening_results}e,f). Majority participants strengthened more often than minority participants, and strengthening was higher when the in-group was equally or more certain than the out-group (for example, 32.4\% versus 16.6\% in the Certainty condition). There was no evidence that the combined treatment increased strengthening beyond Controversy alone. Across disagreement rates, predicted strengthening was highest under Controversy alone, lower when the combined treatment indicated that the in-group was equally or more certain than the out-group, and lowest when the in-group was less certain. Full comparisons are reported in \Cref{app:strengthening-quadrant}.

As a secondary individual-difference finding, higher intellectual humility was associated with both weakening and strengthening, indicating a greater likelihood of moving away from or becoming more confident in an initial judgment. Outcome-specific associations for decision styles are reported in Extended Fig. \ref{tab:full-weakening}, and \ref{tab:full-strengthening}.

Overall, when shown separately, greater disagreement was associated with more weakening and less strengthening, whereas higher in-group certainty showed the opposite pattern and higher out-group certainty was associated with more weakening and less strengthening. In the combined condition, disagreement remained associated with both outcomes, but higher in-group certainty was associated only with less weakening; out-group certainty showed no clear independent association with either outcome. The combined condition did not produce higher rates of weakening or strengthening than Controversy alone.

\subsection*{Exploratory analyses}
\label{sec:deliberations}

\textbf{Written explanations.}
The regression results identify which social signals were associated with judgment change, but not whether participants noticed or accepted those signals. We therefore analyzed participants’ optional written explanations. Of 4,318 judgment entries, 2{,}769 (64\%) included an explanation; after quality filtering, 2,744 remained for analysis (\Cref{subsec:quality_filtering}). Using LLM coding validated against human annotations, we identified whether each explanation mentioned controversy, in-group certainty, or out-group certainty and whether it accepted or rejected the signal as relevant (\Cref{subsec:reasoning}).
For example, one participant whose judgment weakened accepted Controversy: \textit{``The way the audience is split almost 50/50 leaves me a little uncertain either way.''} Another participant maintained their judgment while rejecting it: \textit{``I'm not swayed by other people's opinions.''}

\begin{figure}[!htbp]
    \centering
    \includegraphics[width=0.95\linewidth]{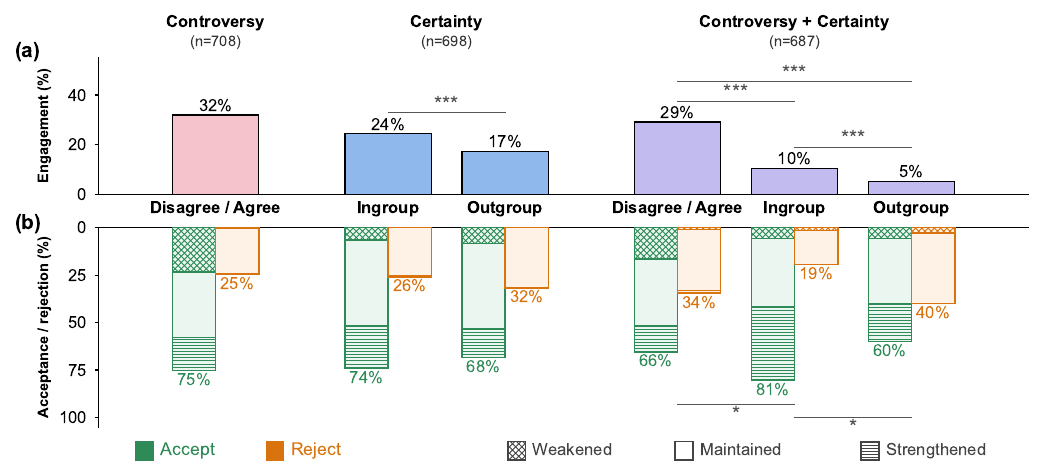}
\caption{
{\bf Engagement with displayed social signals in participants’ written explanations.} {\bf (a)} Percentage of cleaned explanations in each treatment condition that referenced each displayed signal. ($n$) below the bars gives the number of explanations analyzed in that condition. {\bf (b)} Among explanations that referenced a signal, the percentage coded as accepting (green) or rejecting (orange) it as relevant evidence, subdivided by judgment outcome: weakened (cross-hatching), maintained (no hatching), and strengthened (horizontal hatching). Brackets show within-condition pairwise comparisons of reference rates in {\bf (a)} and acceptance rates in {\bf (b)}, based on two-sided Barnard’s exact tests. The Control condition is omitted because it presented no aggregate social information. $^{*}P<0.05$, $^{**}P<0.01$, $^{***}P<0.001$.
}
    \label{fig:deliberation_obs}
\end{figure}

Controversy was mentioned most often (\Cref{fig:deliberation_obs}a): it was mentioned in 32\% of explanations when presented alone and 29\% when presented with Certainty. In the Certainty condition, in-group certainty was mentioned more often than out-group certainty (24\% versus 17\%); the same ordering appeared in the combined condition (10\% versus 5\%). Out-group certainty was usually mentioned alongside another signal rather than independently; detailed co-mention patterns are reported in \Cref{subsec:reasoning}.
Among explanations that mentioned a displayed signal, 60--81\% accepted it as relevant (\Cref{fig:deliberation_obs}b). Explanations rejecting a signal were almost always associated with maintained judgments. Acceptance of Controversy was more often associated with weakening than strengthening, whereas acceptance of in-group certainty was more often associated with strengthening than weakening. Participants’ explanations therefore broadly mirrored the regression results: controversy attracted the most attention, while in-group certainty was more closely associated with strengthening. Because the explanations were optional and observational, they provide exploratory rather than direct evidence about participants’ reasoning.
A broader exploratory classification identified recurring stated reasons for maintaining or changing judgments, but did not provide a conclusive account of why responses changed or remained stable (\ref{subsubsec:reasoning_results}).

\textbf{Dilemma consequences.} (\Cref{subsec:consequence})
Across scenarios ranging from minor etiquette disputes to serious harms, weakening and strengthening rates showed no consistent relationship with severity, affected party, or the involvement of a vulnerable person. Two pairwise differences stood out, but should be interpreted cautiously: greater weakening for dilemmas involving family members than friends, and less strengthening for dilemmas involving a child than no vulnerable person. These exploratory analyses involved multiple unadjusted comparisons and small, uneven groups of dilemmas.

\textbf{Participant feedback.}
At the end of the survey, participants could answer the optional question, ``Any feedback or comments for this survey?''.
After quality filtering and excluding neutral or empty entries, 458 of 2,159 participants (21.2\%) provided meaningful feedback, coded into one or more of eight categories (\Cref{app:overall_feedback}). Among their feedback, 68.6\% described a positive experience and 20.3\% mentioned learning, reflection, or insight, whereas 5.2\% criticized the study design and 2.4\% described a negative experience. Illustrative negative comments questioned the purpose of repeating a judgment without new information in the Control condition or rejected Reddit as an unrepresentative source of opinion. Because feedback was optional, these percentages may not represent the full sample.

\section*{Discussion}
\addcontentsline{toc}{section}{Discussion}

At the broadest level,
moral judgments are responsive to remarkably thin forms of social evidence.
Random assignment to each social-information condition increased both weakening and strengthening relative to the active Control. Within the treatment conditions, disagreement showed the strongest association with updating.
Disagreement was the strongest signal, especially when participants learned that their initial verdict placed them in the minority.
Crucially, the combined display did not simply add the two signals.
When Certainty was shown alone, higher in-group certainty was associated with less weakening and more strengthening, whereas higher out-group certainty showed the reverse pattern. When Controversy was also visible, higher in-group certainty remained associated with less weakening, while out-group certainty had no clear independent association with either outcome.
This pattern is consistent with participants using group certainty in relation to their own position, rather than treating certainty on both sides as equivalent social evidence.

Our findings bridge research on responses to aggregate opinion, which has largely examined numerical proportions in factual tasks \citep{OktarLombrozoGriffiths2024}, with work on confidence communicated between individuals or within small groups \citep{bang2017confidence,Moussaid2013SocialInfluence}. Related research on social influence in moral judgment has instead often used stylized hypothetical scenarios \citep{pryor2019even}. By testing aggregate social signals across 135 varied interpersonal dilemmas drawn from online discussions, the present study extends these traditions to a more content-rich moral setting. The asymmetric pattern in the combined condition is consistent with evidence that people respond more strongly to confidence that supports rather than challenges their prior judgment \citep{Kappes2020}, and connects with research on minority influence \citep{Wood1994}. It also complements recent work presenting factual-answer distributions and group confidence together \citep{HeLienZheng2026} by isolating certainty as a separate signal and showing that its associations with updating differ when controversy is visible.

Whether updating in response to aggregate disagreement and group certainty is epistemically or morally desirable remains an open question. The interpersonal dilemmas studied here are normative questions that do not generally offer the independently verifiable answers available in many factual tasks \citep{SchusterKilov2025MoralDisagreement}. Nor did the experimental summaries provide new arguments, facts, or other first-order evidence about the dilemmas; they provided social information about how others judged them and how certain those judgments appeared. Aggregating judgments can improve accuracy when individual views are independent and diverse \citep{HongPage2004,Surowiecki2004,HeLienZheng2026}, but those conditions do not by themselves establish that an aggregate moral judgment is correct or better justified. Group certainty may signal reliable insight, or instead amplify a shared error \citep{Moscovici1969Minority}. Recognizing disagreement can nevertheless create an opportunity for epistemic self-improvement \citep{christensen2007epistemology}. Participants’ optional feedback, which often described the task as reflective or worthwhile, is compatible with that possibility, but it cannot show that revised judgments became more accurate or better justified.
This ambiguity makes the asymmetric certainty pattern double-edged. Confidence among like-minded others could help preserve well-founded minority dissent, but the same response could shield a mistaken judgment from correction. Our findings establish that these signals prompt updating; they do not establish that the resulting judgments are more accurate or better justified.

Three limitations also point to future research questions. First, the mechanism behind the asymmetric pattern remains unresolved. We explored this question by classifying participants’ optional written explanations into broader types of reasoning, but these retrospective and self-selected accounts could not distinguish among selective attention, conformity, resistance, biased assimilation, and other possible processes (\Cref{subsec:reasoning}). Future studies should use direct measures of attention and information processing, experimentally manipulate candidate mechanisms, or elicit reasoning before and after exposure to the social information.

Second, the present design establishes immediate responses to controlled aggregate summaries in this sample; it does not establish whether the effects generalize to other populations and platforms or persist after repeated exposure. The study used English-language Reddit dilemmas and a predominantly English-speaking sample.
Participants saw concise summaries rather than the richer combination of comments, rankings, and engagement cues found on online platforms. The study also did not compare these summaries with interventions that provide substantive reasons or arguments.
Replication with more diverse content and samples, more platform-like displays, and comparisons with argument-based deliberation could test the generality of these effects; longitudinal designs could test whether change persists.

Finally, our outcomes capture the direction of judgment revision, not whether that revision improved the judgment.
Future work should test whether these displays broaden participants’ consideration of affected parties in moral dilemmas and, in factual tasks with verifiable answers, whether they improve or impair accuracy.

Taken together, the findings show that collective opinion is more than a headcount: people respond both to how divided a group is and to which side appears more certain. For platforms and AI systems that summarize human judgments, representing both features may provide a fuller account of collective opinion without treating either popularity or certainty as a marker of moral correctness.

\vspace{10mm}

\newpage

\section*{Methods}
\phantomsection
\label{sec:methods}
\addcontentsline{toc}{section}{Methods}

We first describe the construction of the \dataset dataset and its social signals, participant recruitment, and the experimental design. We then define the outcomes and variables and report the statistical models and exploratory analyses of participants' written explanations, dilemma consequences, and overall feedback.

\subsection*{\dataset dataset and social-signal construction}
\phantomsection\label{subsec:methods_dataset}

\textbf{Dilemma selection.} We began with the 102,998 AITA posts collected and topic-labeled by \citet{Nguyen2022MappingTI}, covering the subreddit from 2013 to April 2020. Of these posts, 9,998 contained at least 50 top-level comments carrying a YA or NA verdict. From the 47-topic taxonomy, we retained 32 common topics, excluding rare topics (each $<0.8\%$ of the full corpus) and potentially distressing topics such as death, race, and sex. For selection, controversy was the proportion of YA or NA verdicts belonging to the minority group, ranging from 0 (consensus) to 0.5 (an even split). We sampled across this range and oversampled closely divided posts, where the certainty expressed by the two verdict groups could diverge. Specifically, we randomly selected 105 posts with minority proportions between 0.3 and 0.5 and 30 with minority proportions below 0.3. The resulting \dataset dataset contains 135 dilemmas across 32 topics and 54,827 community verdicts. See \Cref{subsec:aita,subsec:datapreparation}.

\textbf{Deriving aggregate controversy and group certainty.} For each selected post, we extracted verdict labels from top-level comments using regular-expression patterns, excluded comments labeled INFO, and grouped YTA and ESH as YA (the original poster was at fault) and NTA and NAH as NA (the original poster was not at fault). We computed controversy as the minority proportion defined above. To derive group certainty, we classified how certain each commenter appeared about their verdict on a four-point ordinal scale. We treated only the highest rating, 4 (certain), as certain; ratings 1--3 were grouped as not certain. The final few-shot, chain-of-thought classifier showed agreement with expert annotations of Krippendorff's $\alpha=0.78$; after dichotomizing the ratings, it achieved F1 $=0.92$ and AUC $=0.88$. Group certainty was the proportion of comments classified as certain within each verdict group. We choose to tally group certainty from the number of comments rather than Reddit votes, because the latter only contain aggregated total at the time of data collection (therefore not reproducible), and its distribution tend to be very skewed to the majority or official verdict due to social cascading effects.
See \Cref{subsec:uncertainty} for the annotation guidelines, validation, and aggregation procedures.

\subsection*{Participants, recruitment, and exclusions}

The human-subject study was approved by the Human Research Ethics Committee at the Australian National University (protocol H/2023/1292) and preregistered on the Open Science Framework on July 18, 2025 (https://osf.io/2hjkt/). Before beginning the study, all participants reviewed the participant information sheet and provided informed consent through Qualtrics (\Cref{subsec:survey_flow}).

We recruited adults through Prolific. Eligible participants were aged 18 or older, fluent in English, and had not previously taken part in the same study or its pilot.
Participants were assigned to one of four
conditions with equal probability and completed only that condition. Within conditions, each participant received two randomly selected dilemmas from the pool of 135.

Across four survey waves, we obtained 2{,}346 survey records: 2{,}171 completed submissions, 158 returned submissions, and 17 timed-out records. Returned submissions included cases in which consent was declined, the participant withdrew, or the task was only partly completed. Of the 2{,}171 completed submissions, 2{,}159 participants passed the preregistered criterion of at least two of three attention checks and formed the final analytic sample. These participants contributed 4{,}318 participant--dilemma observations. Participants received \pounds1.95; the median completion time was 14 minutes (861.5 seconds), equivalent to approximately \pounds8.36 per hour. The final sample was 50.4\% male and 49.5\% female, with a mean age of 38.7 years.
Further information on participant profile is in \Cref{sec:secA4}.

\subsection*{Experimental design}
\label{subsec:expdesign}

We used a four-arm randomized design to test the effects of showing aggregate controversy, group certainty, or both on individual moral judgments. Participants were assigned with equal probability to the Controversy, Certainty, Controversy + Certainty, or Control condition. The single-signal conditions isolate responses to each signal; the combined condition allows us to assess whether responses differ when both signals are available; and the Control condition provides a baseline without social information.

The survey comprised three sequential phases. Phase I introduced the assigned task and display; Phase II contained a three-step moral-judgment task as shown in  \Cref{fig:aggregated_signals}, repeated for two randomly selected dilemmas; and Phase III collected background measures and optional feedback.

\textbf{Phase I: onboarding and training.} After reading the participant information sheet (\Cref{fig:participant_information_sheet}) and providing informed consent, participants were randomized to one of the four conditions and received condition-specific training. The training introduced either the relevant social-information display or the neutral topic-selection task used in Control. Participants then completed a practice trial and the first of three attention checks.
The training material and flow are detailed in \Cref{subsec:survey_flow}.

\textbf{Phase II: repeated moral-judgment task.} For each of two randomly selected dilemmas from the pool of 135, participants completed the same three steps described in \hyperref[sec:expsetup]{Experimental Setup}: (1) they read the dilemma and recorded an initial verdict and confidence rating; (2) they viewed the social information specified by their assigned condition, or completed the Control topic-selection task; and (3) they recorded a final verdict and confidence rating. Participants could optionally explain why they had changed or maintained their response. An attention check followed each dilemma, yielding two additional checks. The presentation format is described below and in \Cref{subsec:survey_flow}.

\textbf{Phase III: post-study surveys and debriefing.} To capture the individual differences mentioned earlier, participants completed a set of background surveys measuring decision-making style (General Decision-Making Style [GDMS] questionnaire \citep{scott1995decision}; see \Cref{subsec:reduce_survey} for the item-reduction procedure used to shorten this questionnaire from 25 to 15 questions), personality traits (Ten-Item Personality Inventory \citep{Gosling2003AVB}), and intellectual humility (Six-Item Intellectual Humility scale \citep{leary2017IH}). The survey concluded with an optional feedback section and a debriefing on the study’s purpose, after which participants were redirected to Prolific for compensation.

Six aspects of the design and measurement merit clarification:

\textbf{Group certainty and participant confidence.} We distinguish two related but different concepts. \textit{Certainty} refers to how sure Reddit commenters appear about their verdict, inferred from comment text and aggregated separately within each verdict group, YA and NA, for each AITA post. \textit{Confidence} refers to how sure survey participants report being about their own verdict, recorded before and after treatment. This distinction is important because certainty is the information shown in the treatment, whereas confidence is a participant-reported outcome. Using a single term would obscure our main result: the effect of group certainty on participant confidence.

\textbf{Design of the treatment infographic.} During Phase I, participants were introduced to the format used to present social information. During each Phase-II dilemma trial, every displayed percentage was represented in three coordinated forms: a 100-icon human array, a numerical percentage, and a short textual description (\Cref{sec:surveyexperiment}).
Verbal and visual information are encoded and remembered through partially independent channels, so presenting both together aids comprehension and recall more than either alone \citep{Paivio1986Mental}.
Presenting the same information visually and verbally was also intended to support comprehension across differences in numeracy and attention \citep{GalesicGarciaRetameroGigerenzer2009,Lipkus2007Numeric}.
The icon array makes a percentage concrete as a proportion of people, while the numerical label preserves the exact value. We used human icons rather than abstract shapes because anthropomorphic icons can attract attention and support recall \citep{ZikmundFisher2014Blocks}; here, they also represent the AITA users whose verdicts and certainty are summarized.

In the Controversy + Certainty condition, the display showed both the distribution of verdicts and the certainty rate within each verdict group. For example, this condition in \Cref{fig:aggregated_signals}(d) could indicate that 48\% judged YA (and agreed with the participant), of whom 88\% were certain, whereas 52\% judged NA (and disagreed with the participant), of whom 65\% were certain.
This layout makes clear that group certainty naturally interacts with verdict position rather than being a single property of the community as a whole.

\textbf{Confidence scale.} We measured participants' confidence on a four-point Likert scale for three reasons. First, it matches our earlier annotation scheme for individual certainty in AITA moral dilemmas (\Cref{subsubsec:individual_uncertainty}). We found the four-point scale more conceptually coherent during discussions among human annotators. It also achieved higher inter-annotator agreement than a five-point scale (Krippendorff's $\alpha$: 0.69 vs. 0.58). This result suggests that a four-point scale is easier for people to comprehend in a more consistent way, and this applies to survey participants as well as trained annotators. Second, the four response options (uncertain, somewhat uncertain, somewhat certain, certain) have clear ordinal meanings that are easy to understand across participants. Third, unlike an odd-point scale, it has no neutral midpoint, reducing central tendency bias by requiring participants to indicate either uncertainty or certainty.

\textbf{Participant judgment format.} In step 1 and step 3 of each judgment task, we asked each participant to first provide their verdict (YA or NA) and then indicate their confidence. One could imagine this being equivalent to one 8-level likert scale including both verdicts and decreasing confidence towards the middle. We choose the current response format to be coherent with the separate dimension of intervention, i.e, controversy and group certainty from Reddit comments.

\textbf{Control task.} In the Control condition, participants labelled each dilemma with a topic rather than viewing information about other people’s verdicts or certainty. This active-control task was designed to maintain engagement with the dilemma while withholding social information \citep{boot2013pervasive}. Topic labels conveyed no information about how other users had judged the dilemma or how certain they were. The Control condition therefore provides a baseline for any judgment changes that occur in the absence of aggregated social information. Total survey duration was similar in the Control and treatment conditions, indicating that the active control also provided comparable time on task (\Cref{subsubsec:time}).

\textbf{Number of dilemmas per participant.} We assigned two dilemmas to each participant to balance time spent on the main judgment task (Phase II) with that required for training and background measures (Phase I and III). With only one dilemma, these fixed components would occupy a large share of the survey; three dilemmas would require more reading and repeated judgments, potentially increasing fatigue and reducing attention. The observed timing was consistent with this choice: participants spent the largest share of the survey on the main task, while their comment engagement and attention-check pass rates suggested that the task remained manageable (\Cref{subsubsec:time}).

\subsection*{Outcome measures and analysis variables}
\phantomsection\label{subsec:methods_vars}

\Cref{tab:variables} provides an overview of the outcome measures, social-information measures, derived social-position categories, and participant-profile measures used in the analyses. The main text introduces these measures; here we specify their coding and the analytic samples.

\textbf{Outcome coding and analysis samples.} We derived the judgment-change outcome from each participant's pre- and post-treatment verdict--confidence pair. A response was \emph{weakened} if the participant changed verdict or retained the verdict with lower confidence, \emph{maintained} if both verdict and confidence were unchanged, and \emph{strengthened} if the verdict was unchanged and confidence increased. Any verdict reversal was classified as weakened regardless of the final confidence rating. These labels describe change relative to the participant's own initial judgment; they do not indicate movement toward the community majority or whether a judgment became morally better justified. For regression, the weakening outcome equalled 1 for weakened responses and 0 otherwise, with all 4{,}318 participant--dilemma observations eligible. The strengthening outcome equalled 1 for strengthened responses and 0 otherwise and was restricted to observations with initial confidence below the scale maximum ($n = 1{,}924$).

\textbf{Participant-relative treatment coding.}
For analysis, we coded the displayed social-information variables separately for each participant--dilemma observation, relative to the participant's initial verdict. \emph{Disagreement rate} was the proportion of AITA verdicts opposing that verdict. \emph{In-group certainty} was the proportion of comments classified as certain within the agreeing verdict group, and \emph{out-group certainty} was the corresponding proportion within the opposing group. Thus, participants giving opposite initial verdicts on the same dilemma received reversed disagreement and certainty values.

For example, suppose a dilemma had 60\% YA and 40\% NA verdicts, with certainty rates of 70\% among YA comments and 50\% among NA comments. For a participant who initially selected YA, disagreement was 40\%, in-group certainty was 70\%, and out-group certainty was 50\%. For a participant who selected NA, the corresponding values were 60\%, 50\%, and 70\%.

\textbf{Derived social-position variables.} For descriptive subgroup comparisons, we also derived binary social-position variables from the displayed information. \emph{In the Majority} indicates that more AITA verdicts agreed than disagreed with the participant’s initial verdict. No dilemma in the \dataset dataset had an exact 50:50 verdict split in its source post. \emph{In $\geq$ Out} indicates that the in-group was equally or more certain than the out-group, whereas \emph{In $<$ Out} indicates that it was less certain. In the example above, the participant selecting YA was \emph{In the Majority} and \emph{In $\geq$ Out}; the participant selecting NA occupied the complementary positions.
Their predictive performance relative to the continuous treatment variables was tested in the model ablation below.

\subsection*{Statistical analysis and regression}

\textbf{Exact rate comparisons and multiple-testing correction.} We used two-sided Barnard’s exact test \citep{barnard1945new} for comparisons of binary outcome rates between independent groups. Barnard’s test evaluates the null hypothesis of equal binomial proportions in a $2\times2$ table without conditioning on the marginal totals.
For each prespecified family of related comparisons, we adjusted $p$-values using the Benjamini--Hochberg procedure \citep{benjamini1995controlling}, controlling the false-discovery rate at $q=.05$. We denote the resulting values $p_{\mathrm{BH}}$. For the overall comparisons in \Cref{fig:weakening_strengthening_rate}, the correction was applied separately across the three treatment-versus-Control tests for weakening and for strengthening. For the \Cref{fig:weakening_strengthening_results} social-position comparisons, correction families are specified in the corresponding figure caption and appendix tables.

\textbf{Regression model specification.} We fitted separate logistic regressions for weakening and strengthening. The preregistered primary specification was a mixed-effects logistic regression with random intercepts for participants and dilemmas.
This is defined in the following equation, where the log-odds of the binary outcome $Y_{ij}$ for participant $i$ responding to dilemma $j$ is represented as a linear combination of treatment and covariate variables.
\begin{align}
\operatorname{logit}\big(\Pr(Y_{ij}=1)\big) &= \,\log\,\left(\frac{\Pr(Y_{ij}=1)}{1-\Pr(Y_{ij}=1)}\right) \label{eq:model-spec-annotated} \\
&= \underbrace{\beta_0 + \beta_c\,c_{ij} }_{\text{shared baseline}} \nonumber\\
&+ \underbrace{ T^{\mathrm{Cont}}_i \left[\, \beta_{\mathrm{Cont}}
   + \beta_{d,\mathrm{Cont}}\, D_{ij} \,\right] }_{\text{Controversy treatment}} \nonumber\\
&+ \underbrace{T^{\mathrm{Cert}}_i \left[\, \beta_{\mathrm{Cert}}\,
   + \beta_{\text{in},\mathrm{Cert}}\,\mathrm{In}_{ij}
   + \beta_{\text{out},\mathrm{Cert}}\,\mathrm{Out}_{ij} \,\right] }_{\text{Certainty treatment}}  \nonumber\\
&+ \underbrace{T^{\mathrm{C+C}}_i \left[\, \beta_{\mathrm{C+C}}\,
   + \beta_{d,\mathrm{C+C}} D_{ij}\,
   + \beta_{\text{in},\mathrm{C+C}}\,\mathrm{In}_{ij}
   + \beta_{\text{out},\mathrm{C+C}}\,\mathrm{Out}_{ij} \,\right] }_{\text{Controversy + Certainty treatment}} \nonumber\\
&+ \underbrace{\mathbf{x}_{ij}^{\top}\boldsymbol{\gamma} }_{\text{selected covariates}} .  \nonumber
\end{align}

Variables are defined in the previous section with their numerical representations summarized here. In the weakening model, $Y_{ij}=1$ for a weakened judgment and $Y_{ij}=0$ for a maintained or strengthened judgment. In the strengthening model, $Y_{ij}=1$ for a strengthened judgment and $Y_{ij}=0$ for a maintained or weakened judgment, within the ceiling-filtered analysis sample.
$c_{ij}$ is participant $i$'s initial confidence on dilemma $j$. $T^{\mathrm{Cont}}_i$, $T^{\mathrm{Cert}}_i$, and $T^{\mathrm{C+C}}_i$ are binary indicators for participant $i$ being in the Controversy, Certainty, or Controversy+Certainty treatment, respectively. $D_{ij}$ is the disagreement rate of Reddit comments for dilemma $j$ against participant $i$. $\mathrm{In}_{ij}$ and $\mathrm{Out}_{ij}$ are in-group and out-group certainty relative to participant $i$'s verdict.
$\mathbf{x}_{ij}$ contains the participant- and dilemma-level covariate blocks retained by model selection (discussed below). Candidate blocks included demographics, personality, decision-making styles, intellectual humility, and dilemma topic.
All non-binary input were rescaled to the $0$--$1$ range using percentage of maximum possible (POMP) scoring \citep{Cohen1999POMP}, i.e., $x_{POMP}=\frac{x-x_{MIN}}{x_{MAX} - x_{MIN}}$, where $x_{MAX}$ and $x_{MIN}$ are the maximum and minimum values according to the definition of variable $x$. For example, for participant confidence $c_{ij}$, $x_{MAX}$ is 4 and $x_{MIN}$ is 1.

There is a shared baseline shift $\beta_0$ and slope $\beta_c$ for participants' initial confidence. $\boldsymbol{\gamma}$ is a slope vector for the selected participant- and dilemma-level covariates, also shared across different treatments and control.
Each treatment-specific intercept $\beta_{\mathrm{Cont}},\beta_{\mathrm{Cert}},\beta_{\mathrm{C+C}}$ represents the baseline shift in weakening or strengthening for that condition. Each treatment-specific slope represents the association with one piece of information shown in that condition - $\beta_{d,*}$ for disagreement rate in the three treatments (denoted as *), as well as $\beta_{\mathrm{In},*}, \beta_{\mathrm{Out},*}$ for the in- and out-group certainty. Note that each treatment-specific intercept and slope were only estimated within the condition that showed the corresponding information, controlled by the binary indicators $T^*_{i}$.

\textbf{Model ablation.}
We used model ablation to identify which model structure, covariates, and representation of the social signals best predicted judgment change on held-out observations. This limits overfitting: more complex participant- or dilemma-level structures and additional covariates can improve fit to observed data without improving performance on new observations~\citep{yarkoni2017choosing}.
All comparisons used 10-fold cross-validation. Participant--dilemma observations were randomly assigned to ten approximately equal folds using a fixed random seed, with each fold serving once as the held-out set. We compared mean log-likelihood, a strictly proper scoring rule that rewards calibrated probabilities without requiring a classification threshold~\citep{gneiting2007strictly}. This approach differed from the preregistration, which specified random intercepts for participants and dilemmas as the primary model.

The ablation targeted three sets of modeling choices.

\textit{1. Model structure.} To test whether accounting for participant or dilemma dependence improved prediction, we compared a generalized linear model (GLM) without dilemma as a factor, a GLM with dilemma as a factor, a dilemma fixed-effects model fitted with \texttt{feglm}, and a dilemma random-intercept model fitted with \texttt{glmmTMB}. The \texttt{feglm} specification used participant-clustered standard errors, and we also tested a participant random intercept in the \texttt{glmmTMB} specification.

\textit{2. Covariate blocks.} To test whether measured participant and dilemma characteristics improved prediction beyond the preregistered treatment terms and initial confidence, we evaluated blocks covering demographics, personality, General Decision-Making Style (GDMS), intellectual humility, and dilemma topic. Because participants were randomly assigned to conditions, adjustment for these covariates was not necessary to estimate the causal effects of treatment assignment. Greedy forward selection retained a block only when it increased held-out mean log-likelihood.

\textit{3. Social signal representation.} To test whether the displayed percentages carried more predictive information than their binary summaries, we compared two mutually exclusive blocks added to the same condition-indicator base model. The continuous block contained disagreement, in-group certainty, and out-group certainty; the binary block contained whether the participant was in majority or minority, and whether the in-group or out-group was more certain.

For both outcomes, the selected specification was the GLM without dilemma as a factor, using the continuous treatment-information block. The weakening model was fitted to all 4{,}318 observations and retained GDMS, intellectual humility, sex, and nationality. The strengthening model was fitted to the ceiling-filtered sample ($n = 1{,}924$) and retained GDMS and intellectual humility. Neither model retained dilemma topic or participant- or dilemma-level effects.

Full model-selection paths, coefficient tables, and signal-representation comparisons are reported in \Cref{app:weakening-regression,app:strengthening-regression,app:weakening-social-position-ablation,app:strengthening-social-position-ablation}.

\subsection*{Analysis of written explanations}

Of 4{,}318 participant--dilemma observations, 2{,}769 included an optional post-judgment written explanation. Quality screening excluded 14 participants and their 25 explanations. This left 2{,}744 written explanations from 1{,}528 participants (\Cref{subsec:quality_filtering}).

\textbf{Willingness to provide an explanation.}
Because written explanations were optional, we modeled whether participants provided one using a multilevel logistic regression with a dilemma-level random intercept. Covariates were selected using the same cross-validated procedure as the weakening and strengthening models. See \Cref{subsec:willingness}.

\textbf{Text classification.} We designed and validated two annotation tools to classify participants' written explanations. Full coding schemes, prompts to large language models (LLM), and validation procedures are reported in \Cref{subsec:reasoning}.

\textit{a. Reasoning expressed in explanations.} We coded eight, non-exclusive kinds of reasoning, grouped into four broader families. Reasons grounded in the participant included appeals to moral principles or intuitions (\textit{moral conviction}) and references to personal experience (\textit{lived experience}). Responses to social information either followed the displayed collective opinion (\textit{conformity}) or explicitly rejected its relevance (\textit{resistance}). Reasons grounded in the dilemma concerned fairness, rights, or responsibility (\textit{fairness}); harm, protection, or emotional effects (\textit{harm}); or situational details, procedures, or communication (\textit{context}). Text that stated a position without giving a substantive reason was coded as \textit{no explanation}. Validation against expert labels on 200 explanations gave macro F1 $=0.69$.

\textit{b. Treatment reference.} For each displayed signal, we first coded whether the explanation referred to disagreement rate, in-group certainty, or out-group certainty. References were then coded as accepting or rejecting the signal as relevant evidence. Validation against expert labels on 150 explanations gave reference-detection accuracy of 0.91--0.95 and three-class macro F1 of 0.75--0.83.

\subsection*{Analysis of participant feedback}

The same participant-level filter was applied to the optional overall feedback, leaving 573 entries. After excluding 115 entries coded as neutral or empty, 458 meaningful responses remained for category analysis. The screening procedure is reported in \Cref{subsec:quality_filtering}.
We coded each response into one or more of eight categories. These covered positive and negative experiences, polite closings, learning or reflection, and interest in online moral communities. Other categories captured appreciation or criticism of the study design and interest in results or future participation. We report category frequencies and common combinations in \Cref{app:overall_feedback}.

\subsection*{Analysis of dilemma consequences}
We annotated the 135 dilemmas in the \dataset dataset along three dimensions. Severity was rated on a 1--7 scale. Affected parties were categorized as the author, family or partners, friends, ongoing contacts, or strangers. Vulnerability was categorized as a child, another vulnerable person, or nobody vulnerable. Severity ratings were averaged across three annotations, while categorical labels were selected by majority vote. Validation against human majority-vote labels gave F1 scores of 0.71--0.92 across tasks. We then described the consequence profile and explored its associations with dilemma-level weakening and strengthening (\Cref{subsec:consequence}).

\backmatter

\newpage
\section*{Extended figures}
\label{sec:extfigures}\phantomsection
\addcontentsline{toc}{section}{Extended figures}

\definecolor{tCtrl}{HTML}{C8C8C8}
\definecolor{tCntr}{HTML}{F4C3CB}
\definecolor{tCert}{HTML}{8FB8ED}
\definecolor{tBoth}{HTML}{C2BBF0}
\definecolor{tCov}{HTML}{F2D862}
\providecommand{\tdot}[1]{\textcolor{#1}{\rule[0.12ex]{1.5ex}{1.5ex}}}
\providecommand{\orci}[2]{
  \ensuremath{[}
  \makebox[2.4em][r]{\ensuremath{#1}}
  \ensuremath{,\;}
  \makebox[2.8em][r]{\ensuremath{#2}}
  \ensuremath{]}
}

\begin{table}[!htbp]
\centering
\renewcommand{\arraystretch}{1.2}
\setlength{\tabcolsep}{4pt}
\small
\begin{tabular}{lccccccc}
\toprule
\textbf{Predictor}
& $\hat{\beta}$ & 95\% CI$_{\beta}$ & SE & $z$ & $p$ & OR & 95\% CI$_{\mathrm{OR}}$ \\
\midrule
\multicolumn{8}{l}{\textbf{Main predictors}} \\[4pt]
\multicolumn{8}{l}{\tdot{tCtrl}~\textit{Pre-treatment}} \\
Intercept
& $-4.577$ & $[-5.659,\;-3.495]$ & $0.552$ & $-8.30$ & ${<}\,2\times10^{-16}$
& $0.01$ & \orci{0.004}{0.03} \\
Initial confidence ($c$)
& $-0.206$ & $[-0.341,\;-0.070]$ & $0.069$ & $-2.98$ & $0.003$
& $0.81$ & \orci{0.71}{0.93} \\[6pt]
\multicolumn{8}{l}{\tdot{tCntr}~\textit{Controversy treatment}} \\
Intercept (Controversy)
& $-1.009$ & $[-1.650,\;-0.369]$ & $0.327$ & $-3.09$ & $0.002$
& $0.36$ & \orci{0.19}{0.69} \\
Disagreement rate
& $4.523$ & $[\phantom{-}3.507,\;\phantom{-}5.538]$ & $0.518$ & $8.73$ & ${<}\,2\times10^{-16}$
& $92.1$ & \orci{33.4}{254.2} \\[6pt]
\multicolumn{8}{l}{\tdot{tCert}~\textit{Certainty treatment}} \\
Intercept (Certainty)
& $1.056$ & $[-0.088,\;\phantom{-}2.200]$ & $0.584$ & $1.81$ & $0.070$
& $2.87$ & \orci{0.91}{9.03} \\
In-group certainty
& $-2.461$ & $[-3.369,\;-1.552]$ & $0.463$ & $-5.31$ & $1.1\times10^{-7}$
& $0.09$ & \orci{0.03}{0.22} \\
Out-group certainty
& $2.056$ & $[\phantom{-}0.939,\;\phantom{-}3.173]$ & $0.570$ & $3.61$ & $3.1\times10^{-4}$
& $7.81$ & \orci{2.56}{23.9} \\[6pt]
\multicolumn{8}{l}{\tdot{tBoth}~\textit{Controversy + Certainty treatment}} \\
Intercept (Both)
& $-0.108$ & $[-1.241,\;\phantom{-}1.025]$ & $0.578$ & $-0.19$ & $0.851$
& $0.90$ & \orci{0.29}{2.79} \\
Disagreement rate
& $3.741$ & $[\phantom{-}2.607,\;\phantom{-}4.876]$ & $0.579$ & $6.47$ & $1.0\times10^{-10}$
& $42.1$ & \orci{13.5}{131.1} \\
In-group certainty
& $-1.663$ & $[-2.609,\;-0.717]$ & $0.483$ & $-3.45$ & $5.7\times10^{-4}$
& $0.19$ & \orci{0.07}{0.49} \\
Out-group certainty
& $0.487$ & $[-0.467,\;\phantom{-}1.441]$ & $0.487$ & $1.00$ & $0.317$
& $1.63$ & \orci{0.63}{4.23} \\[10pt]
\multicolumn{8}{l}{\textbf{Covariates}} \\[4pt]
\multicolumn{8}{l}{\tdot{tCov}~\textit{Decision-making style (GDMS)}} \\
Intuitive
& $-0.022$ & $[-0.161,\;\phantom{-}0.118]$ & $0.071$ & $-0.30$ & $0.761$
& $0.98$ & \orci{0.85}{1.13} \\
Dependent
& $0.287$ & $[\phantom{-}0.137,\;\phantom{-}0.438]$ & $0.077$ & $3.73$ & $1.9\times10^{-4}$
& $1.33$ & \orci{1.15}{1.55} \\
Rational
& $0.072$ & $[-0.128,\;\phantom{-}0.271]$ & $0.102$ & $0.70$ & $0.481$
& $1.07$ & \orci{0.88}{1.31} \\
Avoidant
& $-0.076$ & $[-0.293,\;\phantom{-}0.142]$ & $0.111$ & $-0.68$ & $0.495$
& $0.93$ & \orci{0.75}{1.15} \\
Spontaneous
& $0.298$ & $[\phantom{-}0.112,\;\phantom{-}0.485]$ & $0.095$ & $3.14$ & $0.002$
& $1.35$ & \orci{1.12}{1.63} \\[6pt]
\multicolumn{8}{l}{\tdot{tCov}~\textit{Intellectual humility}} \\
IH
& $0.189$ & $[\phantom{-}0.041,\;\phantom{-}0.337]$ & $0.076$ & $2.50$ & $0.012$
& $1.21$ & \orci{1.04}{1.40} \\
\midrule
\multicolumn{8}{@{}p{\dimexpr\linewidth-2\tabcolsep}@{}}{$N = 4{,}318$. Null deviance: $2988.4$ on $4317$ df. Residual deviance: $2598.5$ on $4294$ df. AIC: $2646.5$.} \\
\bottomrule
\end{tabular}
\caption{Predicting judgment weakening: coefficient estimates, odds ratios (OR), and their 95\% confidence intervals from the best logistic regression model (GLM without dilemma as a factor). Intercept gives the Control baseline, while each Intercept (condition) row gives that condition's intercept shift relative to Control when its displayed social signals equal zero. Coefficient intervals were calculated as $\hat{\beta} \pm 1.96 \times \mathrm{SE}$; OR intervals are the exponentiated coefficient intervals. Covariates were selected by cross-validated greedy forward selection. In addition to the predictors shown, the model includes sex and nationality as control variables, which improved out-of-sample fit; their coefficients under different reference groups are reported in Table~\ref{tab:nationality-weakening}.}
\label{tab:full-weakening}
\end{table}

\begin{table}[!htbp]
\centering
\renewcommand{\arraystretch}{1.2}
\setlength{\tabcolsep}{4pt}
\small
\begin{tabular}{lccccccc}
\toprule
\textbf{Predictor}
& $\hat{\beta}$ & 95\% CI$_{\beta}$ & SE & $z$ & $p$ & OR & 95\% CI$_{\mathrm{OR}}$ \\
\midrule
\multicolumn{8}{l}{\textbf{Main predictors}} \\[4pt]
\multicolumn{8}{l}{\tdot{tCtrl}~\textit{Pre-treatment}} \\
Intercept
& $-3.499$ & $[-4.834,\;-2.164]$ & $0.681$ & $-5.14$ & $2.7\times10^{-7}$
& $0.03$ & \orci{0.01}{0.11} \\
Initial confidence ($c$)
& $-0.646$ & $[-0.855,\;-0.436]$ & $0.107$ & $-6.05$ & $1.5\times10^{-9}$
& $0.52$ & \orci{0.43}{0.65} \\[6pt]
\multicolumn{8}{l}{\tdot{tCntr}~\textit{Controversy treatment}} \\
Intercept (Controversy)
& $3.120$ & $[\phantom{-}2.478,\;\phantom{-}3.763]$ & $0.328$ & $9.52$ & ${<}\,2\times10^{-16}$
& $22.6$ & \orci{11.9}{43.1} \\
Disagreement rate
& $-4.083$ & $[-5.318,\;-2.849]$ & $0.630$ & $-6.48$ & $9.1\times10^{-11}$
& $0.02$ & \orci{0.005}{0.06} \\[6pt]
\multicolumn{8}{l}{\tdot{tCert}~\textit{Certainty treatment}} \\
Intercept (Certainty)
& $0.768$ & $[-0.508,\;\phantom{-}2.044]$ & $0.651$ & $1.18$ & $0.238$
& $2.16$ & \orci{0.60}{7.72} \\
In-group certainty
& $1.821$ & $[\phantom{-}0.589,\;\phantom{-}3.052]$ & $0.628$ & $2.90$ & $0.004$
& $6.17$ & \orci{1.80}{21.2} \\
Out-group certainty
& $-1.212$ & $[-2.155,\;-0.268]$ & $0.481$ & $-2.52$ & $0.012$
& $0.30$ & \orci{0.12}{0.76} \\[6pt]
\multicolumn{8}{l}{\tdot{tBoth}~\textit{Controversy + Certainty treatment}} \\
Intercept (Both)
& $1.877$ & $[\phantom{-}0.252,\;\phantom{-}3.502]$ & $0.829$ & $2.26$ & $0.024$
& $6.53$ & \orci{1.29}{33.2} \\
Disagreement rate
& $-3.450$ & $[-4.750,\;-2.149]$ & $0.663$ & $-5.20$ & $2.0\times10^{-7}$
& $0.03$ & \orci{0.01}{0.12} \\
In-group certainty
& $1.258$ & $[-0.101,\;\phantom{-}2.617]$ & $0.693$ & $1.81$ & $0.070$
& $3.52$ & \orci{0.90}{13.7} \\
Out-group certainty
& $-0.394$ & $[-1.407,\;\phantom{-}0.618]$ & $0.516$ & $-0.76$ & $0.445$
& $0.67$ & \orci{0.24}{1.85} \\[10pt]
\multicolumn{8}{l}{\textbf{Covariates}} \\[4pt]
\multicolumn{8}{l}{\tdot{tCov}~\textit{Decision-making style (GDMS)}} \\
Intuitive
& $0.305$ & $[\phantom{-}0.146,\;\phantom{-}0.465]$ & $0.081$ & $3.75$ & $1.8\times10^{-4}$
& $1.36$ & \orci{1.16}{1.59} \\
Dependent
& $-0.213$ & $[-0.399,\;-0.028]$ & $0.095$ & $-2.26$ & $0.024$
& $0.81$ & \orci{0.67}{0.97} \\
Rational
& $-0.022$ & $[-0.262,\;\phantom{-}0.218]$ & $0.123$ & $-0.18$ & $0.858$
& $0.98$ & \orci{0.77}{1.24} \\
Avoidant
& $0.261$ & $[\phantom{-}0.004,\;\phantom{-}0.519]$ & $0.131$ & $1.99$ & $0.046$
& $1.30$ & \orci{1.00}{1.68} \\
Spontaneous
& $0.209$ & $[-0.018,\;\phantom{-}0.435]$ & $0.116$ & $1.80$ & $0.071$
& $1.23$ & \orci{0.98}{1.55} \\[6pt]
\multicolumn{8}{l}{\tdot{tCov}~\textit{Intellectual humility}} \\
IH
& $0.238$ & $[\phantom{-}0.065,\;\phantom{-}0.411]$ & $0.088$ & $2.70$ & $0.007$
& $1.27$ & \orci{1.07}{1.51} \\
\midrule
\multicolumn{8}{@{}p{\dimexpr\linewidth-2\tabcolsep}@{}}{$N = 1{,}924$ (initial confidence $c < 4$). Null deviance: $1942.6$ on $1923$ df. Residual deviance: $1716.1$ on $1907$ df. AIC: $1750.1$.} \\
\bottomrule
\end{tabular}
\caption{Predicting judgment strengthening: coefficient estimates, odds ratios (OR), and their 95\% confidence intervals from the best logistic regression model (GLM without dilemma as a factor; covariates: GDMS, IH). Intercept gives the Control baseline, while each Intercept (condition) row gives that condition's intercept shift relative to Control when its displayed social signals equal zero. Coefficient intervals were calculated as $\hat{\beta} \pm 1.96 \times \mathrm{SE}$; OR intervals are the exponentiated coefficient intervals.}
\label{tab:full-strengthening}
\end{table}

\begin{table}[htbp]
  \centering
  \begin{tabularx}{\textwidth}{cccc >{\raggedright\arraybackslash}X}
    \toprule
    \textbf{Variable} & \textbf{Treatment} & \textbf{Acceptance} & \textbf{Outcome} & \textbf{Written explanation} \\
    \midrule

    \multirow{6}{*}{\makecell{\textbf{Disagreement} \\ \textbf{rate}}} &
    \cellcolor{colorControversy} Controversy & \cellcolor{colorAccept} Accept & Weakened &
    ... The way the audience is split almost 50/50 leaves me a little uncertain either way. \\
    \cmidrule(lr){3-5}

    & \cellcolor{colorControversy} Controversy & \cellcolor{colorAccept} Accept & Weakened &
    Seeing that this was a controversial post emphasized that point to me and made me feel even more uncertain about my initial judgment. \\
    \cmidrule(lr){3-5}

    & \cellcolor{colorControversy} Controversy & \cellcolor{colorAccept} Accept & Maintained &
    ... most readers agree the individual is not in the wrong here. That further affirms my initial impression. \\
    \cmidrule(lr){3-5}

    & \cellcolor{colorControversy} Controversy & \cellcolor{colorReject} Reject & Maintained &
    ... I do consider other people's positions and arguments with alongside my own, but the decision rests with me. \\
    \cmidrule(lr){2-5}

    & \cellcolor{colorCC} C+C & \cellcolor{colorAccept} Accept & Maintained &
    ... many people strongly disagree with her reasoning. This makes me less confident in my original stance, even though I still believe her choice is valid ... \\
    \cmidrule(lr){3-5}

    & \cellcolor{colorCC} C+C & \cellcolor{colorReject} Reject & Maintained &
    I don't change my mind based on other people's opinions, but I can understand the controversy and different ways of seeing this. \\

    \cmidrule(lr){1-5}

    \multirow{4}{*}{\makecell{\textbf{In-group} \\ \textbf{certainty}}} &
    \cellcolor{colorCertainty} Certainty & \cellcolor{colorAccept} Accept & Maintained &
    I am confident of my position, and it's evident considering the number of people supporting my view. \\
    \cmidrule(lr){3-5}

    & \cellcolor{colorCertainty} Certainty & \cellcolor{colorReject} Reject & Maintained &
    ... I'm not swayed by other people's opinions. \\
    \cmidrule(lr){2-5}

    & \cellcolor{colorCC} C+C & \cellcolor{colorAccept} Accept & Strengthened &
    ..the fact that 98 people agreed changed my decision to `certain', due to `the power of the crowd' dispelling the small amount of uncertainty I had. \\
    \cmidrule(lr){3-5}

    & \cellcolor{colorCC} C+C & \cellcolor{colorAccept} Accept & Strengthened &
    Seeing the large amount of people being certain with their decision of YA also increased my certainty. \\

    \cmidrule(lr){1-5}

    \multirow{3}{*}{\makecell{\textbf{Out-group} \\ \textbf{certainty}}} &
    \cellcolor{colorCertainty} Certainty & \cellcolor{colorAccept} Accept & Weakened &
    ...I am slightly swayed by the very high percentage certainty of those who think the person was not an asshole. \\
    \cmidrule(lr){3-5}

    & \cellcolor{colorCertainty} Certainty & \cellcolor{colorReject} Reject & Maintained &
    For me, my opinion is the same regardless of what the masses think. \\
    \cmidrule(lr){2-5}

    & \cellcolor{colorCC} C+C & \cellcolor{colorAccept} Accept & Weakened &
    High certainty level among those who disagree has made me less confident in my judgement. \\

    \bottomrule
  \end{tabularx}
  \caption{Selected written explanations categorized by explicit mentions of experimental variables (disagreement rate, in-group certainty, or out-group certainty). The table further breaks down participants' reasoning for changing or maintaining their stance by experimental condition and outcome.}
  \label{tab:selected_deliberation_newline}
\end{table}

\clearpage
\section*{Data and code availability}
All analysis code and the data needed to reproduce the figures and statistics reported in the main text are available at \url{https://github.com/ZiyuChen0410/Controversy-and-Group-Certainty}.

\putbib[Reference]
\end{bibunit}

\newpage
\bmhead{Acknowledgements}
We thank Mark Alfano for useful discussion. We also thank Patrik Haslum, Giles Hirst, Siqi Wu, Jenny L. Davis, Pamela Robinson, and Ying Zhang for comments and advice; Yilin Ding and Julian Singh for assistance with data annotation and discussion; and Lydia Lucchesi and Jiamin Lim for testing and providing feedback on the survey.

This work was supported in part by an Australian National University Futures Scheme grant; Australian Research Council grants FT230100563, DP240100506, LP210200818 and DP240100914; and the CSIRO--National Science Foundation (US) AI Research Collaboration Program.

\bmhead{Author contributions}
\begin{itemize}
    \item \textbf{Ziyu Chen:} Conceptualization; Methodology (experimental design, analysis plans, preregistration); Software (experiment implementation); Visualization; Investigation; Formal analysis; Data curation and annotation; Writing (original draft); Writing (review and editing), Project administration.
    \item \textbf{Minjeong Shin:} Formal analysis; Software (website design and implementation); Visualization; Data annotation; Writing (original draft); Writing (review and editing).
    \item \textbf{Tuan Dung Nguyen:} Methodology (experimental design); Formal analysis; Data annotation; Writing (original draft); Writing (review and editing).
    \item \textbf{Colin Klein:} Conceptualization; Methodology (experimental design); Writing (original draft); Writing (review and editing); Funding acquisition.
    \item \textbf{Chenhao Tan:} Conceptualization; Methodology (experimental design); Writing (review and editing).
    \item \textbf{Nick Schuster:} Conceptualization; Data annotation; Writing (review and editing).
    \item \textbf{Nicholas George Carroll:} Conceptualization; Data annotation; Writing (review and editing).
    \item \textbf{Alasdair Tran:} Methodology (experimental design); Data annotation; Writing (review and editing).
    \item \textbf{Lexing Xie:} Conceptualization; Methodology (experimental design and analysis plans); Data annotation; Writing (original draft); Writing (review and editing); Funding acquisition, Project administration.
\end{itemize}

\bmhead{Competing Interests}
All authors of this article declare no competing interest.

\bmhead{Additional Information}
Supplementary information accompanies this manuscript. An interactive overview of the \dataset dataset, study design, and findings is available at \url{https://moralmoments.vercel.app/}.

\newpage
\begin{bibunit}[sn-apacite]
\appendix

\addtocontents{toc}{\protect\setcounter{tocdepth}{3}}
\begin{center}
{\Large\bfseries Supplementary Information\par}
\vspace{0.6em}
{\large\bfseries Controversy and Group Certainty Jointly Shape Everyday Moral Judgments\par}
\end{center}
\vspace{1.5em}
\tableofcontents
\clearpage

\newgeometry{left=2.5cm,right=2.5cm,top=2cm,bottom=2.5cm}

\section{Related work}\label{sec:secA1}

This appendix situates the study within three linked bodies of research. Section~A.1 describes everyday moral judgment as it appears in naturalistic settings and identifies the direct literature on social influence in moral decision-making. Section~A.2 then reviews broader research on how communication, group positions, and aggregate social evidence shape individual judgments across moral and non-moral domains. Section~A.3 examines certainty as a social signal: how it is conceptualized under disagreement, expressed by individuals and groups, and communicated through language. Together, these literatures motivate the present focus on how the distribution of collective moral judgments and the certainty expressed on each side shape individual responses to everyday dilemmas.

\subsection{Moral judgment in everyday contexts}\label{sec:moral}

Moral and social norms concern what people take ought to be done, as well as expectations about what others regard as appropriate \citep{Brennanetal2013}. Computational and behavioral research has made it possible to study such judgments at scale, including in naturally occurring public discussions rather than only in constructed laboratory scenarios. This section introduces empirical setting for everyday moral dilemmas, then distinguishes moral conviction and moralization from the social-information processes examined in the present study.

\subsubsection{Everyday moral dilemmas and online moral norms}\label{subsec:everyday}

Several studies have taken advantage of the scale and access of the internet to describe moral norms. Early work in this vein was the Moral Machine experiment \citep{Awad2018TheMM}, which engages global participants in theoretical moral dilemmas (similar to so-called ``trolley problems'') around autonomous vehicle driving to capture varied ethical priorities across countries. Related experimental research has also developed standardized vignette sets to study particular moral concerns under controlled conditions \citep{clifford2015moral}. Such designs provide valuable control, but they do not capture the narrative detail, interpersonal stakes, and overlapping concerns characteristic of many everyday dilemmas.

Recent studies have increasingly focused on understanding ethical judgments and social norms within an online platform, Reddit, particularly the ``Am I The Asshole'' (AITA) subreddit, where users post personal stories to receive judgments from the community about their stories. For example, \citet{Lourie2020ScruplesAC} introduced SCRUPLES, a dataset containing 625k human evaluations across 32k everyday moral dilemmas extracted from this community. Their findings indicate that a substantial portion of everyday social norms lack a strict consensus, exhibiting high levels of population division. \citet{Nguyen2022MappingTI} used a large corpus from the same community to identify and validate 47 topics in everyday moral dilemmas. Empirical observations showed that most dilemmas combine more than one topic, underscoring the complexity of stakes and judgments in this setting. \citet{Nguyen2023MeasuringMD} developed Mformer to measure moral dimensions in social-media text and applied it to discussions on Reddit and Twitter. Although Mformer does not model community verdicts, it shows how computational methods can characterize the substantive moral content of large-scale online discussion.

Together, these studies establish online discussion as a setting in which people evaluate rich, naturally occurring moral narratives; the moral content and collective evaluation of those narratives can both be studied at scale.

\subsubsection{Computational representations of moral judgment}\label{subsec:computational}

Computational studies have used online moral data both to analyse patterns in collective evaluation and to predict or classify moral judgments. \citet{Candia2022SocialNO} examined demographic correlates of moral judgments on Reddit; \citet{Giorgi2023AuthorAC} analyzed how AITA authors construct themselves as both narrators and characters; and \citet{Haworth2021ClassifyingRI} developed an early classifier of the perceived reasonableness of actions in retold personal events.

Computational work has also sought to represent moral and social norms in machine-readable form. \citet{Jiang2021DelphiTM} developed Delphi to predict ethical judgments across varied everyday scenarios, while Social Chemistry 101 and Moral Stories provide resources for reasoning about social interactions, norms, intentions, actions, and consequences \citep{Forbes2020SocialC1,emelin-etal-2021-moral}. The Moral Integrity Corpus extends this work to ethical dialogue systems \citep{Ziems2022TheMI}.

These projects differ from the present study in their primary goal: they analyse, classify, or generate judgments and normative content, whereas we examine how people respond to information about a community's existing judgments. They nevertheless provide an important backdrop for analysing moral discourse at scale and for retaining disagreement rather than treating a single aggregate label as the whole of a community's view.

\subsubsection{Moral conviction and moralization}\label{subsec:conviction}

Opinions about moral matters can be studied like any other opinion. At the same time, research on moral conviction argues that some attitudes are experienced as matters of objective right and wrong, broadly applicable beyond personal preference, and especially important to their holders \citep{skitka2021psychology}. Such convictions have been associated with intolerance of opposing views, resistance to majority influence, and political or civic engagement \citep{skitka2008moral,aramovich2012opposing,skitka2014social}. Morally convicted attitudes can shape behaviour in domains such as voting, capital punishment, and environmental protection \citep{Jenke2017Vote,skitka2021psychology}.

Related research examines how previously non-moral preferences can become moralized. For example, \citet{feinberg2019understanding} studied the process through which eating meat can acquire moral significance. Research on moral foundations and emotion likewise shows how the content and affective basis of moral concern can differ across people and issues \citep{Graham2009Liberals,Horberg2009Disgust}.

This literature helps explain why moral judgments may sometimes be resistant to social influence. However, the present study does not measure moral conviction, moral foundations, or moralization. It instead measures participants' verdicts and their stated confidence before and after exposure to social information; these should not be interpreted as direct measures of the strength or moral status of an underlying conviction.

\subsubsection{Social influence in moral judgment}\label{subsec:socialinfluence}

A smaller, more directly relevant literature asks how social information affects moral decision-making. \citet{pryor2019even}, for example, showed that even arbitrary social norms can influence moral decisions. Work on moral conviction further suggests that strongly moralized attitudes may resist majority influence \citep{aramovich2012opposing}, while \citet{Jiang2023PeopleCT} found conformity to social norms in decisions involving lives as well as money. Research on moral judgment under uncertainty also shows that the structure of uncertainty can affect how people evaluate moral choices \citep{Ng2023MoralJU}.

These studies establish that moral judgments can be socially responsive, but they differ from the present design in important respects. Existing work commonly uses stylized scenarios, explicit norms, or direct social cues.
This leaves open how people respond to disagreement and expressed certainty when these signals are drawn from naturalistic moral discussion rather than conveyed as explicit norms or stylized social cues.

This distinction also matters for interpreting change. Everyday moral disagreements do not generally offer the independently verifiable answers available in many factual tasks \citep{SchusterKilov2025MoralDisagreement}. Observing a distribution of others' judgments may therefore prompt reflection or revision, but it does not by itself establish that one verdict is correct or better justified. The following section places this moral-literature bridge within broader research on social influence on individual judgment.

\subsection{Social influence on individual judgment}\label{sec:socialinf}

People’s judgments can be shaped by others through several routes: communication that supplies arguments or narratives, visible social norms and group positions, or summaries of what a wider population believes. This section reviews these broader mechanisms across moral and non-moral domains. It first considers influence through communication, then conformity and minority influence, and finally the more specific literature on aggregate social evidence that most directly motivates the present experiment.

\subsubsection{Influence through communication}\label{subsec:infcomm}

Research on opinion change has examined how arguments, interaction, and communication shape judgments over time. In moral contexts, \citet{Strimling2018TheCB} found that the connection between moral positions and moral arguments can drive opinion change. Studies of online discussion likewise show that interaction dynamics and persuasive strategies matter: work on the Reddit \textit{ChangeMyView} community identifies features associated with successful persuasion in good-faith discussion \citep{Tan2016WinningAI}, while related research finds that linguistic phrasing can predict persuasive success in other online settings, such as fundraising appeals \citep{Mitra2014The}.

This computational work builds on a long psychological tradition of attitude change and persuasion. Classic dual-process accounts frame attitude change as a function of how much a person elaborates on a message, and of the strength of the resulting attitude \citep{Petty1986TheEL,Petty1997Attitudes}. Stronger, more important attitudes are more resistant to persuasion \citep{Pomerantz1995Attitude,Zuwerink1996Attitude}, and successfully resisting persuasion can itself increase certainty in the original attitude \citep{Tormala2002What}. Social cues can also shift attitudes as heuristics, independent of argument content, such as an audience's apparent response \citep{Axsom1987Audience} or evidence that others have already complied with a request \citep{Cialdini1999Compliance}.

Recent interventions illustrate the range of communication-based influence. Exposure to people with similar identities but differing political views can reduce opinion polarization \citep{Balietti2021ReducingOP}; AI-mediated dialogue has reduced conspiracy beliefs \citep{costello2024durably}; conversational systems can be persuasive in political discussion \citep{salvi2025conversational}; and AI-assisted deliberation can help participants identify common ground on contentious social issues \citep{tessler2024ai}. These studies typically involve conversational exchange, arguments, narratives, or repeated engagement. They therefore provide an important contrast to the present experiment, in which participants received a terse summary of others' judgments and apparent certainty without new reasons for or factual information about the dilemma.

A related literature examines polarization and depolarization under longer-term social interaction. Group polarization refers to the tendency for group decisions to become more extreme than members' initial inclinations, a phenomenon reviewed by \citet{Myers1976The}. Descriptive work has traced polarization in political communication, including historical changes in the framing of US immigration speeches \citep{Card2022ComputationalAO} and linguistic patterns in online discussion of contentious issues such as abortion, climate change, vaccination, and gun control \citep{Beel2022LinguisticCharacterization,Ojea-QuintanaPolarization22}.

This literature has also examined mechanisms that can sustain or reduce division. Confirmation bias can reinforce existing views \citep{Nickerson1998ConfirmationBA}, while polarized social environments may be shaped by echo chambers, emotionally charged moral communication, and uneven contact between in-groups and out-groups \citep{vanBaar2021ThePM}. Personal experiences can sometimes bridge moral and political divides more effectively than factual arguments \citep{kubin2021personal}, although such narratives can also mislead or oversimplify complex issues \citep{van2021speaking,brady2020mad}. Studies of platform exposure and structured dialogue similarly report mixed effects: like-minded sources on Facebook were prevalent but not themselves polarizing \citep{Nyhan2023LikemindedSO}, exposure to opposing views on social media can increase polarization \citep{Bail2018Exposure}, and a mobile chat platform has reduced political polarization in a structured intervention \citep{Combs2023ReducingPP}.

These studies therefore illuminate influence through exchanged reasons, narratives, and repeated interaction, rather than the interpretation of a summary of others’ views.

\subsubsection{Conformity and minority influence}\label{subsec:conformity}

A second tradition examines how the apparent views of a group shape individual judgment. Social influence has been studied across psychology, sociology, neuroscience, and decision research. Work on social decision-making, for example, examines how social context is integrated into valuation and choice \citep{Ruff2014TheNO,Toelch2015InformationalAN}.

Classic experiments established that people may align their judgments with an apparent group position even when it conflicts with their own perception \citep{asch1955opinions}. Research on obedience further showed how social roles and authority can shape normatively consequential actions \citep{milgram1963behavioral}.
Uncertainty and accountability further qualify the effects of group context. People can be more responsive to group norms when they are uncertain \citep{Smith2007UncertaintyAT}, and people conform to social norms in decisions involving lives as well as money \citep{Jiang2023PeopleCT}. Research on accountability suggests that anticipating evaluation by others can alter the complexity and strategy of reasoning, producing conformity, bolstering, or greater attention to information \citep{tetlock1983accountability,tetlock1989social,Tetlock1989AccountabilityAS}.

Group influence is not limited to majority pressure. A consistent minority can affect majority responses \citep{Moscovici1969Minority}, and a meta-analytic review shows that minority influence depends on features of the source, target, issue, and influence process \citep{Wood1994}. This literature is particularly relevant to the present study because controversy makes a participant's majority or minority position visible, while group certainty may affect whether a minority position appears resilient, persuasive, or both.

Overall, they clarify how visible groups, authorities, and minorities shape judgment, but leave distinct the question of how people interpret anonymous aggregate signals.

\subsubsection{Aggregate social evidence}\label{subsec:aggregate}

A more specific body of work concerns judgments based on aggregate social evidence: information about the distribution of views in a population rather than direct interaction with particular individuals. Such information can function as higher-order evidence about what others believe, even when it supplies no new first-order evidence about the issue itself \citep{oktar2025aggregated}.

Prior research establishes that people learn both from the prevalence of others’ opinions and from the confidence with which those opinions are expressed. Aggregate-opinion experiments have primarily examined the proportion endorsing a claim without presenting the certainty expressed by each side \citep{OktarLombrozoGriffiths2024}. Research on expressed confidence has instead largely used dyadic tasks, in which people communicate confidence during a joint decision or observe the judgment and confidence of one other person \citep{bang2017confidence,Kappes2020}. In the closest earlier integration, participants revised factual estimates after observing another person’s estimate and confidence; models based on these responses produced competing expert and majority effects \citep{Moussaid2013SocialInfluence}. More recently, factual answer distributions were presented together with the average confidence of each side, but without a certainty-only condition or a direct test of whether controversy changes the influence of certainty \citep{HeLienZheng2026}.

The present study extends this body of work in three ways. First, it examines rich, naturally occurring interpersonal moral dilemmas rather than controlled factual estimates. Second, it independently manipulates controversy and group certainty, allowing their separate effects to be distinguished. Third, it represents certainty separately for people whose initial verdict agrees with the participant and for those whose verdict disagrees.
This design makes it possible to test not only whether participants respond to aggregate social evidence, but also how the visibility of disagreement changes the role of certainty.

\subsection{Certainty as social information}\label{sec:certainty}

Certainty can refer to several related but distinct things: a person's private confidence in a judgment, the confidence expressed in communication, the uncertainty of a computational model, or the apparent certainty of a group. These distinctions matter here because the treatment does not reveal whether any individual is objectively correct. Rather, it represents how confidently people on each verdict side appear to hold their collective view. This section reviews conceptual accounts of uncertainty and disagreement, psychological work on confidence in individual and small-group judgment, and research on communicating and measuring certainty in language.

\subsubsection{Moral uncertainty and disagreement}\label{subsec:moraluncertain}

Philosophical work on moral uncertainty asks how a person should act when unsure which moral view is correct. One influential approach treats moral views as graded and recommends choosing actions by weighing each view according to its expected moral value and credibility \citep{macaskill2020moral}. This literature distinguishes uncertainty about the facts of a situation from uncertainty about the relevant moral principle \citep{tarsney2024moral}.

This framing is useful because it treats confidence in moral views as graded rather than all-or-nothing. However, moral uncertainty concerns competing views held by a single agent; it does not directly address how one should respond to other people's confidence in a shared public disagreement. Philosophical work on peer disagreement more directly considers how confidence, evidence, and another person's opposing judgment should bear on one's own view \citep{christensen2007epistemology,vavova2014confidence}.

These accounts clarify the normative stakes of disagreement, but do not themselves predict how people will respond to displayed collective certainty.

\subsubsection{Confidence in individual and group judgment}\label{subsec:confidence}

Psychological research commonly treats certainty as an individual state or trait. Metacognitive accounts examine how people evaluate the reliability of their own knowledge and decisions \citep{fleming2024metacognition}. The literature on attitude certainty examines confidence in one's own opinions, showing that greater certainty is associated with resistance to persuasion and with stronger consistency between attitudes and behaviour \citep{tormala2018attitude}.

Individual certainty can also respond to social cues; when individuals learn that a majority, rather than a minority, of others share their position, they become more confident that their attitude is correct \citep{petrocelli2007unpacking}. Small group studies indicate that dyads often align their confidence levels through verbal interaction \citep{bang2017confidence}, and shared confidence serves as a critical signal for reaching joint decisions \citep{bahrami2012whatfailure}.

The remaining question is how certainty inferred from a large public discussion is used when it is attached to opposing verdict groups.

\subsubsection{Expressing and measuring certainty}\label{subsec:measuring}

Research on uncertainty in machine learning distinguishes aleatoric uncertainty, arising from variability or noise in the data, from epistemic uncertainty, arising from limitations in a model's knowledge \citep{hullermeier2021aleatoric,kendallgal2017uncertainties}.
A related concern is calibration: whether a model's stated confidence corresponds to its accuracy \citep{guo2017calibration}. Recent work has assessed uncertainty in large language models through variation in the semantic content of their answers \citep{farquhar2024detecting} and through direct confidence elicitation \citep{xiong2024llms}.

These computational concepts are not measures of human group certainty. They nevertheless clarify the distinction between uncertainty in a model and certainty expressed by the people whose judgments a model may analyse. In public discussion, people rarely attach an explicit numerical confidence label to each statement. Instead, certainty is often conveyed through linguistic choices such as hedges, modal verbs, and qualifiers.

Early work used hedges as a direct proxy for uncertainty, formalized as a shared task for detecting hedges and their scope in text \citep{farkas2010conll}, and later work shows that certainty in text has both a level and several distinct aspects that a pre-trained language model can recover more fully, though this was demonstrated for single sentences in science communication \citep{pei2021measuring}.

How confident a message sounds changes how it is received. Communicating uncertainty about a fact or a number lowers trust only slightly, and mostly when the uncertainty is stated in words rather than a numeric range \citep{vanderbles2019communicating,vanderbles2020effects}. Expressed confidence also persuades. A confidently stated opinion is treated as more informative, an effect strong enough to sway a group \citep{pulford2018persuasive}.

These findings motivate the present measure of group certainty. Rather than treating certainty as a private feeling or assuming that it directly tracks correctness, the study aggregates linguistic indications of certainty within each verdict group. This makes it possible to examine whether apparent certainty among people who agree or disagree with a participant's initial judgment is associated with different forms of subsequent judgment change.

\section{Data profile and survey experiment}\label{sec:data_methods}

This section details the data and methods behind the \dataset dataset and the survey experiment.
\Cref{subsec:aita} describes the r/AmItheAsshole community that the moral dilemmas are drawn from. \Cref{subsec:datapreparation} details how we selected and profiled the 135 dilemmas within \dataset. \Cref{subsec:uncertainty} describes how we construct controversy and extract and validate individual and group certainty from AITA comments. \Cref{sec:surveyexperiment} details the survey experiment, including recruitment, the three-phase survey procedure, and the background measures collected. \Cref{sec:secA4} profiles the resulting participant sample, including data processing and exclusion criteria.

\subsection{The \textit{r/AmItheAsshole} community}\label{subsec:aita}

\begin{figure}[h]
    \centering
    \begin{subfigure}{0.63\textwidth}
        \centering
        \includegraphics[width=\textwidth]{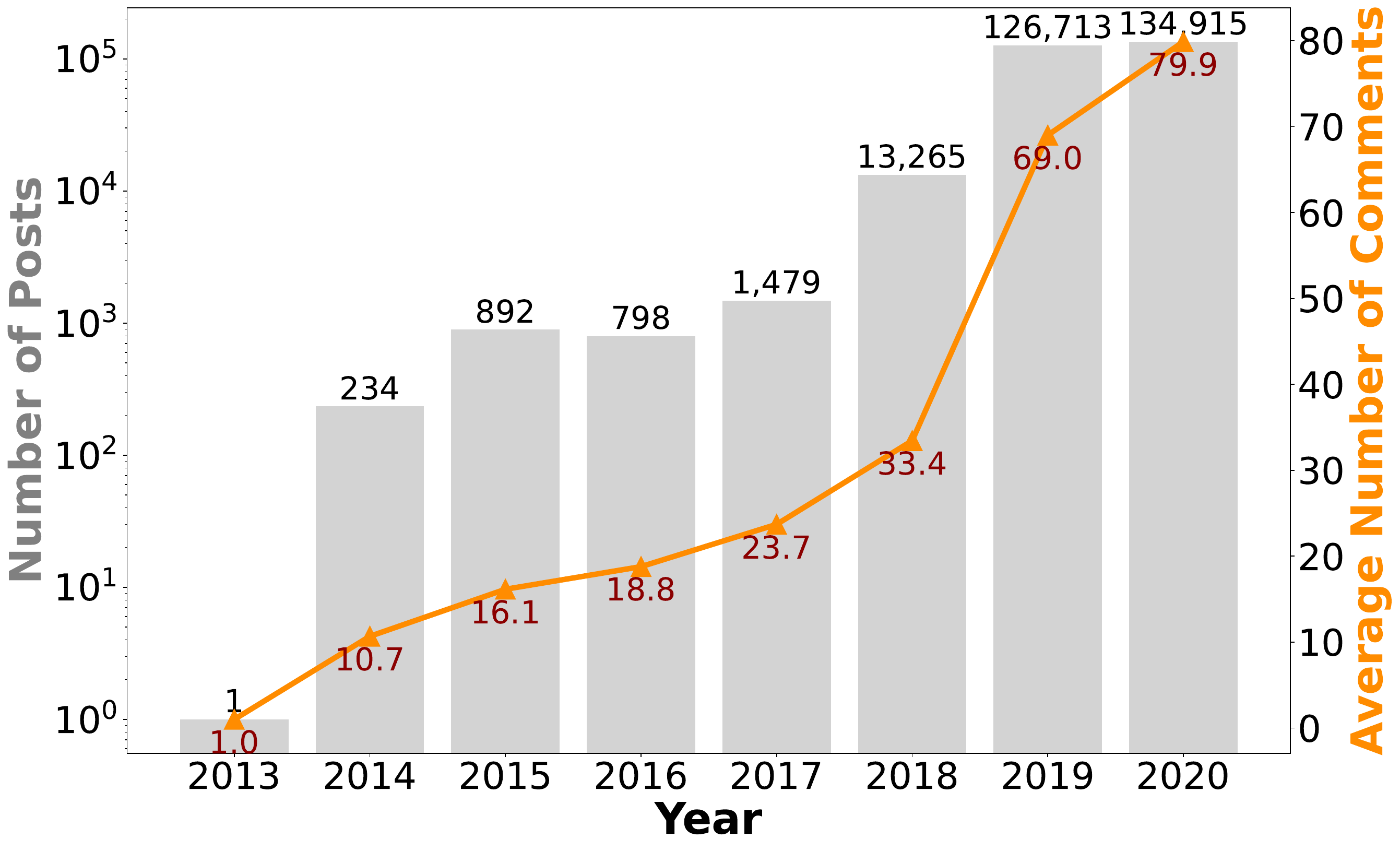}
        \caption{}
        \label{fig:aita_posts}
    \end{subfigure}
    \hfill
    \begin{subfigure}{0.36\textwidth}
        \centering
        \includegraphics[width=\textwidth]{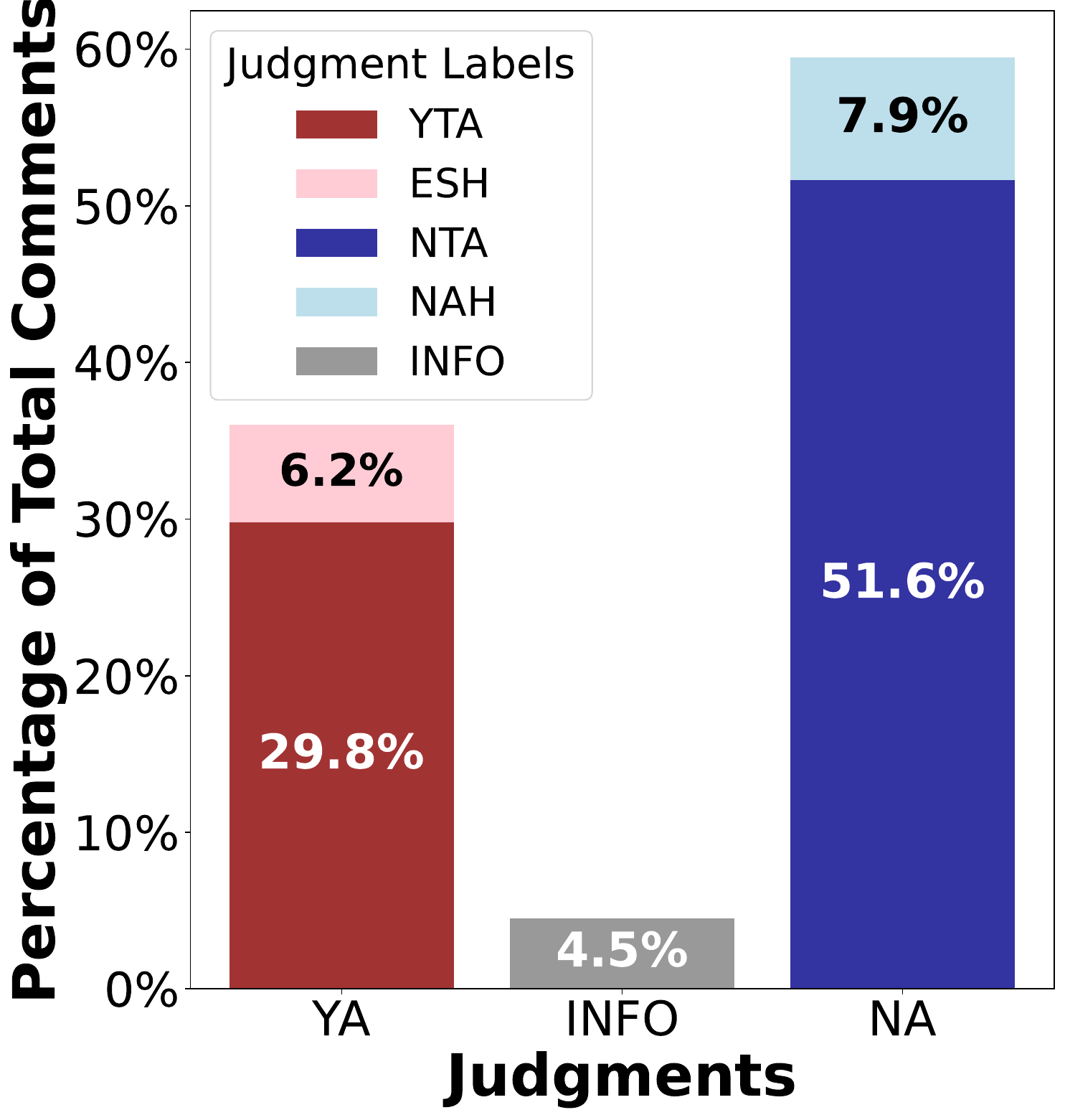}
        \caption{}
        \label{fig:aita_judgments}
    \end{subfigure}
    \caption{Statistics of AITA posts, comments, and verdicts from the source dataset of \citet{Nguyen2022MappingTI}. (a) Number of AITA posts and average comments per post (2013--April 2020). The bars represent the number of posts per year, while the line tracks the average number of comments per post. (b) Distribution of AITA verdicts: over 7 million verdicts were extracted from 15 million top-level comments. After grouping the verdicts into two broader categories, \textbf{YA} (Yes, the OP is at fault, including YTA where the OP is at fault and ESH where everyone involved is at fault) and \textbf{NA} (No, the OP is not at fault, including NTA where the OP is not at fault, and NAH where no one is at fault), \textbf{YA} takes up 36.03\% and \textbf{NA} 59.38\%, while INFO accounts for
4.49\% (see \Cref{subsubsec:aita_structure} for more details).
}
    \label{fig:aita_posts_comments}
\end{figure}

Our moral studies require a large and diverse collection of real-world moral decision scenarios. \textit{r/AmItheAsshole} (AITA) is suitable for this purpose since it provides a rich and large-scale data source of everyday moral dilemmas. AITA is an online Reddit community where a user posts about an interpersonal conflict and receives verdicts from other users. The central question, ``Am I the Asshole?'', in each post revolves around whether the poster was at fault in the situation they have described. From 2014 to 2020, the AITA community showed a notable growth in discussions where the number of posts per year increased from 234 to 134,915 and the average number of comments within each thread per year increased from 10.7 to 79.9 as shown in \Cref{fig:aita_posts}.

AITA has been a source for studying moral situations. \citet{Lourie2020ScruplesAC} curate ``Scruples", a corpus and benchmark consisting of 32,000 everyday moral dilemmas with 625,000 human moral judgments sourced from the AITA subreddit, revealing that many moral and social norms do not have definitive answers, with people often divided in their opinions on these issues. \citet{Candia2022SocialNO} explored how demographic factors impact moral judgments on Reddit, finding that male authors and older individuals are more prone to receiving negative feedback, while discussions centered around relationships and work often attract more supportive responses. \citet{Nguyen2022MappingTI} categorize 100,000 everyday moral dilemmas into 47 distinct topics and find that people perceive most dilemmas as involving a combination of two topics. We use their released dataset, which contains 102,998 posts labeled with 47 topics and over 7 million comments containing verdict label, ranging from 2013 to April 2020.

\subsubsection{The structure of AITA}\label{subsubsec:aita_structure}
\begin{figure}[h]
    \centering
    \includegraphics[width=1\textwidth]{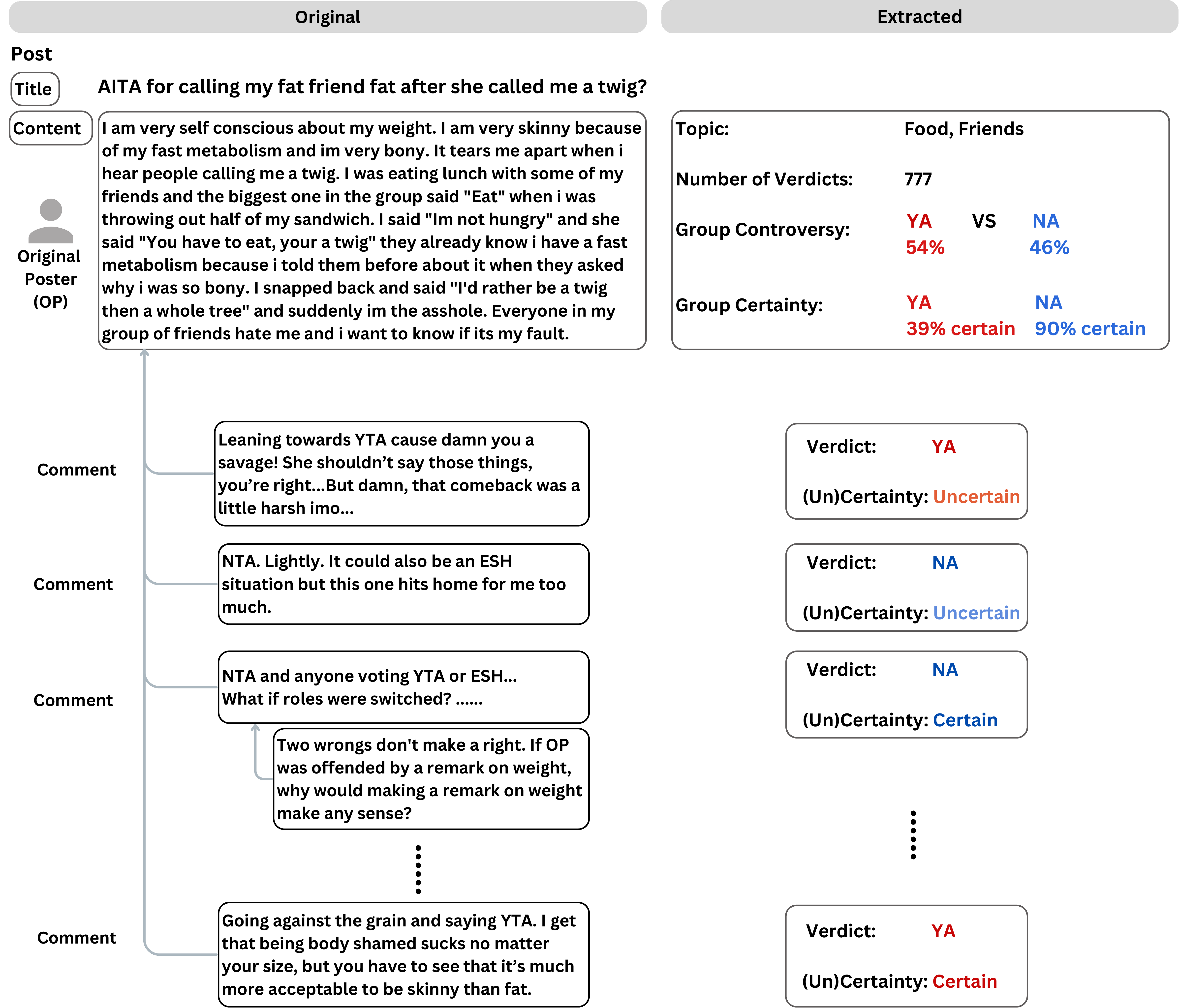}
    \caption{The structure of AITA posts. The left side (Original) shows the flow of a typical AITA thread, including the post's title, content, and comments containing verdicts and reasoning. The right side (Extracted) shows the post topic, number of verdicts, group controversy, group certainty, and each comment's verdict and certainty (see \ref{subsubsec:individual_uncertainty} for details). In this example, the NA minority verdict group is more certain than the YA majority verdict group, showing that controversy and certainty capture distinct features of community responses.}
    \label{fig:aita_structure}
\end{figure}

\Cref{fig:aita_structure} presents an overview of relevant data structures in AITA for our study.

The left side of \Cref{fig:aita_structure} illustrates the original structure of posts and comments. In the AITA community, discussions are organized into \textbf{threads}, each beginning with a post followed by a series of comments. Each \textbf{post} in AITA contains \textbf{title, original poster}, and \textbf{content}. The title provides a brief description of the dilemma, starting with the acronym ``AITA" (Am I The Asshole?) or ``WIBTA" (Would I Be The Asshole?). The original poster refers to the Reddit username of the individual submitting the post. The content includes a detailed narrative of the situation being described.

Each \textbf{comment} is made up of a \textbf{commenter} and \textbf{content} and may either reply to the original post or respond to another comment. \textbf{Top-level comments} are those that directly reply to the original post made by the OP (Original Poster), while non-top-level comments reply to other comments rather than the original post. According to the AITA community's rules, only top-level comments with verdict labels contribute to the final verdict for an AITA post. A verdict label in a reply to another comment does not count. Top-level comments often include a \textbf{verdict label} and reasoning that reflects the user's moral evaluation. The five labels are YTA (You’re the Asshole), where the original poster (OP) is considered at fault; NTA (Not the Asshole), where the OP is not at fault; ESH (Everyone Sucks Here), where all parties involved are deemed at fault; NAH (No Asshole Here), where no party is considered to be at fault; and INFO (More Information Needed), where additional information is required before giving a verdict.

The right side of \Cref{fig:aita_structure} illustrates the extracted structure of posts and comments. For each AITA comment, we extract its verdict and certainty. For each AITA post, we extract its topic and aggregate the number of verdicts, group controversy, and group certainty from top-level verdicts.

\textbf{Verdicts:} We extract verdicts from top-level comments under each selected post using regular-expression patterns that detect the labels. We then combine them into two broader categories: \textbf{YA} (Yes, the OP is at fault) and \textbf{NA} (No, the OP is not at fault). YA includes \textit{YTA} and \textit{ESH}, while NA includes \textit{NTA} and \textit{NAH}. As shown in Figure \ref{fig:aita_judgments}, YA accounts for 36.03\% and NA for 59.38\% of the dataset provided by \citet{Nguyen2022MappingTI}, while INFO accounts for 4.49\%. We remove INFO comments because the commenter has not selected a verdict and is seeking more information.

\textbf{Certainty:} In the AITA
subreddit, individual certainty refers to how confident a commenter is in their chosen verdict. For example, in the post titled ``AITA for calling my fat friend fat after she called me a twig?" as illustrated in \Cref{fig:aita_structure}, some comments are certain, such as ``NTA and anyone voting YTA or ESH... What if roles were switched? ......" while some are uncertain, for instance: ``Leaning towards YTA cause damn you a savage! She shouldn’t say those things, you’re right... But damn, that comeback was a little harsh IMO...". The method of extracting individual certainty is detailed in \ref{subsubsec:individual_uncertainty}.

\textbf{Topic:} The dataset categorizes over 100,000 everyday moral dilemmas into 47 distinct topics via Latent Dirichlet Allocation (LDA) topic modeling, with validations from experts and crowd-sourced workers \citep{Nguyen2022MappingTI}. They also discover that people often identify a pair of two main topics involved in each dilemma: the \textbf{primary topic}, which is the most dominant theme within a post, and the \textbf{secondary topic}, which is the second most prominent theme. For example, in the post titled ``AITA for calling my fat friend fat after she called me a twig?" (illustrated in \Cref{fig:aita_structure}), the primary and secondary topics are food and friends, respectively. In our study, these topic characteristics capture the kinds of moral dilemmas that people discuss in daily life, and they guide our selection of moral situations for the experiment and regression analysis of the experimental results.

\textbf{Number of verdicts:} This is the total number of comments with YA or NA verdict labels for a post and measures its discussion volume (see \Cref{subsec:datapreparation} for details).

\textbf{Group controversy:} The \textbf{majority verdict group} contains the more common binary verdict on an AITA post, and the \textbf{minority verdict group} contains the less common verdict. For instance, in \Cref{fig:aita_structure}, YA is the majority verdict at 54\%, while NA is the minority verdict at 46\%. We define group controversy as the proportion of verdicts in the minority group, ranging from 0 to 0.5. Values near 0.5 indicate a nearly even split, while values near 0 indicate little disagreement. In \Cref{fig:aita_structure}, controversy is 0.46.

\textbf{Group certainty:} This is the percentage of comments classified as certain within a verdict group. In \Cref{fig:aita_structure}, 39\% of comments in the YA majority verdict group are certain, compared with 90\% in the NA minority verdict group. Details on extracting group certainty are provided in \ref{subsubsec:Group_Uncertainty:}.

\subsection{The \dataset dataset}\label{subsec:datapreparation}

\begin{figure}[h]
    \centering
        \begin{subfigure}[t]{1\textwidth}
        \centering
        \includegraphics[width=\textwidth]{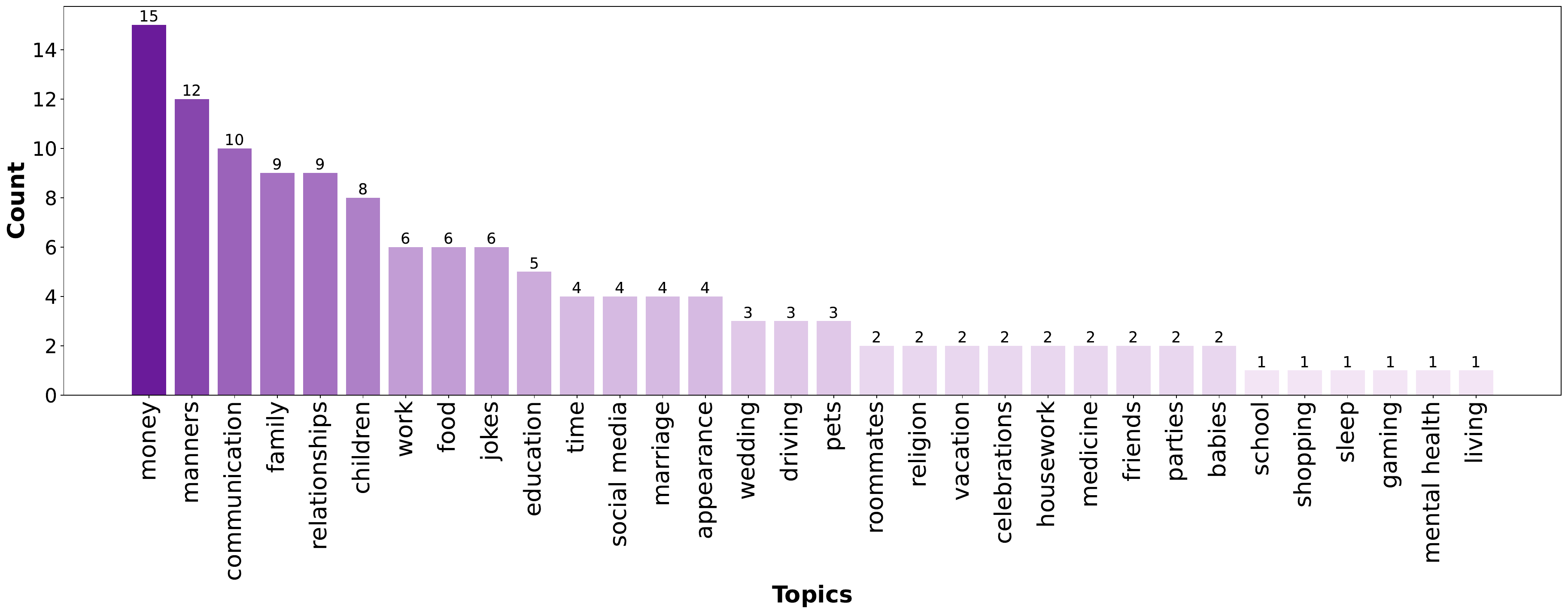}
        \caption{}
        \label{fig:topic_distribution}
    \end{subfigure}

    \vspace{0.5cm}

        \begin{subfigure}[t]{0.48\textwidth}
        \centering
        \includegraphics[width=\textwidth]{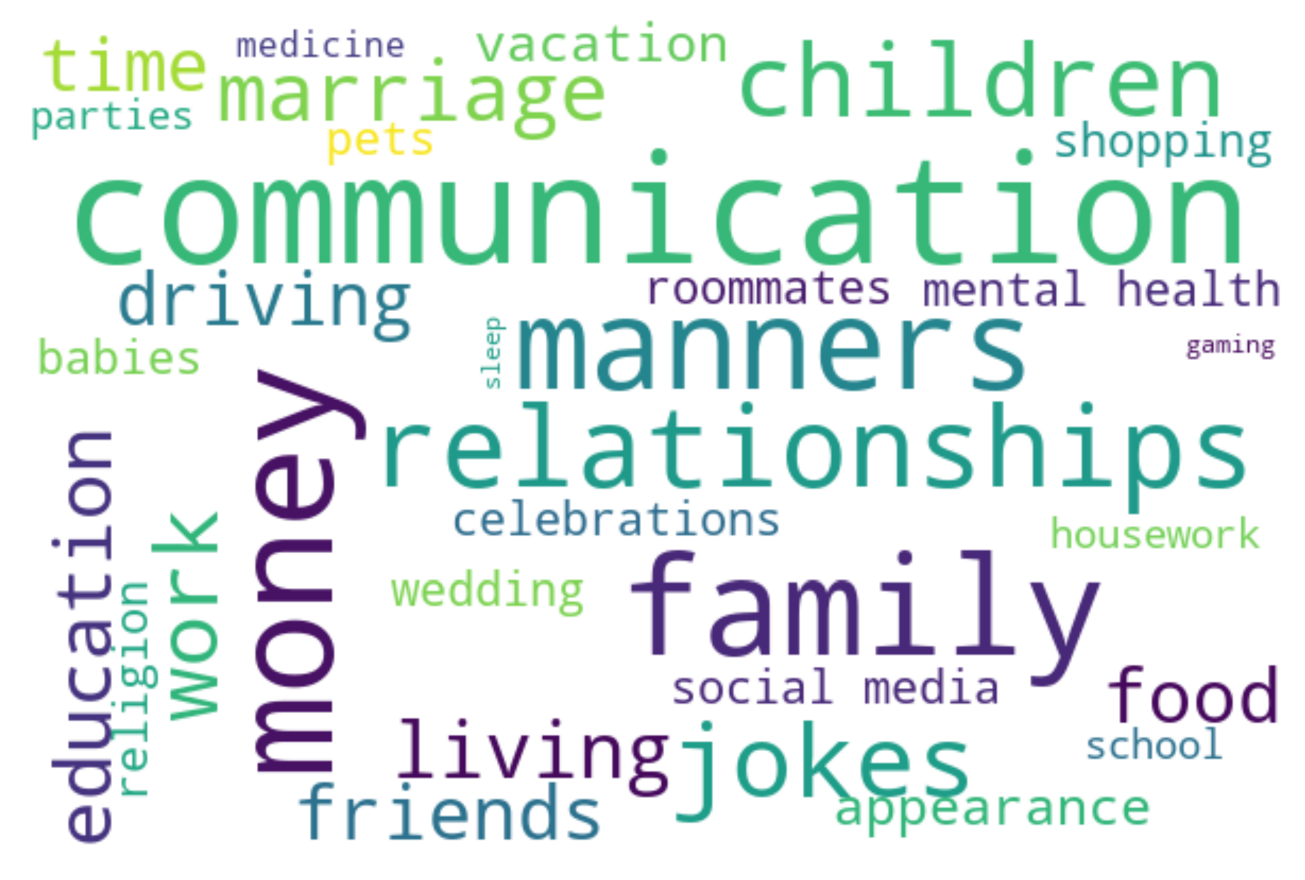}
        \caption{}
        \label{fig:wordcloud_topics}
    \end{subfigure}
    \hfill
    \begin{subfigure}[t]{0.48\textwidth}
        \centering
        \includegraphics[width=\textwidth]{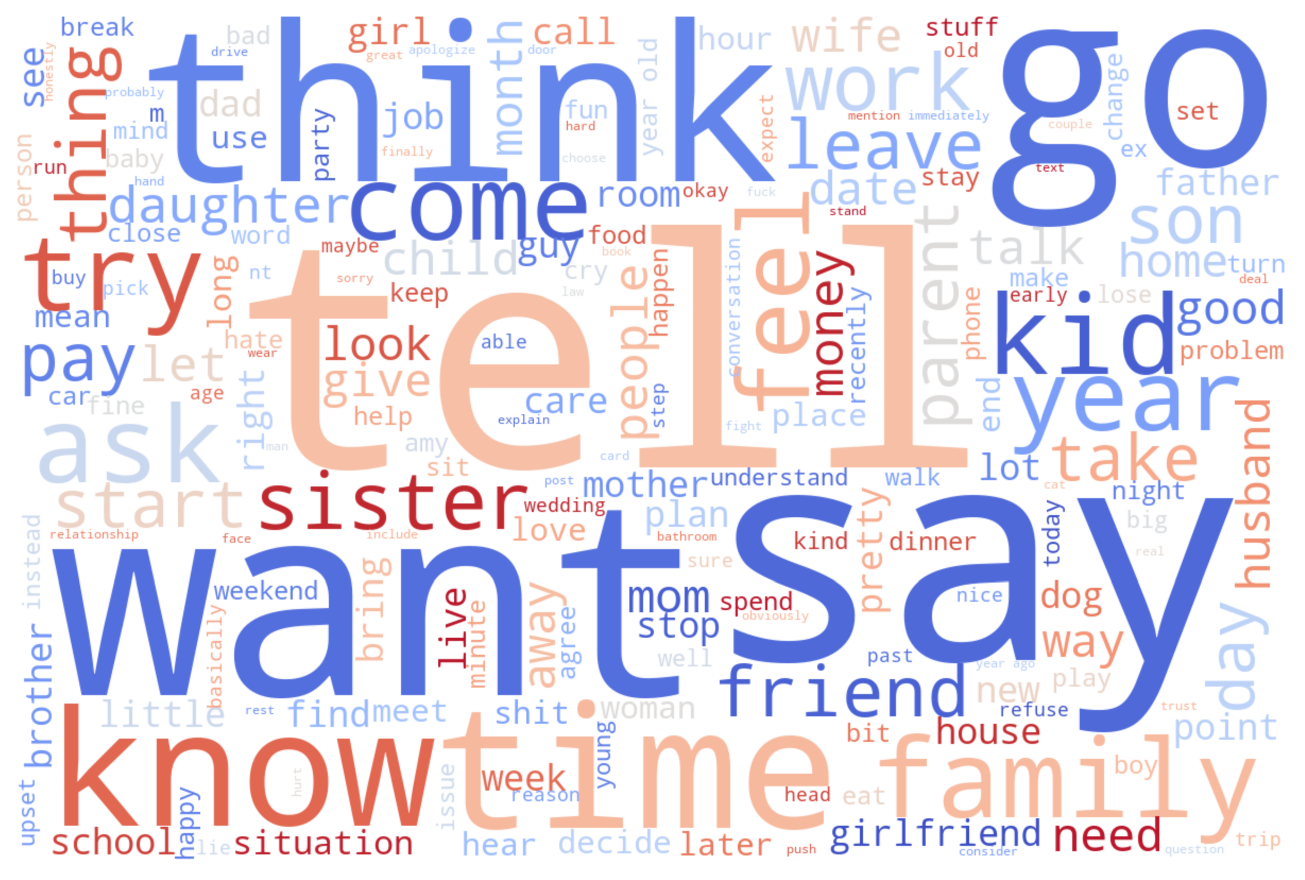}
        \caption{}
        \label{fig:wordcloud_content}
    \end{subfigure}

    \caption{(a) Count of moral situations per primary topic in \dataset dataset: The figure illustrates the count of moral situations represented across 32 primary topics from the AITA community. These topics were chosen based on their frequency in everyday moral dilemmas, excluding rare or potentially discomforting categories. (b) Word cloud visualization representing primary and secondary topics related to 135 moral situations in the \dataset dataset. The size of each word corresponds to its frequency across 135 posts, with larger size indicating more frequently discussed topics. (c) Word cloud of the original text of 135 posts describing moral dilemmas. Larger words indicate commonly used words in the \dataset dataset.}

    \label{fig:combined_figure}
\end{figure}

For the study of everyday moral reasoning, we aim to select a set of moral dilemmas, ensuring the varying levels of moral disagreement, a diverse set of moral dilemma topics, and significant community engagement.
This selection is guided by three main criteria:

\begin{enumerate}
    \item \textbf{Controversy level:} This criterion is the proportion of verdicts in the minority verdict group, ensuring the inclusion of posts with different degrees of moral disagreement, from highly divided to more concentrated verdict distributions.

    \item \textbf{Topic diversity:} This criterion guarantees that the selected dilemmas represent a wide array of moral topics relevant to everyday life. Topics that are rarely discussed or potentially discomforting are excluded.

    \item \textbf{Discussion volume:} This criterion ensures that each selected post has prompted substantial engagement from the community. Specifically, each post must have at least 50 top-level comments with YA or NA verdicts.

\end{enumerate}

First, we applied the \textbf{discussion volume} criterion, resulting in 9,998 posts out of an initial set of 102,998 that met the required threshold of at least 50 top-level comments with YA or NA verdicts. Next, using the \textbf{topic diversity} criterion, 32 were selected from an initial pool of 47 topics, to categorize the posts, focusing on those most frequently occurring and relevant to everyday moral situations (see Figure \ref{fig:topic_distribution}). Specifically, topics that were considered potentially discomforting, including death, race and sex. Topics that appear infrequently (each occurring fewer than 800 times out of 102,998, making up less than 0.8\% of the dataset) are excluded from our analysis, including restaurant, smoking, music, technology, gender, hygiene, breakups, drinking, entertainment, damage, phones, and safety. Finally, we applied the \textbf{controversy level} criterion by random sampling 105 posts with minority-verdict proportions between 0.3 and 0.5, aiming to capture morally ambiguous situations. This range was selected based on the assumption that more controversial dilemmas are more likely to provoke changes in opinion. Additionally, we randomly sampled 30 posts with minority-verdict proportions from 0.0 to 0.3 to represent lower-disagreement contexts.

As shown in Table \ref{tab:posts_stats}, we selected 135 real-life moral dilemmas with 54,827 verdicts across 32 topics and controversy levels from 0 to 0.5 for the survey experiment.
\Cref{fig:wordcloud_topics} counts both the primary and the secondary topic of each post, drawn from the full 47-topic taxonomy of \citet{Nguyen2022MappingTI}. Each post can therefore contribute two words to the cloud. Some of these words may come from secondary topics outside the 32 primary topics shown in \Cref{fig:topic_distribution}. \Cref{fig:wordcloud_content} represents frequently used words within the original text of posts, where words like ``think", ``go", ``tell", ``say" and ``want" appear prominently, revealing how individuals frame and express moral situations in the dataset. The dataset is named as \dataset dataset.

\begin{table}[h]
\centering
\begin{adjustbox}{max width=\textwidth}
\begin{tabularx}{\textwidth}{@{}p{4cm} p{1.2cm} p{1.2cm} p{1.2cm} p{1.2cm} p{1.2cm} p{1.2cm} p{1.2cm}@{}}
\toprule
\textbf{Statistic} & \textbf{min} & \textbf{25\%} & \textbf{50\%} & \textbf{75\%} & \textbf{max} & \textbf{mean} & \textbf{std} \\
\midrule
\textbf{Number of verdicts} & 51.00 & 139.50 & 215.00 & 449.50 & 3950.00 & 406.13 & 535.69 \\
\textbf{Controversy [0.00-0.50]} & 0.01 & 0.30 & 0.36 & 0.45 & 0.50 & 0.34 & 0.13 \\
\bottomrule
\end{tabularx}
\end{adjustbox}
\caption{Statistics of 135 Moral Situations. \dataset dataset}
\label{tab:posts_stats}
\end{table}

\subsection{Extracting controversy and certainty}
\label{subsec:uncertainty}
This section describes how we construct the controversy and certainty signals shown in \Cref{fig:aita_structure}. We first construct controversy from verdict counts and then extract certainty from comment text. We report human agreement, ChatGPT agreement with the human labels, and the controversy and certainty profile of the 135 dilemmas.

\textbf{Constructing controversy.}
For each post, we extracted YA and NA verdicts from top-level comments and calculated controversy as the proportion belonging to the less common verdict group. Controversy therefore ranges from 0 (full consensus) to 0.5 (an even split); see \Cref{subsubsec:aita_structure} for the verdict-group definitions.

\subsubsection{Individual certainty}\label{subsubsec:individual_uncertainty}
Although our final goal is to measure group certainty in post-discussions, it is necessary to quantify individual certainty across all top-level comments within the selected posts so that we can aggregate them into group certainty later on.

\textbf{Defining individual certainty:} AITA users often express different degrees of certainty alongside their verdicts. For example, comments may include "50\% NTA, 50\% YTA. You're NTA for... and YTA for...," "I'm leaning towards YTA because...," or "YTA, YTA, YTA. Every reply you make in this thread confirms it." Here, individual certainty is the confidence a commenter expresses in their verdict on a post.

\textbf{Building annotation guideline and human evaluation set:}
To establish a valid evaluation dataset for ChatGPT's agreement with human experts, we recruit four authors with expertise in data annotation and a deep understanding of the AITA community's discourse. These experts annotate 200 comments from four distinct posts, differing from the 135 selected posts for the survey.

To quantify individual certainty, we develop an annotation guideline using a Likert scale from 1 to 4, supported by detailed explanations and examples for each certainty level to aid annotators in assessing the commenters' confidence levels. We choose this 4-point Likert scale after a preliminary test with a more fine-grained 5-point scale yielded a lower inter-annotator agreement (Krippendorff’s Alpha of 0.58). The 4-point scale improved consistency across annotators. The levels of the 4-point Likert scale are defined as follows:
\begin{itemize}
    \item 1 – uncertain: The commenter cannot make a judgment at all, or chooses conflicting judgments and says they cannot be resolved.
    \item 2 – somewhat uncertain: The commenter leans toward a single judgment but clearly mentions other judgments can be possible or weighted similarly.
    \item 3 – somewhat certain: The commenter chooses and justifies a single judgment without a strong or solid tone.
    \item 4 – certain: The commenter shows a highly confident tone in the chosen judgment.
\end{itemize}

\textbf{Measuring agreement:}
Our goal is to evaluate inter-annotator agreement on ordinal data, focusing on both exact agreement in category assignment and consistency in item ordering. We use \textbf{Krippendorff's Alpha} as our main metric since it is well-suited for ordinal data and it can handle multiple annotators and measure exact agreement on assigned certainty levels.
Mathematically, Krippendorff's Alpha (\(\alpha\)) is defined as:

\[
\alpha = 1 - \frac{D_o}{D_e}
\]

Where \(D_o\) is the observed disagreement among raters, computed based on how frequently raters assign different categories to the same item; \(D_e\) is the expected disagreement, representing the amount of disagreement one would expect if ratings were made randomly, considering the marginal frequencies of the assigned categories. When there is complete agreement among annotators, \(D_o = 0\), and thus \(\alpha = 1\). Conversely, when the observed disagreement equals the expected disagreement due to chance (\(D_o = D_e\)), \(\alpha = 0\), signifying no agreement beyond chance.

In addition to Krippendorff's Alpha, we also employ \textbf{Spearman's Rank Correlation Coefficient} as a supplementary metric. It evaluates the strength and direction of the monotonic relationship between two sets of rankings, which is useful for assessing the consistency in item ordering between pairs of annotators rather than exact category agreement. However, it cannot measure the agreement between multiple annotations when the number of annotators is larger than 2.
Spearman's correlation (\(\rho\)) is defined as:

\[
\rho = \frac{\operatorname{cov} \left( \operatorname{R}[X], \operatorname{R}[Y] \right)}{\sigma_{\operatorname{R}[X]} \sigma_{\operatorname{R}[Y]}}
\]

where \(\operatorname{R}[X_i]\) and \(\operatorname{R}[Y_i]\) are the ranks assigned to the \(i\)-th item by two annotators, \(\operatorname{cov} \left( \operatorname{R}[X], \operatorname{R}[Y] \right)\) is the covariance between the rank variables \(\operatorname{R}[X]\) and \(\operatorname{R}[Y]\), \(\sigma_{\operatorname{R}[X]}\) and \(\sigma_{\operatorname{R}[Y]}\) are the standard deviations of the rank variables for annotators \(X\) and \(Y\).
A Spearman's \(\rho\) of \(+1\) indicates perfect positive correlation (identical rankings), \(0\) signifies no correlation, and \(-1\) indicates perfect negative correlation (inverse rankings).
\begin{table}[h]
\centering
\begin{tabularx}{\textwidth}{@{}p{6cm} p{4cm} p{4cm}@{}}
\toprule
\textbf{Annotators} & \textbf{\(\alpha\)} & \textbf{\(\rho\)} \\
\midrule
(Annotator A, Majority) & 0.85 & 0.87 \\
(Annotator B, Majority) & 0.75 & 0.76 \\
(Annotator C, Majority) & 0.82 & 0.82 \\
(Annotator D, Majority) & 0.84 & 0.84 \\
\midrule
(Annotators A, B, C, D) & \textbf{0.69} & -- \\
\bottomrule
\end{tabularx}
\caption{Inter-Annotator Agreement among Human Experts in Uncertainty Labeling Tasks: Krippendorff's Alpha (\(\alpha\)) and Spearman's Correlation (\(\rho\)) Metrics for Each Annotator Compared to the Majority Vote, along with the Overall Krippendorff's Alpha among All Annotators.}
\label{table:human_agreement}
\end{table}

We first assessed the inter-annotator agreement among four human experts (Annotators A, B, C, and D). \Cref{table:human_agreement} presents the agreement of each annotator compared to the majority vote. The majority vote is determined by selecting the label that at least two annotators agree upon for each comment. In our dataset of 200 comments, every comment has at least two annotators in agreement.  As for cases where there is a tie that two different labels have equal support from the annotators for the same comment, we randomly select one of the two labels as the majority vote.
Annotator A exhibited the highest agreement with the majority, achieving a Spearman's correlation of 0.87 and a Krippendorff's Alpha of 0.85. Annotator B had slightly lower agreement scores, with a Spearman's correlation of 0.76 and a Krippendorff's Alpha of 0.75. Overall, the Krippendorff's Alpha across all annotators was 0.69, indicating a high level of consistency among annotators. This agreement rate surpasses the performance of a similar task, which annotates sentence-level uncertainty in Science Communications and achieves the 0.67\citep{pei2021measuring}.

\textbf{Tuning prompts:} To conduct the annotation of over 50k comments, since recruiting human annotators will be time-consuming and expensive, we leverage ChatGPT, a well-performed model for various annotation tasks in the Natural Language Processing area \citep{ziems2024can}, for this large-scale task. We initially prompt ChatGPT without any examples (zero-shot) but find that its performance does not align closely with human annotations. Due to this mismatch, we explore additional prompting strategies to enhance the model's accuracy.

We tune the prompts in ChatGPT to align with the human performance, by exploring various configurations of methods (zero-shot, zero-shot with chain-of-thought, few-shot and few-shot with chain-of-thought,), models (GPT-3.5 and GPT-4), number of examples per uncertainty level and parameters (temperature set to 0.2 and top p set to 0.1).

By analyzing the trade-offs between using GPT-3.5 Turbo, GPT-4, and GPT-4o, we aim to optimize annotation quality while managing costs.  GPT-3.5 Turbo is significantly more cost-effective, costing approximately \$0.0015 per 1,000 tokens for input and \$0.002 per 1,000 tokens for output. However, it may not perform as effectively as GPT-4 without CoT or examples provided in the annotation. In contrast, GPT-4 offers better alignment with human annotations but at a higher cost—approximately \$0.03 per 1,000 tokens for input and \$0.06 per 1,000 tokens for output. GPT-4o serves as a lighter version of GPT-4, offering a middle ground between cost and performance, priced at approximately \$0.015 per 1,000 tokens for input and \$0.03 per 1,000 tokens for output.

\textbf{Evaluating prompt tuning results:}

\begin{sidewaystable}[htp]
\centering
\begin{tabularx}{0.75\textheight}{@{}l l l | l l l l | l l l l @{}}
\toprule
\textbf{Method} & \textbf{CoT} & \textbf{Model} & \multicolumn{4}{c|}{\textbf{Train}} & \multicolumn{4}{c}{\textbf{Test}} \\
\cmidrule(lr){4-7} \cmidrule(lr){8-11}
& & & $\alpha$ & $\rho$ & AUC & F1 & $\alpha$ & $\rho$ & AUC & F1 \\
\midrule

zero-shot & No & GPT 3.5 Turbo & $0.15 \pm 0.04$ & $0.41 \pm 0.04$ & $0.60 \pm 0.02$ & $0.35 \pm 0.05$ & $0.10 \pm 0.05$ & $0.37 \pm 0.05$ & $0.59 \pm 0.02$ & $0.31 \pm 0.06$ \\

zero-shot & No & GPT 4 & $0.79 \pm 0.02$ & $0.79 \pm 0.02$ & $0.84 \pm 0.02$ & $0.88 \pm 0.02$ & $0.75 \pm 0.02$ & $0.76 \pm 0.02$ & $0.85 \pm 0.02$ & $0.89 \pm 0.02$ \\

zero-shot & No & GPT 4o & $0.71 \pm 0.04$ & $0.77 \pm 0.03$ & $0.82 \pm 0.03$ & $0.80 \pm 0.03$ & $0.73 \pm 0.04$ & $0.79 \pm 0.03$ & $0.86 \pm 0.03$ & $0.84 \pm 0.03$ \\

zero-shot & Yes & GPT 3.5 Turbo & $0.75 \pm 0.04$ & $0.75 \pm 0.04$ & $0.83 \pm 0.02$ & $0.86 \pm 0.02$ & $0.76 \pm 0.04$ & $0.77 \pm 0.04$ & $0.87 \pm 0.02$ & $0.89 \pm 0.02$ \\

zero-shot & Yes & GPT 4 & $0.72 \pm 0.05$ & $0.73 \pm 0.04$ & $0.80 \pm 0.03$ & $0.84 \pm 0.02$ & $0.73 \pm 0.05$ & $0.73 \pm 0.04$ & $0.85 \pm 0.03$ & $0.88 \pm 0.02$ \\

zero-shot & Yes & GPT 4o & $0.73 \pm 0.03$ & $0.77 \pm 0.02$ & $0.84 \pm 0.02$ & $0.83 \pm 0.02$ & $0.72 \pm 0.03$ & $0.78 \pm 0.02$ & $0.85 \pm 0.02$ & $0.83 \pm 0.02$ \\

few-shot & No & GPT 3.5 Turbo & $0.64 \pm 0.03$ & $0.71 \pm 0.02$ & $0.78 \pm 0.02$ & $0.75 \pm 0.03$ & $0.57 \pm 0.03$ & $0.66 \pm 0.02$ & $0.77 \pm 0.02$ & $0.72 \pm 0.03$ \\

few-shot & No & GPT 4 & $0.75 \pm 0.04$ & $0.80 \pm 0.03$ & $0.84 \pm 0.03$ & $0.84 \pm 0.02$ & $0.75 \pm 0.04$ & $0.79 \pm 0.03$ & $0.86 \pm 0.03$ & $0.85 \pm 0.02$ \\

few-shot & No & GPT 4o & $0.79 \pm 0.03$ & $0.80 \pm 0.03$ & $0.86 \pm 0.02$ & $0.88 \pm 0.02$ & $0.79 \pm 0.03$ & $0.80 \pm 0.03$ & $0.87 \pm 0.02$ & $0.89 \pm 0.02$ \\

few-shot & Yes & GPT 3.5 Turbo & $\mathbf{0.79 \pm 0.04}$ & $\mathbf{0.79 \pm 0.04}$ & $\mathbf{0.85 \pm 0.03}$ & $\mathbf{0.90 \pm 0.02}$ & $\mathbf{0.81 \pm 0.04}$ & $\mathbf{0.81 \pm 0.04}$ & $\mathbf{0.88 \pm 0.03}$ & $\mathbf{0.92 \pm 0.02}$ \\

few-shot & Yes & GPT 4 & $0.76 \pm 0.03$ & $0.77 \pm 0.03$ & $0.83 \pm 0.02$ & $0.87 \pm 0.02$ & $0.75 \pm 0.03$ & $0.76 \pm 0.03$ & $0.85 \pm 0.02$ & $0.88 \pm 0.02$ \\

few-shot & Yes & GPT 4o & $0.74 \pm 0.03$ & $0.78 \pm 0.02$ & $0.84 \pm 0.02$ & $0.84 \pm 0.02$ & $0.74 \pm 0.03$ & $0.77 \pm 0.02$ & $0.85 \pm 0.02$ & $0.86 \pm 0.02$ \\

\bottomrule
\end{tabularx}
\caption{Combined agreement between GPT models and humans in both the 4-level certainty labeling task and the binary classification task (certain vs. uncertain). The table examines the impact of the prompting method (zero-shot vs. few-shot), Chain of Thought (CoT) usage, and different GPT models (GPT-3.5 Turbo, GPT-4, GPT-4o) on both training and testing sets. Krippendorff’s Alpha ($\alpha$) and Spearman’s Correlation ($\rho$) are reported for the 4-level certainty labeling Task evaluation, while AUC and F1 Score are reported for the binary classification task. All metrics are presented as mean ± standard deviation.}

\label{table:combined_model_comparison}
\end{sidewaystable}

In this study, we evaluated the robustness of GPT models' agreement via the aforementioned human evaluation dataset using a stratified train-test split repeated across ten iterations, each employing a unique random seed to shuffle the data into balanced 50:50 partitions for training and testing. This stratification preserved a similar distribution of the target variable and uncertainty level across both training and testing sets. We calculate Krippendorff’s alpha and Spearman correlation for each iteration and average the 2 metrics across all splits for train and test separately, and their standard deviations were computed for both training and testing subsets to provide a comprehensive view of variability.

Prompt tuning began by experimenting with \textbf{zero-shot prompting} as a baseline. Zero-shot prompting requires no examples to guide the model, relying solely on the model's inherent capabilities to interpret and annotate the input. We first tested this approach using GPT-4, configuring the model to classify comments based on different certainty levels. As shown in Table \ref{table:combined_model_comparison}, GPT-4's zero-shot performance yielded a \(\alpha\) of 0.79 and \(\rho\) of 0.79 on the training set and a \(\alpha\) of 0.75 and \(\rho\) of 0.76 on the test set. Similarly, GPT-4o performed comparably, achieving a Spearman's correlation of 0.78 and alpha of 0.72. In contrast, GPT-3.5 Turbo performed poorly in zero-shot prompting, with a significantly lower Spearman's correlation of 0.38 and Krippendorff's alpha of 0.14, indicating that GPT-3.5 Turbo struggles without examples.

We then implemented zero-shot prompting combined with \textbf{Chain of Thought (CoT)} reasoning \citep{wei2022chain}, which is a method to enable the model to follow a structured sequence of logical reasoning steps to solve problems. This approach significantly improved GPT-3.5 Turbo's performance, with Spearman's correlation increasing from 0.38 to 0.76 and Krippendorff's alpha rising from 0.14 to 0.75. This suggests that CoT helps less advanced models perform better without examples. For GPT-4 and GPT-4o, zero-shot CoT did not enhance performance and even slightly decreased metrics. GPT-4's Spearman's correlation decreased slightly from 0.78 (zero-shot without CoT) to 0.73 with CoT, while Krippendorff's alpha dropped from 0.76 to 0.72. GPT-4o showed a slight decline as well, with Spearman's correlation of 0.77 and alpha decreasing from 0.72 to 0.71.

Next, we introduced \textbf{few-shot prompting} by providing 14 annotated verdict comments. This led to substantial improvements, especially for GPT-3.5 Turbo. The few-shot approach increased GPT-3.5 Turbo's performance significantly, with a Spearman's correlation of 0.68 and Krippendorff's alpha of 0.60. Few-shot prompting for GPT-4 and GPT-4o also yielded strong results, with GPT-4 scoring 0.79 on Spearman's correlation and 0.75 on Krippendorff's alpha, while GPT-4o reached 0.79 and 0.76, respectively.

We further incorporate \textbf{Chain of Thought (CoT) reasoning} into few-shot prompting to see if it further enhances the model performance. For example, GPT-3.5 Turbo with few-shot and CoT achieved a Spearman's correlation and Krippendorff's alpha of 0.78, demonstrating significantly improved performance. Similarly, GPT-4 and GPT-4o benefited from CoT prompting, achieving Spearman's correlations of 0.78 and 0.76, and Krippendorff's alphas of 0.79 and 0.76, respectively, indicating improved annotation consistency, although the gains were not as pronounced as those seen for GPT-3.5 Turbo.

In summary, while zero-shot prompting yields decent results for GPT-4 and GPT-4o, few-shot prompting with CoT reasoning delivers the best agreement for GPT-3.5 Turbo, making it the final method for large-scale annotation tasks.

\textbf{Performing large-scale annotation via ChatGPT:}
According to the evaluation result, we adopt the few-shot with CoT method with GPT 3.5 Turbo model, temperature at 0.2 and top\_p at 0.1 as the optimal configuration for ChatGPT API request our large-scale annotation task for 54,827 comments under 135 posts. To efficiently manage the large-scale annotation process, we integrated the ChatGPT API into our data pipeline. Comments were batched and sent sequentially to optimize API usage and comply with rate limits. We implemented error handling and logging mechanisms to monitor the process and promptly address any issues related to API requests.

\clearpage
\textbf{The Best Prompt Example:}\begin{tcolorbox}[colback=gray!10, colframe=black, title=Task Prompt, width=1\textwidth, fontupper=\small,breakable]
\textbf{Task:} In the AmITheAsshole subreddit, rate the certainty of the commenter in the chosen judgement, using a Likert scale from 1 to 4.

\textbf{Instructions:} Commenter's Judgment on the Original Poster (OP) can be:

        YTA/YWBTA (You're the Asshole): The OP is to blame, and the other party is not to blame;

        NTA/YWNBTA (Not the Asshole): The OP is NOT to blame, and the other party is;

        ESH (Everyone Sucks Here): Both parties are to blame;

        NAH (No Assholes Here): Neither party is to blame.

\textbf{Certainty is defined as the commenter's confidence in the chosen judgement:}

\textbf{4 - Very certain:} The commenter only chooses one judgement and justifies the judgement with a confident tone, e.g. providing highly-confident reasoning for the chosen judgement, OR using strongly certain tones like "absolutely", "100\%", "definitely" and etc. or strongly emotional words like "WTF" and etc.

\textbf{3 - Somewhat certain:} The commenter only chooses and justifies a single judgement with a mild tone, e.g. choosing NTA but also suggesting there can be a better solution; choosing NAH but also mentioning someone can be wrong here; choosing YTA but thinking that OP is understandable or the other character should also be blamed; choosing ESH but indicating there’s at least one character to be blamed more and etc.

\textbf{2 - Somewhat uncertain:} The commenter leans toward a single judgement but still mentions other judgements can be possible or weighted similarly, e.g. leaning YTA but saying NTA could be possible if some condition occurs.

\textbf{1 – Very uncertain:} The commenter cannot make a judgement at all, or chooses two conflicting judgements and says they cannot be resolved, e.g. Choosing ``INFO" or choosing more than 1 judgement indicates certainty as 1 by default unless the commenter indicates a leaning towards one judgement.

\textbf{Examples per Certainty level are shown below:}

    \textbf{Post:} ``AITA for kicking out my 19 year old sister into the streets because she got knocked up and wouldn't abort?", \textbf{Comment:} ``YTA. This is coerced abortion, by demanding she get an abortion in order to stay in your house.", \textbf{Certainty:} 4;
    (3 more examples) ...

    \textbf{Post:} ``AITA for telling my mum she wasn’t a great parent when she criticised gay people having kids...", Comment: ``NTA. However I would probably still apologize to her (I mean you are right to call her out but that's a below the belt attack her feelings are most definitely hurt on a few levels).", \textbf{Certainty:} 3;
    (2 more examples)  ...

    \textbf{Post:} ``AITA for getting unethical revenge on a sexual harasser?", \textbf{Comment:} ``Mostly NTA. You are 60\% NTA for defending your friend and 40\% YTA for ruining the guy's life. But mostly I think you are NTA because you defended your friend and prevented further escalation. Although, you could’ve exposed him in a different way.", \textbf{Certainty:} 2;
    (2 more examples)  ...

    \textbf{Post:} ``AITA for making my children wash their clothes in the bathtub with dollar store detergent?", \textbf{Comment:} ``50\% YTA, 50\% NTA. Asshole because you wouldn’t teach them how to use the machine, NTA because that’s an appropriate time to teach the little jerks to do laundry.", \textbf{Certainty:} 1;
    (3 more examples)  ...

\textbf{To be labelled:} Post: `\{post\}' Comment: `\{comment\}'

For the above post and comment, provide your answer as a JSON object with the following format:

\{``Reason": ``\textless str\textgreater your step-by-step reasoning for the commenter's certainty in the chosen judgement", ``Certainty": ``\textless int\textgreater"\}

\end{tcolorbox}

\begin{tcolorbox}[colback=yellow!10, colframe=black, title=Assistant:,fontupper=\small, width=1\textwidth]
Reason: \texttt{"[REASON]"} , Certainty: \texttt{"[CERTAINTY]"}.
\end{tcolorbox}

\subsubsection{Group controversy and certainty}\label{subsubsec:Group_Uncertainty:}
\begin{figure}[!htbp]
    \centering
    \begin{subfigure}[b]{0.5\textwidth}
        \centering
        \includegraphics[width=\textwidth]{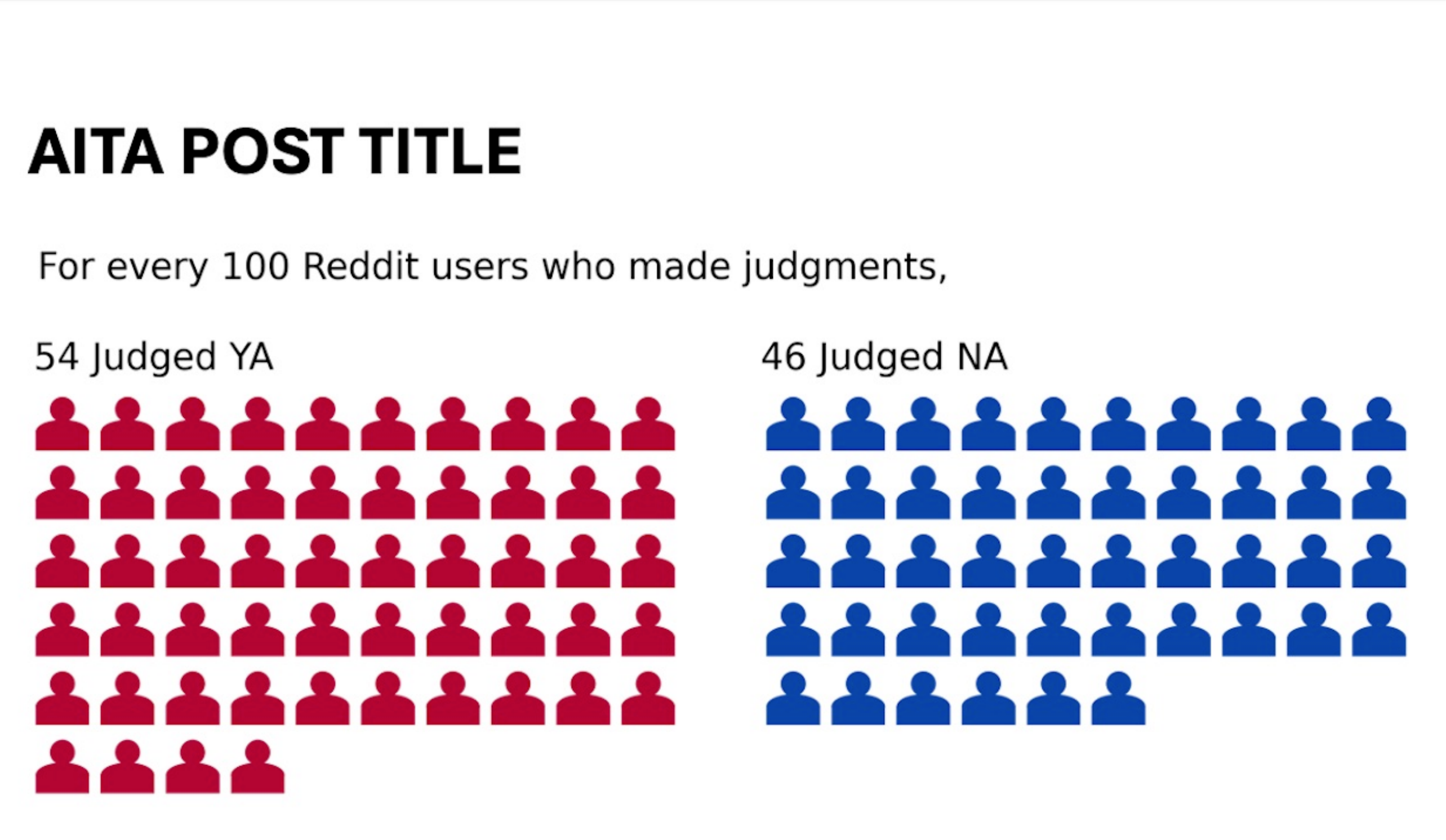}
        \caption{}
        \label{fig:controversy_example}
    \end{subfigure}
    \hfill
    \begin{subfigure}[b]{0.48\textwidth}
        \centering
        \includegraphics[width=\textwidth]{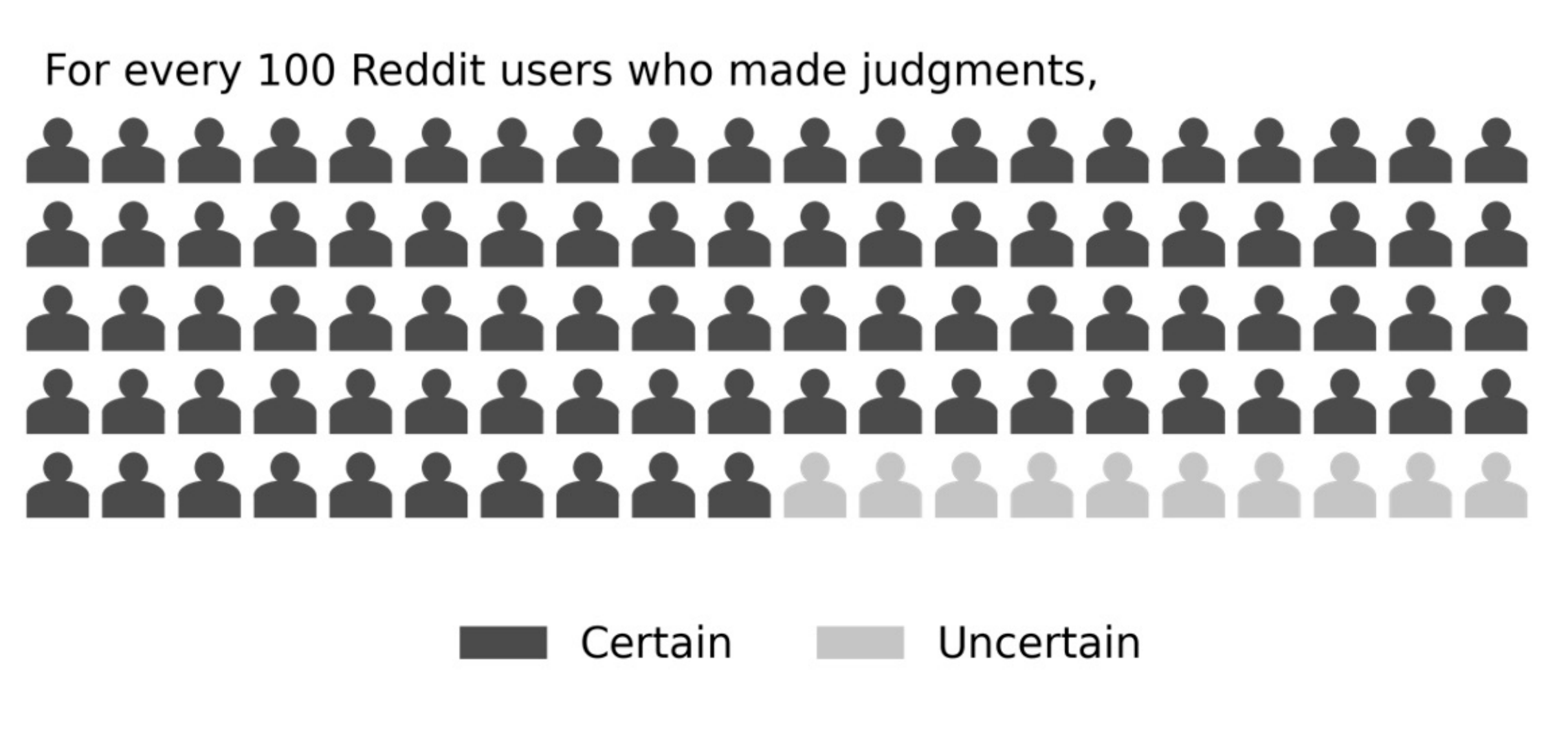}
        \caption{}
        \label{fig:uncertainty_alone}
    \end{subfigure}

    \vskip 1em
    \begin{subfigure}[b]{0.49\textwidth}
        \centering
        \includegraphics[width=\textwidth]{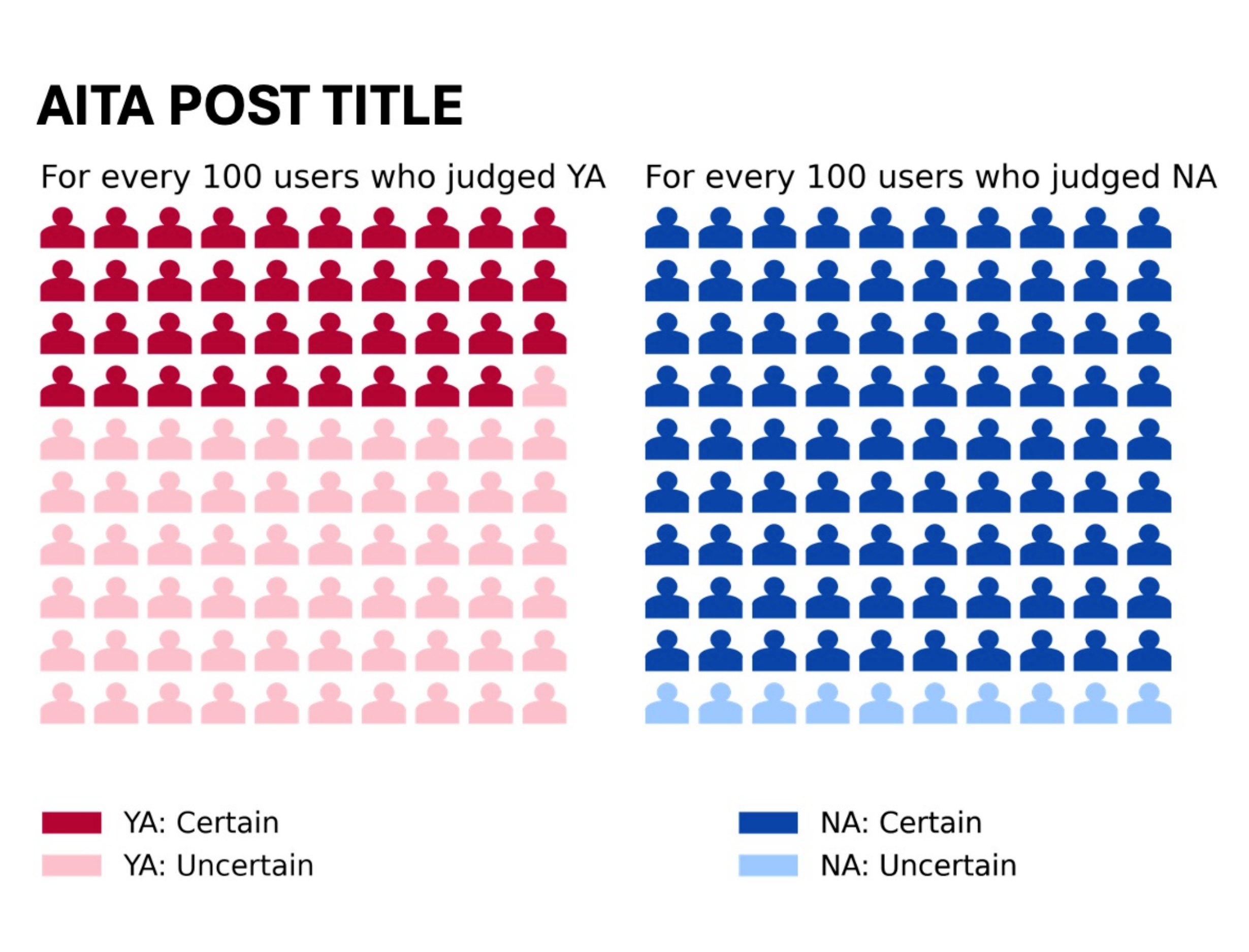}
        \caption{}
        \label{fig:uncertainty_example}
    \end{subfigure}
    \hfill
    \begin{subfigure}[b]{0.49\textwidth}
        \centering
        \includegraphics[width=\textwidth]{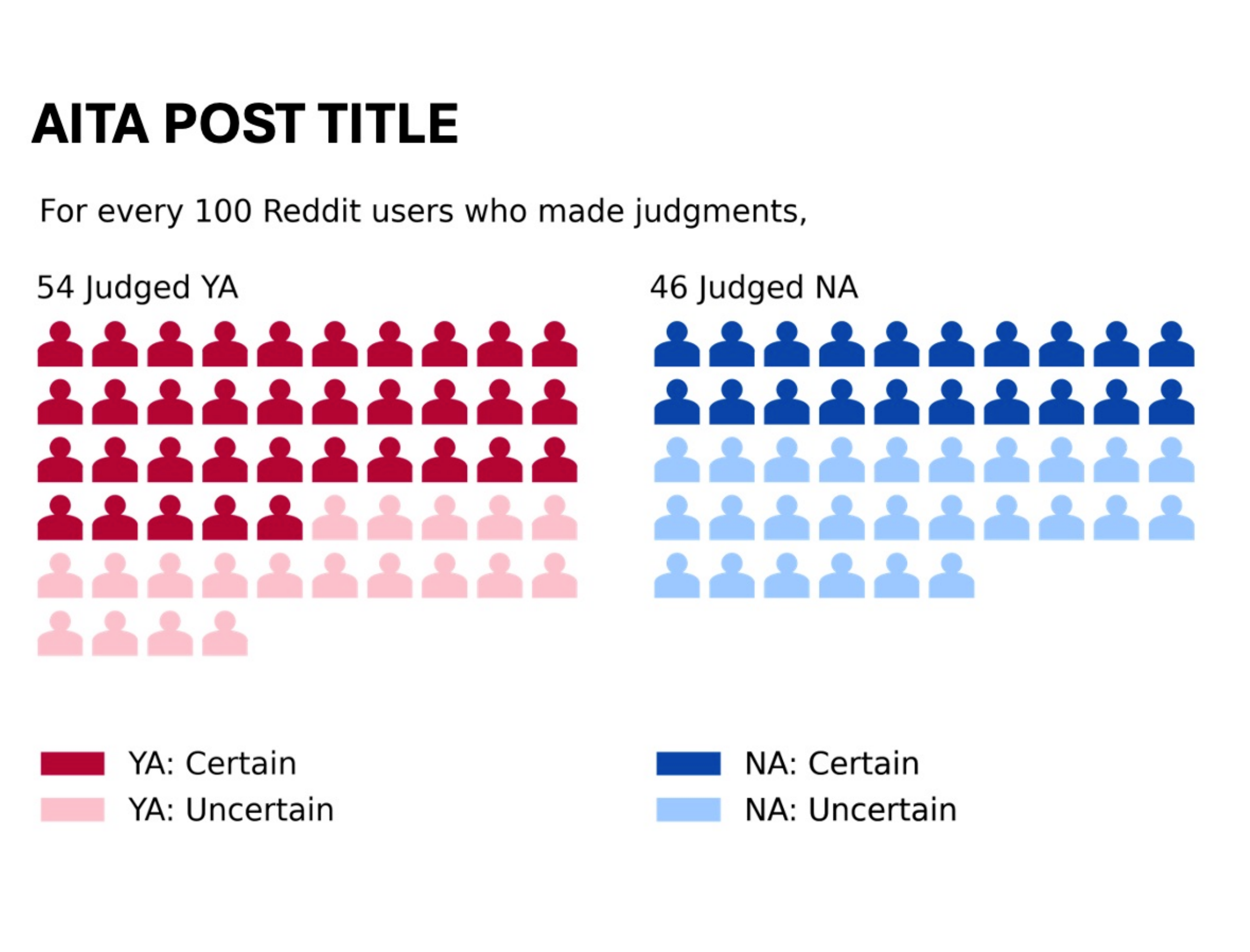}
        \caption{}
        \label{fig:controversy_uncertainty_example}
    \end{subfigure}

    \caption{Visualization of group controversy and certainty. (a) Controversy: the distribution of 100 users' verdicts for a Reddit AITA post, showing the percentages of YA verdicts (red) and NA verdicts (blue). (b) Certainty: the proportions of users classified as certain and uncertain, with darker shades indicating certainty and lighter shades indicating uncertainty. (c) Verdict certainty: the proportions classified as certain and uncertain within each verdict group. (d) Controversy and verdict certainty shown together.}
    \label{fig:group_controversy_certainty_subfigures}
\end{figure}

After annotating individual certainty, we group these levels into binary categories to visualize group certainty. We combine 1 - very uncertain, 2 - somewhat uncertain, and 3 - somewhat certain into a ``not-very-certain" group, and 4 - very certain into a ``very-certain" group. Levels 2 and 3 are grouped because both describe a commenter who leans toward one verdict but remains open to another. For instance, a somewhat uncertain commenter might lean toward YA while allowing that NA could apply under some conditions. A somewhat certain commenter might justify NA in a tentative tone or select YA while acknowledging the OP's perspective. The binary grouping therefore separates highly certain verdicts from more tentative ones.

We evaluated the performance of the models on this binary classification task. As seen in Table \ref{table:combined_model_comparison}, the aggregation does not significantly affect the ranking of model performance, particularly for the few-shot and CoT setup with GPT-3.5 Turbo, which consistently achieved the highest scores. Specifically, GPT-3.5 Turbo with few-shot and CoT prompting yielded the best performance with a test set F1 score of 0.92 and an AUC of 0.88.

We aggregate comment-level verdict certainty into group certainty for each AITA post. Figure \ref{fig:group_controversy_certainty_subfigures} shows examples of group controversy and certainty.

\begin{figure}[!htbp]
    \centering
        \begin{subfigure}[b]{\textwidth}
        \centering
        \includegraphics[width=\textwidth]{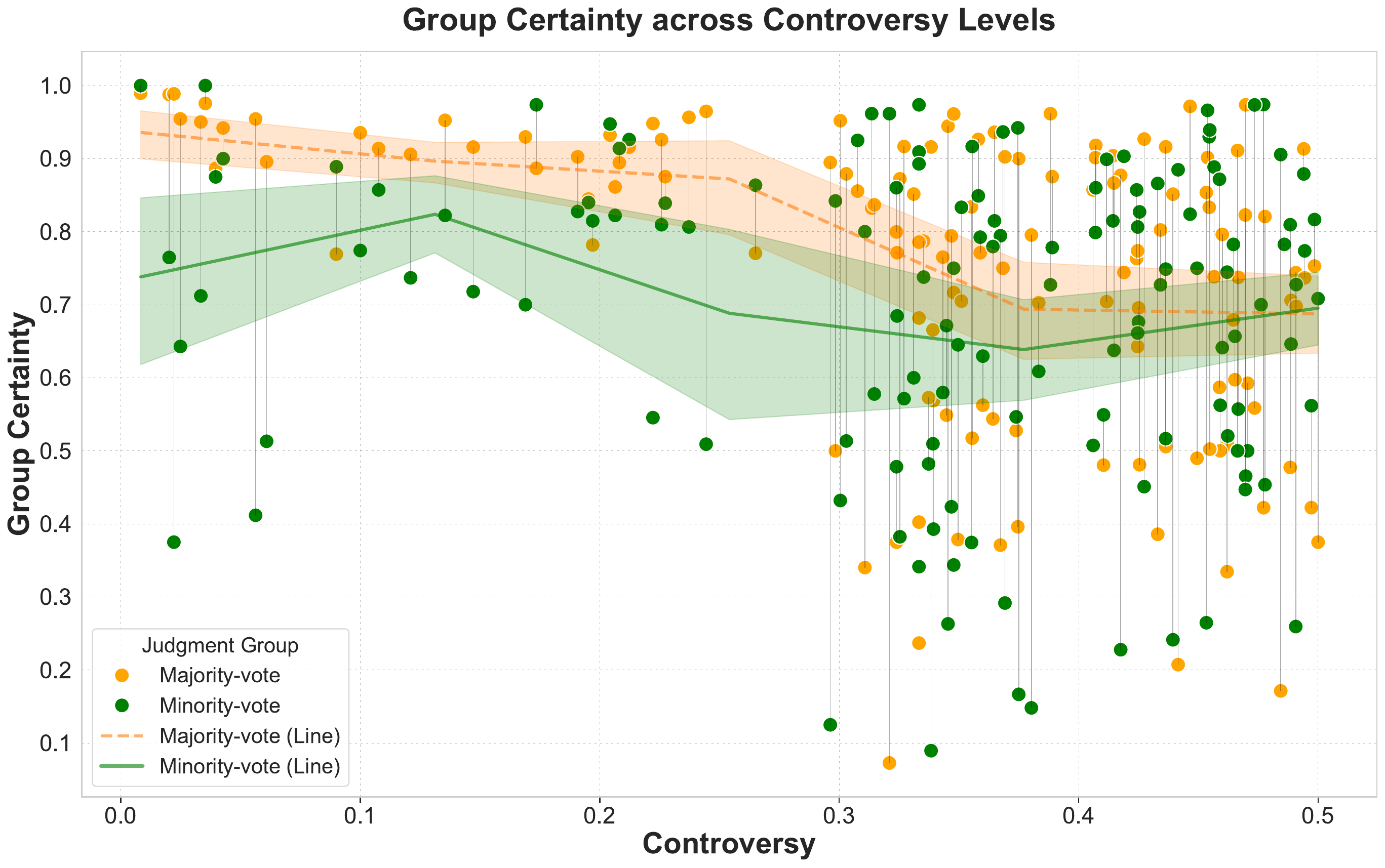}
        \caption{}
        \label{fig:Controversy_certainty}
    \end{subfigure}

        \begin{subfigure}[b]{0.48\textwidth}
        \centering
        \includegraphics[width=\textwidth]{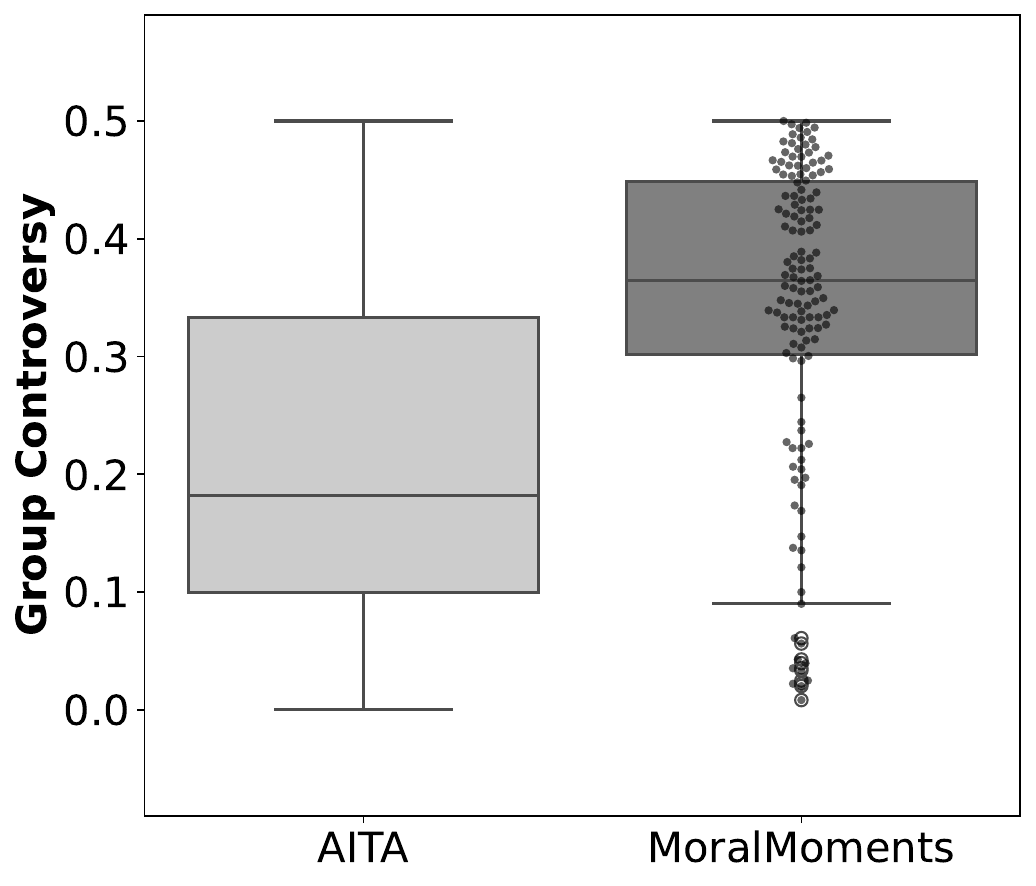}
        \caption{}
        \label{fig:Group_Controversy}
    \end{subfigure}
    \hfill
    \begin{subfigure}[b]{0.50\textwidth}
        \centering
        \includegraphics[width=\textwidth]{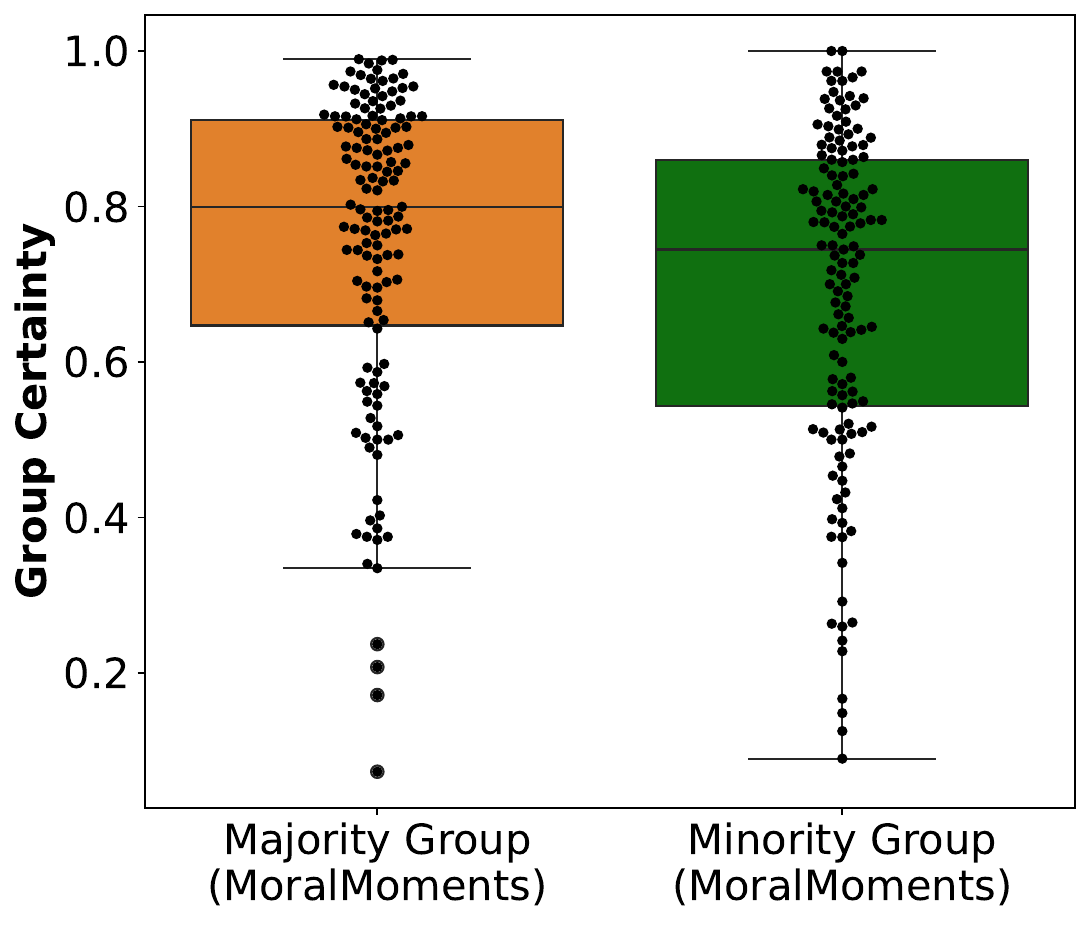}
        \caption{}
        \label{fig:Group_Certainty}
    \end{subfigure}

    \caption{(a) \textbf{Group certainty across controversy levels in the majority and minority verdict groups.} The x-axis shows group controversy, the proportion of verdicts in the minority group, and the y-axis shows the proportion of comments classified as certain within the minority (orange) and majority (green) groups. Each connected pair represents one post in the \dataset{} dataset. Lines show group trends with 95\% confidence intervals. (b) \textbf{Group controversy in the AITA and \dataset{} datasets.} Boxplots and points show the controversy distribution, including the 135 selected posts. (c) \textbf{Group certainty by majority and minority verdict group in the \dataset{} dataset.} Boxplots and points show group-certainty levels for the 135 posts.}
    \label{fig:Controversy_Uncertainty}
\end{figure}

 \Cref{fig:Controversy_Uncertainty} provides an overview of controversy and certainty in the \dataset{} dataset. The paired points show group certainty and the difference between verdict groups for the same post. The majority group contains the more common verdict, while the minority group contains the less common verdict. If YA is the majority verdict, for example, majority-group certainty is calculated from YA comments and minority-group certainty from NA comments.

As to the trend of group certainty over controversy from the line plot is majority group in posts with lower controversy tend to be more certain. Minority groups tend to exhibit more variability but rises slightly when controversy increases from 0.4 to 0.5.

As to the differences between majority and minority group certainty shown in the paired plot, while the majority group generally displays higher certainty compared to the minority group, this is not always the case. As controversy increases, the certainty levels of the majority and minority groups tend to overlap. In some instances, the minority group even surpasses the majority in certainty. The example of this, also referenced in \Cref{fig:aita_structure}, occurs with a controversy of 54:46. In this case, the majority group shows only 39\% certainty, much lower than the 90\% certainty of the minority group.

Certainty levels vary across verdict groups, particularly as controversy increases. The survey experiment tests how these two signals affect moral judgment change.

\lstset{
    language=Python,
    basicstyle=\ttfamily\small,
    breaklines=true,
    columns=fullflexible,
    keepspaces=true,
    showstringspaces=false,
    escapeinside={(*@}{@*)},
    morestring=[b]",
    morecomment=[l]{\#},
    keywordstyle=\color{black},
    stringstyle=\color{red},
    commentstyle=\color{black},
    upquote=true,
    literate={“}{"}1 {”}{"}1 {’}{'}1 {‘}{'}1
             {\%}{\%}1 {\$}{\$}1
}

\subsection{Survey experiment protocol}\label{sec:surveyexperiment}

\begin{figure}[!htbp]
    \centering
    \includegraphics[width=\linewidth]{figure18.pdf}
    \caption{The complete outline of the \dataset online survey. Participants underwent two randomizations: first to one of four experimental conditions (Controversy, Certainty, Controversy$+$Certainty, or Control), and then to two of the 135 dilemmas. In Phase I, participants reviewed the project description, provided consent, learned the concepts used in their assigned condition, completed a practice question, and answered the first attention check. In Phase II, participants evaluated their two assigned dilemmas. For each dilemma, they recorded an initial verdict and confidence, viewed the assigned social information or completed the Control task, and then recorded a final verdict and confidence with optional reasoning. In Phase III, participants completed the decision-making style, personality, and intellectual humility measures, followed by optional feedback and a debriefing. They were then redirected to Prolific for payment.}
    \label{fig:survey_flow}
\end{figure}

The survey experiment tested whether controversy and group certainty, presented separately or together, changed participants' judgments of real-life moral dilemmas.

\paragraph{Ethical approval declarations:}
The ethical aspects of this research have been approved by the Australian National University Human Research Ethics Committee (Protocol H/2023/1292). The methods were carried out under the relevant guidelines and regulations. Informed consent was obtained from all participants before conducting the actual survey.

\paragraph{Recruitment and payment}
Participants were recruited through Prolific. Eligible participants were fluent English speakers aged 18 or older who had not participated in an earlier version of the study. Prolific supplied participant characteristics such as age and gender. Participants received \pounds1.95; the median completion time was 14 minutes (861.5 seconds), equivalent to approximately \pounds8.36 per hour. Attention-check performance did not stop survey completion or payment, but the analysis retained only participants who passed at least two of three checks.

\subsubsection{Detailed survey flow }\label{subsec:survey_flow}
\begin{figure}[htbp]
\centering
\begin{subfigure}[t]{0.48\textwidth}
  \centering
  \fbox{\includegraphics[page=1,width=\linewidth]{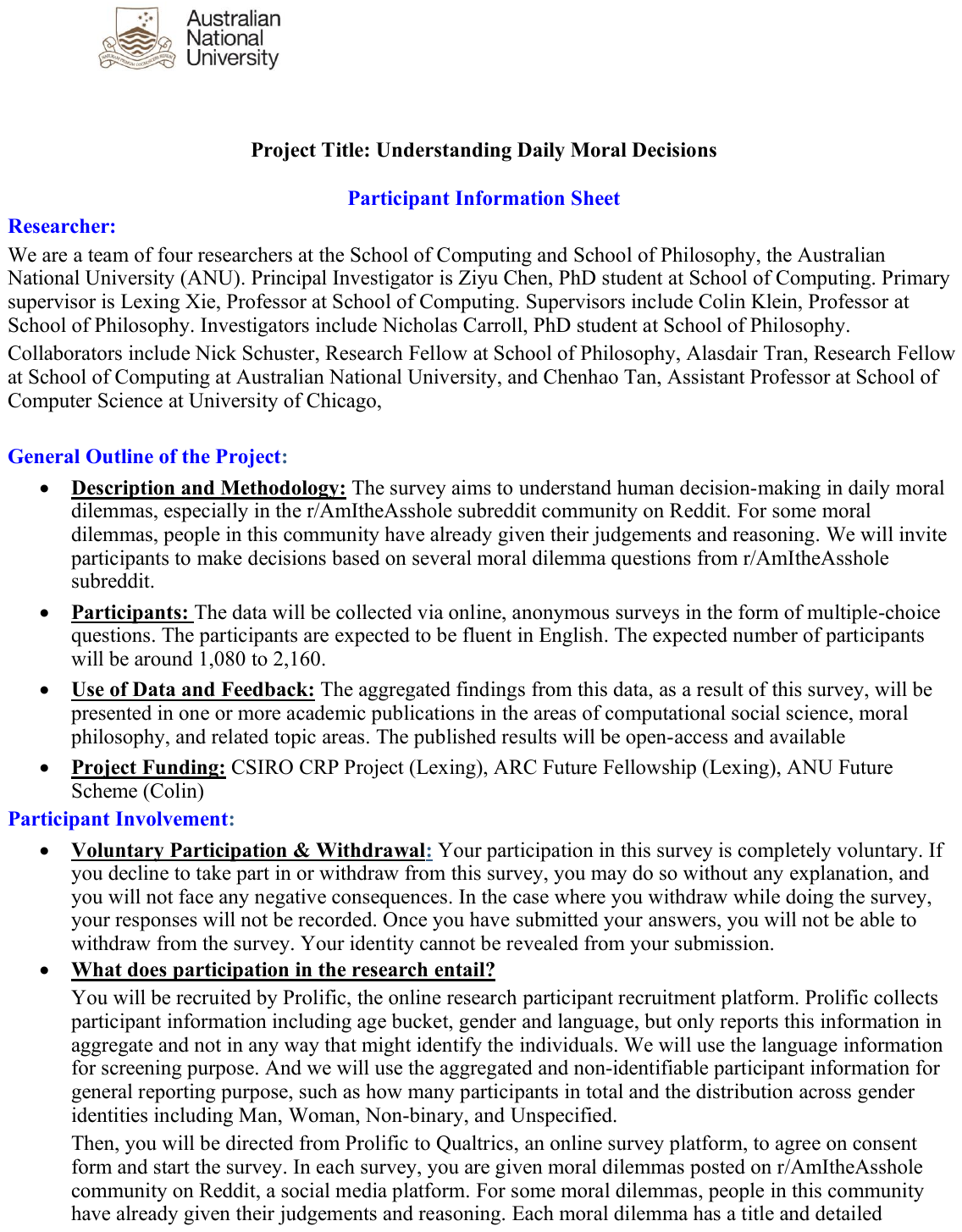}}
  \caption{Page 1}
\end{subfigure}
\hfill
\begin{subfigure}[t]{0.48\textwidth}
  \centering
  \fbox{\includegraphics[page=2,width=\linewidth]{figure19.pdf}}
  \caption{Page 2}
\end{subfigure}

\vspace{0.5em}

\begin{subfigure}[t]{0.48\textwidth}
  \centering
  \fbox{\includegraphics[page=3,width=\linewidth]{figure19.pdf}}
  \caption{Page 3}
\end{subfigure}
\hfill
\begin{subfigure}[t]{0.48\textwidth}
  \centering
  \fbox{\includegraphics[width=\linewidth]{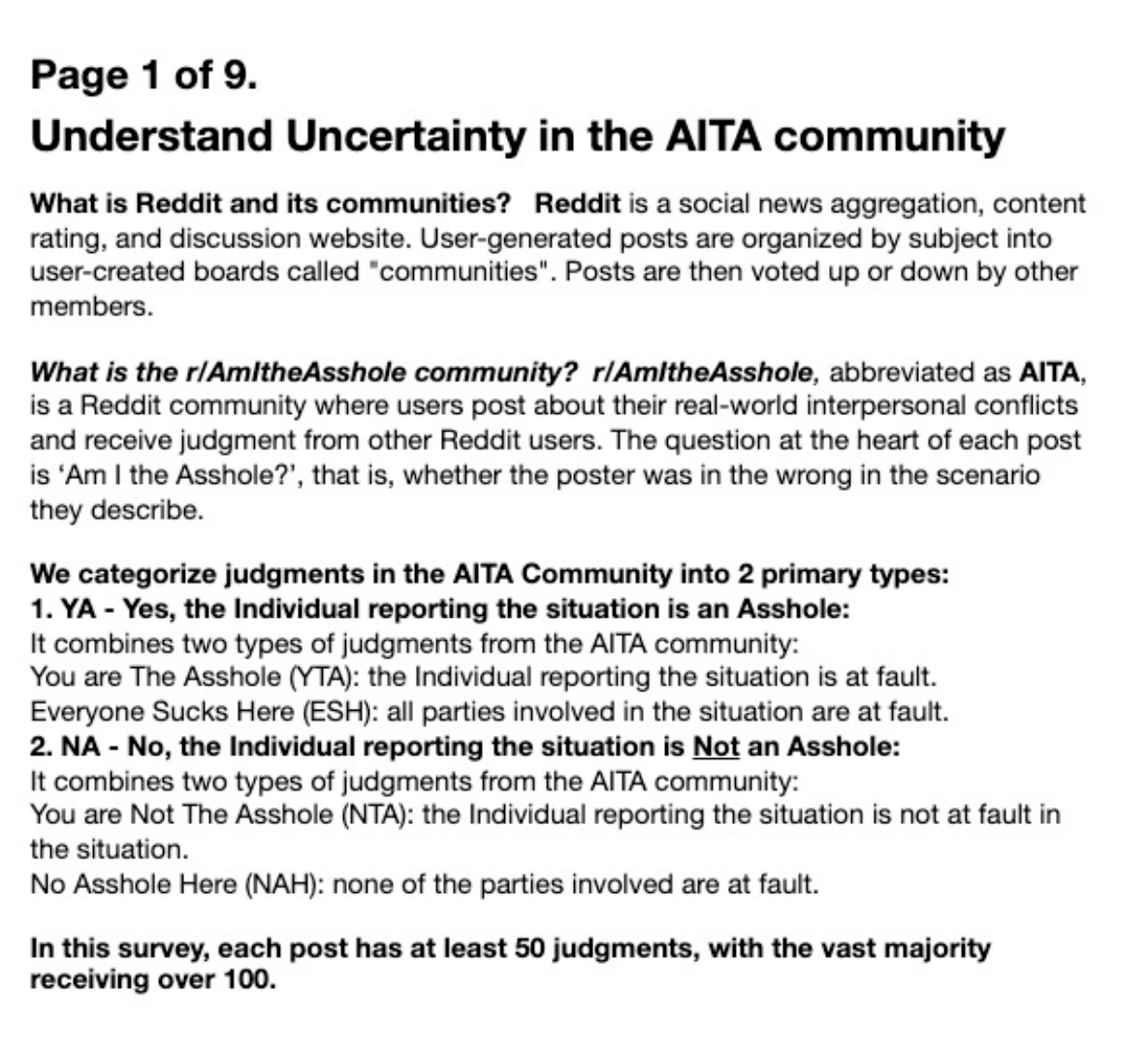}}
  \caption{Background knowledge of AITA}
  \label{fig:background_knowledge}
\end{subfigure}

\caption{Participant information and background material presented during Phase I. Panels (a)--(c) reproduce the three-page participant information sheet shown before consent. Panel (d) introduces Reddit, the AITA community, and the binary verdict groups after consent.}
\label{fig:participant_information_sheet}
\end{figure}

\begin{figure}[htbp]
\centering
\begin{subfigure}{0.48\textwidth}
  \centering
  \fbox{\includegraphics[width=0.82\linewidth]{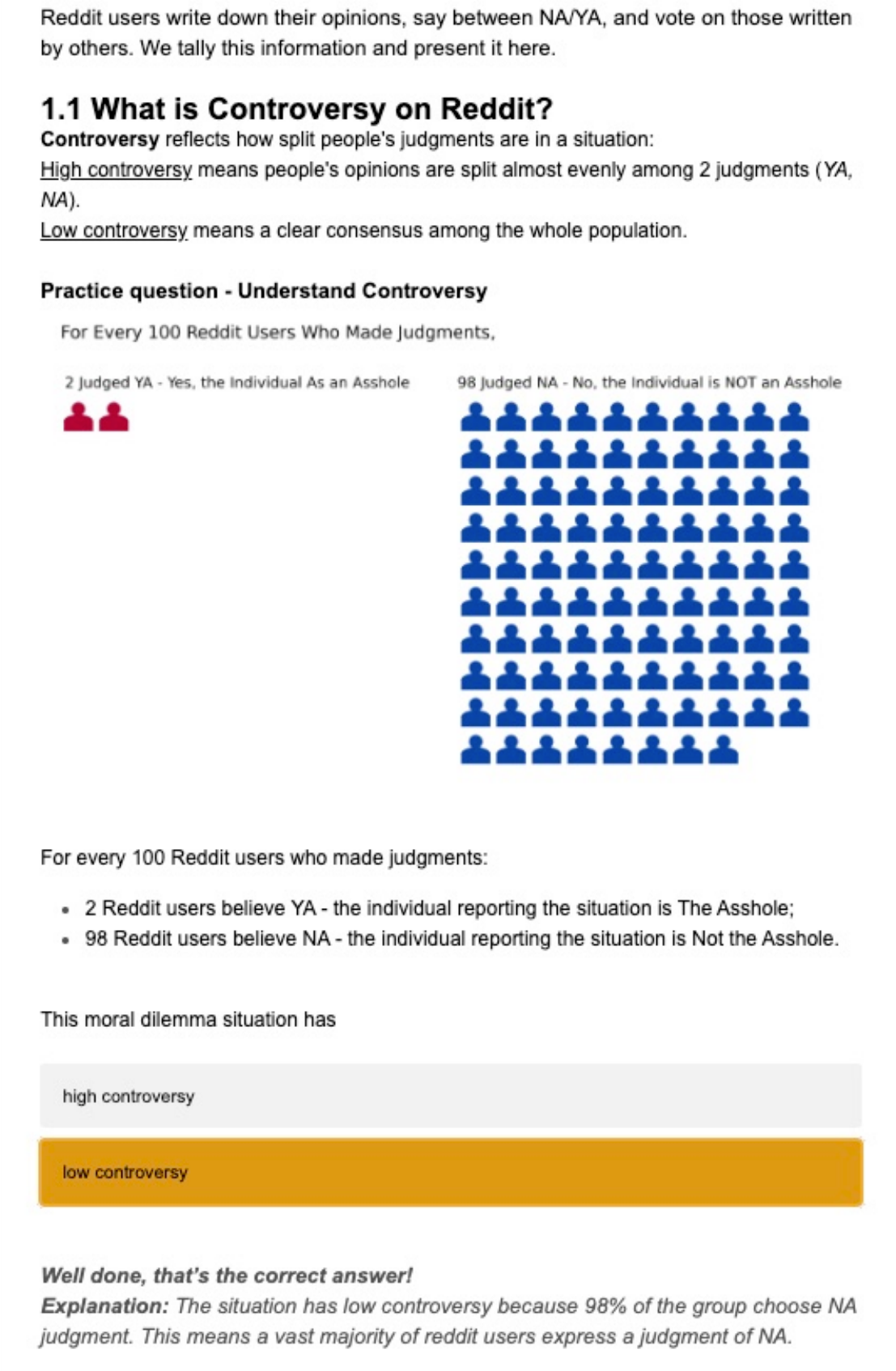}}
  \caption{Controversy treatment: controversy concept}
  \label{fig:training_controversy}
\end{subfigure}
\hfill
\begin{subfigure}{0.48\textwidth}
  \centering
  \fbox{\includegraphics[width=0.82\linewidth]{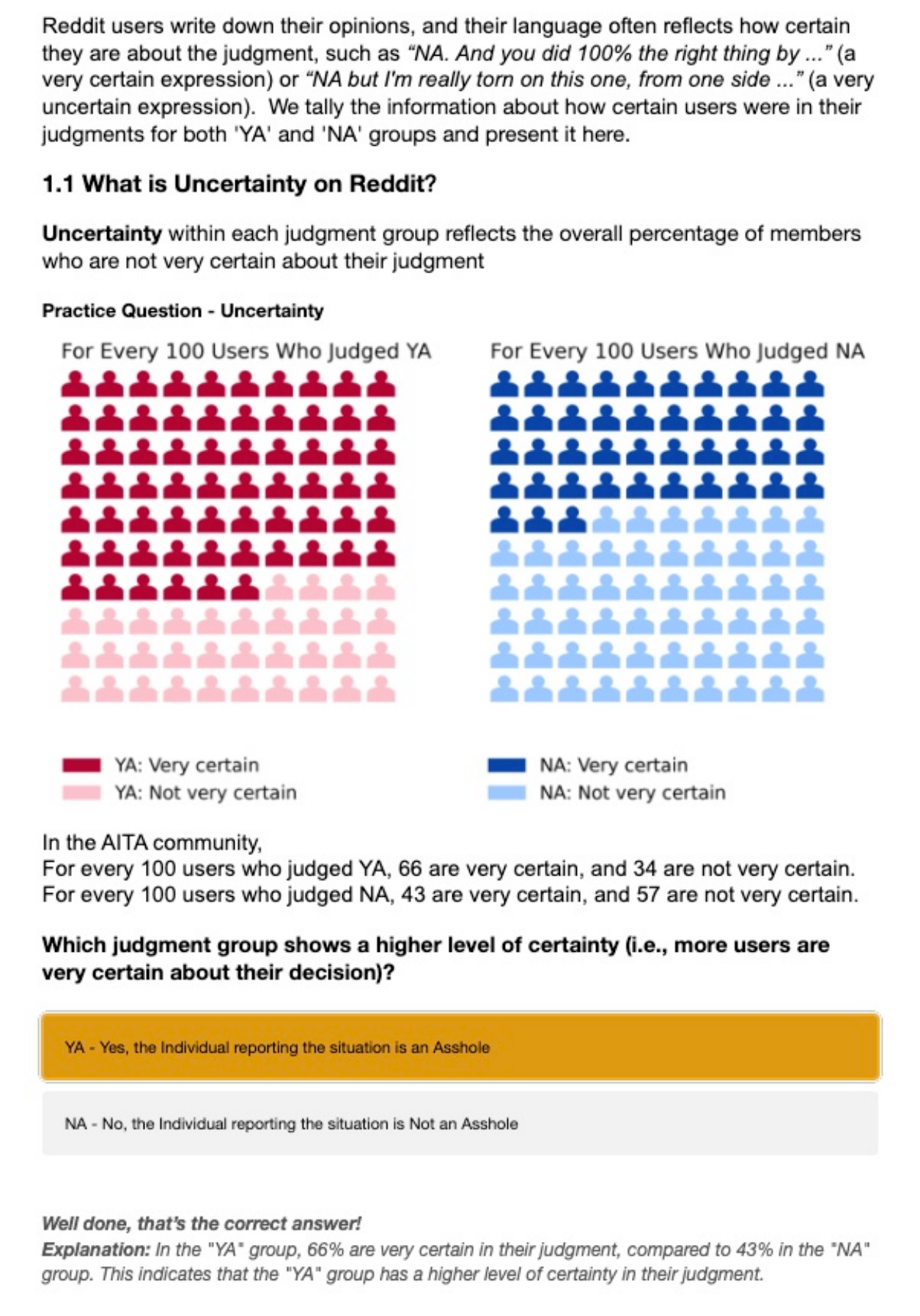}}
  \caption{Certainty treatment: certainty concept}
  \label{fig:training_certainty}
\end{subfigure}

\vspace{1em}

\begin{subfigure}{0.48\textwidth}
  \centering
  \fbox{\includegraphics[width=0.82\linewidth]{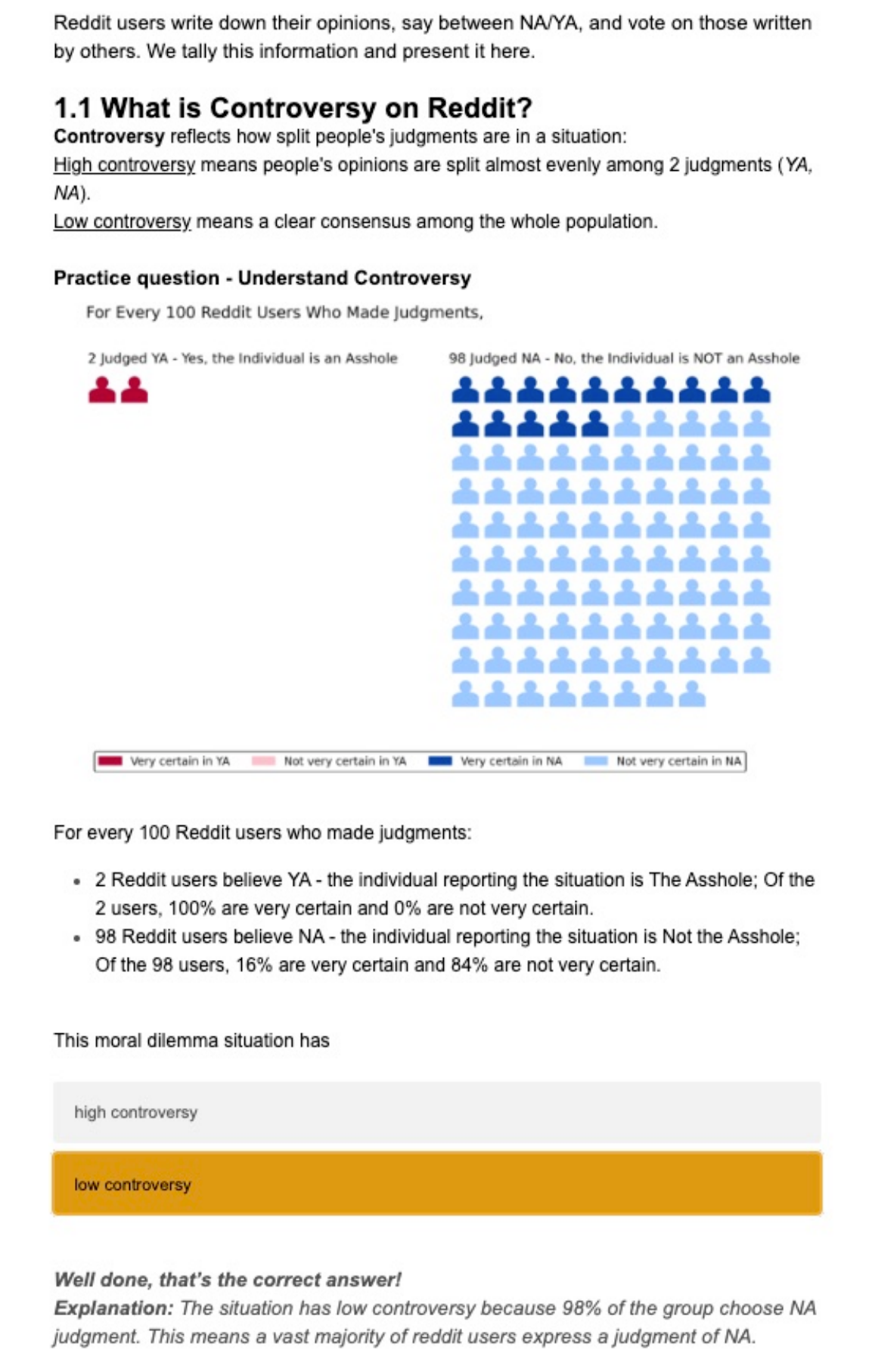}}
  \caption{Controversy + Certainty: controversy concept}
  \label{fig:training_controversy_cc}
\end{subfigure}
\hfill
\begin{subfigure}{0.48\textwidth}
  \centering
  \fbox{\includegraphics[width=0.82\linewidth]{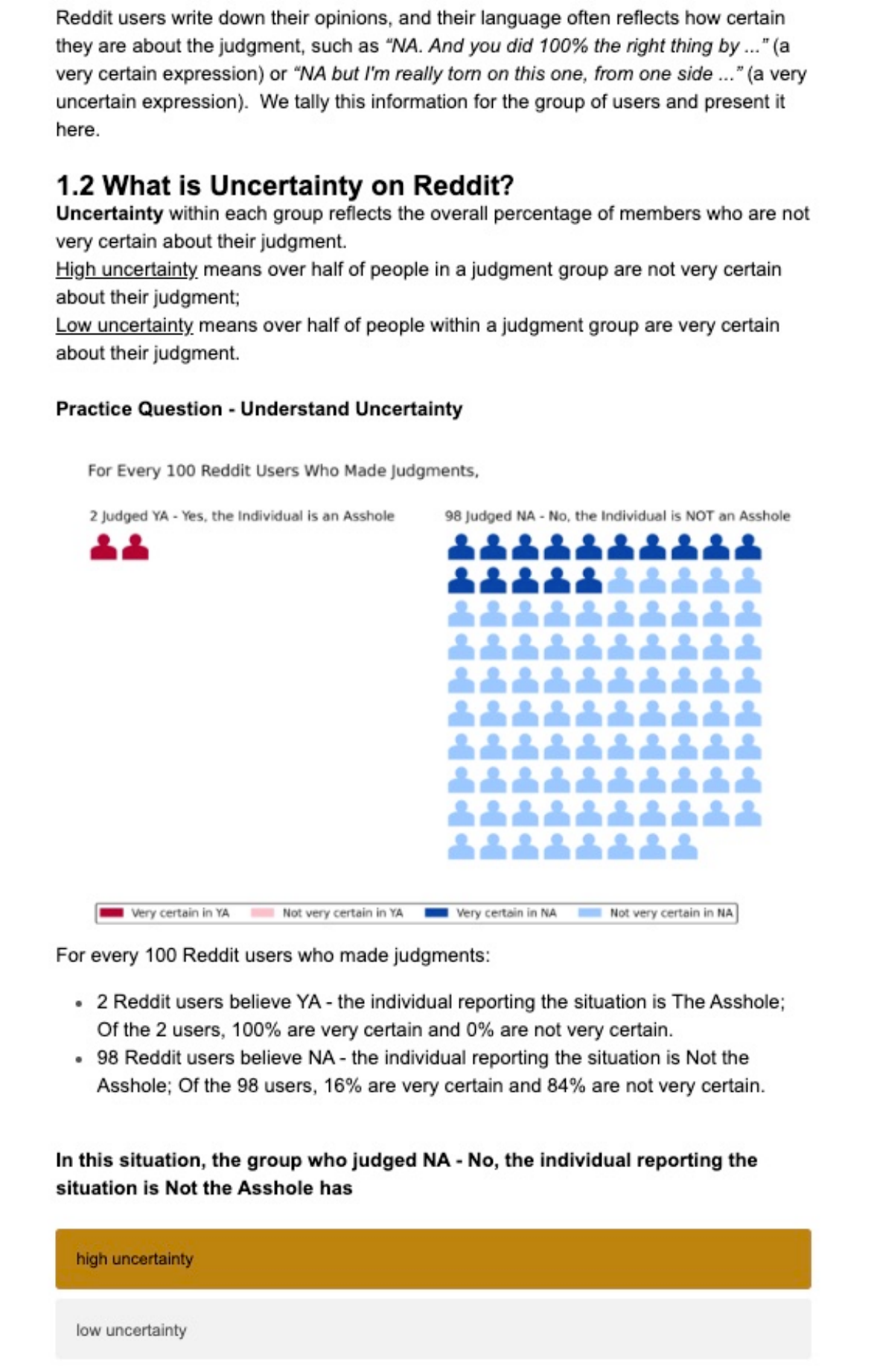}}
  \caption{Controversy + Certainty: certainty concept}
  \label{fig:training_certainty_cc}
\end{subfigure}

\caption{Training screens that introduce the treatment concepts. (a) controversy in the Controversy treatment; (b) certainty in the Certainty treatment; (c) controversy and (d) certainty in the Controversy + Certainty treatment.}
\label{fig:training_concepts}
\end{figure}

\paragraph{Phase I: Project description, Consent and Training}
Participants first read the project description and participant information sheet (\Cref{fig:participant_information_sheet}) and provided consent. Those who consented learned about the AITA community (\Cref{fig:background_knowledge}) and the concepts used in their assigned condition, as shown in \Cref{fig:training_controversy,fig:training_certainty,fig:training_controversy_cc,fig:training_certainty_cc}. They then completed a two-page practice question followed by the first attention check.

\paragraph{Phase II: Treatment and story randomization}

The study used a randomized design within a single Qualtrics survey. Participants were randomized to the Controversy, Certainty, Controversy$+$Certainty, or Control condition and further randomized to two of the 135 dilemmas. For each dilemma, participants recorded an initial verdict and confidence rating.

The Controversy condition showed the percentages of AITA verdicts that agreed and disagreed with the participant's initial verdict. The Certainty condition showed the percentages of certain and uncertain comments separately within the agreeing and disagreeing verdict groups. The Controversy$+$Certainty condition showed both signals, while the Control condition presented a topic-selection task without social information. The signals were presented through percentages, short descriptions, and infographics.

Participants then recorded their final verdict and confidence and could optionally explain why they changed or maintained their response. An attention check followed each dilemma, producing three checks in total when combined with the training check.

\begin{figure}[!htbp]
    \centering
    \begin{subfigure}[t]{0.48\linewidth}
        \centering
        \includegraphics[width=\linewidth]{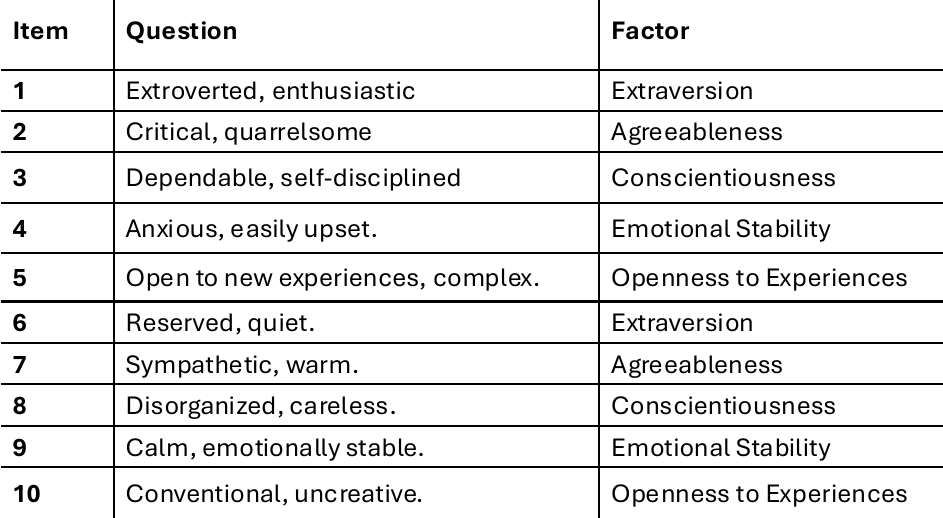}
        \caption{}
        \label{fig:personality_survey}
    \end{subfigure}
    \hfill
    \begin{subfigure}[t]{0.48\linewidth}
        \centering
        \includegraphics[width=\linewidth]{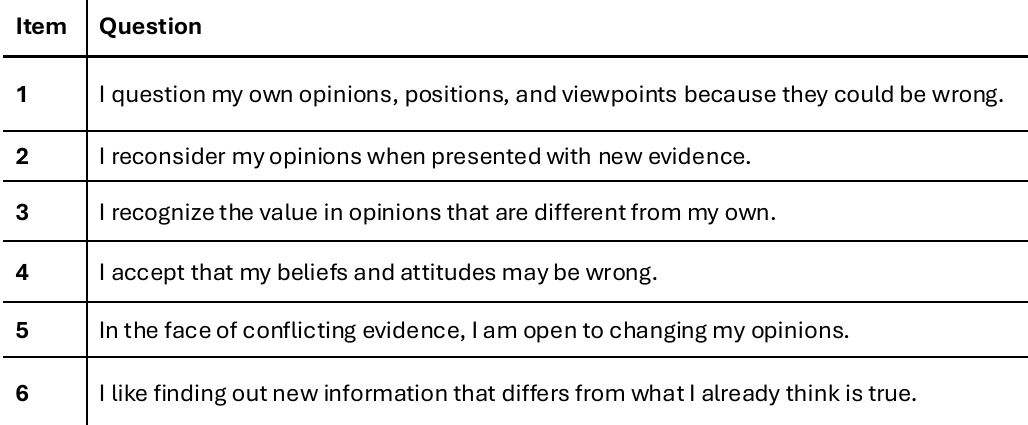}
        \caption{}
        \label{fig:ih_survey}
    \end{subfigure}
    \caption{Survey items for (a) Ten-Item Personality Inventory (TIPI) (b) Intellectual Humility Scale (IH)}
    \label{fig:personality_ih_combined}
\end{figure}

\begin{figure}[!htbp]
    \centering
    \includegraphics[width=1\linewidth]{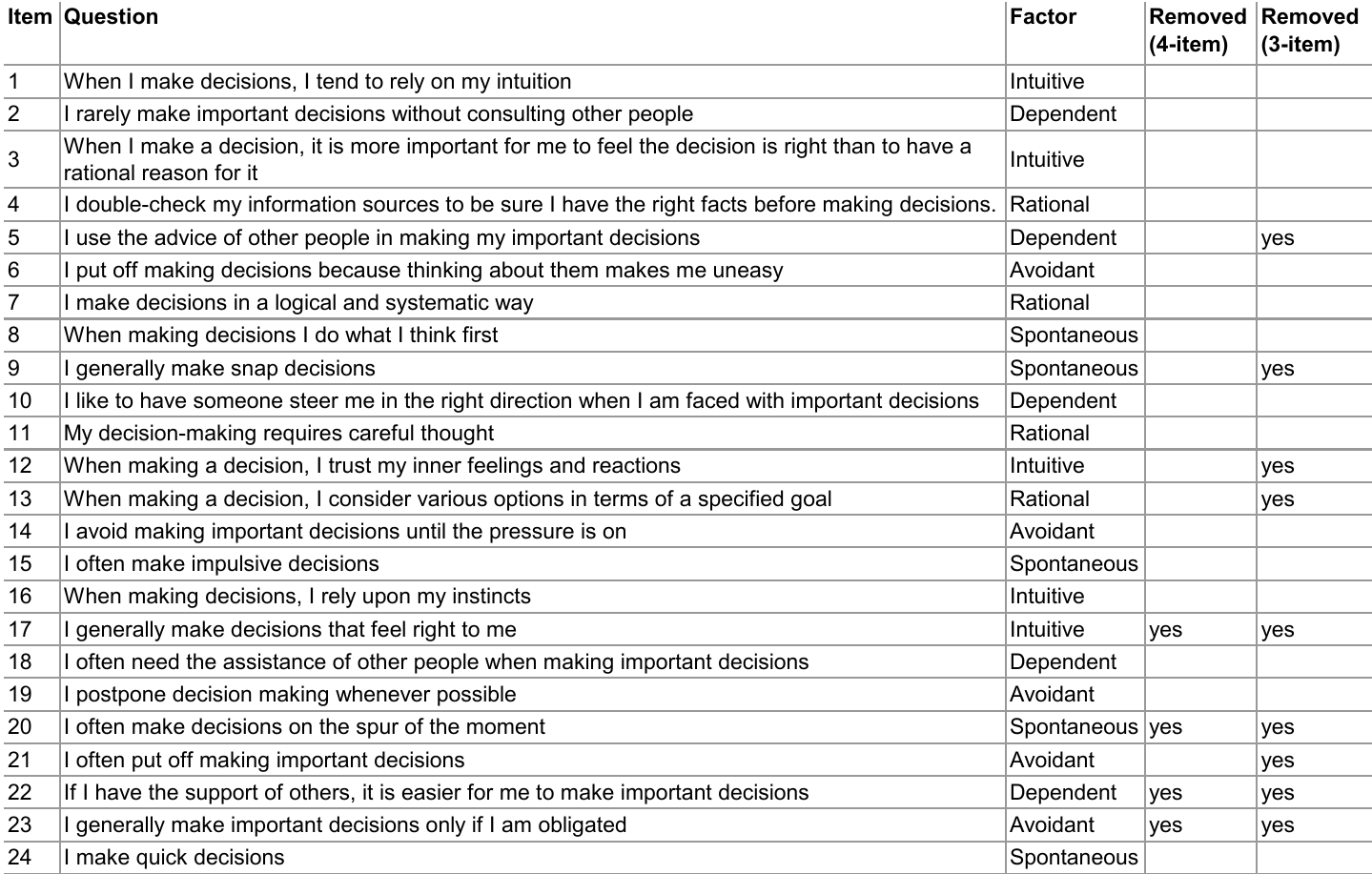}
    \caption{Survey items in the General Decision-Making Style (GDMS) scale. Items 1–25 correspond to the original GDMS version, which includes five items per factor. The “Removed (4-item)” column indicates the version with one item removed per factor to retain the highest variance, resulting in a 4-item scale. The “Removed (3-item)” column shows the version with two items removed per factor, retaining the items that preserve the largest variance in each factor.}
    \label{fig:decision_item}
\end{figure}

\paragraph{Phase III: Background Survey, Feedback, and Conclusion}
After completing both dilemmas, participants completed the shortened 20-item General Decision-Making Style questionnaire \citep{scott1995decision}, shown in \Cref{fig:decision_item}; the Ten-Item Personality Inventory \citep{Gosling2003AVB}, shown in \Cref{fig:personality_survey}; and the six-item Intellectual Humility Scale \citep{leary2017IH}, shown in \Cref{fig:ih_survey}. The survey ended with optional feedback and a debriefing, after which participants were redirected to Prolific for payment. Further details on the GDMS item-reduction procedure are provided in \Cref{subsec:reduce_survey}.

\clearpage
\subsubsection{Reducing survey items in GDMS}\label{subsec:reduce_survey}

We aim to use survey reduction to minimize participant burden in completing the decision-style questionnaire after the treatment while still efficiently capturing the intended psychological constructs. The present study aimed to select optimal subsets of items from an original five-item set for each of five psychological factors—Intuitive, Dependent, Rational, Avoidant, and Spontaneous. The selection was guided by variance retention analysis, initially conducted with 166 participants, and subsequently validated through test-retest reliability involving 80 participants who completed the four-item survey at least one month after participating in the original five-item survey.

\begin{figure}[!htbp]
    \centering
    \includegraphics[width=\linewidth]{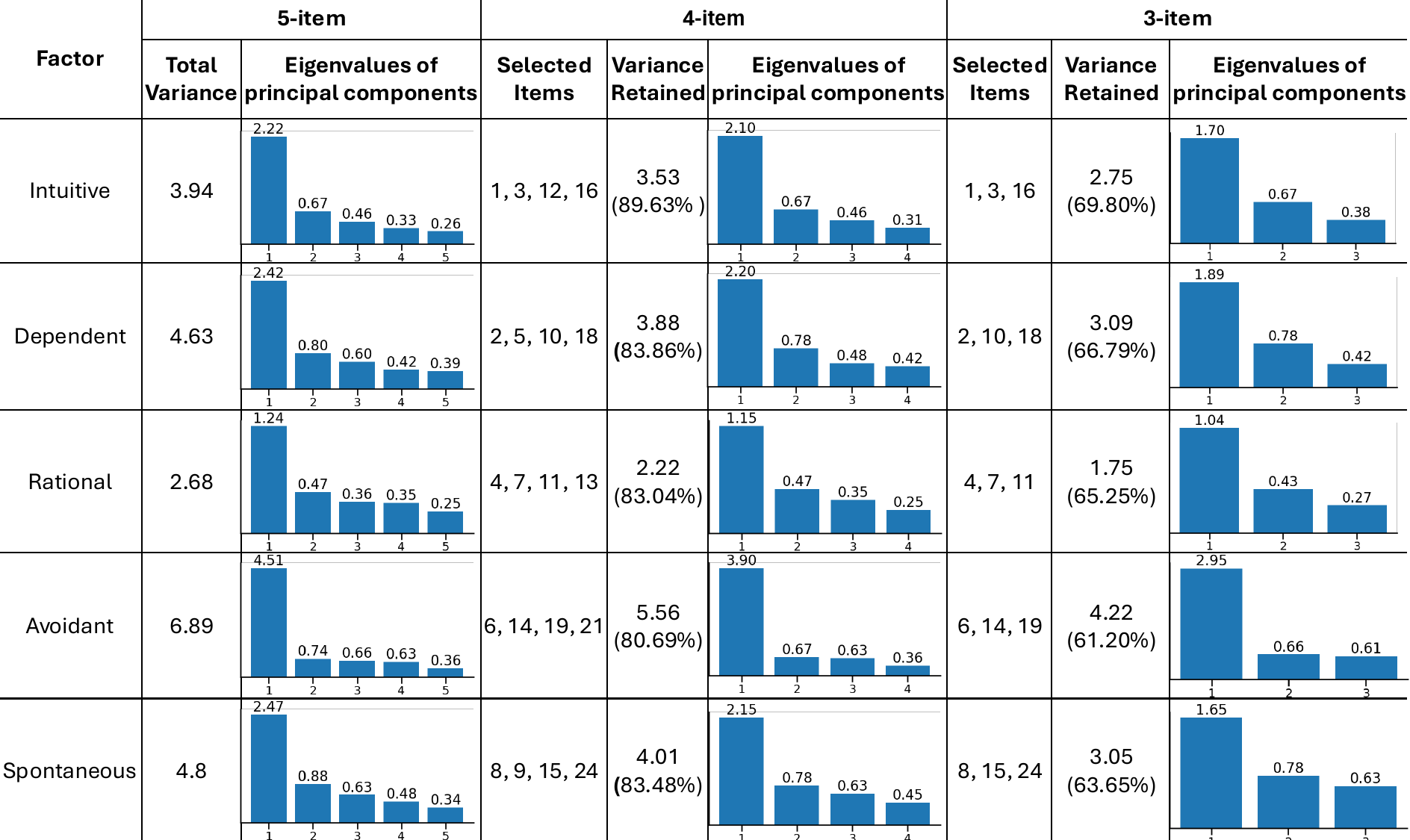}
    \caption{Variance retention analysis for each GDMS factor following item reduction. The figure shows the proportion of variance retained when reducing from the full 5-item scale (original GDMS) to 4-item and 3-item versions. For each factor (Intuitive, Dependent, Rational, Avoidant, Spontaneous), the “Removed (4-item)” column represents the subset with one item removed to maximize retained variance, and the “Removed (3-item)” column reflects two items removed under the same criterion. Item selection is detailed in Figure~\ref{fig:decision_item}.}
    \label{fig:survey_reduce}
\end{figure}

\bigskip

\paragraph{Variance Retention} \label{para:var_retention}
The goal of this analysis is to reduce redundancy in our survey by selecting the optimal subset of four items from the original five for each factor while retaining as much of the original variance as possible (the maximum amount of information). Our procedure leverages the concept of variance retention: by using Principal Component Analysis (PCA), we identify which subset for each factor captures the maximum variance as indicated by the eigenvalues.

\textbf{Step 1 - Covariance Calculation:} To assess redundancy among survey items, we compute the sample covariance matrix, which quantifies the pairwise relationships between items. In our analysis, the covariance matrix is defined as:
\[
S_x = \frac{1}{N} \sum_{i=1}^N (x_i - \bar{x})(x_i - \bar{x})^T,
\]
where \( x_i \) denotes the \(i\)th observation, \( \bar{x} \) is the mean vector over all \(N\) observations, and \(S_x\) encapsulates the covariance structure. This matrix is fundamental because it forms the basis of our eigen-decomposition.

\bigskip

\textbf{Step 2 - Finding maximal variances via PCA:}
PCA identifies the directions (eigenvectors) along which the variance of the data is maximized. Our goal is to find a direction \(u\) (with \(\|u\|=1\)) such that the projected variance is maximized:
\[
\underset{\|u\|=1}{\mathrm{argmax}} \; \frac{1}{N} \sum_{i=1}^N \Bigl((x_i - \bar{x})^T u\Bigr)^2 \;=\; \underset{\|u\|=1}{\mathrm{argmax}} \; u^T S_x\, u.
\]
To solve this, we set up the optimization problem:
\[
\max_{u_1}\; u_1^T S_x\, u_1 \quad \text{subject to} \quad u_1^T u_1 = 1.
\]
Incorporating the constraint using a Lagrange multiplier \(\lambda_1\), the Lagrangian becomes:
\[
\mathcal{L}(u_1,\lambda_1) = u_1^T S_x\, u_1 - \lambda_1 \left(u_1^T u_1 - 1\right).
\]
Taking the derivative with respect to \(u_1\) and setting it to zero gives:
\[
\frac{\partial \mathcal{L}}{\partial u_1} = 2S_x\, u_1 - 2\lambda_1\, u_1 = 0,
\]
which rearranges to the eigenvalue–eigenvector equation:
\[
S_x\, u_1 = \lambda_1\, u_1.
\]
Here, \(u_1\) is an eigenvector of the covariance matrix \(S_x\) and \(\lambda_1\) is the corresponding eigenvalue. The eigenvalue \(\lambda_1\) represents the variance captured by the projection onto \(u_1\); the largest eigenvalue indicates the direction of maximum variance—the first principal component. Additional eigenvectors (and eigenvalues) are computed similarly with the constraint that they remain orthogonal to previous ones.

\bigskip

\textbf{Step 3 - Item reduction in GDMS survey:}
For each factor in the GDMS survey, we evaluated all possible combinations of four items (from the original five). For each subset, the corresponding covariance matrix was computed, eigenvalue analysis was performed, and the total variance was calculated according to step 2. The percentage of variance retained by a four-item subset relative to the original five-item set is given by:
\[
\text{Percentage Variance Retained} = \frac{\sum \lambda_{\text{subset}}}{\sum \lambda_{\text{all}}} \times 100.
\]
This metric directly reflects how well the selected items capture the variance (information) present in the full set.

\bigskip

As shown in \Cref{fig:survey_reduce}, the optimal four-item combinations for each factor were identified as follows: for the Intuitive factor, items 1, 3, 12, and 16 were selected, retaining $89.63\%$ of variance. For the Dependent factor, items 2, 5, 10, and 18 retained $83.86\%$ of variance. The Rational factor's optimal subset included items 4, 7, 11, and 13, retaining $83.04\%$ variance. For the Avoidant factor, the chosen subset was 6, 14, 19, and 21, retaining $80.69\%$ variance. Finally, for the Spontaneous factor, items 8, 9, 15, and 24 retained $83.48\%$ of the original variance.

\paragraph{Pearson Correlation}\label{subsubsec:corr}
We also use Pearson correlation coefficients to compare the performance of the reduced survey constructs (both 4-item and 3-item versions) with the original 5-item version. These correlations provide a quantitative measure of the linear relationship between the scores of the reduced versions and those of the full survey. As indicated in \Cref{4v5_3v5_baseline}, the 4-item version exhibits extremely high correlations (ranging from 0.97 to 0.99) with the original 5-item version, suggesting near-perfect alignment. The 3-item version also shows strong correlations (ranging from 0.92 to 0.96), though these are marginally lower. The extremely low p-values associated with these correlations confirm that the relationships are statistically significant, reinforcing that both the 4-item and 3-item models capture nearly all of the information from the 5-item version.
\begin{table}[!htbp]
\centering

\begin{tabular}{lcccc}
\hline
& \multicolumn{2}{c}{\textbf{4-item vs 5-item}}
& \multicolumn{2}{c}{\textbf{3-item vs 5-item}} \\
\textbf{Factor} & \textbf{Correlation} & \textbf{P-value}
                & \textbf{Correlation} & \textbf{P-value} \\
\hline
Intuitive   & 0.99 & 3.28e--128 & 0.96 & 9.00e--44 \\
Dependent   & 0.98 & 2.17e--109 & 0.94 & 2.87e--39 \\
Rational    & 0.97 & 4.07e--99  & 0.92 & 9.39e--35 \\
Avoidant    & 0.98 & 3.34e--119 & 0.97 & 8.89e--50 \\
Spontaneous & 0.98 & 1.57e--112 & 0.95 & 1.13e--41 \\
\hline
\end{tabular}
\caption{Pearson correlation coefficients and p-values of 4-item and 3-item versus the original 5-item version at baseline (T=0) for each factor.  A total of 166 participants completed the original survey. }\label{4v5_3v5_baseline}
\end{table}
\bigskip

\paragraph{Test-Retest Reliability in the Follow-Up Study}\label{subsubsec:test_retest}

To evaluate the temporal stability of the reduced GDMS constructs, test-retest reliability was assessed using Pearson correlation coefficients between baseline (T=0) and follow-up (T=1) scores across four key comparisons: (1) 4-item (T=1) vs 4-item (T=0), (2) 4-item (T=1) vs 5-item (T=0), (3) 3-item (T=1) vs 3-item (T=0), and (4) 3-item (T=1) vs 5-item (T=0). A total of 80 participants completed the follow-up survey at least one month after the original administration.

\begin{table}[htbp]
\centering
\begin{tabular}{lcccccccc}
\hline
& \multicolumn{2}{c}{\shortstack{4-item (T=1) \\ vs \\ 4-item (T=0)}}
& \multicolumn{2}{c}{\shortstack{4-item (T=1) \\ vs \\ 5-item (T=0)}}
& \multicolumn{2}{c}{\shortstack{3-item (T=1) \\ vs \\ 3-item (T=0)}}
& \multicolumn{2}{c}{\shortstack{3-item (T=1) \\ vs \\ 5-item (T=0)}} \\
\textbf{Factor} & \textbf{Correlation} & \textbf{P-value} & \textbf{Correlation} & \textbf{P-value} & \textbf{Correlation} & \textbf{P-value} & \textbf{Correlation} & \textbf{P-value} \\
\hline
Intuitive   & 0.73 & 1.07e-14 & 0.71 & 6.82e-14 & 0.67 & 1.15e-11 & 0.72 & 2.36e-14 \\
Dependent   & 0.72 & 3.89e-14 & 0.73 & 1.54e-14 & 0.64 & 1.09e-10 & 0.70 & 4.68e-13 \\
Rational    & 0.56 & 5.42e-08 & 0.56 & 6.45e-08 & 0.58 & 1.32e-08 & 0.56 & 4.95e-08 \\
Avoidant    & 0.67 & 5.15e-12 & 0.67 & 9.88e-12 & 0.65 & 3.55e-11 & 0.64 & 9.13e-11 \\
Spontaneous & 0.68 & 3.66e-12 & 0.70 & 5.59e-13 & 0.64 & 1.81e-10 & 0.66 & 2.60e-11 \\
\hline
\end{tabular}
\caption{Test-retest reliability. This compares factor scores from the reduced survey constructs (4-item and 3-item) at follow-up (T=1) with those from baseline (T=0) and the original 5-item survey using Pearson correlation coefficients with p-values. A total of 80 participants completed the follow-up survey. }
\label{tab:4v5_3v5_followup}
\end{table}

As shown in \Cref{tab:4v5_3v5_followup}, the 4-item versions demonstrated strong temporal stability. When compared against themselves across time (T=1 vs T=0), correlations ranged from \(r = 0.56\) to \(r = 0.73\), all with p-values below \(1.1 \times 10^{-14}\). This suggests high consistency in responses over time. Similarly, when compared against the original 5-item versions at baseline, the 4-item scores yielded comparable correlations (\(r = 0.56\) to \(r = 0.73\)), again with p-values below \(1.6 \times 10^{-14}\), indicating strong alignment with the full-length scales.

The 3-item versions showed slightly lower, but still statistically significant, test-retest reliability. Correlations for 3-item (T=1) vs 3-item (T=0) ranged from \(r = 0.54\) to \(r = 0.67\), with p-values below \(1.4 \times 10^{-8}\). When compared with the original 5-item scores at baseline, correlations remained in a similar range (\(r = 0.56\) to \(r = 0.72\), all p $<$ \(2.6 \times 10^{-11}\)).

Both the 4-item and 3-item versions showed a strong linear relationship with the original 5-item constructs at baseline, with only marginal differences in performance. Given that these measures serve as background variables in our study, the 3-item versions provide a practical and efficient alternative—offering acceptable reliability while minimizing survey length.

\paragraph{Different Implications for Measurement Quality (Variance Retention vs Correlation)}\label{subsubsec:var_vs_corr}

Let $\mathbf{x}=(x_1,\dots ,x_p)^{\!\top}$ be the $p$ original items for one
GDMS factor, with sample covariance matrix $\boldsymbol\Sigma$ and eigenvalues
$\lambda_1\ge\cdots\ge\lambda_p>0$.
A reduced set keeps the $m$ items indexed by
$\mathcal I=\{i_1,\dots ,i_m\}$ (\(m<p\)).

\textbf{Variance–retention ratio.}
With $\boldsymbol\Sigma_{\mathcal I}$ the $m\times m$ sub-matrix that matches
the kept items, the fraction of total item variance that survives is
\[
R_{\mathrm{var}}(m)=
\frac{\operatorname{tr}(\boldsymbol\Sigma_{\mathcal I})}
     {\operatorname{tr}(\boldsymbol\Sigma)}
=\frac{\sum_{j=1}^{m}\lambda^{(m)}_j}
       {\sum_{j=1}^{p}\lambda_j},
\]
so any variance unique to the discarded items lowers $R_{\mathrm{var}}$.

\textbf{Correlation between the two composite scores.}
Write both mean scores as
\[
F_{m}= \mathbf w_{m}^{\!\top}\mathbf x,\qquad
F_{p}= \mathbf w_{p}^{\!\top}\mathbf x,
\]
where $\mathbf w_{m}=m^{-1}\mathbf e_{\mathcal I}$ places \(1/m\) on each
retained item and $0$ elsewhere, and
$\mathbf w_{p}=p^{-1}\mathbf 1_{p}$ assigns equal weight to every item.
The Pearson correlation is
\[
\rho_{F_{m},F_{p}}=
\frac{\mathbf w_{m}^{\!\top}\boldsymbol\Sigma\mathbf w_{p}}
     {\sqrt{(\mathbf w_{m}^{\!\top}\boldsymbol\Sigma\mathbf w_{m})
            (\mathbf w_{p}^{\!\top}\boldsymbol\Sigma\mathbf w_{p})}}.
\]
The numerator expands to
\[
\mathbf w_{m}^{\!\top}\boldsymbol\Sigma\mathbf w_{p}
=\frac{1}{mp}\sum_{i\in\mathcal I}\sum_{j=1}^{p}\sigma_{ij},
\]
so it includes \emph{only} covariances that involve items kept
in both composites(3- and 5-item; all covariances among the discarded items disappear.
Therefore $\rho_{F_{m},F_{p}}$ measures the shared structure, whereas
$R_{\mathrm{var}}$ counts every bit of variance, shared or unique.

\textbf{As an illustrative case, consider the \textit{Intuitive} factor.  }
The sample covariance matrix at baseline (\(T = 0\)) is

\[
\boldsymbol\Sigma=
\begin{bmatrix}
0.92 & 0.32 & 0.38 & 0.49 & 0.22\\
0.32 & 1.00 & 0.39 & 0.37 & 0.20\\
0.38 & 0.39 & 0.78 & 0.49 & 0.22\\
0.49 & 0.37 & 0.49 & 0.83 & 0.28\\
0.22 & 0.20 & 0.22 & 0.28 & 0.41
\end{bmatrix},
\qquad
\text{items }(1,3,12,16,17),
\]

the total item variance is the trace
\(\operatorname{tr}(\boldsymbol\Sigma)=0.92+1.00+0.78+0.83+0.41=3.94\).

\medskip
\textbf{Variance retained.}
Keeping the four best items \((1,3,12,16)\) gives
\(\operatorname{tr}(\boldsymbol\Sigma_{(4)})=0.92+1.00+0.78+0.83=3.53\),
so the variance–retention ratio is
\(R_{\text{var}}^{(4)}=3.53/3.94=0.896\;(89.6\%)\).
Keeping the three best items \((1,3,16)\) gives
\(\operatorname{tr}(\boldsymbol\Sigma_{(3)})=0.92+1.00+0.83=2.75\),
hence \(R_{\text{var}}^{(3)}=2.75/3.94=0.698\;(69.8\%)\).

\medskip
\textbf{Correlation between composite scores (3-item vs.\ 5-item).}
Let

\[
F_{5}=\frac{1}{5}(x_{1}+x_{3}+x_{12}+x_{16}+x_{17}),
\qquad
F_{3}=\frac{1}{3}(x_{1}+x_{3}+x_{16}),
\]
so the weight vectors are
\(\mathbf w_{5}=\tfrac15(1,1,1,1,1)^{\!\top}\) and
\(\mathbf w_{3}=\tfrac13(1,1,0,1,0)^{\!\top}\).
Then

\[
\operatorname{Var}(F_{5})=\mathbf w_{5}^{\!\top}\boldsymbol\Sigma\mathbf w_{5}=0.4264,
\quad
\operatorname{Var}(F_{3})=\mathbf w_{3}^{\!\top}\boldsymbol\Sigma\mathbf w_{3}=0.5678,
\]
\[
\operatorname{Cov}(F_{3},F_{5})=\mathbf w_{3}^{\!\top}\boldsymbol\Sigma\mathbf w_{5}=0.4713,
\]
and

\[
\rho_{F_{3},F_{5}}
=\frac{\operatorname{Cov}(F_{3},F_{5})}
       {\sqrt{\operatorname{Var}(F_{3})\,\operatorname{Var}(F_{5})}}
=\frac{0.4713}{\sqrt{0.5678\times0.4264}}
\approx0.96.
\]

\medskip
\textbf{Interpretation.} Keeping three of the five original items retains about \(70\%\) of the \emph{raw} item–level variance, because the two discarded items also carry their own item-specific (unique and error) variance.
The variance–retained statistic is a simple ratio of traces, so any variance that belongs solely to a dropped item is counted as lost, even if that variance is mostly noise. In contrast, the person–level means from the 3-item and 5-item forms remain almost perfectly aligned (\(\rho \approx 0.96\)).
Pearson’s \(r\) reflects only the \emph{shared} variance—i.e.\ the covariance driven by the common latent trait.
Because that trait dominates each item’s variance, averaging three items already recovers nearly all of the signal while down-weighting unshared noise.
Thus, the \(30\%\) reduction in total variance is largely variance that was not common to the retained items and removing it has minimal effect on respondent ranking.

\paragraph{Survey Reduction Decision}
We choose to use the 3-item scales for each GDMS factor, as they offer a strong balance between brevity and measurement quality. Specifically:

\begin{itemize}
    \item The 3-item scales reduce participant burden by forty percent while preserving key psychometric properties. Despite retaining only about \(70\%\) of the total variance compared to the 5-item scales, they maintain very high correlations (\(\rho \ge 0.95\)) with the original 5-item scores at baseline (\(T=0\)) and exhibit stable rank ordering one month later.

    \item The variance lost by dropping two items appears mostly unique to those items and does not substantially affect participant rank ordering at baseline.
    The retained items already capture most of the shared variance defining the latent construct.

    \item The follow-up correlations confirm that the short forms maintain the same relative positions over time as the full form. While the 4-item versions preserve more variance (approximately \(90\%\)) and slightly greater score variability, which may be important in analyses that rely on absolute differences such as regression coefficients, this added variance contributes little to rank-order stability or temporal consistency.
\end{itemize}

\subsubsection{Definition of treatment variables and covariates}

\begin{table}
\centering
\renewcommand{\arraystretch}{1.8}
\normalsize
\begin{tabular}{>{\raggedright\arraybackslash}p{2.2cm}m{2.7cm}m{1cm}>{\raggedright\arraybackslash}p{8.5cm}}
\hline \hline

\multicolumn{4}{c}{\textbf{Treatment Variables}} \\ \hline
\multirow[t]{1}{*}{\textbf{Controversy}} & \( \text{Disagreement}_{j} \) & (0,1) & The percentage of out-group verdicts per post. \\

\multirow[t]{2}{*}{\textbf{Certainty}}
& \( \text{IngroupCertainty}_{j} \) & (0,1) & Group certainty of people agreeing with participants.\\
& \( \text{OutgroupCertainty}_{j} \) & (0,1) & Group certainty of people disagreeing with participants.\\  \hline \hline

\multicolumn{4}{c}{\textbf{Posts Variables}} \\ \hline
\multirow[t]{1}{*}{\textbf{Content}} & \( \text{Post}_{id} \) & \{0,1\}$^{135}$ & Identifier assigned to each post in the dataset. \\
\multirow[t]{1}{*}{\textbf{Topic}} & \( \text{Topic}_{id} \) & \{0,1\}$^{32}$ & Identifier representing the distinct topic of the post. \\ \hline\hline

\multicolumn{4}{c}{\textbf{Participant Variables}} \\ \hline
\multirow[t]{1}{2.5cm}{\textbf{Initial Confidence}} & \( \text{InitialCertainty}_{ij} \) & \{1, 2, 3, 4\} & The participant's initial confidence rating: 1 - uncertain, 2 - somewhat uncertain, 3 - somewhat certain, 4 - certain. \\

\multirow[t]{5}{2.5cm}{\textbf{Big Five Personality}
       \citep{Gosling2003AVB}} & \( \text{Openness}_{i} \) & (0,1) & Curiosity, imagination and openness to new experiences. \\
& \( \text{Conscientiousness}_{i} \) & (0,1) &  Self-discipline, responsibility, and organizational skills. \\
& \( \text{Extraversion}_{i} \) & (0,1) & Sociability and energy in interacting with others. \\
& \( \text{Agreeableness}_{i} \) & (0,1) & Compassion and the tendency to get along well with others. \\
& \( \text{Neuroticism}_{i} \) & (0,1) & Emotional instability and susceptibility to stress. \\

\multirow[t]{5}{2.5cm}{\textbf{General Decision Making Styles       }
    \citep{scott1995decision}} & \( \text{Rational}_{i} \) & (0,1) & Logical and systematic decision-making based on evidence and analysis. \\
& \( \text{Dependent}_{i} \) & (0,1) & Relies on advice and reassurance from others to make decisions. \\
& \( \text{Intuitive}_{i} \) & (0,1) & Decisions driven by instinct, gut feelings, and subjective impressions. \\
& \( \text{Avoidant}_{i} \) & (0,1) & Procrastinates or evades decision-making due to fear or indecisiveness. \\
& \( \text{Spontaneous}_{i} \) & (0,1) & Makes quick, impulsive decisions with minimal deliberation. \\

\multirow[t]{1}{2.5cm}{\textbf{Intellectual Humility} \citep{leary2017IH}} & \( \text{IH}_{i} \) & (0,1) & Openness to the possibility that one's own beliefs could be wrong, and attentiveness to the limits of the evidence. \\

\multirow[t]{4}{2.5cm}{\raggedright\textbf{Demographic Variables} (defined \& provided by Prolific)} & \( \text{Country}_{i} \) & \{0,1\}$^{N}$ & The participant's country of residence or nationality. \\
& \( \text{Age}_{i} \) & $\geq 18$ & The participant's age in years. \\
& \( \text{Gender}_{i} \) & \{0,1\}$^{3}$ & The participant's gender identity: Male, Female, Prefer not to say. \\
\hline

\end{tabular}
\caption{Independent Variables: Treatment, Dilemmas, and Participant Variables}
\label{tab:variables_appendix}
\end{table}

\noindent Table~\ref{tab:variables_appendix} summarizes the independent variables used in our regression analyses, categorized into treatment variables, dilemma-level variables, and participant variables. The \textbf{treatment variables} include $\text{Agreement}_{j}$, $\text{IngroupCertainty}_{j}$, and $\text{OutgroupCertainty}_{j}$. Specifically, $\text{Agreement}_{j}$ measures the percentage of AITA verdicts opposing the focal participant's initial verdict, with higher values indicating greater disagreement. $\text{IngroupCertainty}_{j}$ and $\text{OutgroupCertainty}_{j}$ represent certainty within the AITA verdict groups that agree and disagree with the participant, respectively. These measures allow us to assess how both groups' certainty relates to changes in the participant's judgment.

\noindent The \textbf{dilemma-level variables} consist of $\text{Post}_{id}$ and $\text{Topic}_{id}$. The source-post identifier $\text{Post}_{id}$ uniquely identifies the corresponding dilemma and is represented by 135 dummy variables. Similarly, $\text{Topic}_{id}$ is represented by 32 dummy variables, one for each topic category. Including these identifiers helps account for dilemma- and topic-specific variation when estimating treatment effects and associations with participant characteristics.

\noindent \textbf{Participant variables} include $\text{InitialCertainty}_{ij}$, which records each participant's initial confidence rating. Personality is represented through five traits derived from the TIPI, while decision-making styles are assessed using five GDMS dimensions and one IH score, which captures a person's openness to the possibility that their own beliefs could be wrong, together with their attentiveness to the limits of the evidence behind those beliefs and of their own ability to gather and weigh that evidence \citep{leary2017IH}. These variables aim to examine how individual differences in personality and decision-making affect participants' responses. Detailed measurement formulations are provided in Table~\ref{tab:measurement_formulas}. Demographic variables supplied by Prolific, including $\text{Country}_{i}$, $\text{Age}_{i}$ and $\text{Gender}_{i}$, are also included to control for potential confounding factors related to participants’ backgrounds.

\begin{table}[!htbp]
\renewcommand{\arraystretch}{2}
\centering
\caption{Measurement formulation for GDMS, TIPI, and IH factors}
\begin{tabular}{lll}
\toprule
\textbf{Questionnaire} & \textbf{Factor} & \textbf{Measurement} \\
\midrule
\multirow{5}{*}{GDMS}
& Dependent & $\text{Average}(\text{GDMS}_{2}, \text{GDMS}_{5}, \text{GDMS}_{10}, \text{GDMS}_{18})$ \\
& Intuitive & $\text{Average}(\text{GDMS}_{1}, \text{GDMS}_{3}, \text{GDMS}_{12}, \text{GDMS}_{16})$ \\
& Avoidant & $\text{Average}(\text{GDMS}_{6}, \text{GDMS}_{14}, \text{GDMS}_{19}, \text{GDMS}_{21})$ \\
& Rational & $\text{Average}(\text{GDMS}_{4}, \text{GDMS}_{7}, \text{GDMS}_{11}, \text{GDMS}_{13})$ \\
& Spontaneous & $\text{Average}(\text{GDMS}_{8}, \text{GDMS}_{9}, \text{GDMS}_{15}, \text{GDMS}_{24})$ \\
\midrule
\multirow{5}{*}{TIPI}
& Openness & $\frac{\text{TIPI}_{5} - \text{TIPI}_{10}}{2}$ \\
& Conscientiousness & $\frac{\text{TIPI}_{3} - \text{TIPI}_{8}}{2}$ \\
& Extraversion & $\frac{\text{TIPI}_{1} - \text{TIPI}_{6}}{2}$ \\
& Agreeableness & $\frac{-\text{TIPI}_{2} + \text{TIPI}_{7}}{2}$ \\
& Neuroticism & $\frac{-\text{TIPI}_{4} + \text{TIPI}_{9}}{2}$ \\
\midrule
IH & IH score & $\text{Average}(\text{IH}_{1}, \text{IH}_{2}, \text{IH}_{3}, \text{IH}_{4}, \text{IH}_{5}, \text{IH}_{6})$ \\
\bottomrule
\end{tabular}
\label{tab:measurement_formulas}
\end{table}

\clearpage
\subsection{Survey data processing and profiling}\label{sec:secA4}

Data were collected across four rounds of the online survey, producing 2{,}346 survey records: 2{,}171 completed submissions, 158 returned submissions, and 17 timed-out records. Returned submissions included cases in which participants declined consent, withdrew during the study, or completed only part of the task. Specifically, in the first round, 540 responses were approved, 41 were returned, and 8 timed out; in the second round, 540 were approved and 27 were returned; in the third round, 551 were approved, 41 were returned, and 4 timed out; and in the fourth round, 540 were approved, 49 were returned, and 5 timed out.

Next, we applied an attention-check criterion requiring participants to pass at least two out of three attention-check items.
This filtering step further reduced the sample from 2{,}171 to 2{,}159 valid participants. Exactly 2{,}121 participants passed all three checks, while 38 participants passed two out of three; among these 38 participants, 31 provided a written explanation or feedback comment. We retained these 38 participants because our preregistered criterion specified passing at least two of three attention checks. A single missed item against two correct responses, combined with their willingness to provide optional text, indicates genuine engagement rather than inattentive responding.
The final analytic dataset therefore included 2{,}159 participants with complete demographic and psychological measures. Participant demographics are summarized in \Cref{tab:demographics}, and their decision-making styles, personality traits, and intellectual humility in \Cref{tab:individual-differences}.

\subsubsection{Time distribution and comment engagement}\label{subsubsec:time}
\textbf{Overall completion time.} A histogram of total completion times revealed a median duration of 862.0 seconds, with a 5th percentile cutoff at 406.9 seconds. Of participants below this threshold ($n=108$), 52 participants (48.1\%) nonetheless provided at least one written explanation or feedback comment, suggesting that some low-duration participants were still attentive.

\begin{table}[!htbp]
\centering
\caption{Duration distribution summary.}
\begin{tabular}{lcccccccc}
\toprule
Count & Mean & Min & 5th pct & 25th pct (Q1) & Median & 75th pct (Q3) & 95th pct & Max \\
\midrule
2159.0 & 983.1 & 157.0 & 406.9 & 620.0 & 862.0 & 1203.0 & 1988.2 & 3853.0 \\
\bottomrule
\end{tabular}
\label{tab:duration-summary}
\end{table}

\textbf{Timing across experimental conditions.} To assess whether the active Control task required a similar amount of time as the treatment tasks, we compared total survey duration across conditions. Each participant's total duration covered training, two randomly assigned dilemmas, and the background questionnaire. Median duration was 14.28 minutes in Control and ranged from 14.14 to 14.76 minutes across the three treatment conditions (\Cref{tab:duration-condition}). The interquartile and 5th--95th percentile ranges also overlapped substantially. Thus, the active control was comparable in overall time on task, reducing concern that treatment--control differences reflected total survey duration.

\begin{table}[!htbp]
\centering
\caption{Total survey duration by experimental condition. Each participant contributes one duration. Values are reported in minutes for the final analytic sample.}
\label{tab:duration-condition}
\begin{tabular}{lrrrr}
\toprule
\textbf{Condition} & \textbf{$n$} & \textbf{Mean (SD)} & \textbf{Median [Q1, Q3]} & \textbf{5th--95th percentile} \\
\midrule
Control                 & 535 & 16.67 (8.94) & 14.28 [10.35, 20.79] & 7.11--33.55 \\
Controversy             & 536 & 15.79 (7.79) & 14.14 [10.19, 19.61] & 6.46--30.95 \\
Certainty               & 546 & 16.07 (8.48) & 14.24 [10.22, 19.39] & 6.32--33.27 \\
Controversy $+$ Certainty & 542 & 17.01 (8.95) & 14.76 [10.70, 20.51] & 7.10--34.78 \\
\bottomrule
\end{tabular}
\end{table}

\textbf{Timing across survey sections.} We further separated the total duration into three sections: training, main survey, and background questionnaire. The average durations for these sections were 305.0, 362.7, and 187.8 seconds, respectively (\Cref{tab:section-duration-stats}). The median times ranged from 156 seconds for the background section to 308 seconds for the main survey, suggesting that participants generally spent the most time on the central reasoning task.
\begin{table}[!htbp]
\centering
\caption{Section-wise duration statistics (seconds).}
\begin{tabular}{lrrrrrrrr}
\toprule
Section & Count & Mean & SD & Min & 25th pct (Q1) & Median & 75th pct (Q3) & Max \\
\midrule
Training   & 2159 & 305.0 & 224.6 & 14.7 & 164.3 & 240.8 & 376.3 & 2165.3 \\
Main survey & 2159 & 362.7 & 235.2 & 16.6 & 213.2 & 307.8 & 454.1 & 2537.5 \\
Background  & 2159 & 187.8 & 118.6 & 21.0 & 112.8 & 156.4 & 223.4 & 1372.3 \\
\bottomrule
\end{tabular}
\label{tab:section-duration-stats}
\end{table}

\begin{figure}[!htbp]
\centering
\begin{subfigure}{\linewidth}
    \centering
    \includegraphics[width=0.9\linewidth]{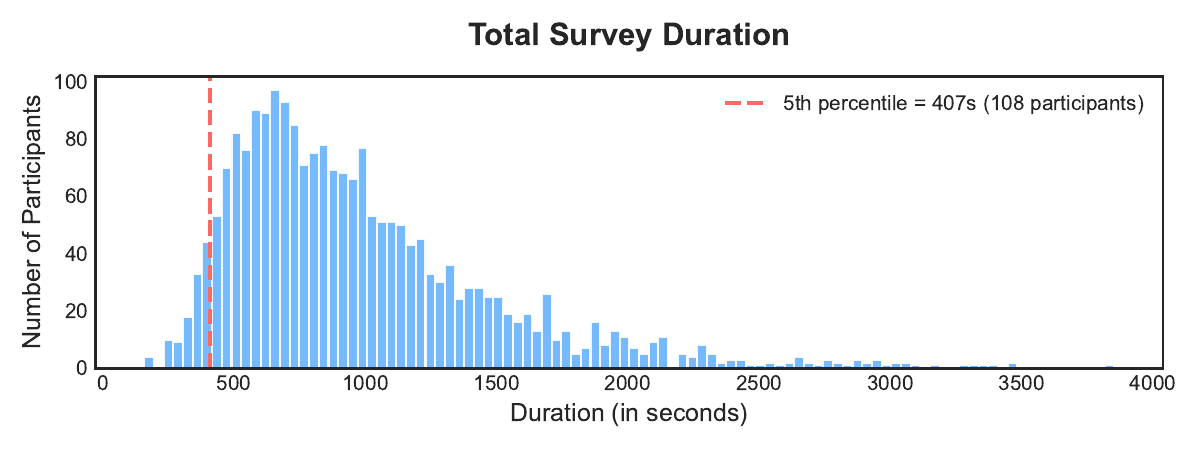}
        \label{fig:duration-total}
\end{subfigure}
\begin{subfigure}{\linewidth}
    \centering
    \includegraphics[width=0.9\linewidth]{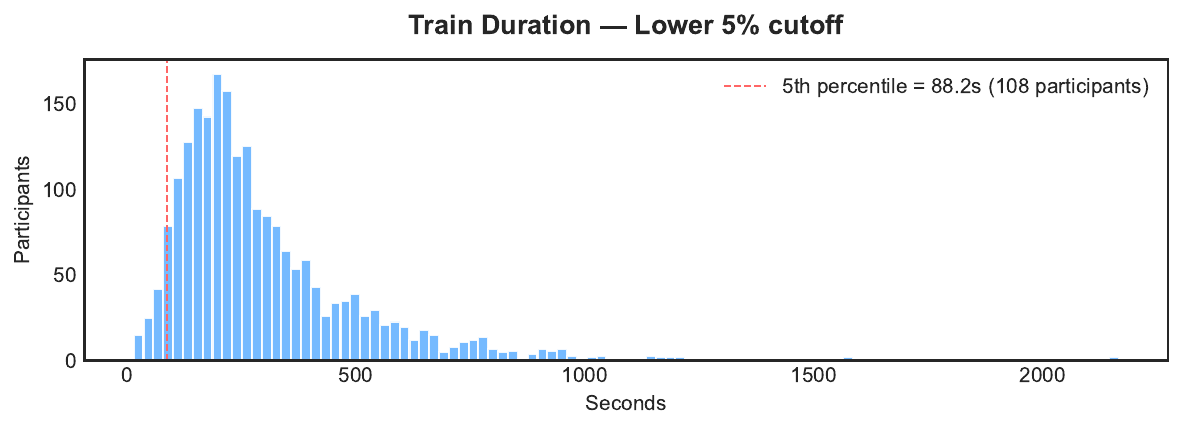}
        \label{fig:duration-train}
\end{subfigure}

\begin{subfigure}{\linewidth}
    \centering
    \includegraphics[width=0.9\linewidth]{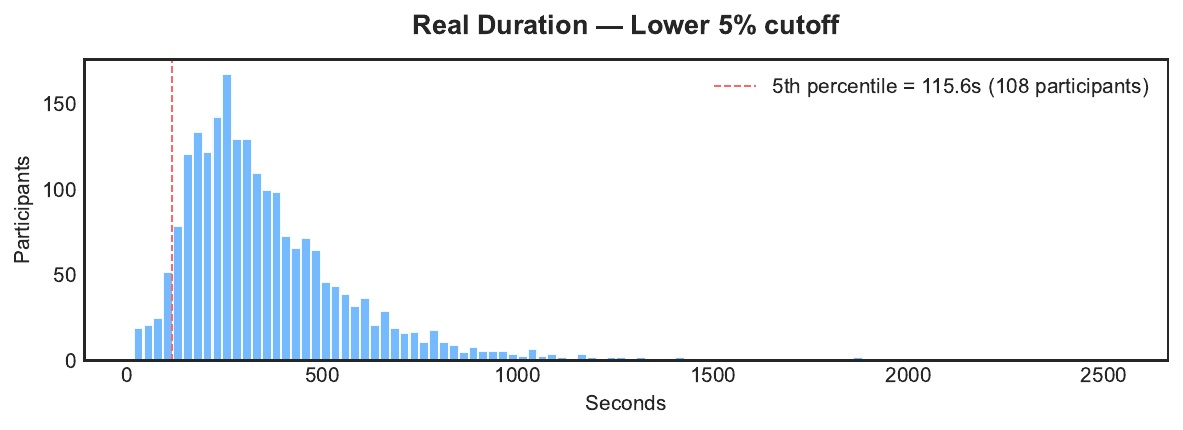}
        \label{fig:duration-survey}
\end{subfigure}

\begin{subfigure}{\linewidth}
    \centering
    \includegraphics[width=0.9\linewidth]{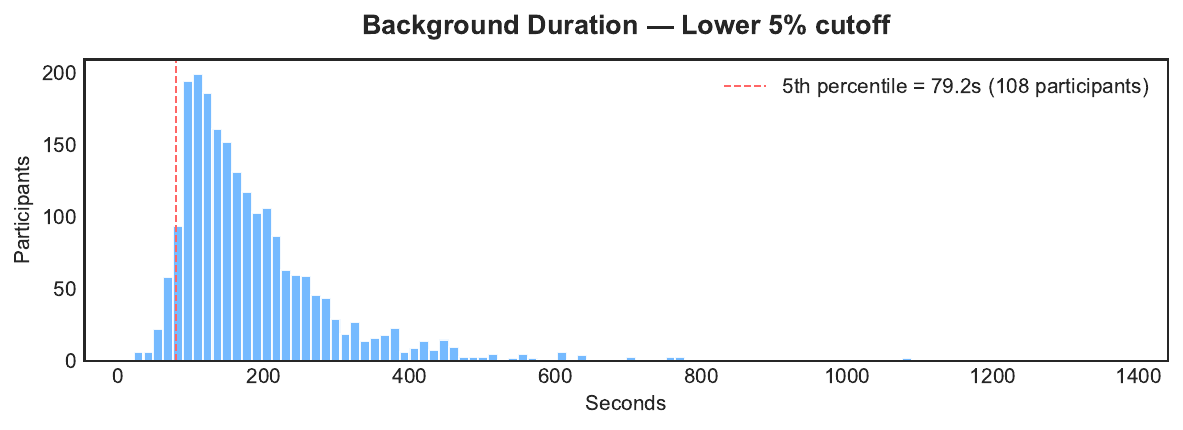}
        \label{fig:duration-background}
\end{subfigure}
\caption{Survey duration distributions and lower 5\% thresholds across sections.}
\label{fig:duration-overview}
\end{figure}

\textbf{Fast completions and written-response engagement.} To identify unusually fast completions, we conducted a 5th percentile analysis for each section (\Cref{tab:section-duration-lower5}). The 5\% cutoff was chosen to flag only the most extreme lower tail of the time distribution while retaining nearly all plausible completions. For instance, the main survey contained two moral-dilemma stories, each typically requiring about one minute to read and respond to; thus, total durations below roughly two minutes were considered implausibly short for genuine engagement. The resulting cutoff points were 88.2 seconds for the training, 115.6 seconds for the main survey, and 79.2 seconds for the background section. Among these fastest 5\% of participants, 57.4\%, 33.3\%, and 55.6\%, respectively, still provided a written explanation or feedback comment. This pattern indicates that shorter completion times did not necessarily correspond to disengagement, though engagement levels varied across different parts of the task.
\begin{table}[!htbp]
\centering
\caption{Fastest 5\% by section: cutoff and written-response presence.}
\begin{tabular}{lrrrr}
\toprule
Section & 5th pct cutoff (s) & Participants below cutoff & With any text & Share with text \\
\midrule
Training    & 88.2  & 108 & 62 & 57.4\% \\
Main survey & 115.6 & 108 & 36 & 33.3\% \\
Background  & 79.2  & 108 & 60 & 55.6\% \\
\bottomrule
\end{tabular}
\label{tab:section-duration-lower5}
\end{table}
\subsubsection{Participants' profile}
This subsection describes the $N = 2{,}159$ participants. Three tables summarize the sample. Table~\ref{tab:demographics} reports age, sex, and ethnicity. Table~\ref{tab:geography} reports country of residence and nationality. Table~\ref{tab:individual-differences} reports decision-making styles, personality traits, and intellectual humility.

Table~\ref{tab:demographics} reports the basic demographics. Age was split into four equal quartiles. The mean age was 38.7 years, with a range of 18 to 83. The sample was close to balanced by sex, at 50.4\% male and 49.5\% female. White participants formed the largest ethnic group (52.9\%), followed by Black (28.7\%) and Asian (8.9\%) participants.

\begin{table}[htbp!]
\centering
\caption{Participant demographic information (N = 2159).}\label{tab:demographics}
\begin{tabular}{lcc}
\toprule
\textbf{Variable} & \textbf{Category} & \textbf{N (\%)} \\
\midrule
\textbf{Age} & 18--27 (Q1) & 540 (25.0) \\
             & 28--35 (Q2) & 540 (25.0) \\
             & 36--47 (Q3) & 540 (25.0) \\
             & 48--83 (Q4) & 539 (25.0) \\
             & \multicolumn{2}{l}{Mean (SD) = 38.7 (13.1), Median = 36, Range = 18--83} \\[4pt]
\textbf{Sex} & Male & 1088 (50.4) \\
             & Female & 1069 (49.5) \\
             & Prefer not to say & 1 (0.0) \\
             & Data expired & 1 (0.0) \\[4pt]
\textbf{Ethnicity} & White & 1143 (52.9) \\
                   & Black & 620 (28.7) \\
                   & Asian & 191 (8.9) \\
                   & Mixed & 117 (5.4) \\
                   & Other & 77 (3.6) \\
                   & Data expired & 11 (0.5) \\
\bottomrule
\end{tabular}
\end{table}

Table~\ref{tab:geography} reports country of residence and nationality. Most participants lived in the United States (39.1\%), South Africa (23.6\%), or the United Kingdom (11.9\%). Nationality followed the same order. The remaining participants were spread across more than thirty countries.

\begin{table}[htbp!]
\centering
\caption{Participant country of residence and nationality (N = 2159).}\label{tab:geography}
\begin{tabular}{lc@{\hskip 1.5cm}lc}
\toprule
\multicolumn{2}{c}{\textbf{Country of residence}} & \multicolumn{2}{c}{\textbf{Nationality}} \\
\cmidrule(r){1-2} \cmidrule(l){3-4}
Category & N (\%) & Category & N (\%) \\
\midrule
United States   & 844 (39.1) & United States   & 808 (37.4) \\
South Africa    & 509 (23.6) & South Africa    & 495 (22.9) \\
United Kingdom  & 256 (11.9) & United Kingdom  & 220 (10.2) \\
Canada          & 149 (6.9)  & Canada          & 132 (6.1)  \\
Australia       & 122 (5.6)  & Australia       & 105 (4.9)  \\
Portugal        & 38 (1.8)   & Mexico          & 40 (1.9)   \\
Mexico          & 37 (1.7)   & Portugal        & 39 (1.8)   \\
Poland          & 27 (1.3)   & Nigeria         & 37 (1.7)   \\
Italy           & 27 (1.3)   & Italy           & 32 (1.5)   \\
New Zealand     & 18 (0.8)   & Poland          & 29 (1.3)   \\
Spain           & 16 (0.7)   & New Zealand     & 17 (0.8)   \\
Germany         & 15 (0.7)   & Germany         & 15 (0.7)   \\
Kenya           & 12 (0.6)   & Kenya           & 13 (0.6)   \\
Greece          & 11 (0.5)   & Greece          & 12 (0.6)   \\
Chile           & 9 (0.4)    & Spain           & 12 (0.6)   \\
Brazil          & 8 (0.4)    & India           & 11 (0.5)   \\
Netherlands     & 7 (0.3)    & Brazil          & 10 (0.5)   \\
Ireland         & 6 (0.3)    & Zimbabwe        & 9 (0.4)    \\
India           & 6 (0.3)    & Chile           & 9 (0.4)    \\
Latvia          & 5 (0.2)    & Netherlands     & 8 (0.4)    \\
Hungary         & 5 (0.2)    & China           & 8 (0.4)    \\
Czech Republic  & 4 (0.2)    & Ireland         & 6 (0.3)    \\
Sweden          & 4 (0.2)    & Latvia          & 6 (0.3)    \\
France          & 4 (0.2)    & Czech Republic  & 5 (0.2)    \\
Slovenia        & 3 (0.1)    & Philippines     & 5 (0.2)    \\
Austria         & 3 (0.1)    & Turkey          & 5 (0.2)    \\
Denmark         & 3 (0.1)    & Hungary         & 5 (0.2)    \\
Belgium         & 2 (0.1)    & France          & 4 (0.2)    \\
Israel          & 2 (0.1)    & Iran            & 4 (0.2)    \\
Japan           & 1 (0.0)    & Sweden          & 4 (0.2)    \\
Morocco         & 1 (0.0)    & Austria         & 3 (0.1)    \\
Argentina       & 1 (0.0)    & Slovenia        & 3 (0.1)    \\
Switzerland     & 1 (0.0)    & Denmark         & 3 (0.1)    \\
Malaysia        & 1 (0.0)    & Vietnam         & 3 (0.1)    \\
Finland         & 1 (0.0)    & Other ($n \leq 2$ each) & 28 (1.3) \\
Estonia         & 1 (0.0)    & Data expired    & 5 (0.2)    \\
\bottomrule
\end{tabular}
\end{table}

Table~\ref{tab:individual-differences} reports the individual-difference measures on a 0 to 1 scale. These cover the five General Decision-Making Style scales, the Big Five traits, and intellectual humility. The mean intellectual humility score was 0.71. The table also gives the median and the 25th and 75th percentiles for each measure.

\begin{table}
\centering
\caption{Participants' decision-making styles, personality traits, and intellectual humility (0--1 scale, N = 2159).}
\label{tab:individual-differences}
\begin{tabular}{lcccc}
\toprule
\textbf{Variable} & \textbf{Mean} & \textbf{25\%} & \textbf{Median} & \textbf{75\%} \\
\midrule
\textit{\textbf{Decision-making styles (GDMS)}} & & & & \\
Intuitive & 0.62 & 0.50 & 0.67 & 0.75 \\
Dependent & 0.41 & 0.25 & 0.42 & 0.58 \\
Rational & 0.55 & 0.42 & 0.50 & 0.67 \\
Avoidant & 0.66 & 0.58 & 0.67 & 0.75 \\
Spontaneous & 0.44 & 0.33 & 0.42 & 0.58 \\[4pt]
\textit{\textbf{Big Five personality traits}} & & & & \\
Extraversion & 0.41 & 0.17 & 0.42 & 0.58 \\
Agreeableness & 0.71 & 0.58 & 0.75 & 0.83 \\
Conscientiousness & 0.76 & 0.58 & 0.83 & 0.92 \\
Emotional Stability & 0.66 & 0.50 & 0.67 & 0.83 \\
Openness & 0.71 & 0.58 & 0.75 & 0.92 \\[4pt]
\textit{\textbf{Intellectual humility}} & & & & \\
IH & 0.71 & 0.58 & 0.71 & 0.83 \\
\bottomrule
\end{tabular}
\end{table}

\clearpage
\section{Results}
\label{app:results}
This section provides the full statistical detail behind the results summarized in the main text. \Cref{app:overall} reports the complete rate comparisons behind whether and how participants changed their minds, including the three-way classification of weakened, maintained, and strengthened responses used throughout the main text. \Cref{app:weakening} and \Cref{app:strengthening} give the model-selection paths and coefficient tables behind the weakening and strengthening results, respectively, together with the social-position breakdowns and robustness checks comparing continuous and binary representations of social position. \Cref{app:j_c_change} reports the two preregistered models for verdict change and confidence change as originally specified, and shows that they agree with the combined weakening and strengthening framework used in the main text. Finally, \Cref{sec:deliberation_analysis} reports exploratory analyses of participants' written explanations, dilemma consequences, and overall survey feedback.

\subsection{Overall judgment changes}
\label{app:overall}

This section reports the full rate comparisons behind the main-text result on whether people change their minds. It has two parts. Section~\ref{app:judgment-confidence-overall} separates weakening into its two preregistered component outcomes. These are verdict change, and confidence change among participants who kept their verdict. Section~\ref{app:weakening-strengthening-overall} then combines these outcomes into the three-way classification of weakened, maintained, and strengthened responses used throughout the main text.

All pairwise comparisons of rates in this appendix use Barnard's exact test \citep{barnard1945new}, an unconditional exact test for $2 \times 2$ contingency tables. Unlike Fisher's exact test, Barnard's test does not condition on the marginal totals, which gives it greater power when comparing two independent proportions. Two-sided tests are used throughout. An earlier version of this analysis used one-sided tests with the alternative set in the direction of the observed difference; because that direction was chosen after seeing the data, the resulting $p$-values were anticonservative, and we report two-sided tests instead.

To address multiple comparisons, $p$-values are corrected with the Benjamini--Hochberg false discovery rate (FDR) procedure \citep{benjamini1995controlling} at $q = .05$. The correction is applied within families of related comparisons. Each family is defined where its table appears. BH-adjusted values, denoted $p_{\mathrm{BH}}$, are reported throughout. Significance thresholds are $^{*}\ p < .05$, $^{**}\ p < .01$, and $^{***}\ p < .001$. For Tables~\ref{tab:barnard_results},\ref{tab:confidence_given_judgment_change}, \ref{tab:weakening_rate}, and \ref{tab:strengthening_rate}, the correction family is the three treatments against control.

\subsubsection{Verdict and confidence change rate}\label{app:judgment-confidence-overall}
This part reports the rate of each component outcome across treatments, relative to control. Table~\ref{tab:barnard_results} covers verdict change, and confidence change among participants who kept their verdict. Table~\ref{tab:confidence_given_judgment_change} then examines confidence change within the subset who did change their verdict.

\begin{table}[ht]
\centering
\caption{Barnard exact test results for verdict change and confidence change among participants who did not change their verdict across treatments, relative to control.}
\label{tab:barnard_results}
\small
\begin{tabular}{llrrrrr}
\toprule
Outcome & Treatment & $n$ & $N$ & Rate (\%) & $p_{\mathrm{raw}}$ & $p_{\mathrm{BH}}$ \\
\midrule
\multirow{4}{*}{Verdict change}
 & Control              &  34 & 1070 &  3.18 & ---                & ---                \\
 & Certainty            &  72 & 1092 &  6.59 & $2.35 \times 10^{-4}$ & $3.53 \times 10^{-4}$ \\
 & Controversy          &  95 & 1072 &  8.86 & $3.47 \times 10^{-8}$ & $1.04 \times 10^{-7}$ \\
 & Controversy+Certainty &  56 & 1084 &  5.17 & $2.17 \times 10^{-2}$ & $2.17 \times 10^{-2}$ \\
\midrule
\multirow{4}{*}{Confidence decrease}
 & Control              &  23 & 1036 &  2.22 & ---                & ---                \\
 & Certainty            &  44 & 1020 &  4.31 & $7.65 \times 10^{-3}$ & $7.65 \times 10^{-3}$ \\
 & Controversy          &  75 &  977 &  7.68 & $1.28 \times 10^{-8}$ & $3.83 \times 10^{-8}$ \\
 & Controversy+Certainty &  75 & 1028 &  7.30 & $5.86 \times 10^{-8}$ & $8.78 \times 10^{-8}$ \\
\midrule
\multirow{4}{*}{Confidence increase}
 & Control              &  40 & 1036 &  3.86 & ---                & ---                \\
 & Certainty            & 118 & 1020 & 11.57 & $5.34 \times 10^{-11}$ & $8.02 \times 10^{-11}$ \\
 & Controversy          & 126 &  977 & 12.90 & $1.59 \times 10^{-13}$ & $4.78 \times 10^{-13}$ \\
 & Controversy+Certainty & 107 & 1028 & 10.41 & $7.13 \times 10^{-9}$ & $7.13 \times 10^{-9}$ \\
\bottomrule
\end{tabular}
\end{table}

Table~\ref{tab:barnard_results} reports the three component outcomes. All three treatments raised the rate of verdict change relative to control (3.2\%): Certainty 6.6\%, Controversy 8.9\%, and Controversy+Certainty 5.2\%. Among participants who kept their verdict, confidence decreases also rose relative to control (2.2\%): Certainty 4.3\%, Controversy 7.7\%, and Controversy+Certainty 7.3\%. Confidence increases rose as well, from 3.9\% under control to 11.6\% (Certainty), 12.9\% (Controversy), and 10.4\% (Controversy+Certainty). Every treatment comparison was significant.

\begin{table}[htbp]
\centering
\caption{Confidence change among participants who changed their verdict, across treatments and control. Barnard exact tests (two-sided) compare each treatment to control; $N$ refers to the count of verdict changes; $p$-values are BH-corrected across treatments within each column. No comparison reaches significance.}
\label{tab:confidence_given_judgment_change}
\small
\begin{tabular}{lrrrrrrrr}
\toprule
Treatment & $N$ & Conf. dec. (\%) & Conf. stable (\%) & Conf. inc. (\%) & $p_{\mathrm{raw}}^{\mathrm{dec}}$ & $p_{\mathrm{BH}}^{\mathrm{dec}}$ & $p_{\mathrm{raw}}^{\mathrm{inc}}$ & $p_{\mathrm{BH}}^{\mathrm{inc}}$ \\
\midrule
Control              & 34 &  8 (23.5) & 20 (58.8) &  6 (17.6) & ---  & ---   & ---  & ---   \\
Certainty            & 72 &  8 (11.1) & 46 (63.9) & 18 (25.0) & .107 & .320  & .415 & .733  \\
Controversy          & 95 & 21 (22.1) & 53 (55.8) & 21 (22.1) & .878 & .878  & .733 & .733  \\
Controversy+Certainty & 56 & 15 (26.8) & 33 (58.9) &  8 (14.3) & .790 & .878  & .713 & .733  \\
\bottomrule
\end{tabular}
\end{table}
Table~\ref{tab:confidence_given_judgment_change} then looks at confidence among the smaller set of participants who did change their verdict. Here the treatments did not differ from control. No comparison reached significance in either direction.

\subsubsection{Weakening and strengthening rate}
\label{app:weakening-strengthening-overall}

This part combines the two component outcomes into a single three-way classification. We group each participant's verdict and confidence changes into one of three types. A response is \textbf{\emph{maintained}} when verdict and confidence both stay the same. A response is \textbf{\emph{weakened}} when the participant changes their verdict, or keeps the same verdict but reports lower confidence. Lower confidence without a reversal shows growing doubt that has not yet crossed the threshold of a verdict change. A response is \textbf{\emph{strengthened}} when the participant keeps the same verdict and reports higher confidence. The classification follows a two-step logic. The first step records whether verdict changed. The second step applies only when verdict did not change, and records whether confidence moved on the scale from $-3$ to $+3$. Tables~\ref{tab:weakening_rate} and \ref{tab:strengthening_rate} report the rate of each type by treatment, for the full sample to analyze weakening rate and for the subset used to analyze strengthening rate.

\begin{table}[ht]
\centering
\caption{Belief change rates by treatment for the full sample ($N = 4{,}318$). Columns give the percentage of responses that weakened, were maintained, or strengthened. The test compares each treatment's weakening rate to control.}
\label{tab:weakening_rate}
\small
\begin{tabular}{lrrrrrrrrrr}
\toprule
 & & \multicolumn{2}{c}{Weakened} & \multicolumn{2}{c}{Maintained} & \multicolumn{2}{c}{Strengthened} & & \\
\cmidrule(lr){3-4} \cmidrule(lr){5-6} \cmidrule(lr){7-8}
Treatment & $N$ & $n$ & \% & $n$ & \% & $n$ & \% & $p_{\mathrm{raw}}$ & $p_{\mathrm{BH}}$ \\
\midrule
Control              & 1070 &  57 &  5.3 & 973 & 90.9 &  40 &  3.7 & ---                    & ---                    \\
Certainty            & 1092 & 116 & 10.6 & 858 & 78.6 & 118 & 10.8 & $5.79 \times 10^{-6}$  & $5.79 \times 10^{-6}$  \\
Controversy          & 1072 & 170 & 15.9 & 776 & 72.4 & 126 & 11.8 & $2.17 \times 10^{-15}$ & $6.52 \times 10^{-15}$ \\
Controversy+Certainty & 1084 & 131 & 12.1 & 846 & 78.0 & 107 &  9.9 & $2.76 \times 10^{-8}$  & $4.13 \times 10^{-8}$  \\
\bottomrule
\end{tabular}
\vspace{2pt}
\noindent\textit{Note.} $p$ values are from Barnard exact tests comparing each treatment's weakening rate to control, with Benjamini--Hochberg correction.
\end{table}
Table~\ref{tab:weakening_rate} reports the three-way rates for the full sample. Weakening rose under all three treatments relative to control (5.3\%): Certainty 10.6\%, Controversy 15.9\%, and Controversy+Certainty 12.1\%. Every treatment comparison was significant. Controversy produced the most weakening. The combined condition produced less weakening than Controversy alone (12.1\% against 15.9\%). Maintenance fell in step, from 90.9\% under control to 72.4\% under Controversy.

\begin{table}[ht]
\centering
\caption{Belief change rates by treatment for the strengthening subset ($n = 1{,}924$), restricted to participants whose initial confidence was below the ceiling ($c < 4$). Columns give the percentage of responses that weakened, were maintained, or strengthened. The test compares each treatment's strengthening rate to control.}
\label{tab:strengthening_rate}
\small
\begin{tabular}{lrrrrrrrrrr}
\toprule
 & & \multicolumn{2}{c}{Weakened} & \multicolumn{2}{c}{Maintained} & \multicolumn{2}{c}{Strengthened} & & \\
\cmidrule(lr){3-4} \cmidrule(lr){5-6} \cmidrule(lr){7-8}
Treatment & $N$ & $n$ & \% & $n$ & \% & $n$ & \% & $p_{\mathrm{raw}}$ & $p_{\mathrm{BH}}$ \\
\midrule
Control              & 463 &  26 &  5.6 & 397 & 85.7 &  40 &  8.6 & ---                    & ---                    \\
Certainty            & 461 &  64 & 13.9 & 279 & 60.5 & 118 & 25.6 & $6.86 \times 10^{-12}$ & $1.03 \times 10^{-11}$ \\
Controversy          & 487 &  84 & 17.2 & 277 & 56.9 & 126 & 25.9 & $2.28 \times 10^{-12}$ & $6.85 \times 10^{-12}$ \\
Controversy+Certainty & 513 &  74 & 14.4 & 332 & 64.7 & 107 & 20.9 & $9.71 \times 10^{-8}$  & $9.71 \times 10^{-8}$  \\
\bottomrule
\end{tabular}
\vspace{2pt}
\noindent\textit{Note.} $p$ values are from Barnard exact tests comparing each treatment's strengthening rate to control, with Benjamini--Hochberg correction. Sample restricted to participants with initial confidence $c < 4$.
\end{table}

Table~\ref{tab:strengthening_rate} reports the same three-way rates for the subset with initial confidence below the ceiling ($c < 4$). This restriction is needed because participants already at the top of the scale cannot strengthen. In this subset, strengthening rose under all three treatments relative to control (8.6\%): Certainty 25.6\%, Controversy 25.9\%, and Controversy+Certainty 20.9\%. Every treatment comparison was significant.

\clearpage
\setcounter{secnumdepth}{4}
\subsection{When do judgments weaken?}
\label{app:weakening}

This appendix gives the full analysis behind the weakening results in the main text. The common regression specification is reported in Methods (\Cref{eq:model-spec-annotated}), together with the model-ablation rationale. The evaluation metric and cross-validated selection procedure shared across the weakening, strengthening (\Cref{app:strengthening-regression}), verdict-change (\Cref{app:judgmentchange}), and confidence-change (\Cref{app:confidencechange}) models are detailed in \ref{app:weakening-selection-details}. The weakening selection results are reported in \ref{app:weakening-ablation}, followed by the selected-model coefficients in \ref{app:weakening-coefficients}. We close with social-position comparisons and a robustness check testing whether continuous social signals predict judgment change better than their binary summaries (\Cref{app:weakening-quadrant,app:weakening-social-position-ablation}).

\subsubsection{Weakening regression details}
\label{app:weakening-regression}

The outcome, treatment, pretreatment, and individual-trait variables used in the regression are defined in Table~\ref{tab:variables_appendix} and Table~\ref{tab:measurement_formulas}.
We applied the logistic specification in \Cref{eq:model-spec-annotated} to the binary weakening outcome. A response was weakened when the participant changed their verdict or retained it with lower confidence. Every response had a defined weakening outcome, so the model was fitted to all $N = 4{,}318$ participant--dilemma observations. The coefficients were estimated separately from those in the strengthening model.

\paragraph{Model-selection details}\label{app:weakening-selection-details}
The four candidate model specifications and their rationale are described in Methods. Here we provide the evaluation metric and detailed covariate-block selection procedure used for all outcome models.

\textbf{Evaluation metric.} We used the mean log-likelihood on held-out observations as the selection metric. For each validation-fold observation $i$ with outcome $y_i \in \{0, 1\}$ and predicted probability $\hat{p}_i$, the observation-level log-likelihood is:
\begin{equation}
\ell_i = y_i \log \hat{p}_i + (1 - y_i) \log(1 - \hat{p}_i).
\label{eq:obs-loglik}
\end{equation}
The metric is the average of $\ell_i$ across all held-out observations. Because $\ell_i$ is always negative and equals zero only when the model assigns probability 1 to the outcome that actually occurred, a higher (less negative) value indicates better out-of-sample prediction. The score penalizes both false confidence ($y_i = 0$, $\hat{p}_i$ large) and missed events ($y_i = 1$, $\hat{p}_i$ small) symmetrically. It is a strictly proper scoring rule~\citep{gneiting2007strictly}, meaning that a model can only improve its score through better-calibrated probability estimates, not through systematic over- or under-prediction. We chose this metric over classification accuracy because accuracy requires an arbitrary decision threshold and discards calibration information. Our interest is in how well covariates improve the estimation of weakening probabilities, not in binary classification performance.

\textbf{Cross Validation and Ablation.}
We used 10-fold cross-validation to select which covariate blocks improve prediction on held-out data. Overfitting is a well-documented concern in behavioral research, where models with many covariates can fit training data closely but fail to generalize to new observations~\citep{yarkoni2017choosing}.
The ablation proceeded in four steps, applied identically to each of the four model specifications.

\textit{Step 1: Single-block addition.} Each candidate covariate block was added to the base model one at a time, independently of the others. The candidate blocks were: Big Five personality traits (Openness, Conscientiousness, Extraversion, Agreeableness, Emotional Stability), GDMS subscales (Intuitive, Dependent, Rational, Avoidant, Spontaneous), intellectual humility (IH), Age, Sex, dilemma topic, and nationality or country of residence. Two versions of the nationality/country variable were tested: a grouped version retaining only groups with $n > 100$ observations and collapsing the rest into an ``Other'' category, and a raw version retaining all individual groups. This step measures the marginal contribution of each block in isolation and does not determine the final model.

\textit{Step 2: Nationality/country filter.} Among the nationality and country blocks, only the grouped version ($n > 100$) with the highest $\Delta\bar{\ell}$ was retained for subsequent steps. The raw versions were dropped because they consistently reduced out-of-sample fit, likely due to overfitting on small groups. Across all four model specifications, Nationality (grouped, $n > 100$) outperformed Country (grouped, $n > 100$).

\textit{Step 3: Greedy forward selection.} Starting from the base model, every remaining candidate block was tested on top of the current accepted set at each step. The block with the largest positive $\Delta\bar{\ell}$ was accepted. Blocks that did not improve fit at a given step were not discarded but were retained in the candidate pool and retested at subsequent steps, because a block that is unhelpful alone may become useful in the presence of other covariates. When no single remaining block improved fit, all remaining blocks were tested jointly as a combined final step. If the combined step also failed to improve fit, the procedure stopped.

\textit{Step 4: Best model identification.} The step with the highest mean log-likelihood across the entire greedy path was identified as the best model. This is not necessarily the last accepted step, because the combined addition can sometimes reduce fit. The best model was then re-fit on the full data. For the \texttt{feglm} specification, standard errors were clustered by participant in the final fit.

\paragraph{Ablation results}\label{app:weakening-ablation}

\textbf{Single-block addition.} Table~\ref{tab:ablation-single-weakening} reports the marginal contribution of each covariate block when added individually to the base model. We denote the change in mean log-likelihood relative to the base model as $\Delta\bar{\ell}$, where a positive value indicates improved out-of-sample prediction. GDMS was the strongest single predictor across all four specifications ($\Delta\bar{\ell} \approx 0.007$, rank 1 in every model). IH ranked second or fourth depending on the specification ($\Delta\bar{\ell} \approx 0.001$). Age and Nationality (grouped, $n > 100$) showed small positive gains of similar magnitude. Sex and Personality contributed little in isolation ($\Delta\bar{\ell} < 0.001$). Three blocks consistently reduced out-of-sample fit. Topic hurt prediction in all specifications where it was testable ($\Delta\bar{\ell} \approx -0.011$). Raw country and raw nationality variables produced large negative $\Delta\bar{\ell}$ values ($-0.034$ to $-0.093$), indicating overfitting on small nationality groups. In the \texttt{glmmTMB} specification, adding a participant-level random intercept produced the worst single-block result ($\Delta\bar{\ell} = -0.462$), confirming that this component overfits the data.

\begin{table}[htbp]
\centering
\caption{Single-block ablation for weakening: $\Delta\bar{\ell}$ (rank) for each covariate block across model specifications, where $\Delta\bar{\ell}$ is the change in 10-fold cross-validated mean log-likelihood relative to the base model. The base row shows $\bar{\ell}$ (SE) of the base model. All other rows show $\Delta\bar{\ell}$ relative to base, with rank in parentheses. Higher (less negative) values indicate better out-of-sample fit.}
\label{tab:ablation-single-weakening}
\small
\begin{tabular}{l cccc}
\toprule
 & GLM & GLM & \texttt{feglm} & \texttt{glmmTMB} \\
 & (no dilemma factor) & (dilemma factor) & (dilemma FE) & (dilemma RE) \\
\midrule
Base $\bar{\ell}$ (SE)
  & $-0.315$ $(0.007)$ & $-0.396$ $(0.029)$ & $-0.333$ $(0.008)$ & $-0.315$ $(0.008)$ \\
\addlinespace
\multicolumn{5}{l}{\textit{Covariate block ($\Delta\bar{\ell}$, rank)}} \\
\addlinespace
GDMS
  & $+0.0072$ (1) & $+0.0069$ (1) & $+0.0070$ (1) & $+0.0073$ (1) \\
IH
  & $+0.0010$ (2) & $+0.0009$ (4) & $+0.0009$ (4) & $+0.0009$ (2) \\
Age
  & $+0.0009$ (3) & $+0.0011$ (3) & $+0.0013$ (2) & $+0.0009$ (3) \\
Nationality ($n{>}100$)
  & $+0.0008$ (4) & $+0.0012$ (2) & $+0.0012$ (3) & $+0.0008$ (4) \\
Country ($n{>}100$)
  & $+0.0007$ (5) & $+0.0008$ (5) & $+0.0009$ (5) & $+0.0006$ (5) \\
Sex
  & $+0.0002$ (6) & $+0.0000$ (7) & $+0.0000$ (7) & $+0.0002$ (6) \\
Personality
  & $+0.0001$ (7) & $+0.0002$ (6) & $+0.0002$ (6) & $+0.0001$ (7) \\
Topic
  & $-0.0106$ (9) & $\phantom{+}0.0000$ (8) & --- & $-0.0105$ (9) \\
Country (raw)
  & $-0.0336$ (10) & $-0.0404$ (10) & $-0.0340$ (9) & $-0.0520$ (10) \\
Nationality (raw)
  & $-0.0641$ (11) & $-0.0797$ (11) & $-0.0647$ (10) & $-0.0932$ (11) \\
RE: participant
  & --- & --- & --- & $-0.462\phantom{0}$ (12) \\
\bottomrule
\end{tabular}

\vspace{4pt}
\raggedright\footnotesize
\textit{Note.} GLM = generalized linear model with a binomial family and logit link. GLM without dilemma as a factor excludes dilemma indicators; GLM with dilemma as a factor includes dummy-coded dilemma indicators. \texttt{feglm} (dilemma FE) = fixed-effects logistic regression with dilemma absorbed via iterative demeaning. \texttt{glmmTMB} (dilemma RE) = generalized linear mixed model with dilemma as a random intercept. Nationality/Country ($n{>}100$) = grouped version retaining only groups with more than 100 observations. ``---'' = block not applicable for that specification. Topic is not available for \texttt{feglm} because it is absorbed into the dilemma fixed effect.
\end{table}

\textbf{Greedy forward selection.} Table~\ref{tab:ablation-greedy-weakening} reports the greedy forward selection path for each specification. In the GLM without dilemma as a factor (panel a), GDMS entered first ($\Delta\bar{\ell} = +0.0072$), followed by IH ($+0.0006$), Sex ($+0.0004$), and Nationality ($+0.0005$). The combined addition of the three remaining blocks (Topic, Personality, Age) reduced fit. The best model was reached at step 4 ($\bar{\ell} = -0.3062$, SE $= 0.0091$).
In the \texttt{glm} with dilemma as a factor, GDMS again entered first ($\Delta\bar{\ell} = +0.0069$). Nationality entered second ($+0.0006$) and IH entered third ($+0.0004$). The remaining four blocks (Topic, Personality, Age, Sex) were jointly rejected. The best model was reached at step 3 ($\bar{\ell} = -0.3881$, SE $= 0.0295$). The higher baseline loss compared with the GLM without dilemma as a factor reflects the cost of estimating 134 dilemma dummy coefficients from limited within-dilemma data in each training fold.
In the \texttt{feglm} specification with dilemma fixed effects, GDMS entered first ($\Delta\bar{\ell} = +0.0070$), followed by Nationality ($+0.0005$), Sex ($+0.0005$), and IH ($+0.0003$). Personality and Age were jointly rejected. The best model was reached at step 4 ($\bar{\ell} = -0.3249$, SE $= 0.0097$).
In the \texttt{glmmTMB} specification with dilemma as a random intercept, the selection path matched the GLM without dilemma as a factor exactly: GDMS, IH, Sex, Nationality, in the same order and with nearly identical $\Delta\bar{\ell}$ values. The best model was reached at step 4 ($\bar{\ell} = -0.3063$, SE $= 0.0092$). The combined final step, which included Topic, Personality, Age, and a participant-level random intercept, produced a large drop ($\Delta\bar{\ell} = -0.300$), driven almost entirely by the participant random intercept.

In summary, the best model across all specifications was the GLM without dilemma as a factor at step 4 (GDMS + IH + Sex + Nationality; $\bar{\ell} = -0.3062$, SE $= 0.0091$). The \texttt{glmmTMB} model reached a nearly identical best-step value ($\bar{\ell} = -0.3063$, SE $= 0.0092$), and its dilemma-level random effect variance was effectively zero (Variance $= 4.77 \times 10^{-9}$), confirming that the random intercept adds no predictive value. The coefficient estimates from the \texttt{glmmTMB} fit matched those of the GLM without dilemma as a factor to at least three decimal places.

\begin{table}[htbp]
\centering
\caption{Greedy forward selection for weakening: covariate block added at each step, resulting mean log-likelihood $\bar{\ell}$ (SE), and step-wise $\Delta\bar{\ell}$. Steps marked with $\checkmark$ were accepted; steps marked with $\times$ were rejected. The best model is the step with the highest $\bar{\ell}$.}
\label{tab:ablation-greedy-weakening}
\small
\begin{tabular}{cl cccc}
\toprule
Step & Block added & $\bar{\ell}$ & SE & $\Delta\bar{\ell}$ & Accepted \\
\midrule
\multicolumn{6}{l}{\textit{(a) GLM without dilemma as a factor}} \\
\addlinespace
0 & Base & $-0.3152$ & $0.0075$ & --- & --- \\
1 & GDMS & $-0.3079$ & $0.0081$ & $+0.0072$ & $\checkmark$ \\
2 & IH & $-0.3074$ & $0.0085$ & $+0.0006$ & $\checkmark$ \\
3 & Sex & $-0.3070$ & $0.0087$ & $+0.0004$ & $\checkmark$ \\
4 & Nationality ($n{>}100$) & $-0.3062$ & $0.0091$ & $+0.0005$ & $\checkmark$ \\
5 & combined(Topic, Pers., Age) & $-0.3064$ & $0.0092$ & $-0.0002$ & $\times$ \\
\addlinespace
\multicolumn{6}{l}{\textit{(b) GLM with dilemma as factor}} \\
\addlinespace
0 & Base & $-0.3960$ & $0.0286$ & --- & --- \\
1 & GDMS & $-0.3891$ & $0.0289$ & $+0.0069$ & $\checkmark$ \\
2 & Nationality ($n{>}100$) & $-0.3885$ & $0.0296$ & $+0.0006$ & $\checkmark$ \\
3 & IH & $-0.3881$ & $0.0295$ & $+0.0004$ & $\checkmark$ \\
4 & combined(Topic, Pers., Age, Sex) & $-0.3889$ & $0.0297$ & $-0.0008$ & $\times$ \\
\addlinespace
\multicolumn{6}{l}{\textit{(c) \texttt{feglm} with dilemma fixed effects}} \\
\addlinespace
0 & Base & $-0.3330$ & $0.0084$ & --- & --- \\
1 & GDMS & $-0.3260$ & $0.0086$ & $+0.0070$ & $\checkmark$ \\
2 & Nationality ($n{>}100$) & $-0.3255$ & $0.0093$ & $+0.0005$ & $\checkmark$ \\
3 & Sex & $-0.3250$ & $0.0095$ & $+0.0005$ & $\checkmark$ \\
4 & IH & $-0.3249$ & $0.0097$ & $+0.0003$ & $\checkmark$ \\
5 & combined(Pers., Age) & $-0.3254$ & $0.0098$ & $-0.0005$ & $\times$ \\
\addlinespace
\multicolumn{6}{l}{\textit{(d) \texttt{glmmTMB} with dilemma random intercept}} \\
\addlinespace
0 & Base & $-0.3153$ & $0.0075$ & --- & --- \\
1 & GDMS & $-0.3080$ & $0.0082$ & $+0.0073$ & $\checkmark$ \\
2 & IH & $-0.3074$ & $0.0085$ & $+0.0006$ & $\checkmark$ \\
3 & Sex & $-0.3071$ & $0.0087$ & $+0.0004$ & $\checkmark$ \\
4 & Nationality ($n{>}100$) & $-0.3063$ & $0.0092$ & $+0.0004$ & $\checkmark$ \\
5 & combined(Topic, Pers., Age, RE:part.) & $-0.6063$ & $0.0898$ & $-0.3000$ & $\times$ \\
\bottomrule
\end{tabular}

\vspace{4pt}
\raggedright\footnotesize
\textit{Note.} ``combined(...)'' denotes the joint addition of all remaining blocks that individually failed to improve fit. Pers.\ = Big Five Personality traits. RE:part.\ = participant-level random intercept. SE is the standard error of $\bar{\ell}$ across the 10 cross-validation folds.
\end{table}

\paragraph{Best-model coefficients}\label{app:weakening-coefficients}

\begin{table}[htbp]
\centering
\caption{Pairwise nationality contrasts from the best weakening model.
Each cell shows $\hat{\beta}$ (SE), $p$ for the row nationality
relative to the column nationality. The matrix is antisymmetric:
$\hat{\beta}_{A \text{ vs.\ } B} = -\hat{\beta}_{B \text{ vs.\ } A}$.
Treatment, initial-confidence, GDMS, and IH coefficients are reported in Extended Data Table~\ref{tab:full-weakening}; sex coefficients are summarized in the text.}
\label{tab:nationality-weakening}
\setlength{\tabcolsep}{3pt}
\begin{tabular}{l cccccc}
\toprule
& Aus. & Can. & S.Afr. & UK & US & Other \\
\midrule
Aus.
  & ---
  & \makecell{$0.55$ ($0.35$) \\ $p = .113$}
  & \makecell{$0.80$ ($0.30$) \\ $p = .009^{**}$}
  & \makecell{$0.40$ ($0.33$) \\ $p = .225$}
  & \makecell{$0.59$ ($0.29$) \\ $p = .046^{*}$}
  & \makecell{$0.32$ ($0.31$) \\ $p = .300$} \\[6pt]
Can.
  & \makecell{$-0.55$ ($0.35$) \\ $p = .113$}
  & ---
  & \makecell{$0.25$ ($0.23$) \\ $p = .291$}
  & \makecell{$-0.15$ ($0.27$) \\ $p = .570$}
  & \makecell{$0.03$ ($0.22$) \\ $p = .879$}
  & \makecell{$-0.23$ ($0.24$) \\ $p = .336$} \\[6pt]
S.Afr.
  & \makecell{$-0.80$ ($0.30$) \\ $p = .009^{**}$}
  & \makecell{$-0.25$ ($0.23$) \\ $p = .291$}
  & ---
  & \makecell{$-0.40$ ($0.21$) \\ $p = .055^{\dagger}$}
  & \makecell{$-0.21$ ($0.15$) \\ $p = .143$}
  & \makecell{$-0.48$ ($0.17$) \\ $p = .004^{**}$} \\[6pt]
UK
  & \makecell{$-0.40$ ($0.33$) \\ $p = .225$}
  & \makecell{$0.15$ ($0.27$) \\ $p = .570$}
  & \makecell{$0.40$ ($0.21$) \\ $p = .055^{\dagger}$}
  & ---
  & \makecell{$0.19$ ($0.19$) \\ $p = .333$}
  & \makecell{$-0.08$ ($0.21$) \\ $p = .709$} \\[6pt]
US
  & \makecell{$-0.59$ ($0.29$) \\ $p = .046^{*}$}
  & \makecell{$-0.03$ ($0.22$) \\ $p = .879$}
  & \makecell{$0.21$ ($0.15$) \\ $p = .143$}
  & \makecell{$-0.19$ ($0.19$) \\ $p = .333$}
  & ---
  & \makecell{$-0.27$ ($0.16$) \\ $p = .086^{\dagger}$} \\[6pt]
Other
  & \makecell{$-0.32$ ($0.31$) \\ $p = .300$}
  & \makecell{$0.23$ ($0.24$) \\ $p = .336$}
  & \makecell{$0.48$ ($0.17$) \\ $p = .004^{**}$}
  & \makecell{$0.08$ ($0.21$) \\ $p = .709$}
  & \makecell{$0.27$ ($0.16$) \\ $p = .086^{\dagger}$}
  & --- \\
\bottomrule
\end{tabular}
\vspace{4pt}

\raggedright\footnotesize
\textit{Note.} $^{\dagger}p < .10$, $^{*}p < .05$, $^{**}p < .01$.
Values are log-odds differences; positive values indicate higher
weakening odds for the row nationality. $p$-values are not adjusted
for multiple comparisons because these contrasts are exploratory.
\end{table}

Extended Data Table~\ref{tab:full-weakening} reports the treatment, initial-confidence, GDMS, and IH coefficients from the best weakening model (GLM without dilemma as a factor; GDMS + IH + Sex + Nationality). The outcome is the binary indicator for whether a participant's judgment was weakened (verdict changed or confidence decreased without verdict change). Among treatment variables, disagreement rate in the controversy had a strong positive effect on weakening ($\beta = 4.52$, $p < 0.001$), indicating that exposure to a higher share of opposing verdicts increased the probability of judgment weakening. In-group certainty in the certainty condition reduced weakening ($\beta = -2.46$, $p < 0.001$), while out-group certainty increased it ($\beta = 2.06$, $p < 0.001$). In the combined condition, disagreement rate again increased weakening ($\beta = 3.74$, $p < 0.001$), and in-group certainty again reduced it ($\beta = -1.66$, $p < 0.001$). Initial confidence was negatively associated with weakening ($\beta = -0.21$, $p = 0.003$). Among individual traits, Dependent decision-making style ($\beta = 0.29$, $p < 0.001$) and Spontaneous style ($\beta = 0.30$, $p = 0.002$) were the strongest predictors. IH had a positive effect ($\beta = 0.19$, $p = 0.012$). Relative to female participants, male participants had higher weakening odds ($\beta = 0.253$, SE $= 0.106$, $z = 2.38$, $p = 0.017$, OR $= 1.29$, 95\% CI $[1.05, 1.59]$).

Table~\ref{tab:nationality-weakening} reports nationality coefficients under all six possible reference groups. South Africa showed the highest weakening rate. It was higher than Australia ($\beta = 0.80$, $p = 0.009$), the Other group ($\beta = 0.48$, $p = 0.004$), and the United Kingdom at trend level ($\beta = 0.40$, $p = 0.055$). The United States also showed elevated weakening relative to Australia ($\beta = 0.58$, $p = 0.046$) and a trend relative to the Other group ($\beta = 0.27$, $p = 0.086$). No contrast between Canada, the United Kingdom, the United States, or the Other group reached significance. These results indicate that the nationality effect is primarily driven by South Africa, with a secondary contribution from the United States. All non-nationality coefficients are identical across reference groups because re-leveling only re-partitions the nationality intercepts.

\subsubsection{Treatment and social position}\label{app:weakening-quadrant}

Table~\ref{tab:quadrant_within} reports the within-treatment comparisons. Under Controversy, holding the minority position produced much more weakening than holding the majority position (27.3\% against 7.9\%). Under Certainty, weakening was higher when the in-group was less certain than the out-group (18.5\% against 5.3\%). Both contrasts were significant. Under Controversy+Certainty, the minority position again weakened more than the majority position in both certainty cells (18.1\% against 4.0\% when the in-group was equally or more certain than the out-group, and 20.3\% against 11.7\% when it was less certain). The certainty contrast was significant only within the majority position (11.7\% against 4.0\%). It was not significant within the minority position (20.3\% against 18.1\%).
\begin{table}[ht]
\centering
\caption{Within-treatment comparisons of weakening rates (two-sided Barnard exact tests, BH-corrected within each treatment). Controversy is compared on disagreement only, and Certainty on the ingroup--outgroup certainty contrast only, because each manipulates one dimension. Controversy+Certainty is compared across the full quadrant. A dash marks the dimension that was collapsed. Control is a single reference cell (5.3\%, $N = 1070$) and is not split.}
\label{tab:quadrant_within}
\small
\begin{tabular}{lllrrrrrr}
\toprule
Treatment & Comparison & Controlled & $N_1$ & Rate 1 (\%) & $N_2$ & Rate 2 (\%) & $p_{\mathrm{raw}}$ & $p_{\mathrm{BH}}$ \\
\midrule
Controversy & Maj. vs Min. & --- & 633 & 7.9 & 439 & 27.3 & $8.68 \times 10^{-18}$ & $8.68 \times 10^{-18}$ \\
\midrule
Certainty & In$\geq$Out vs In$<$Out & --- & 655 & 5.3 & 437 & 18.5 & $3.70 \times 10^{-12}$ & $3.70 \times 10^{-12}$ \\
\midrule
\multirow{4}{*}{Contr.+Cert.}
 & Maj. vs Min. & In$\geq$Out & 403 &  4.0 & 249 & 18.1 & $1.76 \times 10^{-9}$ & $7.03 \times 10^{-9}$ \\
 & Maj. vs Min. & In$<$Out    & 205 & 11.7 & 227 & 20.3 & $1.62 \times 10^{-2}$ & $2.15 \times 10^{-2}$ \\
 & In$\geq$Out vs In$<$Out & Maj. & 403 &  4.0 & 205 & 11.7 & $2.73 \times 10^{-4}$ & $5.46 \times 10^{-4}$ \\
 & In$\geq$Out vs In$<$Out & Min. & 249 & 18.1 & 227 & 20.3 & $5.64 \times 10^{-1}$ & $5.64 \times 10^{-1}$ \\
\bottomrule
\end{tabular}
\end{table}

\begin{table}[htbp]
\centering
\caption{Cross-treatment comparisons of weakening rates (two-sided Barnard exact tests, BH-corrected within each treatment pair, separated by the middle rule). Each single-manipulation treatment is collapsed to one cell along its manipulated dimension ($N_1$, Rate 1) and compared against each matching Controversy+Certainty subcell ($N_2$, Rate 2).}
\label{tab:quadrant_cross}
\small
\begin{tabular}{llrrrrrr}
\toprule
Single-treatment cell & C+C treatment cell & $N_1$ & Rate 1 (\%) & $N_2$ & Rate 2 (\%) & $p_{\mathrm{raw}}$ & $p_{\mathrm{BH}}$ \\
\midrule
\multirow{2}{*}{Controversy, Majority} & Maj., In$\geq$Out & \multirow{2}{*}{633} & \multirow{2}{*}{7.9}  & 403 &  4.0 & $1.16 \times 10^{-2}$ & $2.32 \times 10^{-2}$ \\
                                       & Maj., In$<$Out    &                      &                       & 205 & 11.7 & $9.56 \times 10^{-2}$ & $9.56 \times 10^{-2}$ \\
\multirow{2}{*}{Controversy, Minority} & Min., In$\geq$Out & \multirow{2}{*}{439} & \multirow{2}{*}{27.3} & 249 & 18.1 & $6.35 \times 10^{-3}$ & $2.32 \times 10^{-2}$ \\
                                       & Min., In$<$Out    &                      &                       & 227 & 20.3 & $4.70 \times 10^{-2}$ & $6.26 \times 10^{-2}$ \\
\midrule
\multirow{2}{*}{Certainty, In$\geq$Out} & Maj., In$\geq$Out & \multirow{2}{*}{655} & \multirow{2}{*}{5.3}  & 403 &  4.0 & $4.04 \times 10^{-1}$ & $5.38 \times 10^{-1}$ \\
                                        & Min., In$\geq$Out &                      &                       & 249 & 18.1 & $1.06 \times 10^{-8}$ & $4.24 \times 10^{-8}$ \\
\multirow{2}{*}{Certainty, In$<$Out}    & Maj., In$<$Out    & \multirow{2}{*}{437} & \multirow{2}{*}{18.5} & 205 & 11.7 & $2.97 \times 10^{-2}$ & $5.94 \times 10^{-2}$ \\
                                        & Min., In$<$Out    &                      &                       & 227 & 20.3 & $6.07 \times 10^{-1}$ & $6.07 \times 10^{-1}$ \\
\bottomrule
\end{tabular}
\end{table}

Table~\ref{tab:quadrant_cross} reports the cross-treatment comparisons. Controversy alone produced more weakening than Controversy+Certainty in three of the four subcells: the majority cell in which the in-group was equally or more certain than the out-group (7.9\% against 4.0\%), the corresponding minority cell (27.3\% against 18.1\%), and the minority cell in which the in-group was less certain (27.3\% against 20.3\%). The first two were significant; the third fell just short after correction ($p_{\mathrm{BH}} = .063$). The two conditions did not differ in the majority cell with the in-group less certain. The Certainty comparison was mixed. Adding controversy raised weakening sharply in the minority cell in which the in-group was equally or more certain than the out-group (18.1\% against 5.3\%), the only significant contrast in that family. In the majority cell with the in-group less certain, Certainty alone produced numerically more weakening than the combined condition (18.5\% against 11.7\%), but this did not survive correction ($p_{\mathrm{BH}} = .059$). The other two cells did not differ. Across the cells that differed, adding certainty to controversy lowered weakening rather than raising it.

\subsubsection{Regression robustness: continuous treatment variables versus binary social-position variables}
\label{app:weakening-social-position-ablation}

The primary weakening regression represents the displayed social information using continuous disagreement, in-group certainty, and out-group certainty. We tested whether the selected representation depended on retaining these continuous values. We derived two binary variables for each participant--dilemma observation: \emph{In the Minority} indicated that more than 50\% of AITA verdicts opposed the participant's initial verdict, and \emph{In $\geq$ Out} indicated that the in-group was equally or more certain than the out-group. Lower in-group certainty was coded as \emph{In $<$ Out}.

Within the GLM without dilemma as a factor, we added either the continuous or binary representation to the same base model containing initial confidence and the Controversy, Certainty, and Controversy$+$Certainty condition indicators. The continuous block contained the condition-specific disagreement, in-group-certainty, and out-group-certainty slopes used in the primary model. The binary block replaced these slopes with minority status in the Controversy condition, relative certainty in the Certainty condition, and both binary variables in the combined condition. The mutually exclusive blocks were evaluated using the same 10-fold cross-validation procedure and held-out mean log-likelihood used for the primary model selection.

\Cref{tab:weakening-social-position-representation-ablation} shows that both representations improved prediction over the base model. The binary social-position block increased held-out mean log-likelihood by $0.0182$, compared with $0.0227$ for the continuous treatment-information block. The continuous block therefore produced a larger gain ($+0.0045$) and was retained. Continuing the greedy ablation from this block reproduced the primary covariate selection and final held-out score reported in \Cref{tab:ablation-greedy-weakening} ($\bar{\ell}=-0.3062$). The binary variables capture the broad positions used in the descriptive comparisons, but the continuous variables preserve variation in the displayed percentages.

\begin{table}[htbp]
\centering
\caption{Robustness of the signal representation for weakening in the GLM without dilemma as a factor. The binary social-position and continuous treatment-information blocks were added separately to the same condition-indicator base model. $\bar{\ell}$ is the 10-fold cross-validated mean log-likelihood, SE is its standard error across folds, and $\Delta\bar{\ell}$ is the gain over the base model. Higher values indicate better held-out prediction.}
\label{tab:weakening-social-position-representation-ablation}
\small
\begin{tabular}{lccc}
\toprule
Signal representation & $\bar{\ell}$ & SE & $\Delta\bar{\ell}$ \\
\midrule
Base model & $-0.338$ & $0.0083$ & --- \\
Binary social position & $-0.319$ & $0.0079$ & $+0.0182$ \\
Continuous treatment information & $-0.315$ & $0.0075$ & $+0.0227$ \\
\bottomrule
\end{tabular}
\end{table}

\subsection{When do judgments strengthen?}
\label{app:strengthening}

This part gives the full analysis behind the strengthening results in the main text. The common regression specification is reported in Methods (\Cref{eq:model-spec-annotated}), and the detailed model-selection procedure is reported with the weakening analysis (\ref{app:weakening-selection-details}). The strengthening selection results are reported in \ref{app:strengthening-ablation}, followed by the selected-model coefficients in \ref{app:strengthening-coefficients}. We close with social-position comparisons and a robustness check testing whether continuous social signals predict judgment change better than their binary summaries (\Cref{app:strengthening-quadrant,app:strengthening-social-position-ablation}).

\subsubsection{Strengthening regression: model selection and estimates}
\label{app:strengthening-regression}

We applied the logistic specification in \Cref{eq:model-spec-annotated} to the binary strengthening outcome. A response was strengthened when the participant retained their verdict and reported higher confidence. The analysis was restricted to observations with initial confidence below the ceiling ($c < 4$), because confidence could not otherwise increase. This yielded $N = 1{,}924$ participant--dilemma observations, of which 391 (20.3\%) were strengthened. The coefficients were estimated separately from those in the weakening model.

\paragraph{Ablation results}\label{app:strengthening-ablation}

\textbf{Single-block addition.} Table~\ref{tab:ablation-single-strengthening} reports the marginal contribution of each covariate block for the strengthening outcome. GDMS was again the strongest single predictor across all three specifications ($\Delta\bar{\ell} \approx 0.003$--$0.005$, rank 1). IH ranked second ($\Delta\bar{\ell} \approx 0.001$). No other block produced a positive $\Delta\bar{\ell}$ in the GLM or \texttt{glmmTMB} specifications. In the \texttt{feglm} specification, Age and Sex showed small positive gains ($\Delta\bar{\ell} < 0.001$). Nationality and Country (both grouped and raw) consistently reduced fit across all specifications, unlike in the weakening analysis where grouped Nationality improved prediction. Topic produced the largest negative $\Delta\bar{\ell}$ in the GLM ($-0.050$) and \texttt{glmmTMB} ($-0.064$). The participant-level random intercept in \texttt{glmmTMB} also reduced fit ($\Delta\bar{\ell} = -0.005$).

\begin{table}[htbp]
\centering
\caption{Single-block ablation for strengthening: $\Delta\bar{\ell}$ (rank) for each covariate block across model specifications. The base row shows $\bar{\ell}$ (SE) of the base model. Restricted to participants with initial confidence $c < 4$ ($N = 1{,}924$).}
\label{tab:ablation-single-strengthening}
\small
\begin{tabular}{l ccc}
\toprule
 & GLM & \texttt{feglm} & \texttt{glmmTMB} \\
 & (no dilemma factor) & (dilemma FE) & (dilemma RE) \\
\midrule
Base $\bar{\ell}$ (SE)
  & $-0.462$ $(0.012)$ & $-0.509$ $(0.017)$ & $-0.463$ $(0.012)$ \\
\addlinespace
\multicolumn{4}{l}{\textit{Covariate block ($\Delta\bar{\ell}$, rank)}} \\
\addlinespace
GDMS
  & $+0.0045$ (1) & $+0.0026$ (1) & $+0.0046$ (1) \\
IH
  & $+0.0012$ (2) & $+0.0007$ (2) & $+0.0012$ (2) \\
Personality
  & $+0.0000$ (3) & $-0.0009$ (6) & $+0.0002$ (3) \\
Age
  & $-0.0000$ (5) & $+0.0002$ (3) & $-0.0000$ (5) \\
Sex
  & $-0.0002$ (6) & $+0.0001$ (4) & $-0.0002$ (6) \\
Nationality ($n{>}100$)
  & $-0.0006$ (7) & $-0.0021$ (7) & $-0.0006$ (7) \\
Country ($n{>}100$)
  & $-0.0010$ (8) & $-0.0029$ (8) & $-0.0009$ (8) \\
Topic
  & $-0.0501$ (9) & --- & $-0.0644$ (10) \\
Country (raw)
  & $-0.0571$ (10) & $-0.0462$ (9) & $-0.0838$ (11) \\
Nationality (raw)
  & $-0.103\phantom{0}$ (11) & $-0.0782$ (10) & $-0.146\phantom{0}$ (12) \\
RE: participant
  & --- & --- & $-0.0052$ (9) \\
\bottomrule
\end{tabular}

\vspace{4pt}
\raggedright\footnotesize
\textit{Note.} Specifications are defined in Table~\ref{tab:ablation-single-weakening}. The GLM with dilemma as a factor is omitted for the strengthening analysis because its results parallel the GLM without dilemma as a factor. ``---'' = block not applicable for that specification.
\end{table}

\textbf{Greedy forward selection.} Table~\ref{tab:ablation-greedy-strengthening} reports the greedy forward selection path for each specification. All three specifications selected the same two blocks in the same order: GDMS first, then IH. No further block improved fit. In the GLM (panel a), the best model was reached at step 2 ($\bar{\ell} = -0.4567$, SE $= 0.0116$). In the \texttt{feglm} specification (panel b), the best model was also reached at step 2 ($\bar{\ell} = -0.5059$, SE $= 0.0165$). In the \texttt{glmmTMB} specification (panel c), the best model was reached at step 2 ($\bar{\ell} = -0.4569$, SE $= 0.0116$). The \texttt{glmmTMB} coefficient estimates matched those of the GLM to at least three decimal places, and the dilemma-level random effect variance was again effectively zero (Variance $= 6.76 \times 10^{-9}$).

Compared to the weakening analysis, the strengthening model selected fewer covariates. Sex and Nationality, which improved weakening prediction, did not improve strengthening prediction. Only GDMS and IH generalized across both outcomes.

\begin{table}[htbp]
\centering
\caption{Greedy forward selection for strengthening: covariate block added at each step, resulting mean log-likelihood $\bar{\ell}$ (SE), and step-wise $\Delta\bar{\ell}$. Restricted to participants with initial confidence $c < 4$ ($N = 1{,}924$).}
\label{tab:ablation-greedy-strengthening}
\small
\begin{tabular}{cl cccc}
\toprule
Step & Block added & $\bar{\ell}$ & SE & $\Delta\bar{\ell}$ & Accepted \\
\midrule
\multicolumn{6}{l}{\textit{(a) GLM without dilemma as a factor}} \\
\addlinespace
0 & Base & $-0.4619$ & $0.0121$ & --- & --- \\
1 & GDMS & $-0.4575$ & $0.0119$ & $+0.0045$ & $\checkmark$ \\
2 & IH & $-0.4567$ & $0.0116$ & $+0.0013$ & $\checkmark$ \\
3 & combined(Topic, Pers., Age, Sex, Nat., Cntry.) & $-0.4590$ & $0.0113$ & $-0.0022$ & $\times$ \\
\addlinespace
\multicolumn{6}{l}{\textit{(b) \texttt{feglm} with dilemma fixed effects}} \\
\addlinespace
0 & Base & $-0.5094$ & $0.0174$ & --- & --- \\
1 & GDMS & $-0.5068$ & $0.0173$ & $+0.0026$ & $\checkmark$ \\
2 & IH & $-0.5059$ & $0.0165$ & $+0.0007$ & $\checkmark$ \\
3 & combined(Pers., Age, Sex, Cntry., Nat.) & $-0.5095$ & $0.0161$ & $-0.0036$ & $\times$ \\
\addlinespace
\multicolumn{6}{l}{\textit{(c) \texttt{glmmTMB} with dilemma random intercept}} \\
\addlinespace
0 & Base & $-0.4630$ & $0.0121$ & --- & --- \\
1 & GDMS & $-0.4583$ & $0.0120$ & $+0.0046$ & $\checkmark$ \\
2 & IH & $-0.4569$ & $0.0116$ & $+0.0013$ & $\checkmark$ \\
3 & combined(Topic, Pers., Age, Sex, Nat., Cntry., RE:part.) & $-0.4610$ & $0.0133$ & $-0.0041$ & $\times$ \\
\bottomrule
\end{tabular}

\vspace{4pt}
\raggedright\footnotesize
\textit{Note.} Nat.\ = Nationality ($n{>}100$). Cntry.\ = Country ($n{>}100$). Other abbreviations as in Table~\ref{tab:ablation-greedy-weakening}. SE is the standard error of $\bar{\ell}$ across the 10 cross-validation folds.
\end{table}

\paragraph{Best-model coefficients}\label{app:strengthening-coefficients}

Extended Data Table~\ref{tab:full-strengthening} reports the full coefficient estimates from the best strengthening model (GLM without dilemma as a factor; GDMS + IH). The outcome is the binary indicator for whether a participant's judgment was strengthened (verdict unchanged and confidence increased), restricted to participants with initial confidence below ceiling ($c < 4$).

The disagreement rate in the Controversy condition  was strongly negative ($\beta = -4.08$, $p < 0.001$, OR $= 0.02$), indicating that higher disagreement reduced the probability of strengthening. This is the mirror of the weakening result, where disagreement increased weakening. In-group certainty in the Certainty condition increased strengthening ($\beta = 1.82$, $p = 0.004$, OR $= 6.17$), while out-group certainty decreased it ($\beta = -1.21$, $p = 0.012$, OR $= 0.30$). Both effects reversed relative to weakening. In the combined condition, disagreement rate again reduced strengthening ($\beta = -3.45$, $p < 0.001$). Initial confidence had a stronger negative association with strengthening ($\beta = -0.65$, $p < 0.001$, OR $= 0.52$) than with weakening ($\beta = -0.21$), consistent with a ceiling effect: participants with lower confidence are more likely to strengthen their judgments.

The GDMS coefficients also showed notable reversals. Intuitive style was not significant for weakening but was the strongest GDMS predictor of strengthening ($\beta = 0.31$, $p < 0.001$, OR $= 1.36$). Dependent style was positive for weakening ($\beta = 0.29$) but negative for strengthening ($\beta = -0.21$, $p = 0.024$, OR $= 0.81$). Avoidant style, which was not significant for weakening, had a positive effect on strengthening ($\beta = 0.26$, $p = 0.046$, OR $= 1.30$). Spontaneous style, which was a strong predictor of weakening ($\beta = 0.30$, $p = 0.002$), showed only a trend for strengthening ($\beta = 0.21$, $p = 0.071$). IH had a positive effect on both outcomes (weakening: $\beta = 0.19$, $p = 0.012$; strengthening: $\beta = 0.24$, $p = 0.007$), making it the only covariate with a consistent direction across both forms of belief change.

\clearpage
\subsubsection{Strengthening by social position}\label{app:strengthening-quadrant}

Table~\ref{tab:quadrant_within_reinf} reports the within-treatment comparisons. Under Controversy, holding the majority position produced much more strengthening than holding the minority position (35.1\% against 14.4\%). Under Certainty, strengthening was higher when the in-group was equally or more certain than the out-group (32.4\% against 16.6\%). Both contrasts were significant. Under Controversy+Certainty, the majority position again strengthened more than the minority position in both certainty cells (34.9\% against 10.9\% when the in-group was equally or more certain than the out-group, and 22.5\% against 9.7\% when it was less certain). Both contrasts were significant. The certainty contrast was significant only within the majority position (34.9\% against 22.5\%). It was not significant within the minority position (10.9\% against 9.7\%).

\begin{table}[ht]
\centering
\caption{Within-treatment comparisons of strengthening rates (two-sided Barnard exact tests, BH-corrected within each treatment). The sample is restricted to participants whose initial confidence left room to increase. Controversy is compared on disagreement only, and Certainty on the ingroup and outgroup certainty contrast only, because each manipulates one dimension. Controversy+Certainty is compared across the full quadrant. A dash marks the dimension that was collapsed. Control is a single reference cell (8.6\%, $N = 463$) and is not split.}
\label{tab:quadrant_within_reinf}
\small
\begin{tabular}{lllrrrrrr}
\toprule
Treatment & Comparison & Controlled & $N_1$ & Rate 1 (\%) & $N_2$ & Rate 2 (\%) & $p_{\mathrm{raw}}$ & $p_{\mathrm{BH}}$ \\
\midrule
Controversy & Maj. vs Min. & --- & 271 & 35.1 & 216 & 14.4 & $2.02 \times 10^{-7}$ & $2.02 \times 10^{-7}$ \\
\midrule
Certainty & In$\geq$Out vs In$<$Out & --- & 262 & 32.4 & 199 & 16.6 & $1.12 \times 10^{-4}$ & $1.12 \times 10^{-4}$ \\
\midrule
\multirow{4}{*}{Contr.+Cert.}
 & Maj. vs Min. & In$\geq$Out & 169 & 34.9 & 129 & 10.9 & $1.56 \times 10^{-6}$ & $6.23 \times 10^{-6}$ \\
 & Maj. vs Min. & In$<$Out    & 102 & 22.5 & 113 &  9.7 & $1.05 \times 10^{-2}$ & $2.11 \times 10^{-2}$ \\
 & In$\geq$Out vs In$<$Out & Maj. & 169 & 34.9 & 102 & 22.5 & $3.30 \times 10^{-2}$ & $4.39 \times 10^{-2}$ \\
 & In$\geq$Out vs In$<$Out & Min. & 129 & 10.9 & 113 &  9.7 & $8.01 \times 10^{-1}$ & $8.01 \times 10^{-1}$ \\
\bottomrule
\end{tabular}
\end{table}

Table~\ref{tab:quadrant_cross_reinf} reports the cross-treatment comparisons. Controversy alone did not differ from Controversy+Certainty in any of the four subcells. Controversy produced numerically higher strengthening in every cell. No contrast reached significance after correction. The Certainty comparison was mostly null. Certainty alone produced much more strengthening than the combined condition in the minority cell in which the in-group was equally or more certain than the out-group (32.4\% against 10.9\%). This was the only significant cross-treatment contrast. The other three cells did not differ. Across the comparisons, combining the two signals never produced more strengthening than the stronger single signal.

\begin{table}[ht]
\centering
\caption{Cross-treatment comparisons of strengthening rates (two-sided Barnard exact tests, BH-corrected within each treatment pair, separated by the middle rule). Each single-manipulation treatment is collapsed to one cell along its manipulated dimension ($N_1$, Rate 1) and compared against each matching Controversy+Certainty subcell ($N_2$, Rate 2).}
\label{tab:quadrant_cross_reinf}
\small
\begin{tabular}{llrrrrrr}
\toprule
Single-treatment cell & C+C treatment cell & $N_1$ & Rate 1 (\%) & $N_2$ & Rate 2 (\%) & $p_{\mathrm{raw}}$ & $p_{\mathrm{BH}}$ \\
\midrule
\multirow{2}{*}{Controversy, Majority} & Maj., In$\geq$Out & \multirow{2}{*}{271} & \multirow{2}{*}{35.1} & 169 & 34.9 & $9.81 \times 10^{-1}$ & $9.81 \times 10^{-1}$ \\
                                       & Maj., In$<$Out    &                      &                       & 102 & 22.5 & $2.09 \times 10^{-2}$ & $8.35 \times 10^{-2}$ \\
\multirow{2}{*}{Controversy, Minority} & Min., In$\geq$Out & \multirow{2}{*}{216} & \multirow{2}{*}{14.4} & 129 & 10.9 & $4.22 \times 10^{-1}$ & $5.63 \times 10^{-1}$ \\
                                       & Min., In$<$Out    &                      &                       & 113 &  9.7 & $2.67 \times 10^{-1}$ & $5.34 \times 10^{-1}$ \\
\midrule
\multirow{2}{*}{Certainty, In$\geq$Out} & Maj., In$\geq$Out & \multirow{2}{*}{262} & \multirow{2}{*}{32.4} & 169 & 34.9 & $6.05 \times 10^{-1}$ & $6.05 \times 10^{-1}$ \\
                                        & Min., In$\geq$Out &                      &                       & 129 & 10.9 & $4.75 \times 10^{-6}$ & $1.90 \times 10^{-5}$ \\
\multirow{2}{*}{Certainty, In$<$Out}    & Maj., In$<$Out    & \multirow{2}{*}{199} & \multirow{2}{*}{16.6} & 102 & 22.5 & $2.24 \times 10^{-1}$ & $2.99 \times 10^{-1}$ \\
                                        & Min., In$<$Out    &                      &                       & 113 &  9.7 & $9.66 \times 10^{-2}$ & $1.93 \times 10^{-1}$ \\
\bottomrule
\end{tabular}
\end{table}

\subsubsection{Regression robustness: continuous treatment variables versus binary social-position variables}
\label{app:strengthening-social-position-ablation}

We repeated the representation check in \Cref{app:weakening-social-position-ablation} for strengthening. Within the GLM without dilemma as a factor, the continuous and binary blocks were added separately to the same base model containing initial confidence and the three condition indicators, then compared using 10-fold cross-validation. The binary block contained minority status in the Controversy condition, relative certainty in the Certainty condition, and both variables in the combined condition.

\Cref{tab:strengthening-social-position-representation-ablation} shows that both representations improved prediction over the base model. The binary social-position block increased held-out mean log-likelihood by $0.0181$, compared with $0.0223$ for the continuous treatment-information block. The continuous block therefore produced a larger gain ($+0.0042$) and was retained. Continuing the greedy ablation from this block reproduced the primary covariate selection and final held-out score reported in \Cref{tab:ablation-greedy-strengthening} ($\bar{\ell}=-0.4567$). As for weakening, the continuous variables performed better while preserving variation in the displayed percentages.

\begin{table}[htbp]
\centering
\caption{Robustness of the signal representation for strengthening in the GLM without dilemma as a factor. The binary social-position and continuous treatment-information blocks were added separately to the same condition-indicator base model. $\bar{\ell}$ is the 10-fold cross-validated mean log-likelihood, SE is its standard error across folds, and $\Delta\bar{\ell}$ is the gain over the base model. Higher values indicate better held-out prediction.}
\label{tab:strengthening-social-position-representation-ablation}
\small
\begin{tabular}{lccc}
\toprule
Signal representation & $\bar{\ell}$ & SE & $\Delta\bar{\ell}$ \\
\midrule
Base model & $-0.485$ & $0.0087$ & --- \\
Binary social position & $-0.467$ & $0.0107$ & $+0.0181$ \\
Continuous treatment information & $-0.462$ & $0.0121$ & $+0.0223$ \\
\bottomrule
\end{tabular}
\end{table}

\setcounter{secnumdepth}{3}
\clearpage
\subsection{Preregistered regression models for verdict and confidence change}
\label{app:j_c_change}
Our preregistration specified verdict change and confidence change as two separate outcomes. We fit both preregistered models and report them in full in \Cref{app:judgmentchange,app:confidencechange}. Their treatment effects agree with the combined regression outcomes of weakening and strengthening. The disagreement rate, in-group certainty, and out-group certainty act in the same direction and with similar significance across both the separate models and the combined outcome.

\subsubsection{Verdict-change regression}\label{app:judgmentchange}
Before presenting the equations, we define the coefficients and quantities used in the models. Let $\beta_0, \beta_1, \beta_2, \ldots$ denote the coefficients in the regression models for RQ1. These coefficients are specific to each model and are not shared across treatment conditions, even when the variables share a name. Let
\begin{itemize}
    \item $y_{ij}^b$ and $y_{ij}$ be the verdicts before and after treatment, which are 1 if participant $i$ chooses YA for dilemma $j$ and 0 if they choose NA;
    \item $t_{ij}^b$ and $t_{ij}$ be the certainty levels before and after treatment, which are both between 1 and 4;
    \item $C_{ij}$ be the binary indicator for the controversy treatment;
    \item $U_{ij}$ be the binary indicator for the certainty treatment;
    \item $B_{ij}$ be the binary indicator for the controversy and certainty treatment;
    \item $\mathbf{x}_{i}$ be the vector of all pre-treatment covariates for participant $i$, which include their demographic information, as well as big-five personality and decision style variables;
    \item $\text{Disagree}_{ij}$ is the percentage of Reddit users that made the opposite verdict to participant $i$'s initial verdict for dilemma $j$;
    \item $\text{InConf}_{ij}$ and (respectively $\text{OutConf}_{ij}$) be the percentage of Reddit users who were certain about the same verdict (resp. the opposite verdict) as participant $i$'s verdict for dilemma $j$.
\end{itemize}
The reported model for each outcome includes a selected subvector of $\mathbf{x}_i$. We write this selected subvector as $\mathbf{x}_i^{(J)}$ for verdict change.

Let $p_{ij}$ be the probability that participant $i$ changes their initial verdict for dilemma $j$, that is, $y_{ij} \neq y_{ij}^b$. We fit the following logistic regression:
\begin{align}
    \text{logit}(p_{ij})
    =\,& \tau_0 + \tau_1 t_{ij}^b + \boldsymbol{\tau_2}^T \mathbf{x}_i^{(J)} \nonumber \\
    & + \alpha_1 C_{ij} + \alpha_2 (C_{ij} \times \text{Disagree}_{ij}) \nonumber \\
    & + \beta_1 U_{ij} + \beta_2 (U_{ij} \times \text{InConf}_{ij}) + \beta_3 (U_{ij} \times \text{OutConf}_{ij}) \nonumber \\
    & + \gamma_1 B_{ij} + \gamma_2 (B_{ij} \times \text{Disagree}_{ij}) + \gamma_{3} (B_{ij} \times \text{InConf}_{ij}) + \gamma_{4} (B_{ij} \times \text{OutConf}_{ij}).
    \label{eq:causal_effect_judgment}
\end{align}
The selected covariate vector $\mathbf{x}_i^{(J)}$ contains the GDMS decision-style subscales, intellectual humility, sex, nationality (grouped to nationalities with more than 100 participants), and age. \textbf{\textit{1) Ablation}} and \textbf{\textit{2) Best Model - Full Coefficient Table}} below reports the selection path and the full coefficient table.

\paragraph{1) Ablation.}
\textbf{Single-block addition.} Table~\ref{tab:ablation-single-judgmentchange} reports the marginal contribution of each covariate block when added individually to the base model. We denote the change in mean log-likelihood relative to the base model as $\Delta\bar{\ell}$, where a positive value indicates improved out-of-sample prediction. GDMS was the strongest single predictor in three of the four specifications ($\Delta\bar{\ell}\approx0.004$). Grouped Nationality ($n{>}100$) ranked first in the GLM with dilemma factors, and grouped Country ($n{>}100$) ranked first in the \texttt{feglm} specification. IH, Age, and Sex produced only very small gains ($\Delta\bar{\ell}<0.001$). Personality contributed little and slightly reduced prediction in most specifications. Three blocks consistently reduced out-of-sample fit. Topic hurt prediction in all specifications where it was testable ($\Delta\bar{\ell}\approx-0.02$). Raw Country and raw Nationality variables produced substantial negative $\Delta\bar{\ell}$ values ($-0.012$ to $-0.070$), indicating overfitting on small nationality groups. In the \texttt{glmmTMB} specification, adding a participant-level random intercept produced the worst single-block result ($\Delta\bar{\ell}=-0.748$), confirming that this component strongly overfits the data.

\begin{table}[htbp]

\centering

\caption{Single-block ablation for verdict change: $\Delta\bar{\ell}$ (rank) for each covariate block across model specifications, where $\Delta\bar{\ell}$ is the change in 10-fold cross-validated mean log-likelihood relative to the base model. The base row shows $\bar{\ell}$ (SE) of the base model. All other rows show $\Delta\bar{\ell}$ relative to base, with rank in parentheses. Higher (less negative) values indicate better out-of-sample fit.}

\label{tab:ablation-single-judgmentchange}

\small

\begin{tabular}{l cccc}

\toprule

& GLM & GLM & \texttt{feglm} & \texttt{glmmTMB} \\

& (no post) & (post factor) & (post FE) & (post RE) \\

\midrule

Base $\bar{\ell}$ (SE)

& $-0.2065$ $(0.0109)$ & $-0.4060$ $(0.0371)$ & $-0.2087$ $(0.0113)$ & $-0.2070$ $(0.0110)$ \\

\addlinespace

\multicolumn{5}{l}{\textit{Covariate block ($\Delta\bar{\ell}$, rank)}} \\

\addlinespace

GDMS

& $+0.0044$ (1) & $+0.0043$ (2) & $+0.0037$ (3) & $+0.0043$ (1) \\

IH

& $+0.0006$ (4) & $+0.0005$ (4) & $-0.0003$ (8) & $+0.0006$ (4) \\

Age

& $+0.0002$ (5) & $+0.0004$ (5) & $+0.0001$ (5) & $+0.0001$ (5) \\

Nationality ($n{>}100$)

& $+0.0038$ (2) & $+0.0044$ (1) & $+0.0042$ (2) & $+0.0038$ (2) \\

Country ($n{>}100$)

& $+0.0032$ (3) & $+0.0035$ (3) & $+0.0046$ (1) & $+0.0032$ (3) \\

Sex

& $+0.0001$ (6) & $-0.0001$ (8) & $-0.0000$ (7) & $+0.0001$ (6) \\

Personality

& $-0.0009$ (8) & $-0.0006$ (9) & $+0.0008$ (4) & $-0.0009$ (8) \\

Topic

& $-0.0222$ (9) & $\phantom{+}0.0000$ (7) & --- & $-0.0255$ (9) \\

Country (raw)

& $-0.0301$ (10) & $-0.0155$ (10) & $-0.0120$ (9) & $-0.0434$ (10) \\

Nationality (raw)

& $-0.0508$ (11) & $-0.0318$ (11) & $-0.0378$ (10) & $-0.0703$ (11) \\

RE: participant

& --- & --- & --- & $-0.7476$ (12) \\

\bottomrule

\end{tabular}

\vspace{4pt}

\raggedright\footnotesize

\textit{Note.} GLM = generalized linear model with a binomial family and logit link. GLM without dilemma as a factor excludes dilemma indicators; GLM with dilemma as a factor includes dummy-coded dilemma indicators. \texttt{feglm} (dilemma FE) = fixed-effects logistic regression with dilemma absorbed via iterative demeaning. \texttt{glmmTMB} (dilemma RE) = generalized linear mixed model with dilemma as a random intercept. Nationality/Country ($n{>}100$) = grouped version retaining only groups with more than 100 observations. ``---'' = block not applicable for that specification. Topic is not available for \texttt{feglm} because it is absorbed into the dilemma fixed effect.

\end{table}

\textbf{Greedy forward selection.} Table~\ref{tab:ablation-greedy-judgmentchange} reports the greedy forward selection path for each specification. In the GLM without dilemma as a factor (panel a), GDMS entered first ($\Delta\bar{\ell}=+0.0044$), followed by Nationality ($+0.0019$), Sex ($+0.0009$), IH ($+0.0001$), and Age ($+0.0001$). The combined addition of the two remaining blocks (Topic, Personality) reduced fit. The best model was reached at step 5 ($\bar{\ell}=-0.1992$, SE $=0.0102$). In the \texttt{glm} with dilemma as a factor, Nationality entered first ($\Delta\bar{\ell}=+0.0044$), followed by GDMS ($+0.0022$) and Sex ($+0.0005$). The remaining four blocks (Topic, Personality, IH, Age) were jointly rejected. The best model was reached at step 3 ($\bar{\ell}=-0.3988$, SE $=0.0365$). The higher baseline loss compared with the GLM without dilemma as a factor reflects the cost of estimating 134 dilemma dummy coefficients from limited within-dilemma data in each training fold. In the \texttt{feglm} specification with dilemma fixed effects, Country entered first ($\Delta\bar{\ell}=+0.0046$), followed by GDMS ($+0.0024$) and Sex ($+0.0010$). Personality, IH, and Age were jointly rejected. The best model was reached at step 3 ($\bar{\ell}=-0.2007$, SE $=0.0109$). In the \texttt{glmmTMB} specification with dilemma as a random intercept, the selection path closely matched the GLM without dilemma as a factor. GDMS entered first ($\Delta\bar{\ell}=+0.0043$), followed by Nationality ($+0.0019$), Sex ($+0.0008$), IH ($+0.0001$), and Age ($+0.0001$). The combined final step, which included Topic, Personality, and a participant-level random intercept, produced a large drop ($\Delta\bar{\ell}=-0.733$), driven almost entirely by the participant random intercept.

In summary, the best model across all specifications was the GLM without dilemma as a factor at step 5 (GDMS + Nationality + Sex + IH + Age; $\bar{\ell}=-0.1992$, SE $=0.0102$). The \texttt{glmmTMB} model reached a nearly identical best-step value ($\bar{\ell}=-0.1997$, SE $=0.0103$), suggesting that the random effects add little predictive value. Across specifications, GDMS and grouped Country/Nationality variables emerged as the most consistent predictors of verdict change, whereas Topic, raw nationality variables, and participant-level random effects consistently impaired out-of-sample prediction.

\begin{table}[htbp]

\centering

\caption{Greedy forward selection for verdict change: covariate block added at each step, resulting mean log-likelihood $\bar{\ell}$ (SE), and step-wise $\Delta\bar{\ell}$. Steps marked with $\checkmark$ were accepted; steps marked with $\times$ were rejected. The best model is the step with the highest $\bar{\ell}$.}

\label{tab:ablation-greedy-judgmentchange}

\small

\begin{tabular}{cl cccc}

\toprule

Step & Block added & $\bar{\ell}$ & SE & $\Delta\bar{\ell}$ & Accepted \\

\midrule

\multicolumn{6}{l}{\textit{(a) GLM without dilemma as a factor}} \\

\addlinespace

0 & Base & $-0.2065$ & $0.0109$ & --- & --- \\

1 & GDMS & $-0.2021$ & $0.0105$ & $+0.0044$ & $\checkmark$ \\

2 & Nationality ($n{>}100$) & $-0.2002$ & $0.0105$ & $+0.0019$ & $\checkmark$ \\

3 & Sex & $-0.1993$ & $0.0103$ & $+0.0009$ & $\checkmark$ \\

4 & IH & $-0.1992$ & $0.0104$ & $+0.0001$ & $\checkmark$ \\

5 & Age & $-0.1992$ & $0.0102$ & $+0.0001$ & $\checkmark$ \\

6 & combined(Topic, Pers.) & $-0.2005$ & $0.0100$ & $-0.0013$ & $\times$ \\

\addlinespace

\multicolumn{6}{l}{\textit{(b) GLM with dilemma as factor}} \\

\addlinespace

0 & Base & $-0.4060$ & $0.0371$ & --- & --- \\

1 & Nationality ($n{>}100$) & $-0.4015$ & $0.0365$ & $+0.0044$ & $\checkmark$ \\

2 & GDMS & $-0.3994$ & $0.0364$ & $+0.0022$ & $\checkmark$ \\

3 & Sex & $-0.3988$ & $0.0365$ & $+0.0005$ & $\checkmark$ \\

4 & combined(Topic, Pers., IH, Age) & $-0.3990$ & $0.0363$ & $-0.0002$ & $\times$ \\

\addlinespace

\multicolumn{6}{l}{\textit{(c) \texttt{feglm} with dilemma fixed effects}} \\

\addlinespace

0 & Base & $-0.2087$ & $0.0113$ & --- & --- \\

1 & Country ($n{>}100$) & $-0.2041$ & $0.0112$ & $+0.0046$ & $\checkmark$ \\

2 & GDMS & $-0.2017$ & $0.0109$ & $+0.0024$ & $\checkmark$ \\

3 & Sex & $-0.2007$ & $0.0109$ & $+0.0010$ & $\checkmark$ \\

4 & combined(Pers., IH, Age) & $-0.2008$ & $0.0108$ & $-0.0002$ & $\times$ \\

\addlinespace

\multicolumn{6}{l}{\textit{(d) \texttt{glmmTMB} with dilemma random intercept}} \\

\addlinespace

0 & Base & $-0.2070$ & $0.0110$ & --- & --- \\

1 & GDMS & $-0.2026$ & $0.0107$ & $+0.0043$ & $\checkmark$ \\

2 & Nationality ($n{>}100$) & $-0.2007$ & $0.0106$ & $+0.0019$ & $\checkmark$ \\

3 & Sex & $-0.1999$ & $0.0104$ & $+0.0008$ & $\checkmark$ \\

4 & IH & $-0.1998$ & $0.0105$ & $+0.0001$ & $\checkmark$ \\

5 & Age & $-0.1997$ & $0.0103$ & $+0.0001$ & $\checkmark$ \\

6 & combined(Topic, Pers., RE:part.) & $-0.9328$ & $0.1731$ & $-0.7331$ & $\times$ \\

\bottomrule

\end{tabular}

\vspace{4pt}

\raggedright\footnotesize

\textit{Note.} ``combined(...)'' denotes the joint addition of all remaining blocks that individually failed to improve fit. Pers.\ = Big Five Personality traits. RE:part.\ = participant-level random intercept. SE is the standard error of $\bar{\ell}$ across the 10 cross-validation folds.

\end{table}

\paragraph{2) Best Model - Full Coefficient Table}

\begin{table}[!htbp]
\centering
\renewcommand{\arraystretch}{1.2}
\setlength{\tabcolsep}{5pt}
\resizebox{\linewidth}{!}{
\begin{tabular}{lcccccc}
\toprule
\textbf{Predictor}
& $\hat{\beta}$ & SE & $z$ & $p$ & OR & 95\% CI \\
\midrule
\multicolumn{7}{l}{\textbf{Main predictors}} \\[4pt]
\textit{Pre-treatment} \\
(Intercept)
& $-4.744$ & $0.654$ & $-7.26$ & $3.9\times10^{-13}$
& $0.009$ & $[0.002,\;0.03]$ \\
Initial confidence ($c$)
& $-1.873$ & $0.252$ & $-7.42$ & $1.2\times10^{-13}$
& $0.15$ & $[0.09,\;0.25]$ \\[6pt]
\textit{Controversy treatment} \\
Controversy
& $-1.163$ & $0.424$ & $-2.75$ & $0.006$
& $0.31$ & $[0.14,\;0.72]$ \\
Disagreement rate (controversy:disagree)
& $4.461$ & $0.645$ & $6.92$ & $4.6\times10^{-12}$
& $86.59$ & $[24.47,\;306.45]$ \\[6pt]
\textit{Certainty treatment} \\
Certainty
& $0.596$ & $0.716$ & $0.83$ & $0.405$
& $1.81$ & $[0.45,\;7.37]$ \\
In-group certainty (certainty:inconf)
& $-1.380$ & $0.585$ & $-2.36$ & $0.018$
& $0.25$ & $[0.08,\;0.79]$ \\
Out-group certainty (certainty:outconf)
& $1.672$ & $0.672$ & $2.49$ & $0.013$
& $5.32$ & $[1.43,\;19.88]$ \\[6pt]
\textit{Controversy + Certainty treatment} \\
Both
& $-0.260$ & $0.801$ & $-0.33$ & $0.745$
& $0.77$ & $[0.16,\;3.71]$ \\
Disagreement rate (disagree:both)
& $4.000$ & $0.858$ & $4.66$ & $3.2\times10^{-6}$
& $54.60$ & $[10.16,\;293.36]$ \\
In-group certainty (inconf:both)
& $-1.781$ & $0.698$ & $-2.55$ & $0.011$
& $0.17$ & $[0.04,\;0.66]$ \\
Out-group certainty (outconf:both)
& $-0.086$ & $0.710$ & $-0.12$ & $0.903$
& $0.92$ & $[0.23,\;3.69]$ \\[10pt]
\multicolumn{7}{l}{\textbf{Covariates}} \\[4pt]
\textit{Decision-making style (GDMS)} \\
Intuitive
& $0.144$ & $0.386$ & $0.37$ & $0.709$
& $1.15$ & $[0.54,\;2.46]$ \\
Dependent
& $0.534$ & $0.400$ & $1.33$ & $0.182$
& $1.71$ & $[0.78,\;3.74]$ \\
Rational
& $1.045$ & $0.534$ & $1.96$ & $0.050$
& $2.84$ & $[1.00,\;8.10]$ \\
Avoidant
& $-0.115$ & $0.573$ & $-0.20$ & $0.841$
& $0.89$ & $[0.29,\;2.74]$ \\
Spontaneous
& $1.178$ & $0.494$ & $2.38$ & $0.017$
& $3.25$ & $[1.23,\;8.56]$ \\[6pt]
\textit{Demographics} \\
Nationality (Base: Australia) \\
\hspace{3mm}Canada
& $0.359$ & $0.492$ & $0.73$ & $0.466$
& $1.43$ & $[0.55,\;3.76]$ \\
\hspace{3mm}South Africa
& $1.207$ & $0.425$ & $2.84$ & $0.005$
& $3.34$ & $[1.45,\;7.68]$ \\
\hspace{3mm}United Kingdom
& $-0.049$ & $0.482$ & $-0.10$ & $0.919$
& $0.95$ & $[0.37,\;2.45]$ \\
\hspace{3mm}United States
& $0.560$ & $0.413$ & $1.36$ & $0.175$
& $1.75$ & $[0.78,\;3.94]$ \\
\hspace{3mm}Other
& $0.389$ & $0.433$ & $0.90$ & $0.369$
& $1.48$ & $[0.63,\;3.45]$ \\
Sex (Base: Female) \\
\hspace{3mm}Male
& $0.430$ & $0.140$ & $3.07$ & $0.002$
& $1.54$ & $[1.17,\;2.02]$ \\
\hspace{3mm}Other
& $-11.977$ & $423.661$ & $-0.03$ & $0.977$
& $0.00$ & -- \\
Age (scaled)
& $0.098$ & $0.074$ & $1.32$ & $0.186$
& $1.10$ & $[0.95,\;1.28]$ \\[6pt]
\textit{Intellectual humility} \\
IH
& $0.644$ & $0.399$ & $1.61$ & $0.107$
& $1.90$ & $[0.87,\;4.17]$ \\
\midrule
\multicolumn{7}{l}{$N = 4{,}318$. Null deviance: $1948.6$ on $4317$ df. Residual deviance: $1669.7$ on $4293$ df. AIC: $1719.7$.} \\
\bottomrule
\end{tabular}
}
\caption{Predicting verdict change: coefficient estimates, odds ratios (OR), and 95\% confidence intervals from the best logistic regression model (GLM without dilemma as a factor; covariates: GDMS, Nationality, Sex, IH, Age). The control condition serves as the baseline: when all treatment indicators equal zero, the intercept gives the log-odds of verdict change for a control-condition participant with all continuous/scaled predictors at zero. OR $= \exp(\hat{\beta})$. CI $= \exp(\hat{\beta} \pm 1.96 \times \text{SE})$.}
\label{tab:coefficients-judgment-change}
\end{table}

Table~\ref{tab:coefficients-judgment-change} reports the full coefficient estimates from the best verdict change model (GLM without dilemma as a factor; GDMS + Nationality + Sex + IH + Age). The outcome is the binary indicator for whether a participant changed their verdict.

The treatment effects show that structural social information strongly drives verdict revisions. The disagreement rate in the Controversy condition was strongly positive ($\beta = 4.46$, $p < 0.001$, OR $= 86.59$), indicating that higher disagreement massively increased the probability of changing one's verdict. Similarly, in the combined condition, the disagreement rate significantly elevated verdict change odds ($\beta = 4.00$, $p < 0.001$, OR $= 54.60$). When participants were exposed to certainty information alone, higher in-group certainty significantly reduced the odds of changing verdicts ($\beta = -1.38$, $p = 0.018$, OR $= 0.25$), whereas higher out-group certainty had the opposite effect, increasing verdict changes ($\beta = 1.67$, $p = 0.013$, OR $= 5.32$). This pattern is reinforced in the combined condition for in-group certainty ($\beta = -1.78$, $p = 0.011$, OR $= 0.17$), though the out-group certainty effect vanishes when paired with controversy ($\beta = -0.09$, $p = 0.903$). Notably, initial confidence had a powerful negative association with verdict change ($\beta = -1.87$, $p < 0.001$, OR $= 0.15$), confirming that highly confident individuals are deeply resistant to modifying their positions.

Among the individual difference covariates, the GDMS subscales yielded distinct profiles. Spontaneous decision-making style was a strong and significant positive predictor of shifting verdicts ($\beta = 1.18$, $p = 0.017$, OR $= 3.25$), while a Rational style showed a strong positive trend on the margin of significance ($\beta = 1.05$, $p = 0.050$, OR $= 2.84$). Intuitive, Dependent, and Avoidant styles did not significantly influence verdict change. Demographic factors also revealed localized variations: South African participants were substantially more likely to change verdicts compared to the Australian baseline ($\beta = 1.21$, $p = 0.005$, OR $= 3.34$), and male participants showed higher odds of changing their minds relative to females ($\beta = 0.43$, $p = 0.002$, OR $= 1.54$). Neither intellectual humility (IH) nor age reached statistical significance within this broader model structure.

\subsubsection{Confidence-change regression}\label{app:confidencechange}

We model how confidence moved among participants who kept their initial verdict. This conditioning separates the two preregistered outcomes cleanly. A verdict change is already counted as weakening by the verdict change model. The confidence change model then asks, for the responses where verdict held, whether confidence changed.

We define the quantities used below. The $\beta$ coefficients are specific to this model and are not shared with the verdict change model, even where variables share a name. Let
\begin{itemize}
    \item $y_{ij}^b$ and $y_{ij}$ be the verdicts before and after treatment, which are 1 if participant $i$ chooses YA for dilemma $j$ and 0 if they choose NA;
    \item $t_{ij}^b$ and $t_{ij}$ be the certainty levels before and after treatment, both between 1 and 4;
    \item $C_{ij}$, $U_{ij}$, and $B_{ij}$ be the binary indicators for the controversy, certainty, and combined treatments;
    \item $\mathbf{x}_{i}$ be the vector of pre-treatment covariates for participant $i$, covering demographics, big-five personality, and decision style;
    \item $\text{Disagree}_{ij}$ be the percentage of Reddit users who made the opposite verdict to participant $i$'s initial verdict for dilemma $j$;
    \item $\text{InConf}_{ij}$ and $\text{OutConf}_{ij}$ be the percentages of Reddit users who were certain about the same verdict and the opposite verdict, respectively.
\end{itemize}

The model is fit on the subset of responses where the participant kept their initial verdict, that is, $y_{ij} = y_{ij}^b$. The outcome is the signed change in certainty within this subset, $\Delta t_{ij} = t_{ij} - t_{ij}^b$. Negative values denote reduced confidence. Positive values denote increased confidence. We fit the following linear model:
\begin{align}
    \Delta t_{ij}
    =\,& \tau_0 + \tau_1 t_{ij}^b + \boldsymbol{\tau_2}^T \mathbf{x}_i^{(C)} \nonumber \\
    & + \alpha_1 C_{ij} + \alpha_2 (C_{ij} \times \text{Disagree}_{ij}) \nonumber \\
    & + \beta_1 U_{ij} + \beta_2 (U_{ij} \times \text{InConf}_{ij}) + \beta_3 (U_{ij} \times \text{OutConf}_{ij}) \nonumber \\
    & + \gamma_1 B_{ij} + \gamma_2 (B_{ij} \times \text{Disagree}_{ij}) + \gamma_{3} (B_{ij} \times \text{InConf}_{ij}) + \gamma_{4} (B_{ij} \times \text{OutConf}_{ij}) + \varepsilon_{ij}.
    \label{eq:causal_effect_confidence}
\end{align}
Here $\mathbf{x}_i^{(C)}$ is the selected covariate subvector. It contains the GDMS decision-style subscales and the big-five personality traits, chosen by cross-validated forward selection. Part \textbf{\textit{1) Ablation}} reports the selection path. Part \textbf{\textit{2) Best Model -- Full Coefficient Table}} reports the full coefficient table.

\paragraph{1) Ablation.}
\textbf{Single-block addition.} Table~\ref{tab:ablation-single-confidencechange} reports the marginal contribution of each covariate block when added individually to the base model. We denote the change in mean log-likelihood relative to the base model as $\Delta\bar{\ell}$, where a positive value indicates improved out-of-sample prediction. The mean log-likelihoods are positive for this outcome because the response is continuous. The Gaussian density can exceed one when the residual variance is small. This contrasts with the logistic models for verdict change, where per-observation log-likelihoods are always negative. GDMS was the strongest decision-relevant block. It ranked first in the \texttt{feols} specification and second in the other three ($\Delta\bar{\ell}\approx0.003$ to $0.004$). Personality ranked second or third in every specification ($\Delta\bar{\ell}\approx0.0015$ to $0.0017$). This differs from verdict change, where Personality reduced fit. For confidence change, Personality improved out-of-sample prediction in all four specifications. IH, Age, and Sex produced near-zero or slightly negative gains ($|\Delta\bar{\ell}|<0.001$). Grouped Country and Nationality ($n{>}100$) gave very small positive gains. Raw Nationality ranked first in the three specifications without dilemma fixed effects ($\Delta\bar{\ell}\approx+0.007$). It reversed sign under \texttt{feols} ($\Delta\bar{\ell}=-0.009$). This reversal indicates that the raw geographic effect was not stable across specifications. Raw Country reduced fit in all specifications ($\Delta\bar{\ell}\approx-0.003$). Topic reduced fit in the specifications where it was testable ($\Delta\bar{\ell}\approx-0.01$). In the \texttt{glmmTMB} specification, the participant-level random intercept reduced fit ($\Delta\bar{\ell}=-0.006$). The raw geographic variables overfit small groups and did not replicate under dilemma fixed effects. We therefore excluded them from the greedy candidate pool, following the same rule applied to verdict change.

\begin{table}[htbp]

\centering

\caption{Single-block ablation for confidence change: $\Delta\bar{\ell}$ (rank) for each covariate block across model specifications, where $\Delta\bar{\ell}$ is the change in 10-fold cross-validated mean log-likelihood relative to the base model. The base row shows $\bar{\ell}$ (SE) of the base model. All other rows show $\Delta\bar{\ell}$ relative to base, with rank in parentheses. Higher (less negative) values indicate better out-of-sample fit.}

\label{tab:ablation-single-confidencechange}

\small

\begin{tabular}{l cccc}

\toprule

& GLM & GLM & \texttt{feols} & \texttt{glmmTMB} \\

& (no post) & (post factor) & (post FE) & (post RE) \\

\midrule

Base $\bar{\ell}$ (SE)

& $0.6047$ $(0.0176)$ & $0.5844$ $(0.0180)$ & $0.5844$ $(0.0180)$ & $0.6047$ $(0.0176)$ \\

\addlinespace

\multicolumn{5}{l}{\textit{Covariate block ($\Delta\bar{\ell}$, rank)}} \\

\addlinespace

GDMS

& $+0.0037$ (2) & $+0.0027$ (2) & $+0.0027$ (1) & $+0.0037$ (2) \\

Personality

& $+0.0017$ (3) & $+0.0015$ (3) & $+0.0015$ (2) & $+0.0017$ (3) \\

IH

& $-0.0001$ (7) & $-0.0002$ (9) & $-0.0002$ (7) & $-0.0001$ (7) \\

Age

& $-0.0003$ (9) & $-0.0005$ (10) & $-0.0005$ (8) & $-0.0003$ (9) \\

Nationality ($n{>}100$)

& $+0.0000$ (5) & $+0.0001$ (5) & $+0.0001$ (4) & $+0.0000$ (5) \\

Country ($n{>}100$)

& $+0.0002$ (4) & $+0.0003$ (4) & $+0.0003$ (3) & $+0.0002$ (4) \\

Sex

& $-0.0001$ (8) & $-0.0001$ (8) & $-0.0001$ (6) & $-0.0001$ (8) \\

Topic

& $-0.0098$ (11) & $\phantom{+}0.0000$ (6) & --- & $-0.0101$ (12) \\

Country (raw)

& $-0.0027$ (10) & $-0.0033$ (11) & $-0.0033$ (9) & $-0.0027$ (10) \\

Nationality (raw)

& $+0.0072$ (1) & $+0.0070$ (1) & $-0.0092$ (10) & $+0.0071$ (1) \\

RE: participant

& --- & --- & --- & $-0.0057$ (11) \\

\bottomrule

\end{tabular}

\vspace{4pt}

\raggedright\footnotesize

\textit{Note.} GLM = generalized linear model with a Gaussian family and identity link. GLM without dilemma as a factor excludes dilemma indicators; GLM with dilemma as a factor includes dummy-coded dilemma indicators. \texttt{feols} (dilemma FE) = fixed-effects linear model with dilemma absorbed via iterative demeaning. \texttt{glmmTMB} (dilemma RE) = generalized linear mixed model with dilemma as a random intercept. Nationality/Country ($n{>}100$) = grouped version retaining only groups with more than 100 observations. ``---'' = block not applicable for that specification. Topic is not available for \texttt{feols} because it is absorbed into the dilemma fixed effect.

\end{table}

\textbf{Greedy forward selection.} Table~\ref{tab:ablation-greedy-confidencechange} reports the greedy forward selection path for each specification. The path was identical across all four. GDMS entered first ($\Delta\bar{\ell}\approx+0.003$ to $+0.004$). Personality entered second ($\Delta\bar{\ell}\approx+0.0002$ to $+0.0003$). No further block improved fit. The best model in every specification was base + GDMS + Personality. The GLM without dilemma as a factor reached $\bar{\ell}=0.6086$ (SE $=0.0175$). The \texttt{glmmTMB} specification reached the same value. The two dilemma-absorbing specifications reached lower values ($\bar{\ell}=0.5873$), because estimating dilemma effects costs predictive accuracy in each training fold. Topic and the participant-level random intercept were the most harmful rejected blocks. Adding Topic dropped $\bar{\ell}$ to $0.5983$ in the GLM without dilemma as a factor and to $0.5969$ in \texttt{glmmTMB}. Adding the participant random intercept dropped $\bar{\ell}$ to $0.6040$.

In summary, the best model across all specifications was the GLM without dilemma as a factor (base + GDMS + Personality; $\bar{\ell}=0.6086$, SE $=0.0175$). The \texttt{glmmTMB} model matched this value. Its dilemma-level random intercept variance was effectively zero ($1.39\times10^{-11}$). The random effect therefore added no predictive value. We selected the GLM without dilemma as a factor as the final model. This matched the choice made for verdict change. Across specifications, GDMS and Personality emerged as the only blocks that improved out-of-sample prediction, whereas Topic, raw geographic variables, and participant-level random effects consistently impaired it.

\begin{table}[htbp]

\centering

\caption{Greedy forward selection for confidence change: covariate block added at each step, resulting mean log-likelihood $\bar{\ell}$ (SE), and step-wise $\Delta\bar{\ell}$. Steps marked with $\checkmark$ were accepted; steps marked with $\times$ were rejected. The best model is the step with the highest $\bar{\ell}$.}

\label{tab:ablation-greedy-confidencechange}

\small

\begin{tabular}{cl cccc}

\toprule

Step & Block added & $\bar{\ell}$ & SE & $\Delta\bar{\ell}$ & Accepted \\

\midrule

\multicolumn{6}{l}{\textit{(a) GLM without dilemma as a factor}} \\

\addlinespace

0 & Base & $0.6047$ & $0.0176$ & --- & --- \\

1 & GDMS & $0.6084$ & $0.0168$ & $+0.0037$ & $\checkmark$ \\

2 & Personality & $0.6086$ & $0.0175$ & $+0.0002$ & $\checkmark$ \\

  & IH & $0.6086$ & $0.0175$ & $-0.0000$ & $\times$ \\

  & Sex & $0.6085$ & $0.0175$ & $-0.0002$ & $\times$ \\

  & Age & $0.6084$ & $0.0175$ & $-0.0002$ & $\times$ \\

  & Country ($n{>}100$) & $0.6078$ & $0.0175$ & $-0.0008$ & $\times$ \\

  & Topic & $0.5983$ & $0.0178$ & $-0.0104$ & $\times$ \\

\addlinespace

\multicolumn{6}{l}{\textit{(b) GLM with dilemma as factor}} \\

\addlinespace

0 & Base & $0.5844$ & $0.0180$ & --- & --- \\

1 & GDMS & $0.5870$ & $0.0172$ & $+0.0027$ & $\checkmark$ \\

2 & Personality & $0.5873$ & $0.0181$ & $+0.0003$ & $\checkmark$ \\

  & Topic & $0.5873$ & $0.0181$ & $\phantom{+}0.0000$ & $\times$ \\

  & IH & $0.5872$ & $0.0182$ & $-0.0001$ & $\times$ \\

  & Sex & $0.5871$ & $0.0181$ & $-0.0002$ & $\times$ \\

  & Age & $0.5870$ & $0.0181$ & $-0.0003$ & $\times$ \\

  & Country ($n{>}100$) & $0.5865$ & $0.0182$ & $-0.0008$ & $\times$ \\

\addlinespace

\multicolumn{6}{l}{\textit{(c) \texttt{feols} with dilemma fixed effects}} \\

\addlinespace

0 & Base & $0.5844$ & $0.0180$ & --- & --- \\

1 & GDMS & $0.5870$ & $0.0172$ & $+0.0027$ & $\checkmark$ \\

2 & Personality & $0.5873$ & $0.0181$ & $+0.0003$ & $\checkmark$ \\

  & IH & $0.5872$ & $0.0182$ & $-0.0001$ & $\times$ \\

  & Sex & $0.5871$ & $0.0181$ & $-0.0002$ & $\times$ \\

  & Age & $0.5870$ & $0.0181$ & $-0.0003$ & $\times$ \\

  & Country ($n{>}100$) & $0.5865$ & $0.0182$ & $-0.0008$ & $\times$ \\

\addlinespace

\multicolumn{6}{l}{\textit{(d) \texttt{glmmTMB} with dilemma random intercept}} \\

\addlinespace

0 & Base & $0.6047$ & $0.0176$ & --- & --- \\

1 & GDMS & $0.6084$ & $0.0168$ & $+0.0037$ & $\checkmark$ \\

2 & Personality & $0.6086$ & $0.0175$ & $+0.0002$ & $\checkmark$ \\

  & IH & $0.6086$ & $0.0175$ & $-0.0000$ & $\times$ \\

  & Sex & $0.6085$ & $0.0175$ & $-0.0002$ & $\times$ \\

  & Age & $0.6084$ & $0.0175$ & $-0.0002$ & $\times$ \\

  & Country ($n{>}100$) & $0.6078$ & $0.0175$ & $-0.0008$ & $\times$ \\

  & RE: participant & $0.6040$ & $0.0193$ & $-0.0047$ & $\times$ \\

  & Topic & $0.5969$ & $0.0179$ & $-0.0117$ & $\times$ \\

\bottomrule

\end{tabular}

\vspace{4pt}

\raggedright\footnotesize

\textit{Note.} Steps 0 to 2 were accepted. The remaining rows show every block tested at the third iteration, ordered by $\Delta\bar{\ell}$. None improved on the step-2 best, so selection stopped. RE: participant = participant-level random intercept. SE is the standard error of $\bar{\ell}$ across the 10 cross-validation folds.

\end{table}

\paragraph{2) Best Model - Full Coefficient Table}

\begin{table}[!htbp]
\centering
\renewcommand{\arraystretch}{1.2}
\setlength{\tabcolsep}{5pt}
\resizebox{\linewidth}{!}{
\begin{tabular}{lccccc}
\toprule
\textbf{Predictor}
& $\hat{\beta}$ & SE & $t$ & $p$ & 95\% CI \\
\midrule
\multicolumn{6}{l}{\textbf{Main predictors}} \\[4pt]
\textit{Pre-treatment} \\
(Intercept)
& $0.137$ & $0.017$ & $7.92$ & $3.1\times10^{-15}$ & $[0.103,\;0.171]$ \\
Initial confidence ($c$)
& $-0.220$ & $0.009$ & $-24.73$ & $<10^{-16}$ & $[-0.238,\;-0.203]$ \\[6pt]
\textit{Controversy treatment} \\
Controversy
& $0.096$ & $0.011$ & $8.65$ & $<10^{-16}$ & $[0.074,\;0.117]$ \\
Disagreement rate (controversy:disagree)
& $-0.186$ & $0.021$ & $-8.66$ & $<10^{-16}$ & $[-0.228,\;-0.144]$ \\[6pt]
\textit{Certainty treatment} \\
Certainty
& $-0.024$ & $0.023$ & $-1.04$ & $0.297$ & $[-0.069,\;0.021]$ \\
In-group certainty (certainty:inconf)
& $0.108$ & $0.022$ & $4.98$ & $6.7\times10^{-7}$ & $[0.065,\;0.150]$ \\
Out-group certainty (certainty:outconf)
& $-0.050$ & $0.019$ & $-2.64$ & $0.008$ & $[-0.088,\;-0.013]$ \\[6pt]
\textit{Controversy + Certainty treatment} \\
Both
& $0.015$ & $0.027$ & $0.56$ & $0.578$ & $[-0.038,\;0.068]$ \\
Disagreement rate (disagree:both)
& $-0.127$ & $0.022$ & $-5.83$ & $5.8\times10^{-9}$ & $[-0.170,\;-0.085]$ \\
In-group certainty (inconf:both)
& $0.084$ & $0.023$ & $3.66$ & $2.6\times10^{-4}$ & $[0.039,\;0.130]$ \\
Out-group certainty (outconf:both)
& $-0.030$ & $0.018$ & $-1.62$ & $0.105$ & $[-0.066,\;0.006]$ \\[10pt]
\multicolumn{6}{l}{\textbf{Covariates}} \\[4pt]
\textit{Decision-making style (GDMS)} \\
Intuitive
& $0.038$ & $0.011$ & $3.61$ & $3.1\times10^{-4}$ & $[0.017,\;0.059]$ \\
Dependent
& $-0.046$ & $0.013$ & $-3.60$ & $3.2\times10^{-4}$ & $[-0.072,\;-0.021]$ \\
Rational
& $-0.003$ & $0.017$ & $-0.17$ & $0.864$ & $[-0.036,\;0.030]$ \\
Avoidant
& $0.024$ & $0.018$ & $1.36$ & $0.175$ & $[-0.011,\;0.060]$ \\
Spontaneous
& $0.013$ & $0.016$ & $0.82$ & $0.412$ & $[-0.018,\;0.043]$ \\[6pt]
\textit{Big Five personality} \\
Openness
& $0.003$ & $0.011$ & $0.29$ & $0.771$ & $[-0.018,\;0.025]$ \\
Conscientiousness
& $0.000$ & $0.012$ & $0.01$ & $0.995$ & $[-0.023,\;0.023]$ \\
Extraversion
& $-0.013$ & $0.008$ & $-1.60$ & $0.110$ & $[-0.029,\;0.003]$ \\
Agreeableness
& $0.017$ & $0.011$ & $1.46$ & $0.144$ & $[-0.006,\;0.039]$ \\
Emotional Stability
& $0.023$ & $0.010$ & $2.23$ & $0.026$ & $[0.003,\;0.044]$ \\
\midrule
\multicolumn{6}{l}{$N = 4{,}061$. Null deviance: $82.99$ on $4060$ df. Residual deviance: $69.40$ on $4040$ df. AIC: $-4956.8$. Residual variance: $0.0172$.} \\
\bottomrule
\end{tabular}
}
\caption{Predicting confidence change: coefficient estimates, $t$ statistics, $p$ values, and 95\% confidence intervals from the best generalized linear model with a Gaussian family and identity link (GLM without dilemma as a factor; covariates: GDMS, Big Five personality). The control condition serves as the baseline: when all treatment indicators equal zero, the intercept gives the expected confidence change for a control-condition participant with all continuous predictors at zero. CI $= \hat{\beta} \pm 1.96 \times \text{SE}$.}
\label{tab:coefficients-confidence-change}
\end{table}

Table~\ref{tab:coefficients-confidence-change} reports the full coefficient estimates from the best confidence change model. The model is a generalized linear model with a Gaussian family and identity link, without dilemma as a factor. The covariates are GDMS and Big Five personality. The outcome is the signed change in confidence. Negative values denote reduced confidence. Positive values denote increased confidence.

The treatment effects show that the two information types moved confidence in opposite directions. The disagreement rate in the Controversy condition reduced confidence ($\beta = -0.186$, $p < 0.001$). Higher disagreement produced larger reductions in confidence. In the Certainty condition, in-group certainty raised confidence ($\beta = 0.108$, $p < 0.001$), whereas out-group certainty lowered it ($\beta = -0.050$, $p = 0.008$). The combined condition reproduced part of this pattern. Disagreement still reduced confidence ($\beta = -0.127$, $p < 0.001$). In-group certainty still raised confidence ($\beta = 0.084$, $p < 0.001$). Out-group certainty no longer mattered ($\beta = -0.030$, $p = 0.105$). Out-group certainty therefore dropped out when paired with controversy. This non-additive pattern matches the result reported for verdict change. Initial confidence had a strong negative association with confidence change ($\beta = -0.220$, $p < 0.001$). Participants who started with high confidence had little room to rise and tended to fall.

Among the covariates, two decision styles stood out. An Intuitive style raised confidence ($\beta = 0.038$, $p < 0.001$). A Dependent style lowered confidence ($\beta = -0.046$, $p < 0.001$). These two styles moved confidence in opposite directions. Emotional Stability raised confidence by a small amount ($\beta = 0.023$, $p = 0.026$). The remaining decision styles (Rational, Avoidant, Spontaneous) did not reach significance. The remaining Big Five traits (Openness, Conscientiousness, Extraversion, Agreeableness) also did not reach significance.

\clearpage
\setcounter{secnumdepth}{4}
\subsection{Exploratory analyses: written explanations, dilemma consequences, and participant feedback}\label{sec:deliberation_analysis}

This section provides full details for the exploratory analyses summarized in the main text and Methods. We first describe quality filtering for participants' optional written explanations (\Cref{subsec:quality_filtering}). Because these explanations were optional, we next model who chose to provide one (\Cref{subsec:willingness}). We then classify the explanations, first by general reasoning for change and then by whether participants referenced and accepted the displayed treatment information (\Cref{subsec:reasoning}). Next, we describe the consequences at stake in the dilemmas and explore their associations with judgment change (\Cref{subsec:consequence}). We close with participants' optional feedback about the survey (\Cref{app:overall_feedback}).

\subsubsection{Quality filtering of written explanations}\label{subsec:quality_filtering}
We clean the written explanations in three steps. We first check text length. We then apply two checks on the wording of the responses. The first wording check draws on an earlier manual review. The second uses an automated detector. Each step is described below.

We start with text length, since automatically generated answers tend to be longer and more uniform than human ones. We compare the length distribution of 2,769 human-written texts with texts produced by an LLM for the same questions and treatments. Human responses are short and varied. LLM samples are more uniform and much longer. Table~\ref{tab:length_stats} shows the character and word count distributions for each source.
\begin{table}[!htbp]
\centering
\caption{Length distribution of participant and LLM-generated written explanations}
\label{tab:length_stats}
\begin{tabular}{llrrrrrrr}
\toprule
Source & Type & mean & std & min & 25\% & 50\% & 75\% & max \\
\midrule
\multirow{2}{*}{Human}
       & Characters & 136.71 & 116.89 & 1   & 62  & 105 & 172 & 1391 \\
       & Words      & 25.23  & 21.31  & 0   & 11  & 19  & 32  & 268  \\
\midrule
\multirow{2}{*}{LLM}
       & Characters & 342.85 & 98.48  & 118 & 268 & 340 & 411 & 712  \\
       & Words      & 53.26  & 15.71  & 17  & 42  & 53  & 64  & 117  \\
\bottomrule
\end{tabular}
\end{table}

Length alone cannot catch every case, so we add two further checks on the wording of the responses. The first check builds on an earlier manual pass. In that pass, about 18 of 727 responses (2.48\%) looked suspicious. More than 2,000 responses remained unread, and manual review cannot scale to that number. We therefore turn to an automated tool. The second check uses GPTZero \citep{2026gptzero} to estimate the AI probability of each participant's text. For each participant, we merge all responses into a single string (1,542 participants in total) and process it with the GPTZero API. Table~\ref{tab:gptzero_stats} gives the probability distribution across all participants with non-empty text.
\begin{table}[!htbp]
\centering
\caption{GPTZero-estimated AI probability distribution (participant-level)}
\label{tab:gptzero_stats}
\begin{tabular}{lrrrrrrr}
\toprule
 mean & std & min & 25\% & 50\% & 75\% & max \\
\midrule
 0.0169 & 0.1042 & 0.0000 & 0.000005 & 0.000015 & 0.000086 & 0.999997 \\
\bottomrule
\end{tabular}
\end{table}

GPTZero flagged 21 of the 1,542 participants who provided an explanation (1.4\%). These had an AI probability above 0.5. We then inspected these cases by hand. Seven of them were very short, under 250 characters. This is well below the typical LLM length range, and they did not read like AI-generated text. We kept these seven and removed the remaining 14 participants, who had both a high AI probability and long text.

This leaves 1,528 participants with 2,744 written explanations. All subsequent analyses use this cleaned set.

We also record how often participants provided an explanation in each condition.
Of the 4,318 responses, 2,744 (63.5\%) included a written explanation.
The rate was similar across the four conditions.
It was 60.8\% in Control (651 of 1,070), 66.0\% in Controversy (708 of 1,072), 63.9\% in Certainty (698 of 1,092), and 63.4\% in Controversy + Certainty (687 of 1,084).

\subsubsection{Willingness to provide a written explanation}\label{subsec:willingness}

We ask who chooses to provide a written explanation. Of the 4,318 responses, 2,744 (63.5\%) included an explanation and 1,574 (36.5\%) did not.

\paragraph{Model Specification}\label{app:willingness}

We model whether a participant chose to provide a written explanation. The outcome is binary. It equals one when participant $i$ wrote an explanation for dilemma $j$, and zero otherwise. We denote it $D_{ij}$. The model uses the same treatment terms and pre-treatment covariates as the verdict change and confidence change models. We do not restate those definitions here. The treatment indicators $C_{ij}$, $U_{ij}$, and $B_{ij}$, the disagreement rate $\text{Disagree}_{ij}$, and the in-group and out-group certainty terms $\text{InConf}_{ij}$ and $\text{OutConf}_{ij}$ are defined in \Cref{app:confidencechange}. The $\beta$ coefficients below are specific to this model and are not shared with the other models, even where variables share a name.

The model is a multilevel logistic regression with a random intercept for dilemma. Let $u_j \sim \mathcal{N}(0, \sigma^2_{\text{dilemma}})$ be the dilemma-level random intercept, let $t_{ij}$ be the post-treatment confidence level, and let $\mathbf{x}_i^{(D)}$ be the selected covariate subvector. We fit the following model:
\begin{align}
    \operatorname{logit}\Pr(D_{ij}=1)
    =\,& \theta_0 + u_j + \boldsymbol{\theta}^{T}\mathbf{x}_i^{(D)} + \lambda\, t_{ij} \nonumber \\
    & + \alpha_1 C_{ij} + \alpha_2 (C_{ij} \times \text{Disagree}_{ij}) \nonumber \\
    & + \beta_1 U_{ij} + \beta_2 (U_{ij} \times \text{InConf}_{ij}) + \beta_3 (U_{ij} \times \text{OutConf}_{ij}) \nonumber \\
    & + \gamma_1 B_{ij} + \gamma_2 (B_{ij} \times \text{Disagree}_{ij}) + \gamma_{3} (B_{ij} \times \text{InConf}_{ij}) + \gamma_{4} (B_{ij} \times \text{OutConf}_{ij}).
    \label{eq:willingness}
\end{align}
The treatment terms and the dilemma random intercept form the base model. The covariate subvector $\mathbf{x}_i^{(D)}$ was chosen by cross-validated forward selection. Part \textbf{\textit{1) Ablation}} reports the selection path. Part \textbf{\textit{2) Best Model -- Full Coefficient Table}} reports the full coefficient table.

\subparagraph{1) Ablation.}
\textbf{Single-block addition.} Table~\ref{tab:ablation-single-willingness} reports the marginal contribution of each covariate block when added on its own to the base model. We ran the sweep in two specifications: a single-level GLM and a \texttt{glmmTMB} model with a dilemma-level random intercept. We denote the change in mean log-likelihood relative to base as $\Delta\bar{\ell}$, where a positive value indicates improved out-of-sample prediction. The mean log-likelihoods are negative because the outcome is binary. Age gave the largest single-block gain in both specifications ($\Delta\bar{\ell}=+0.0077$ in the GLM, $+0.0072$ in \texttt{glmmTMB}). GDMS ranked second in both. Grouped Nationality ($n{>}100$) ranked third in both. The two specifications differed on Personality. It reduced fit on its own in the GLM ($\Delta\bar{\ell}=-0.0007$) but improved fit in \texttt{glmmTMB} ($\Delta\bar{\ell}=+0.0010$). Raw Country and raw Nationality reduced fit sharply in both specifications. We excluded them from the greedy candidate pool. This followed the same rule applied to the verdict change and confidence change models.

\begin{table}[htbp]
\centering
\caption{Single-block ablation for willingness to provide a written explanation: $\Delta\bar{\ell}$ (rank) for each covariate block when added on its own to the base model, where $\Delta\bar{\ell}$ is the change in 10-fold cross-validated mean log-likelihood relative to base. The base row shows $\bar{\ell}$ (SE). All other rows show $\Delta\bar{\ell}$ relative to base, with rank in parentheses. Higher (less negative) values indicate better out-of-sample fit.}
\label{tab:ablation-single-willingness}
\small
\begin{tabular}{l cc}
\toprule
& GLM & \texttt{glmmTMB} \\
& (single level) & (dilemma RE) \\
\midrule
Base $\bar{\ell}$ (SE) & $-0.6567$ $(0.0032)$ & $-0.6531$ $(0.0030)$ \\
\addlinespace
\multicolumn{3}{l}{\textit{Covariate block ($\Delta\bar{\ell}$, rank)}} \\
\addlinespace
Age                       & $+0.0077$ (1)  & $+0.0072$ (1)  \\
GDMS                      & $+0.0068$ (2)  & $+0.0038$ (2)  \\
Nationality ($n{>}100$)   & $+0.0014$ (3)  & $+0.0016$ (3)  \\
Personality               & $-0.0007$ (10) & $+0.0010$ (4)  \\
IH                        & $+0.0000$ (6)  & $+0.0011$ (5)  \\
Sex                       & $+0.0004$ (4)  & $+0.0002$ (6)  \\
Post-treatment confidence & $+0.0001$ (5)  & $+0.0002$ (7)  \\
Change type               & $-0.0000$ (7)  & $-0.0004$ (8)  \\
Country ($n{>}100$)       & $-0.0001$ (8)  & $-0.0007$ (9)  \\
Pre-treatment confidence  & $-0.0002$ (9)  & $-0.0002$ (10) \\
Nationality (raw)         & $-0.0093$ (11) & $-0.0173$ (11) \\
Country (raw)             & $-0.0347$ (12) & $-0.0571$ (12) \\
\bottomrule
\end{tabular}
\vspace{4pt}
\raggedright\footnotesize
\textit{Note.} The GLM column is a true single-block sweep (each block added alone to base). The \texttt{glmmTMB} column reports the round-1 marginal gains from the greedy run, which are equivalent to single-block additions at the first step. The two specifications agreed on the strongest blocks (Age, GDMS, grouped Nationality). They differed on Personality, which reduced fit on its own in the GLM but improved fit in \texttt{glmmTMB}. Raw geographic variables reduced fit sharply in both and were excluded from the greedy candidate pool.
\end{table}

\textbf{Greedy forward selection.} Table~\ref{tab:ablation-greedy-willingness} reports the greedy forward selection path for both specifications. The two paths accepted the same seven blocks in the same order: Age, GDMS, grouped Nationality, Personality, IH, Sex, and post-treatment confidence. Pre-treatment confidence and change type were the first rejected blocks in both. The \texttt{glmmTMB} model reached a higher cross-validated log-likelihood than the GLM at every step. The final values were $\bar{\ell}=-0.6380$ for \texttt{glmmTMB} and $\bar{\ell}=-0.6403$ for the GLM. The estimated dilemma-level variance was $0.026$ (SD $=0.16$), so the dilemma grouping carried information. We therefore selected the \texttt{glmmTMB} model with the dilemma random intercept as the final model for its better out-of-sample fit. This differs from the confidence change model, where the dilemma random-intercept variance was near zero and the GLM was selected instead.

\begin{table}[htbp]
\centering
\caption{Greedy forward selection for willingness to provide a written explanation: covariate block added at each step, resulting mean log-likelihood $\bar{\ell}$ (SE), and step-wise $\Delta\bar{\ell}$. Steps marked with $\checkmark$ were accepted; steps marked with $\times$ were rejected. The best model is the step with the highest $\bar{\ell}$.}
\label{tab:ablation-greedy-willingness}
\small
\begin{tabular}{cl cccc}
\toprule
Step & Block added & $\bar{\ell}$ & SE & $\Delta\bar{\ell}$ & Accepted \\
\midrule
\multicolumn{6}{l}{\textit{(a) Single-level GLM}} \\
\addlinespace
0 & Base & $-0.6567$ & $0.0032$ & --- & --- \\
1 & Age & $-0.6491$ & $0.0040$ & $+0.0077$ & $\checkmark$ \\
2 & GDMS & $-0.6444$ & $0.0053$ & $+0.0047$ & $\checkmark$ \\
3 & Nationality ($n{>}100$) & $-0.6429$ & $0.0053$ & $+0.0015$ & $\checkmark$ \\
4 & Personality & $-0.6417$ & $0.0050$ & $+0.0012$ & $\checkmark$ \\
5 & IH & $-0.6407$ & $0.0053$ & $+0.0010$ & $\checkmark$ \\
6 & Sex & $-0.6405$ & $0.0054$ & $+0.0003$ & $\checkmark$ \\
7 & Post-treatment confidence & $-0.6403$ & $0.0054$ & $+0.0001$ & $\checkmark$ \\
  & Pre-treatment confidence & $-0.6405$ & $0.0055$ & $-0.0002$ & $\times$ \\
  & Change type & $-0.6407$ & $0.0056$ & $-0.0003$ & $\times$ \\
\addlinespace
\multicolumn{6}{l}{\textit{(b) Multilevel with dilemma random intercept}} \\
\addlinespace
0 & Base & $-0.6531$ & $0.0030$ & --- & --- \\
1 & Age & $-0.6459$ & $0.0039$ & $+0.0072$ & $\checkmark$ \\
2 & GDMS & $-0.6421$ & $0.0049$ & $+0.0038$ & $\checkmark$ \\
3 & Nationality ($n{>}100$) & $-0.6405$ & $0.0049$ & $+0.0016$ & $\checkmark$ \\
4 & Personality & $-0.6395$ & $0.0047$ & $+0.0010$ & $\checkmark$ \\
5 & IH & $-0.6384$ & $0.0051$ & $+0.0011$ & $\checkmark$ \\
6 & Sex & $-0.6381$ & $0.0052$ & $+0.0002$ & $\checkmark$ \\
7 & Post-treatment confidence & $-0.6380$ & $0.0052$ & $+0.0002$ & $\checkmark$ \\
  & Pre-treatment confidence & $-0.6381$ & $0.0052$ & $-0.0002$ & $\times$ \\
  & Change type & $-0.6383$ & $0.0054$ & $-0.0004$ & $\times$ \\
  & Country ($n{>}100$) & $-0.6387$ & $0.0054$ & $-0.0007$ & $\times$ \\
  & Nationality (raw) & $-0.6553$ & $0.0116$ & $-0.0173$ & $\times$ \\
  & Country (raw) & $-0.6950$ & $0.0215$ & $-0.0571$ & $\times$ \\
\bottomrule
\end{tabular}
\vspace{4pt}
\raggedright\footnotesize
\textit{Note.} The two specifications accepted the same seven blocks in the same order. The multilevel model reached a higher $\bar{\ell}$ at every step, so the dilemma random intercept was retained. SE is the standard error of $\bar{\ell}$ across the 10 cross-validation folds.
\end{table}

\subparagraph{2) Best Model - Full Coefficient Table}

\begin{table}[!htbp]
\centering
\renewcommand{\arraystretch}{0.9}
\setlength{\tabcolsep}{5pt}
\resizebox{\linewidth}{!}{
\begin{tabular}{lccccc}
\toprule
\textbf{Predictor} & $\hat{\beta}$ & SE & $z$ & $p$ & 95\% CI \\
\midrule
\multicolumn{6}{l}{\textbf{Main predictors}} \\[4pt]
\textit{Baseline} \\
(Intercept)
& $0.910$ & $0.407$ & $2.24$ & $0.025$ & $[0.112,\;1.708]$ \\
Post-treatment confidence ($t$)
& $0.098$ & $0.048$ & $2.04$ & $0.042$ & $[0.004,\;0.191]$ \\[6pt]
\textit{Controversy treatment} \\
Controversy
& $-0.124$ & $0.177$ & $-0.70$ & $0.484$ & $[-0.471,\;0.223]$ \\
Disagreement rate (controversy:disagree)
& $0.846$ & $0.342$ & $2.48$ & $0.013$ & $[0.177,\;1.516]$ \\[6pt]
\textit{Certainty treatment} \\
Certainty
& $0.587$ & $0.372$ & $1.58$ & $0.114$ & $[-0.142,\;1.316]$ \\
In-group certainty (certainty:inconf)
& $-0.309$ & $0.344$ & $-0.90$ & $0.370$ & $[-0.984,\;0.366]$ \\
Out-group certainty (certainty:outconf)
& $-0.322$ & $0.308$ & $-1.05$ & $0.296$ & $[-0.926,\;0.282]$ \\[6pt]
\textit{Controversy + Certainty treatment} \\
Both
& $0.644$ & $0.434$ & $1.49$ & $0.137$ & $[-0.206,\;1.494]$ \\
Disagreement rate (disagree:both)
& $-0.015$ & $0.342$ & $-0.04$ & $0.965$ & $[-0.685,\;0.655]$ \\
In-group certainty (inconf:both)
& $-0.452$ & $0.369$ & $-1.23$ & $0.221$ & $[-1.175,\;0.271]$ \\
Out-group certainty (outconf:both)
& $-0.247$ & $0.297$ & $-0.83$ & $0.407$ & $[-0.829,\;0.336]$ \\[10pt]
\multicolumn{6}{l}{\textbf{Covariates}} \\[4pt]
\textit{Decision-making style (GDMS)} \\
Intuitive
& $0.164$ & $0.045$ & $3.69$ & $2.3\times10^{-4}$ & $[0.077,\;0.252]$ \\
Rational
& $-0.390$ & $0.067$ & $-5.82$ & $5.8\times10^{-9}$ & $[-0.522,\;-0.259]$ \\
Dependent
& $-0.096$ & $0.051$ & $-1.87$ & $0.061$ & $[-0.196,\;0.005]$ \\
Avoidant
& $0.057$ & $0.072$ & $0.79$ & $0.429$ & $[-0.084,\;0.197]$ \\
Spontaneous
& $-0.072$ & $0.062$ & $-1.17$ & $0.242$ & $[-0.194,\;0.049]$ \\[6pt]
\textit{Big Five personality} \\
Openness
& $0.041$ & $0.031$ & $1.33$ & $0.184$ & $[-0.020,\;0.101]$ \\
Conscientiousness
& $-0.074$ & $0.032$ & $-2.32$ & $0.021$ & $[-0.136,\;-0.011]$ \\
Extraversion
& $0.017$ & $0.022$ & $0.75$ & $0.452$ & $[-0.027,\;0.060]$ \\
Agreeableness
& $-0.026$ & $0.032$ & $-0.83$ & $0.406$ & $[-0.088,\;0.036]$ \\
Emotional Stability
& $-0.063$ & $0.029$ & $-2.19$ & $0.029$ & $[-0.119,\;-0.006]$ \\[6pt]
\textit{Intellectual humility} \\
IH
& $0.174$ & $0.048$ & $3.63$ & $2.9\times10^{-4}$ & $[0.080,\;0.268]$ \\[6pt]
\textit{Demographics} \\
Age (scaled)
& $0.273$ & $0.038$ & $7.20$ & $6.2\times10^{-13}$ & $[0.199,\;0.347]$ \\
Sex: Male
& $-0.165$ & $0.069$ & $-2.41$ & $0.016$ & $[-0.300,\;-0.031]$ \\
Sex: Other
& $-0.519$ & $1.018$ & $-0.51$ & $0.610$ & $[-2.514,\;1.476]$ \\
Nationality: Australia
& $0.431$ & $0.166$ & $2.59$ & $0.009$ & $[0.105,\;0.756]$ \\
Nationality: Canada
& $0.243$ & $0.146$ & $1.67$ & $0.096$ & $[-0.043,\;0.528]$ \\
Nationality: South Africa
& $0.216$ & $0.101$ & $2.14$ & $0.032$ & $[0.018,\;0.413]$ \\
Nationality: United Kingdom
& $0.477$ & $0.123$ & $3.89$ & $9.9\times10^{-5}$ & $[0.237,\;0.718]$ \\
Nationality: Other
& $0.139$ & $0.096$ & $1.45$ & $0.146$ & $[-0.048,\;0.327]$ \\
\midrule
\multicolumn{6}{l}{Random intercept (post): variance $=0.026$, SD $=0.160$.} \\
\multicolumn{6}{l}{$N = 4{,}318$ responses across $135$ dilemmas. AIC $=5509.6$. BIC $=5707.1$. logLik $=-2723.8$.} \\
\bottomrule
\end{tabular}
}
\caption{Predicting willingness to provide a written explanation: coefficient estimates, $z$ statistics, $p$ values, and 95\% confidence intervals from the best model (multilevel logistic regression with a dilemma-level random intercept; covariates selected by cross-validated forward selection: GDMS, Big Five personality, IH, age, sex, grouped nationality). Estimates are log-odds. Sex uses Female as the reference category. Nationality uses the omitted group as the reference category. CI $= \hat{\beta} \pm 1.96 \times \text{SE}$.}
\label{tab:coefficients-willingness}
\end{table}

Table~\ref{tab:coefficients-willingness} reports the coefficient estimates from the best willingness model. The model is a multilevel logistic regression with a dilemma-level random intercept. The outcome equals one when a participant provided a written explanation.

Age was the strongest predictor. Older participants provided explanations more often ($\beta = 0.273$, $p < 0.001$). Two decision styles moved in opposite directions. An Intuitive style raised the chance of providing an explanation ($\beta = 0.164$, $p < 0.001$). A Rational style lowered it ($\beta = -0.390$, $p < 0.001$). Intellectual humility raised it ($\beta = 0.174$, $p < 0.001$). Higher Conscientiousness lowered it ($\beta = -0.074$, $p = 0.021$). Higher Emotional Stability also lowered it ($\beta = -0.063$, $p = 0.029$).

Sex and nationality served as controls. Male participants provided explanations less often than female participants ($\beta = -0.165$, $p = 0.016$). Participants from Australia, the United Kingdom, and South Africa provided explanations more often than the reference group, the United States (all $p < 0.05$). Post-treatment confidence had a small positive association with providing an explanation ($\beta = 0.098$, $p = 0.042$).

The treatment terms were mostly not significant. The three treatment indicators did not predict whether participants provided an explanation (Controversy $p = 0.484$; Certainty $p = 0.114$; Controversy + Certainty $p = 0.137$). Only one treatment term reached significance. Under the Controversy condition, a higher disagreement rate raised the chance of providing an explanation ($\beta = 0.846$, $p = 0.013$). The certainty slopes and the combined-condition slopes were not significant. Treatment assignment alone therefore did not drive explanation rates. Stable individual differences did most of the work. The dilemma-level random intercept had a variance of $0.026$ (SD $=0.16$), so baseline explanation rates varied modestly across the 135 dilemmas.

\clearpage
\subsubsection{Written explanations}\label{subsec:reasoning}

We classify the cleaned written explanations using two coding tasks. The first labels the kind of reasoning each explanation gives for change. The second labels whether the explanation refers to each treatment variable and whether it accepts that variable as relevant evidence. We describe annotation and validation in \ref{subsubsec:annotation}, followed by the general reasoning results in \ref{subsubsec:reasoning_results} and the treatment-reference results in \ref{subsec:tr_results}.

\paragraph{Annotation and validation}\label{subsubsec:annotation}

We annotate the cleaned written explanations for two coding tasks.
The first task assigns each explanation to one or more reasoning subtypes.
The second task records the participant's stance toward each treatment variable.
We describe each task and its validation in turn.

\subparagraph{Task 1: General reasoning for change}
To analyze the content of participants' written explanations ($n$ = 2,744), we construct a categorization scheme that captures the main forms of reasoning observed in these texts. Our goal is to create a set of categories that is expressive enough to cover the range of explanations while still being compact and reliable to code at scale.

We begin with an initial scheme generated through LLM-assisted exploration across a diverse subset of responses. This produces fifteen fine-grained reasoning patterns, which serve as a starting point for building a structured and interpretable hierarchy. Through iterative grouping, we merge related items and organize them into a two-level scheme. The top level contains four broad types of reasoning: lack of meaningful explanation, personal-based values or experience, social-influence-based reasoning, and situation-based reasoning. The second level refines these into specific subtypes that reflect the linguistic cues and themes present in the data. \textit{Moral conviction} and \textit{Lived experience} fall under internal reasoning. \textit{Conformity} and \textit{Resistance} represent external social influence. \textit{Fairness / Rights / Responsibility}, \textit{Harm / Protection / Emotional Impact}, and \textit{Context / Procedure / Communication} correspond to situation-based reasoning. Finally, textual responses with no further explanation are assigned to the \textit{NO-EXPLANATION} subtype. Including a \textit{NO-EXPLANATION} subtype is important because it prevents forcing a classification when no reasoning is provided and keeps the coding accurate. The absence of an explanation is also informative: it marks cases where participants do not engage in reflective reasoning and helps us later compare whether treatments, relative to control, prompt more substantive explanations.
Our human validation on more than one hundred explanations shows that this scheme captures all observed patterns, and no additional categories are required.

To annotate the full dataset, we use the structured prompt shown in \textit{Deliberation annotation prompt}, which guides the model to classify all subtypes simultaneously. The prompt provides background for the task, instructions for consistent coding, definitions for each subtype, and a JSON response format. We use GPT-4o mini with top\_p and temperature set to 0 for annotation.

\begin{tcolorbox}[colframe=black, colback=white, boxrule=0.5pt,
                  title=Deliberation annotation prompt, breakable]
\label{tab:deliberation_prompt}
\small
You will be shown a participant's written deliberation about whether and why they changed their judgment or confidence after viewing treatment information.

\medskip
\textbf{\textless background\textgreater} \\
Participants first view an ``AITA'' (Am I the Asshole) post that includes a title and story. They then make an initial judgment and give a confidence level. After this, they see treatment information about how the Reddit community judged the situation. The treatment varies by condition:
\begin{itemize}[nosep, leftmargin=1.5em]
  \item Controversy: the percentage of people who agree with the participant's judgment.
  \item Certainty: the percentage of people who are certain among those who agree, and the percentage who are certain among those who disagree.
  \item Controversy + Certainty: both controversy and certainty information.
  \item Control: no treatment information is shown.
\end{itemize}
After viewing (or not viewing) this information, participants may change their judgment or confidence. They then write a short deliberation explaining the reason for their decision. \\
\textbf{\textless/background\textgreater}

\medskip
\textbf{\textless instructions\textgreater} \\
Your task is to determine whether each reasoning subtype is present in the deliberation.

\medskip
Binary classification: \\
0 = not relevant \\
1 = relevant

\medskip
Rules: \\
- Classify every subtype (do not skip). \\
- Do not infer motives that are not stated; classify only what is explicitly mentioned or clearly implied. \\
- Mark a subtype as 1 (relevant) if it appears as a meaningful part of the reasoning, even if brief. \\
- If the deliberation provides no meaningful explanation, set ``0\_NO\_EXPLANATION'' = 1 and set all other subtypes to 0.

Return only the JSON object requested below. Do not add explanations. \\
\textbf{\textless/instructions\textgreater}

\medskip
\textbf{\textless coding\_scheme\textgreater}

\medskip
0.\ NO-EXPLANATION \\
No meaningful explanation. \\
Examples: ``Nothing changed.'' ``Same as before.'' ``No reason to change.'' \\[4pt]

1a.\ Moral conviction (INTERNAL REASONING) \\
Internal certainty or personal principles/values as justification. \\
Examples: ``I'm sure about my answer.'' ``My values don't change.'' \\[4pt]

1b.\ Lived experience (INTERNAL REASONING) \\
Past experiences used as justification. \\
Examples: ``As a parent, I understand this.'' ``I've been through something similar.'' \\[4pt]

2a.\ Conformity (EXTERNAL SOCIAL INFLUENCE) \\
Crowd opinions (treatment information) influences judgment or confidence (agreement increases confidence; disagreement creates doubt/change). \\
Examples: ``Since most people agreed, I feel more confident.'' ``Lots of people disagreed, so I became unsure.'' \\[4pt]

2b.\ Resistance (EXTERNAL SOCIAL INFLUENCE) \\
Explicit rejection/dismissal of crowd opinions (``don't'', ``won't'', ``irrelevant'', ``doesn't matter''). \\
Examples: ``I don't care what others think.'' ``The votes don't influence me.'' \\[4pt]

3a.\ Fairness / Rights / Responsibility (SITUATION-BASED MORAL REASONING) \\
Entitlement, obligations, rules, equality. \\
Examples: ``Both siblings should pay equally.'' ``It is her property, so it is her choice.'' \\[4pt]

3b.\ Harm / Protection / Emotional Impact (SITUATION-BASED MORAL REASONING) \\
Preventing harm, safety, emotional damage. \\
Examples: ``This could harm the child emotionally.'' ``Safety should come first.'' \\[4pt]

3c.\ Context / Procedure / Communication (SITUATION-BASED MORAL REASONING) \\
Intent, tone, missing info, how it was done. \\
Examples: ``The intention was good but execution was poor.'' ``It depends; we need more context.'' \\[4pt]

\textbf{\textless/coding\_scheme\textgreater}

\medskip
\textbf{\textless deliberation\textgreater} \\
post title: '\{post\_title\}' \\
treatment: '\{treatment\}' \\
initial judgment and confidence: '\{j\}', '\{c\}' \\
post-treatment judgment and confidence: '\{j\_2\}', '\{c\_2\}' \\
deliberation: '\{deliberation\}' \\
\textbf{\textless/deliberation\textgreater}

\medskip
\textbf{\textless response\_format\textgreater} \\
Return a JSON object in exactly this format, map each subtype name to 0 or 1: \\
\textbf{\textless/response\_format\textgreater}

\medskip
\begin{verbatim}
{
    "0_NO_EXPLANATION": "<0 or 1>",
    "1a_MORAL_CONVICTION": "<0 or 1>",
    "1b_LIVED_EXPERIENCE": "<0 or 1>",
    "2a_CONFORMITY": "<0 or 1>",
    "2b_RESISTANCE": "<0 or 1>",
    "3a_FAIRNESS_RIGHTS_RESPONSIBILITY": "<0 or 1>",
    "3b_HARM_PROTECTION_EMOTIONAL_IMPACT": "<0 or 1>",
    "3c_CONTEXT_PROCEDURE_COMMUNICATION": "<0 or 1>"
}
\end{verbatim}

\end{tcolorbox}

\begin{table}[!htbp]
\centering
\begin{tabular}{lrr}
\hline
\textbf{Subtype} & \textbf{Frequency} & \textbf{Percentage (\%)} \\
\hline
0\_NO\_EXPLANATION & 221 & 8.05 \\
1a\_MORAL\_CONVICTION & 1654 & 60.28 \\
1b\_LIVED\_EXPERIENCE & 109 & 3.97 \\
2a\_CONFORMITY & 358 & 13.05 \\
2b\_RESISTANCE & 229 & 8.35 \\
3a\_FAIRNESS\_RIGHTS\_RESPONSIBILITY & 950 & 34.62 \\
3b\_HARM\_PROTECTION\_EMOTIONAL\_IMPACT & 598 & 21.79 \\
3c\_CONTEXT\_PROCEDURE\_COMMUNICATION & 1007 & 36.70 \\
\hline
\end{tabular}
\caption{Frequency and percentage of each annotated explanation subtype.}
\label{tab:deliberation_frequency}
\end{table}

\textbf{Coverage evaluation.} We next describe the distribution of subtypes after annotation. The frequencies are summarized in \Cref{tab:deliberation_frequency}. Every explanation contains at least one interpretable element, further indicating that the coding scheme captures all observed patterns. The most common subtype is moral conviction, which appears in more than half of all explanations (60.28\%). Situation-based reasoning is also widespread: contextual or procedural considerations appear in 36.70\% of texts, and fairness, rights, or responsibility appear in 34.62\%. Harm-related reasoning appears in 21.79\% of cases. Social influence appears in two distinct forms: conformity (13.05\%) and resistance (8.35\%), showing that participants differ in whether they align with or reject community feedback. Personal lived experience is relatively rare (3.97\%), and 8.05\% of explanations contain no meaningful explanation. Together, the distribution suggests that participants draw on a mixture of internal principles, situational interpretation, and reactions to social information when explaining their decisions.

\textbf{Alignment evaluation.} To assess annotation quality, we randomly sampled 200 explanations from the full dataset and one expert manually annotated them without access to the LLM annotations. We then compared the expert labels against the GPT-4o mini labels using multilabel classification metrics. \Cref{tab:alignment} reports precision, recall, and F1 for each subtype. The macro-averaged F1 score is 0.69, and the weighted-averaged F1 score is 0.68. Subtypes related to social influence show the strongest alignment: conformity (F1 = 0.87) and resistance (F1 = 0.85). Harm-related reasoning (F1 = 0.73) and fairness (F1 = 0.72) also show acceptable agreement. Moral conviction (F1 = 0.59) and contextual reasoning (F1 = 0.62) show moderate agreement, likely because these subtypes require more subjective judgment about whether a statement reflects a personal principle or a situational interpretation. All subtypes substantially exceed a random baseline (macro F1 = 0.19), confirming that the LLM annotations capture meaningful structure in the written explanations.

\begin{table}[!htbp]
\centering
\caption{Alignment between GPT-4o mini and expert annotations on a random sample of 200 written explanations. The random baseline reports the expected F1 under independent random assignment at observed base rates.}
\label{tab:alignment}
\small
\begin{tabular}{lrrrr}
\toprule
\textbf{Subtype} & \textbf{Precision} & \textbf{Recall} & \textbf{F1} & \textbf{Random Baseline F1} \\
\midrule
No explanation     & 0.40 & 0.89 & 0.55 & 0.05 \\
Moral conviction   & 0.46 & 0.81 & 0.59 & 0.35 \\
Lived experience   & 0.83 & 0.45 & 0.59 & 0.06 \\
Conformity         & 0.89 & 0.86 & 0.87 & 0.14 \\
Resistance         & 1.00 & 0.73 & 0.85 & 0.08 \\
Fairness           & 0.66 & 0.79 & 0.72 & 0.30 \\
Harm               & 0.82 & 0.65 & 0.73 & 0.25 \\
Context            & 0.63 & 0.60 & 0.62 & 0.33 \\
\midrule
Macro average      & 0.71 & 0.72 & 0.69 & 0.19 \\
Weighted average   & 0.67 & 0.73 & 0.68 & --- \\
\bottomrule
\end{tabular}
\end{table}

\subparagraph{Task 2: Treatment reference and stance}\label{subsec:tr_prompt_eval}
We code each explanation for its stance toward the three treatment variables: disagreement rate, ingroup certainty, and outgroup certainty. Stance is how the participant treats the information, not whether the verdict or confidence changed. Each variable receives one of three labels. A label of 0 means the participant did not refer to the variable, or was not shown it. A label of 1 means the participant referred to the variable and treated it as legitimate, relevant evidence, whether or not the final judgment changed. A label of $-1$ means the participant referred to the variable and rejected its relevance or validity. We annotate every explanation with an LLM using the prompt shown in \textit{Treatment-reference annotation prompt}.

\begin{tcolorbox}[colframe=black, colback=white, boxrule=0.5pt,
                  title=Treatment-reference annotation prompt, breakable]
\label{tab:tr_prompt}
\small
You will be shown a participant's written deliberation explaining whether and why they changed their judgment or confidence after viewing social information about a Reddit AITA post.

\medskip
\textbf{\textless background\textgreater} \\
Each participant first read an AITA post and gave an initial judgment and confidence rating. They were then shown one or more pieces of treatment information about how other Reddit users responded, and asked to provide a final judgment, final confidence rating, and a written deliberation explaining their reasoning.

\medskip
Judgment labels:
\begin{itemize}[nosep, leftmargin=1.5em]
  \item NA = No, the individual reporting the situation is NOT an asshole.
  \item YA = Yes, the individual reporting the situation IS an asshole.
\end{itemize}

The three treatment variables are:
\begin{itemize}[nosep, leftmargin=1.5em]
  \item disagreement\_rate: the percentage of Reddit users who DISAGREE with the participant's initial judgment. (May be null if the participant was not shown this.)
  \item ingroup\_certainty: among Reddit users who AGREE with the participant's initial judgment, the percentage who expressed certainty. (May be null.)
  \item outgroup\_certainty: among Reddit users who DISAGREE with the participant's initial judgment, the percentage who expressed certainty. (May be null.)
\end{itemize}
\textbf{\textless/background\textgreater}

\medskip
\textbf{\textless task\textgreater} \\
For each of the three treatment variables, classify the participant's STANCE toward that piece of information. Stance is how the participant treats the information. It is not whether their judgment or confidence changed. Output one label per variable.

\medskip
Labels:
\begin{itemize}[nosep, leftmargin=1.5em]
  \item 0 = The participant does NOT reference this information, OR was not shown this variable (input value is null).
  \item 1 = ACCEPT. The participant references this information and treats it as legitimate, relevant evidence that bears on the question. This holds whether or not the information changed the final judgment or confidence.
  \item $-1$ = REJECT. The participant references this information and rejects its relevance or validity.
\end{itemize}
\textbf{\textless/task\textgreater}

\medskip
\textbf{\textless rules\textgreater}
\begin{enumerate}[nosep, leftmargin=1.5em]
  \item Classify all three variables (disagreement\_rate, ingroup\_certainty, outgroup\_certainty). Do not skip any.
  \item If a variable's input value is null, output 0 for that variable.
  \item The label measures stance, not movement. Judge stance only from the deliberation text. Do not read it off the final judgment or confidence numbers.
  \item ``Reference'' includes paraphrases and indirect mentions, not only literal numbers. For example, ``most people seemed to agree with me'' references disagreement\_rate.
  \item Output 0 when the participant does not reference the information, or the variable was not shown (null).
  \item Output 1 (accept) when the participant treats the information as legitimate, relevant evidence. A participant who takes the information seriously but keeps their original verdict is still 1.
  \item Output $-1$ (reject) when the participant rejects the information's relevance or validity. For example, ``Reddit users are biased,'' ``I do not care what the crowd thinks,'' or ``the percentages do not matter to me.''
\end{enumerate}
\textbf{\textless/rules\textgreater}

\medskip
\textbf{\textless input\textgreater}
\begin{verbatim}
initial_judgment: {j}
initial_confidence: {c}
final_judgment: {j_2}
final_confidence: {c_2}
treatment information shown to participant:
  disagreement_rate: {disagree}   (null if not shown)
  ingroup_certainty: {inconf}     (null if not shown)
  outgroup_certainty: {outconf}   (null if not shown)
deliberation: {deliberation}
\end{verbatim}
\textbf{\textless/input\textgreater}

\medskip
\textbf{\textless output\_format\textgreater} \\
Return a single JSON object with this exact schema, and nothing else:
\begin{verbatim}
{
  "disagreement_rate": <-1 | 0 | 1>,
  "ingroup_certainty": <-1 | 0 | 1>,
  "outgroup_certainty": <-1 | 0 | 1>
}
\end{verbatim}
\textbf{\textless/output\_format\textgreater}
\end{tcolorbox}

\begin{table}[!htbp]
\centering
\caption{Reference detection (label 0 versus non-zero). Precision, recall, and F1 are reported for the referenced class. Accuracy and macro-F1 are over both classes.}
\label{tab:tr_detection}
\small
\begin{tabular}{lrrrrrr}
\toprule
\textbf{Variable} & \textbf{n} & \textbf{Precision} & \textbf{Recall} & \textbf{F1} & \textbf{Accuracy} & \textbf{Macro F1} \\
\midrule
Disagreement rate  & 100 & 0.91 & 0.87 & 0.89 & 0.95 & 0.93 \\
Ingroup certainty  & 100 & 0.87 & 0.80 & 0.83 & 0.92 & 0.89 \\
Outgroup certainty & 100 & 0.80 & 0.67 & 0.73 & 0.91 & 0.84 \\
\bottomrule
\end{tabular}
\end{table}

\begin{table}[!htbp]
\centering
\caption{Stance classification (three classes: $-1$ reject, 0 none, 1 accept) against expert labels.}
\label{tab:tr_stance}
\small
\begin{tabular}{lrrrr}
\toprule
\textbf{Class} & \textbf{Precision} & \textbf{Recall} & \textbf{F1} & \textbf{Support} \\
\midrule
\multicolumn{5}{l}{\textit{Disagreement rate} ($n = 100$)} \\[2pt]
reject ($-1$)        & 0.83 & 0.71 & 0.77 & 7  \\
none ($0$)            & 0.96 & 0.97 & 0.97 & 77 \\
accept ($1$)           & 0.75 & 0.75 & 0.75 & 16 \\
Macro average      & 0.85 & 0.81 & 0.83 & 100 \\[4pt]
\multicolumn{5}{l}{\textit{Ingroup certainty} ($n = 100$)} \\[2pt]
reject ($-1$)        & 0.80 & 0.50 & 0.62 & 8  \\
none ($0$)             & 0.94 & 0.96 & 0.95 & 75 \\
accept ($1$)          & 0.83 & 0.88 & 0.86 & 17 \\
Macro average      & 0.86 & 0.78 & 0.81 & 100 \\[4pt]
\multicolumn{5}{l}{\textit{Outgroup certainty} ($n = 100$)} \\[2pt]
reject ($-1$)        & 0.80 & 0.50 & 0.62 & 8  \\
none ($0$)             & 0.93 & 0.96 & 0.95 & 82 \\
accept ($1$)          & 0.70 & 0.70 & 0.70 & 10 \\
Macro average      & 0.81 & 0.72 & 0.75 & 100 \\
\bottomrule
\end{tabular}
\end{table}

\textbf{LLM Annotation.} We annotate every explanation with \texttt{gpt-5.4-mini}, using the prompt shown in \textit{Treatment-reference annotation prompt}. We use the default sampling settings (temperature $= 1$, \texttt{top\_p} unset), because the GPT-5 reasoning models accept only the default temperature of 1 through the API and return an error for any other value.

\textbf{Evaluation.} To check the annotations, one expert labeled a random sample of 150 explanations, without access to the model labels. The sample was balanced across the three treatments, with about 50 explanations from each. Each variable appears in two of the treatments. Disagreement rate appears in Controversy and Controversy + Certainty. Ingroup and outgroup certainty appear in Certainty and Controversy + Certainty. Each variable was therefore present in about 100 of the sampled explanations. We compare the model labels against the expert labels in two ways. First, we treat the task as detection: did the participant refer to the variable at all (label 0 versus a non-zero label)? Second, we keep all three classes (reject ($-1$), none ($0$), accept ($1$)).

Table~\ref{tab:tr_detection} reports the detection results on whether the participant refers to the variable at all. The model finds references reliably. Accuracy is 0.91 to 0.95 across the three variables. The F1 score for the referenced class is 0.89 for disagreement rate, 0.83 for ingroup certainty, and 0.73 for outgroup certainty.
Table~\ref{tab:tr_stance} reports the three-class results. Each label records whether the participant refers to the variable ($0$ if not), and, if they do, whether they accept it as relevant evidence ($1$) or reject it as evidence ($-1$). The macro-F1 score is 0.83 for disagreement rate, 0.81 for ingroup certainty, and 0.75 for outgroup certainty. The \textit{accept} class reaches F1 from 0.70 to 0.86, and the \textit{reject} class from 0.62 to 0.77.

Overall, the model detects variable references reliably and recovers the accept and reject stances well enough to annotate the full set. We therefore apply this prompt to all 2,744 explanations.

\paragraph{Results: General reasoning for change}\label{subsubsec:reasoning_results}
\begin{figure}[!htbp]
    \centering
    \includegraphics[width=\linewidth]{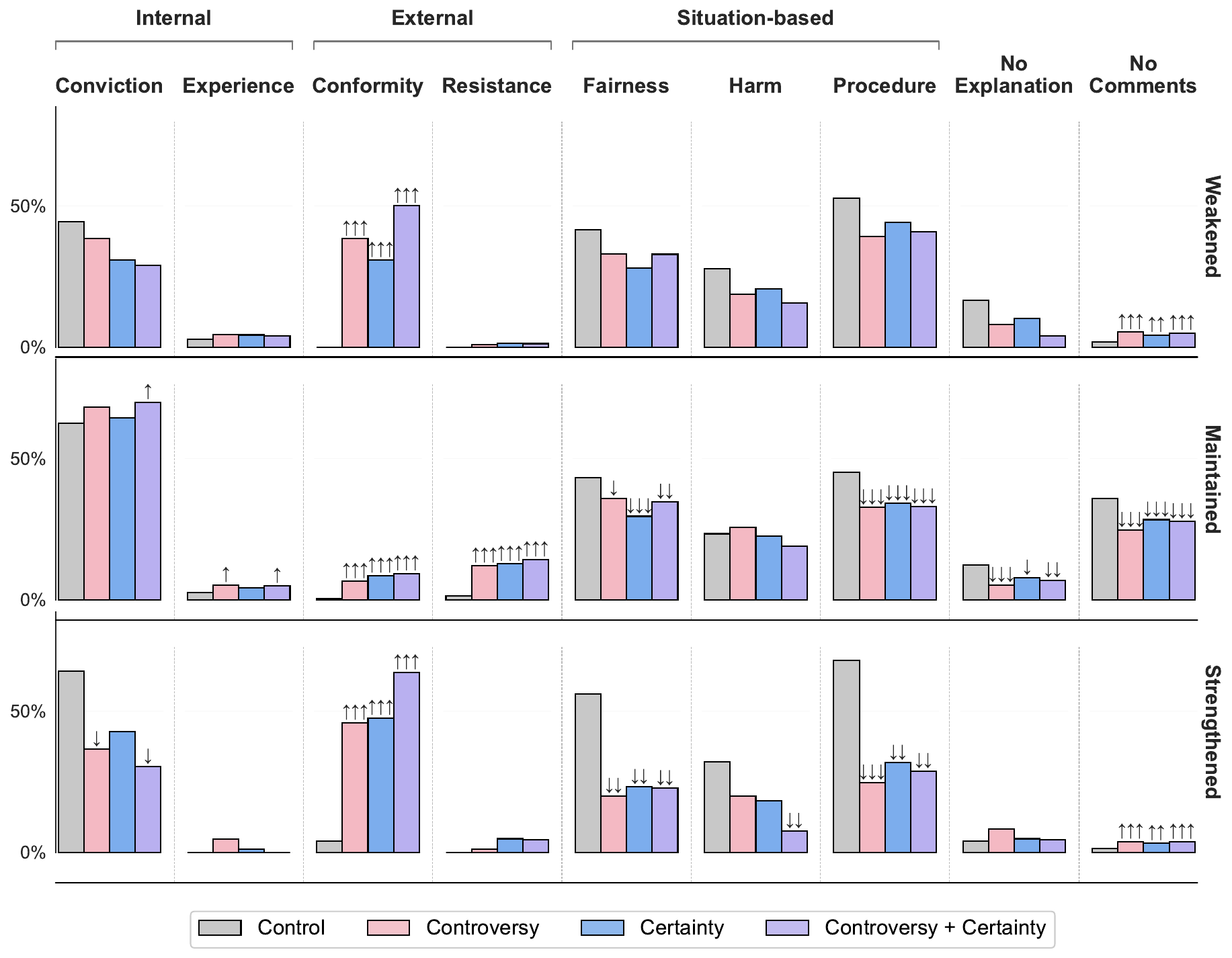}
    \caption{\textbf{General reasons for change.}
    \textbf{(a)},
    Among those who provided an explanation, proportions of reasoning types occurring across
    Control, Controversy, Certainty, and Controversy + Certainty by change types. The bottom row shows participants who did not provide an explanation. Each bar shows the proportion of participants using a given reasoning type within each change type.
    \textbf{Top arrows} indicate significant differences between each experimental treatment and control (↑: Treatment $>$ Control; ↓: Control $>$ Treatment).
    Significance markers (* $p < .05$, ** $p < .01$, *** $p < .001$) are based on two-sided Barnard's exact tests with Benjamini--Hochberg FDR correction across the three treatment-versus-control comparisons, applied separately for each change type and reasoning subtype. Comparisons of responses without written explanations were corrected separately within each change type.}
    \label{fig:deliberation_results}
\end{figure}

\Cref{fig:deliberation_results} shows the reasoning profiles for each change type across treatments. Proportions are computed within each treatment among responses that included a written explanation (control: $n = 651$; Controversy: $n = 708$; Certainty: $n = 698$; Controversy + Certainty: $n = 687$). Each explanation was coded into one or more of the eight subtypes defined in \Cref{subsubsec:annotation}, organized into four groups (mean $= 1.87$ subtypes per comment, median $= 2$, $N = 2{,}744$ comments). The overall subtype distribution is reported in \Cref{tab:deliberation_frequency}.

\textbf{Reasoning comparison.}
To assess whether reasoning-type proportions differed between each experimental treatment and the control condition, we used two-sided Barnard's exact tests \citep{barnard1945new}. Barnard's exact test is an unconditional exact test for $2\times2$ contingency tables; unlike Fisher's exact test, it does not condition on the marginal totals, which gives it greater power when comparing two independent proportions. Comparisons were conducted separately within each of the three change types (weakened, maintained, strengthened).

To address multiple comparisons, the Benjamini--Hochberg (BH) false discovery rate (FDR) correction \citep{benjamini1995controlling} was applied per change type per subtype across the three experimental treatments, controlling the expected rate of false discoveries at $q = .05$. BH-adjusted values, denoted $p_{\mathrm{BH}}$, are reported throughout, with significance thresholds shown as $^{*}\ p < .05$, $^{**}\ p < .01$, and $^{***}\ p < .001$.

We test whether each treatment produced a different reasoning distribution from control, separately within each change type. Results are presented in \Cref{tab:cross_treatment}.

\textbf{Responses without written explanations.} Before examining reasoning subtypes, we compared the proportion of responses without a written explanation. Across all three treatments, significantly more participants who weakened or strengthened their position did so without providing an explanation compared with Control (all $p_{\mathrm{BH}} < .01$). Conversely, the proportion of maintained responses without an explanation was significantly lower in all treatments than in Control (all $p_{\mathrm{BH}} < .001$). This pattern suggests that treatment information prompted more active engagement: participants who changed were more likely to act without explaining, while those who held firm were more likely to articulate a reason.

\textbf{Weakened.} Among participants who weakened their position, the only significant cross-treatment difference was conformity, which was absent in the Control condition (0.0\%) but appeared in 38.4\% of Controversy explanations, 30.9\% of Certainty explanations, and 50.0\% of Controversy + Certainty explanations (all $p_{\mathrm{BH}} < .001$). No other subtype differed significantly between any treatment and Control after correction. This indicates that when treatment information led participants to weaken their position, conformity to the community signal was the primary reasoning pathway that distinguished them from Control participants who weakened for other reasons.

\textbf{Maintained.} The maintained group showed the most widespread shifts in reasoning composition. Conformity and resistance both increased significantly across all three treatments relative to control (conformity: 6.7--9.4\% vs.\ 0.5\%; resistance: 12.1--14.3\% vs.\ 1.4\%; all $p_{\mathrm{BH}} < .001$). These two subtypes reflect opposite reactions to the same social information: some participants cited community agreement as reinforcing their position, while others explicitly rejected the community signal. Moral conviction increased significantly only in the Controversy + Certainty condition (69.9\% vs.\ 62.4\%, $p_{\mathrm{BH}} = .023$). At the same time, situation-based reasoning declined. Fairness decreased significantly in all three treatments (29.6--35.8\% vs.\ 43.2\%; all $p_{\mathrm{BH}} < .05$), as did contextual and procedural reasoning (32.7--34.1\% vs.\ 45.1\%; all $p_{\mathrm{BH}} < .001$). No-explanation responses also decreased significantly in all treatments (5.3--7.8\% vs.\ 12.4\%; all $p_{\mathrm{BH}} < .05$). Harm-related reasoning showed no significant differences. Together, these results suggest that treatment information shifted reasoning among maintainers away from situational engagement and toward social-relational and conviction-based justifications.

\textbf{Strengthened.} The strengthened group showed the sharpest compositional shift. Conformity increased from 4.0\% in control to 45.9--63.6\% across treatments (all $p_{\mathrm{BH}} < .001$). Moral conviction declined in the Controversy and Controversy + Certainty conditions (36.5\% and 30.3\% vs.\ 64.0\%; both $p_{\mathrm{BH}} < .05$), but not under Certainty alone. Fairness dropped from 56.0\% to 20.0--23.2\% (all $p_{\mathrm{BH}} < .01$), and contextual reasoning dropped from 68.0\% to 24.7--31.7\% (all $p_{\mathrm{BH}} < .01$). Harm-related reasoning declined significantly only in the Controversy + Certainty condition (7.6\% vs.\ 32.0\%, $p_{\mathrm{BH}} = .009$). These results indicate that participants who strengthened in the control did so by re-engaging with the moral content of the scenario, whereas participants in the treatment who strengthened were driven primarily by social conformity.

\textbf{Summary.} Across all three change types, the most consistent effect of treatment information was the activation of conformity-based reasoning, which was nearly absent in the control condition. Among maintainers and strengtheners, this activation co-occurred with reduced reliance on situation-based reasoning. The pattern suggests that social signals do not simply add a new input to the reasoning process; they reshape its composition, substituting social reference for moral and situational engagement.

\begin{table}[!htbp]
\centering
\caption{Cross-treatment comparisons of reasoning subtype proportions
and rates of responses without written explanations (each experimental treatment versus control).
Two-sided Barnard exact tests with Benjamini--Hochberg correction across the three treatment-versus-control comparisons, applied separately for each change type and subtype. Comparisons of responses without written explanations were corrected separately within each change type. Significance: $^{*}p < .05$, $^{**}p < .01$,
$^{***}p < .001$.}
\label{tab:cross_treatment}
\scriptsize
\setlength{\tabcolsep}{3.5pt}
\begin{tabular}{l
  rr@{\hskip 2pt}r@{\hskip 4pt}c
  rr@{\hskip 2pt}r@{\hskip 4pt}c
  rr@{\hskip 2pt}r@{\hskip 4pt}c}
\toprule
& \multicolumn{4}{c}{Controversy}
& \multicolumn{4}{c}{Certainty}
& \multicolumn{4}{c}{Contr.+Cert.} \\
\cmidrule(lr){2-5}\cmidrule(lr){6-9}\cmidrule(lr){10-13}
Subtype
& Exp & Ctrl & $p_{\mathrm{BH}}$ &
& Exp & Ctrl & $p_{\mathrm{BH}}$ &
& Exp & Ctrl & $p_{\mathrm{BH}}$ & \\
\midrule
\multicolumn{13}{l}{\textit{Panel A: Weakened}} \\[4pt]
Moral conviction    & .38 & .44 & $.582$               &          & .31 & .44 & $.274$               &          & .29 & .44 & $.274$               &          \\
Lived experience    & .04 & .03 & $.835$               &          & .04 & .03 & $.835$               &          & .04 & .03 & $.835$               &          \\
Conformity          & .38 & .00 & $3.74\times10^{-5}$ & $^{***}$ & .31 & .00 & $1.55\times10^{-4}$ & $^{***}$ & .50 & .00 & $6.31\times10^{-7}$ & $^{***}$ \\
Resistance          & .01 & .00 & $.729$               &          & .01 & .00 & $.729$               &          & .01 & .00 & $.729$               &          \\
Fairness            & .33 & .42 & $.415$               &          & .28 & .42 & $.415$               &          & .33 & .42 & $.415$               &          \\
Harm                & .19 & .28 & $.414$               &          & .21 & .28 & $.427$               &          & .16 & .28 & $.414$               &          \\
Context             & .39 & .53 & $.417$               &          & .44 & .53 & $.417$               &          & .41 & .53 & $.417$               &          \\[2pt]
No explanation      & .08 & .17 & $.212$               &          & .10 & .17 & $.417$               &          & .04 & .17 & $.0635$              &          \\
No written explanation & .05 & .02 & $7.46\times10^{-5}$ & $^{***}$ & .04 & .02 & $1.31\times10^{-3}$ & $^{**}$  & .05 & .02 & $1.42\times10^{-4}$ & $^{***}$ \\[4pt]
\multicolumn{13}{l}{\quad\scriptsize Subtype $N$: Contr.\ = 112, Cert.\ = 68, C+C = 76, Control = 36.} \\[2pt]
\multicolumn{13}{l}{\quad\scriptsize No-written-explanation $N$: Contr.\ = 1072, Cert.\ = 1092, C+C = 1084, Control = 1070.} \\[4pt]
\midrule
\multicolumn{13}{l}{\textit{Panel B: Maintained}} \\[4pt]
Moral conviction    & .68 & .62 & $.0741$              &          & .64 & .62 & $.490$               &          & .70 & .62 & $.0230$              & $^{*}$   \\
Lived experience    & .05 & .03 & $.0483$              & $^{*}$   & .04 & .03 & $.124$               &          & .05 & .03 & $.0483$              & $^{*}$   \\
Conformity          & .07 & .01 & $1.60\times10^{-8}$ & $^{***}$ & .09 & .01 & $4.47\times10^{-11}$& $^{***}$ & .09 & .01 & $6.79\times10^{-12}$& $^{***}$ \\
Resistance          & .12 & .01 & $2.23\times10^{-13}$& $^{***}$ & .13 & .01 & $3.16\times10^{-14}$& $^{***}$ & .14 & .01 & $3.90\times10^{-16}$& $^{***}$ \\
Fairness            & .36 & .43 & $.0125$              & $^{*}$   & .30 & .43 & $5.25\times10^{-6}$ & $^{***}$ & .35 & .43 & $.00506$             & $^{**}$  \\
Harm                & .26 & .23 & $.590$               &          & .23 & .23 & $.768$               &          & .19 & .23 & $.202$               &          \\
Context             & .33 & .45 & $5.08\times10^{-5}$ & $^{***}$ & .34 & .45 & $1.68\times10^{-4}$ & $^{***}$ & .33 & .45 & $5.08\times10^{-5}$ & $^{***}$ \\[2pt]
No explanation      & .05 & .12 & $1.33\times10^{-4}$ & $^{***}$ & .08 & .12 & $.0118$              & $^{*}$   & .07 & .12 & $.00336$             & $^{**}$  \\
No written explanation & .25 & .36 & $7.34\times10^{-8}$ & $^{***}$ & .28 & .36 & $2.28\times10^{-4}$ & $^{***}$ & .28 & .36 & $9.78\times10^{-5}$ & $^{***}$ \\[4pt]
\multicolumn{13}{l}{\quad\scriptsize Subtype $N$: Contr.\ = 511, Cert.\ = 548, C+C = 545, Control = 590.} \\[2pt]
\multicolumn{13}{l}{\quad\scriptsize No-written-explanation $N$: Contr.\ = 1072, Cert.\ = 1092, C+C = 1084, Control = 1070.} \\[4pt]
\midrule
\multicolumn{13}{l}{\textit{Panel C: Strengthened}} \\[4pt]
Moral conviction    & .36 & .64 & $.0334$              & $^{*}$   & .43 & .64 & $.0630$              &          & .30 & .64 & $.0101$              & $^{*}$   \\
Lived experience    & .05 & .00 & $.936$               &          & .01 & .00 & $1.00$               &          & .00 & .00 & $1.00$               &          \\
Conformity          & .46 & .04 & $2.92\times10^{-4}$ & $^{***}$ & .48 & .04 & $2.92\times10^{-4}$ & $^{***}$ & .64 & .04 & $1.91\times10^{-6}$ & $^{***}$ \\
Resistance          & .01 & .00 & $.735$               &          & .05 & .00 & $.477$               &          & .05 & .00 & $.477$               &          \\
Fairness            & .20 & .56 & $3.19\times10^{-3}$ & $^{**}$  & .23 & .56 & $3.35\times10^{-3}$ & $^{**}$  & .23 & .56 & $3.35\times10^{-3}$ & $^{**}$  \\
Harm                & .20 & .32 & $.228$               &          & .18 & .32 & $.226$               &          & .08 & .32 & $.00913$             & $^{**}$  \\
Context             & .25 & .68 & $6.69\times10^{-4}$ & $^{***}$ & .32 & .68 & $1.42\times10^{-3}$ & $^{**}$  & .29 & .68 & $1.17\times10^{-3}$ & $^{**}$  \\[2pt]
No explanation      & .08 & .04 & $.963$               &          & .05 & .04 & $.963$               &          & .05 & .04 & $.963$               &          \\
No written explanation & .04 & .01 & $.000801$            & $^{***}$ & .03 & .01 & $.00375$             & $^{**}$  & .04 & .01 & $.000801$            & $^{***}$ \\[4pt]
\multicolumn{13}{l}{\quad\scriptsize Subtype $N$: Contr.\ = 85, Cert.\ = 82, C+C = 66, Control = 25.} \\[2pt]
\multicolumn{13}{l}{\quad\scriptsize No-written-explanation $N$: Contr.\ = 1072, Cert.\ = 1092, C+C = 1084, Control = 1070.} \\
\bottomrule
\end{tabular}
\end{table}
\clearpage
\paragraph{Results: Treatment reference and acceptance}\label{subsec:tr_results}
In this part, we report two things: how often participants reference each treatment variable, and, among those who do, whether they accept or reject it as relevant evidence. \Cref{tab:tr_attention} reports the reference and acceptance counts, and \Cref{tab:tr_tests} reports the pairwise Barnard tests behind the comparisons below. \Cref{tab:tr_tests_clustered} tests whether these pairwise differences remain after accounting for multiple signal codes within explanations and repeated responses from the same participant.

We first look at how often comments mention each treatment variable they were shown (the ``Referenced'' columns of \Cref{tab:tr_attention}). When Controversy is presented alone, the disagreement rate is mentioned in 31.9\% of comments. In the Certainty treatment, in-group certainty (24.4\%) is mentioned more often than out-group certainty (17.2\%), and the difference is significant. When the two are presented together, the mention rate drops for every variable: the disagreement rate (29.1\%), in-group certainty (10.5\%), and out-group certainty (5.1\%). The same ordering holds, with the disagreement rate mentioned most, then in-group certainty, then out-group certainty, and all three differences are significant. When both signals are shown, attention concentrates on the disagreement rate.
\begin{table}[!htbp]
\centering
\caption{Attention to and acceptance of treatment variables. ``Referenced'' gives the count and percentage of shown participants who mentioned the variable. ``Among engagers'' gives the accept or reject count and its share of engagers. The last three columns are the outcome split within each stance and sum to 100. Pairwise tests are reported in \Cref{tab:tr_tests}.}
\label{tab:tr_attention}
\small
\setlength{\tabcolsep}{4.5pt}
\begin{tabular}{llrrlrrrrr}
\toprule
\multirow{2}{*}{\textbf{Condition}} & \multirow{2}{*}{\textbf{Variable}}
 & \multicolumn{2}{c}{\textbf{Referenced}} & \multirow{2}{*}{\textbf{Stance}}
 & \multicolumn{2}{c}{\textbf{Among engagers}}
 & \multicolumn{3}{c}{\textbf{Outcome within stance (\%)}} \\
\cmidrule(lr){3-4}\cmidrule(lr){6-7}\cmidrule(lr){8-10}
 & & n & \% & & n & \% & Weak. & Maint. & Streng. \\
\midrule

\multirow{2}{*}{Controversy ($N=708$)}
  & \multirow{2}{*}{Disagreement} & \multirow{2}{*}{226} & \multirow{2}{*}{31.9}
    & Accept & 170 & 75.2 & 31.2 & 45.9 & 22.9 \\
  & & & & Reject & 56 & 24.8 & 1.8 & 96.4 & 1.8 \\
\midrule

\multirow{4}{*}{Certainty ($N=698$)}
  & \multirow{2}{*}{Ingroup} & \multirow{2}{*}{170} & \multirow{2}{*}{24.4}
    & Accept & 126 & 74.1 & 8.7 & 61.1 & 30.2 \\
  & & & & Reject & 44 & 25.9 & 0.0 & 97.7 & 2.3 \\
\cmidrule(l){2-10}
  & \multirow{2}{*}{Outgroup} & \multirow{2}{*}{120} & \multirow{2}{*}{17.2}
    & Accept & 82 & 68.3 & 12.2 & 65.9 & 22.0 \\
  & & & & Reject & 38 & 31.7 & 0.0 & 100.0 & 0.0 \\
\midrule

\multirow{6}{*}{Contr. + Cert. ($N=687$)}
  & \multirow{2}{*}{Disagreement} & \multirow{2}{*}{200} & \multirow{2}{*}{29.1}
    & Accept & 131 & 65.5 & 25.2 & 53.4 & 21.4 \\
  & & & & Reject & 69 & 34.5 & 2.9 & 92.8 & 4.3 \\
\cmidrule(l){2-10}
  & \multirow{2}{*}{Ingroup} & \multirow{2}{*}{72} & \multirow{2}{*}{10.5}
    & Accept & 58 & 80.6 & 6.9 & 44.8 & 48.3 \\
  & & & & Reject & 14 & 19.4 & 7.1 & 92.9 & 0.0 \\
\cmidrule(l){2-10}
  & \multirow{2}{*}{Outgroup} & \multirow{2}{*}{35} & \multirow{2}{*}{5.1}
    & Accept & 21 & 60.0 & 9.5 & 57.1 & 33.3 \\
  & & & & Reject & 14 & 40.0 & 7.1 & 92.9 & 0.0 \\
\bottomrule
\end{tabular}
\end{table}

\begin{table}[!htbp]
\centering
\caption{Pairwise tests for reference rate and acceptance rate. All tests are two-sided Barnard exact tests. Significance: $^{*}p < .05$, $^{**}p < .01$, $^{***}p < .001$; n.s.\ = not significant.}
\label{tab:tr_tests}
\small
\begin{tabular}{lllr l}
\toprule
\textbf{Measure} & \textbf{Condition} & \textbf{Comparison} & \textbf{p} & \textbf{Sig.} \\
\midrule
\multirow{4}{*}{Reference rate}
 & Certainty       & Ingroup vs.\ Outgroup       & $.000988$ & $^{***}$ \\
 & Contr. + Cert.  & Disagreement vs.\ Ingroup   & $3.33 \times 10^{-18}$ & $^{***}$ \\
 & Contr. + Cert.  & Disagreement vs.\ Outgroup  & $9.65 \times 10^{-33}$ & $^{***}$ \\
 & Contr. + Cert.  & Ingroup vs.\ Outgroup       & $.000205$ & $^{***}$ \\
\midrule
\multirow{4}{*}{Acceptance rate}
 & Certainty       & Ingroup vs.\ Outgroup       & $.343$  & n.s.     \\
 & Contr. + Cert.  & Disagreement vs.\ Ingroup   & $.0176$ & $^{*}$   \\
 & Contr. + Cert.  & Disagreement vs.\ Outgroup  & $.612$  & n.s.     \\
 & Contr. + Cert.  & Ingroup vs.\ Outgroup       & $.0237$ & $^{*}$   \\
\bottomrule
\end{tabular}
\end{table}

\paragraph{Participant-clustered regression robustness check.}
The Barnard tests above compare the observed proportions but do not account for dependence between signal codes from the same explanation or between the two dilemma responses from each participant. We therefore repeated the eight pairwise comparisons using binomial logistic regressions in signal-level long format. For reference-rate comparisons, the binary outcome indicated whether an explanation referenced each signal. For acceptance-rate comparisons, the analysis was restricted to explanations referencing the signal, and the outcome indicated whether the signal was accepted rather than rejected. Each model included an indicator distinguishing the two signals being compared, with robust standard errors clustered by participant. Tests were two-sided, with Benjamini--Hochberg correction applied jointly across the eight contrasts.

\Cref{tab:tr_tests_clustered} shows that all four reference-rate contrasts remained significant. The acceptance-rate results also supported greater acceptance of in-group than out-group certainty in both the Certainty and combined conditions. In the combined condition, in-group certainty was accepted more often than disagreement, whereas acceptance of disagreement and out-group certainty did not differ. Thus, accounting for repeated responses preserved the main ordering of attention and acceptance while providing additional evidence for the in-group versus out-group acceptance contrast in the Certainty condition.

\begin{table}[!htbp]
\centering
\caption{Participant-clustered regression robustness checks for the reference-rate and acceptance-rate comparisons in \Cref{tab:tr_tests}. Each row reports a separate binomial logistic regression contrasting the first named signal with the second; a positive coefficient indicates a higher rate for the first signal. Standard errors were clustered by participant. All tests were two-sided, with Benjamini--Hochberg correction applied jointly across the eight contrasts.}
\label{tab:tr_tests_clustered}
\scriptsize
\setlength{\tabcolsep}{3.5pt}
\renewcommand{\arraystretch}{1.15}
\begin{tabularx}{\textwidth}{@{}ll>{\hsize=1.4\hsize\linewidth=\hsize\raggedright\arraybackslash}X>{\hsize=.8\hsize\linewidth=\hsize\centering\arraybackslash}X>{\hsize=.8\hsize\linewidth=\hsize\centering\arraybackslash}Xcc@{}}
\toprule
Measure & Condition & Comparison & $\beta$ [95\% CI] & OR [95\% CI] & $p_{\mathrm{raw}}$ & $p_{\mathrm{BH}}$ \\
\midrule
\multirow{4}{*}{Reference rate}
 & Certainty & Ingroup vs. Outgroup
 & $0.439$ [$0.302$, $0.575$] & $1.551$ [$1.353$, $1.778$] & $3.08 \times 10^{-10}$ & $8.22 \times 10^{-10}$ \\
 & Contr. + Cert. & Disagreement vs. Ingroup
 & $1.255$ [$1.022$, $1.488$] & $3.508$ [$2.778$, $4.429$] & $5.06 \times 10^{-26}$ & $2.03 \times 10^{-25}$ \\
 & Contr. + Cert. & Disagreement vs. Outgroup
 & $2.035$ [$1.679$, $2.391$] & $7.650$ [$5.358$, $10.923$] & $4.14 \times 10^{-29}$ & $3.32 \times 10^{-28}$ \\
 & Contr. + Cert. & Ingroup vs. Outgroup
 & $0.780$ [$0.484$, $1.076$] & $2.181$ [$1.622$, $2.932$] & $2.40 \times 10^{-7}$ & $4.80 \times 10^{-7}$ \\
\midrule
\multirow{4}{*}{Acceptance rate}
 & Certainty & Ingroup vs. Outgroup
 & $0.283$ [$0.098$, $0.468$] & $1.327$ [$1.103$, $1.597$] & $.00277$ & $.00369$ \\
 & Contr. + Cert. & Disagreement vs. Ingroup
 & $-0.780$ [$-1.346$, $-0.214$] & $0.458$ [$0.260$, $0.807$] & $.00691$ & $.00790$ \\
 & Contr. + Cert. & Disagreement vs. Outgroup
 & $0.236$ [$-0.427$, $0.898$] & $1.266$ [$0.653$, $2.455$] & $.486$ & $.486$ \\
 & Contr. + Cert. & Ingroup vs. Outgroup
 & $1.016$ [$0.615$, $1.417$] & $2.762$ [$1.850$, $4.124$] & $6.77 \times 10^{-7}$ & $1.08 \times 10^{-6}$ \\
\bottomrule
\end{tabularx}
\end{table}

We next break down the comments that engage with a variable by whether they accept or reject it, and by whether the judgment changed (the ``Among engagers'' and outcome columns of \Cref{tab:tr_attention}). Across all treatments, most engaging comments accept the information (60\% to 81\%). Among the comments that reject it, almost all maintained their judgment. For the disagreement rate, acceptance dominates in both treatments where it appears (Controversy: 75.2\%; Controversy + Certainty: 65.5\%). Among comments that accept the disagreement rate, more lead to a weakened than a strengthened judgment (Controversy: 31.2\% versus 22.9\%; Controversy + Certainty: 25.2\% versus 21.4\%). In-group certainty draws the highest acceptance among the variables (Certainty: 74.1\%; Controversy + Certainty: 80.6\%). Among comments that accept in-group certainty, more lead to a strengthened than a weakened judgment (Certainty: 30.2\% versus 8.7\%; Controversy + Certainty: 48.3\% versus 6.9\%).
\begin{figure}[!htbp]
    \centering
    \includegraphics[width=\linewidth]{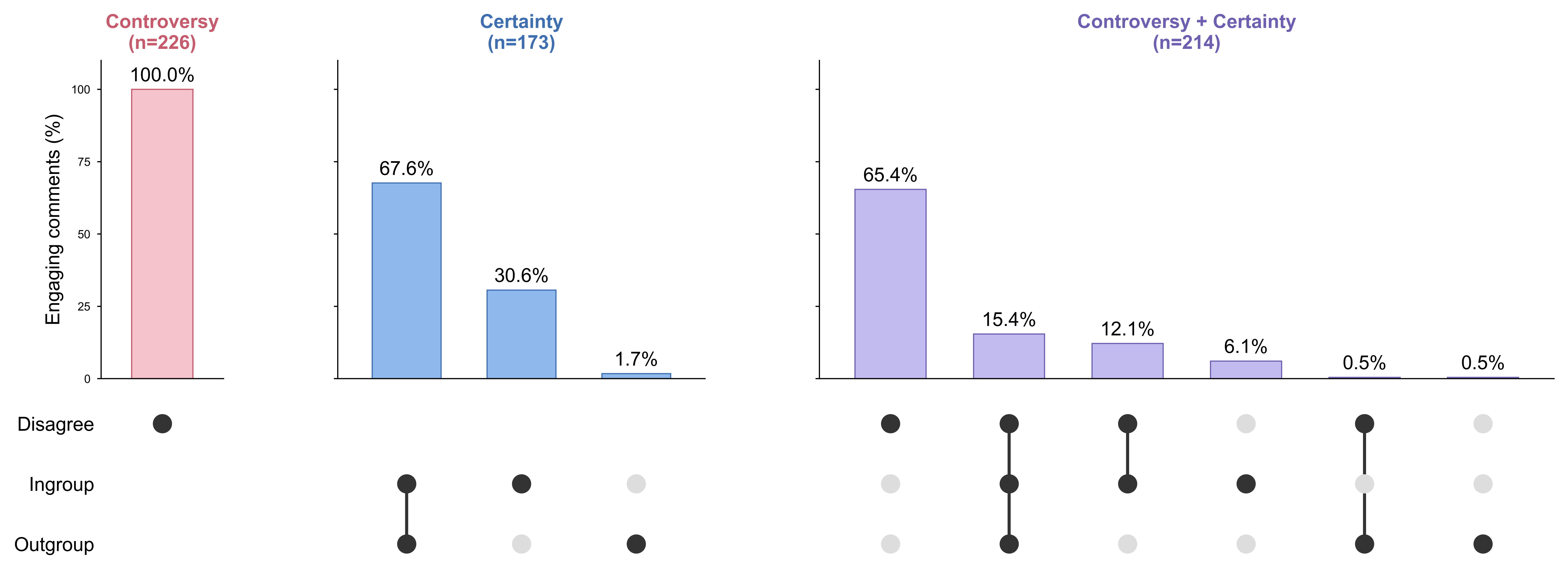}
    \caption{\textbf{Co-occurrence of treatment variables in written explanations.}}
    \label{fig:cooccurence}
\end{figure}

Comments often reference more than one signal (\Cref{fig:cooccurence}). The Certainty treatment is dominated by the co-mention of in-group and out-group certainty (68\%). In the Controversy + Certainty treatment, participants mostly mention the disagreement rate alone (65\%) or the disagreement rate paired with in-group certainty (28\%). Across both treatments, out-group certainty is rarely discussed on its own.

Overall, these patterns point to three things. First, attention mostly follows the disagreement rate and in-group certainty. Under Certainty, in-group certainty draws more notice than out-group certainty. When both signals appear together, attention narrows further onto the disagreement rate and then in-group certainty, and out-group certainty is rarely mentioned. Second, most participants who engage with a signal accept it as relevant evidence, and those who reject a signal almost always hold their judgment. Third, the signals push in opposite directions: the disagreement rate tends to move participants away from their initial judgment, while in-group certainty tends to reinforce it.

\subsubsection{Dilemma consequences}\label{subsec:consequence}
The 135 dilemmas differ in their severity, affected parties, and involvement of vulnerable people. We annotated these features, validated the annotations against human labels (\Cref{subsubsec:consequence_annotation}), described their distribution, and explored whether they were associated with weakening or strengthening. This part is exploratory rather than confirmatory.

\paragraph{Annotation and validation}\label{subsubsec:consequence_annotation}

We run two coding tasks on the post text (title and body).
The first rates the severity of the dilemma.
The second records who is affected and whether any affected person is vulnerable.
Both tasks are blind to all outcome data: the model sees only the post, never the Reddit tallies or our participants' responses.
Each post is annotated three times independently.
The three runs are then aggregated, by a rule that differs between the two tasks because their outputs are of different kinds; we state each rule below.
We use \texttt{gpt-5.6-luna} with the default sampling settings, because the GPT-5 reasoning models accept only the default temperature through the API.

\subparagraph{Task 1: Severity}
Severity is the seriousness of what is at stake for the people involved, independent of who is at fault.
We rate it on a 1--7 scale with the odd points anchored, shown in the prompt below.
An earlier version of this rubric also rated impact asymmetry, reversibility, harm type, and affected party in a single pass; we retired it because those dimensions were largely collinear with severity, two harm labels were near-constant, and its affected-party scheme was single-select and so could not represent dilemmas affecting more than one group.
Because the output is a single ordinal rating, we aggregate the three runs by taking their \emph{mean}, which yields a continuous score on the 1--7 scale.

\begin{tcolorbox}[colframe=black, colback=white, boxrule=0.5pt,
                  title=Severity annotation prompt, breakable]
\label{tab:severity_prompt}
\small
You will be shown a post from a public online forum in which an author describes an interpersonal conflict and asks readers whether they behaved wrongly. Your task is to rate the STAKES of the situation described.

\medskip
\textbf{\textless background\textgreater} \\
These posts come from the Reddit forum ``Am I the Asshole'' (AITA). An author narrates a conflict from their own point of view and asks the community to judge them; readers reply with a verdict.

\medskip
We are studying how the seriousness of a scenario relates to how readers respond to it. That requires a rating of what is at stake in each scenario, independent of who is at fault. Two scenarios can be equally hard to judge while differing enormously in what is at stake, and it is that difference this rating captures. \\
\textbf{\textless/background\textgreater}

\medskip
\textbf{\textless task\textgreater} \\
Rate the scenario on one dimension.

\medskip
\texttt{severity\_rating}: integer 1--7. How serious are the stakes for the people involved --- how much is actually on the line, materially or emotionally, in the situation as a whole?
\begin{itemize}[nosep, leftmargin=1.5em]
  \item 1 = trivial: a matter of taste, etiquette or mild annoyance; forgotten within days
  \item 3 = moderate: real but bounded --- hurt feelings, a spoiled occasion, small sums of money
  \item 5 = serious: substantial and lasting --- significant money, a damaged close relationship, sustained distress
  \item 7 = grave: major and life-altering --- physical safety, health, a child's welfare, custody, financial ruin, permanent rupture of a family
\end{itemize}
\textbf{\textless/task\textgreater}

\medskip
\textbf{\textless rules\textgreater}
\begin{enumerate}[nosep, leftmargin=1.5em]
  \item Rate the stakes, not the blame. Do not state or imply a verdict on who is in the right, and do not let a view about who is at fault influence the rating.
  \item Rate the situation as described in the text. Do not speculate about facts that are not stated, and do not penalise the author for narrating from their own point of view.
  \item Only the odd values of the scale are described above. When a scenario falls between two described points, use the even value in between (2, 4 or 6).
  \item If the post body is missing or was removed, rate from the title alone.
\end{enumerate}
\textbf{\textless/rules\textgreater}

\medskip
\textbf{\textless output\_format\textgreater} \\
Return a single JSON object with this exact schema, and nothing else:
\begin{verbatim}
{
  "severity_rating": <integer 1-7>
}
\end{verbatim}
\textbf{\textless/output\_format\textgreater}
\end{tcolorbox}

\subparagraph{Task 2: Affected party and vulnerability}
Consequences rarely fall on one person, so we code the affected party as multi-select over five categories that follow the author's expanding social circle: the author (OP), family and romantic partners, friends, ongoing contacts such as co-workers or neighbours, and strangers.
A separate single-select question records whether anyone vulnerable is affected: a child, another vulnerable person (elderly, ill, or disabled), or nobody in particular.
The two questions are independent by design.
A post can name family in the first and a child in the second.
Both outputs are categorical, so we aggregate the three runs by \emph{majority vote} rather than by averaging.
The two questions need slightly different rules because one is multi-select and the other is not.
For affected party, each of the five categories is voted on separately: a category enters the post's set if at least two of the three runs select it.
For vulnerability, the post takes the modal label across runs, which with three runs means the label given by at least two of them.
A post whose three runs give three different vulnerability labels would have no majority and be dropped; in practice this never occurs, and 126 of 135 posts are unanimous.

\begin{tcolorbox}[colframe=black, colback=white, boxrule=0.5pt,
                  title=Affected-party and vulnerability annotation prompt, breakable]
\label{tab:party_prompt}
\small
You will be shown a post from a public online forum in which an author describes an interpersonal conflict and asks readers whether they behaved wrongly. Your task is to identify who is affected by the situation described, not to judge who is right.

\medskip
\textbf{\textless background\textgreater} \\
These posts come from the Reddit forum ``Am I the Asshole'' (AITA). An author narrates a conflict from their own point of view and asks the community to judge them; readers reply with a verdict. We are not asking you to judge who's right or wrong, and not asking how interesting or dramatic the story is. We are asking about consequences --- who bears the cost of what happens in the situation described. \\
\textbf{\textless/background\textgreater}

\medskip
\textbf{\textless task\textgreater} \\
Answer two separate questions.

\medskip
\texttt{affected\_party}: array of one or more strings. Select every group that is genuinely affected by the situation described, following the author's (OP's) expanding social circle, drawn only from:
\begin{itemize}[nosep, leftmargin=1.5em]
  \item \texttt{op} = OP themself bears the consequence
  \item \texttt{family} = a parent, sibling, spouse, child, or other family member; or a romantic partner
  \item \texttt{friends} = a friend of OP
  \item \texttt{ongoing\_contact} = someone OP interacts with regularly but who isn't family or a friend (e.g.\ a co-worker, neighbour, roommate)
  \item \texttt{stranger} = someone outside OP's regular life
\end{itemize}
Select every group that's genuinely affected --- a post can easily involve both \texttt{op} and a family member, for instance. This question is only about OP's relationship to the affected person.

\medskip
\texttt{vulnerable}: exactly one string. Are there vulnerable person(s) affected? Choose the single best answer, drawn only from:
\begin{itemize}[nosep, leftmargin=1.5em]
  \item \texttt{child} = a child or children --- one or more minors are affected
  \item \texttt{other\_vulnerable} = other vulnerable person(s), i.e.\ the elderly, or those with a disability or illness --- not a minor
  \item \texttt{none} = no person is particularly vulnerable
\end{itemize}
This is a separate, independent question from \texttt{affected\_party} --- a post can select \texttt{family} for \texttt{affected\_party} and \texttt{child} for \texttt{vulnerable} at the same time (e.g.\ OP's own kid, or an elderly parent). If a post involves both a child and an elderly relative, pick whichever one is more central to the post's consequences --- only one answer is allowed here. \\
\textbf{\textless/task\textgreater}

\medskip
\textbf{\textless rules\textgreater}
\begin{enumerate}[nosep, leftmargin=1.5em]
  \item Answer both questions. Do not skip either.
  \item Judge who is affected and how vulnerable they are, not who is at fault. Do not state or imply a verdict on who is in the right.
  \item Base your answer on the situation as described in the text. Do not speculate about facts that are not stated.
  \item If the post body is missing or was removed, answer from the title alone.
\end{enumerate}
\textbf{\textless/rules\textgreater}

\medskip
\textbf{\textless output\_format\textgreater} \\
Return a single JSON object with this exact schema, and nothing else:
\begin{verbatim}
{
  "affected_party": [<one or more allowed strings>],
  "vulnerable": "<one allowed string>"
}
\end{verbatim}
\textbf{\textless/output\_format\textgreater}
\end{tcolorbox}

\textbf{Across-run reliability.}
Annotations were stable across the three independent model runs. Severity ratings showed high absolute agreement across the three runs, measured using a two-way random-effects, absolute-agreement intraclass correlation coefficient for a single run ($\operatorname{ICC}(2,1)=0.96$). The three runs gave identical severity ratings for 98 of 135 dilemmas; for the remaining 37, the highest and lowest ratings differed by only one severity level. Vulnerability labels were unanimous for 126 of 135 dilemmas, while 287 of the 301 affected-party selections retained by majority vote were unanimous and 14 were selected in two of three runs.

\textbf{Human validation.}
Three human annotators per task coded a random sample of 50 posts; the severity annotators also compared 68 post pairs. Agreement between individual annotators and the human majority-vote labels was high, with mean F1 scores of $0.91$ for affected party, $0.83$ for any vulnerable person, $0.88$ for a child, and $0.88$ for pairwise severity. Against these labels, the model achieved F1 scores of $0.92$, $0.71$, $0.78$, and $0.85$, respectively (\Cref{tab:consequence_interrater,tab:consequence_validation}).

\begin{table}[!htbp]
\centering
\caption{Inter-annotator reliability: each annotator against the human majority-vote label for that task. Annotators are numbered within each task, so \textnormal{A1} in one task is not the same person as \textnormal{A1} in another.}
\label{tab:consequence_interrater}
\small
\begin{tabular}{llrrr}
\toprule
\textbf{Task} & \textbf{Annotator} & \textbf{Precision} & \textbf{Recall} & \textbf{F1} \\
\midrule
\multirow{4}{*}{Affected party}
 & A1   & 0.87 & 0.98 & 0.92 \\
 & A2   & 0.90 & 0.93 & 0.92 \\
 & A3   & 0.97 & 0.86 & 0.91 \\
 & \textit{mean} & \textit{0.91} & \textit{0.92} & \textit{0.91} \\
\midrule
\multirow{4}{*}{Vulnerable (any)}
 & A1   & 1.00 & 0.67 & 0.80 \\
 & A2   & 1.00 & 0.87 & 0.93 \\
 & A3   & 0.63 & 1.00 & 0.77 \\
 & \textit{mean} & \textit{0.88} & \textit{0.84} & \textit{0.83} \\
\midrule
\multirow{4}{*}{Vulnerable (child)}
 & A1   & 1.00 & 0.64 & 0.78 \\
 & A2   & 1.00 & 0.93 & 0.96 \\
 & A3   & 0.82 & 1.00 & 0.90 \\
 & \textit{mean} & \textit{0.94} & \textit{0.86} & \textit{0.88} \\
\midrule
\multirow{4}{*}{Pairwise severity}
 & A1   & 0.86 & 0.86 & 0.86 \\
 & A2   & 0.92 & 0.97 & 0.94 \\
 & A3   & 0.88 & 0.78 & 0.83 \\
 & \textit{mean} & \textit{0.89} & \textit{0.87} & \textit{0.88} \\
\bottomrule
\end{tabular}
\end{table}

\begin{table}[!htbp]
\centering
\caption{Validation: LLM against the same human majority-vote labels used in \Cref{tab:consequence_interrater}.}
\label{tab:consequence_validation}
\small
\begin{tabular}{lrrr}
\toprule
\textbf{Task} & \textbf{Precision} & \textbf{Recall} & \textbf{F1} \\
\midrule
Affected party        & 0.87 & 0.98 & 0.92 \\
Vulnerable (any)      & 0.56 & 1.00 & 0.71 \\
Vulnerable (child)    & 0.64 & 1.00 & 0.78 \\
Pairwise severity     & 0.88 & 0.81 & 0.85 \\
\bottomrule
\end{tabular}
\end{table}

\paragraph{Results: the consequence profile of dilemmas}\label{subsubsec:consequence_profile}

\begin{figure}[!htbp]
\centering
\includegraphics[width=\linewidth]{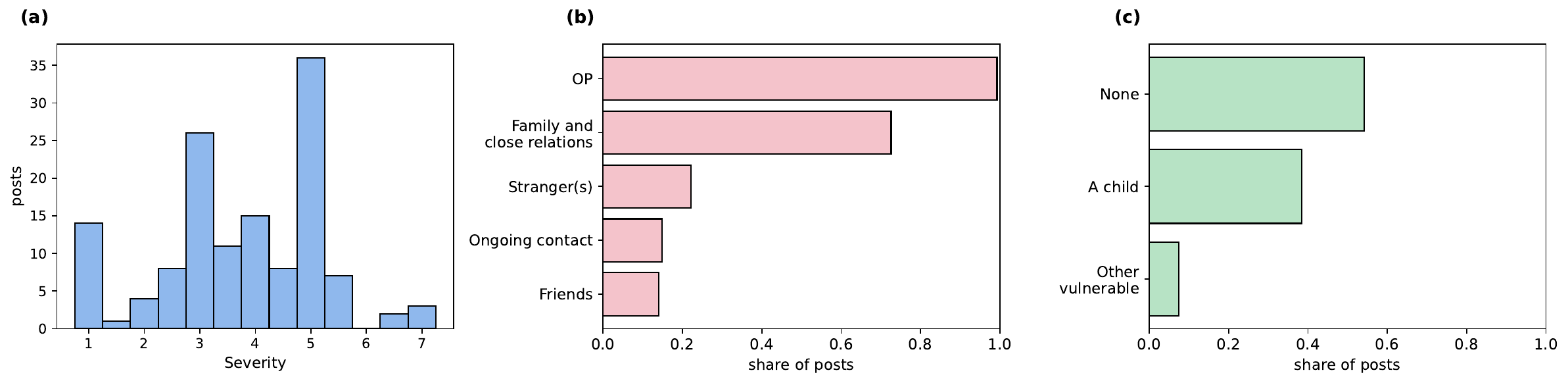}
\caption{\textbf{Profile of the 135 annotated dilemmas.}
\textbf{(a)} Distribution of severity (mean of three LLM runs, 1--7).
\textbf{(b)} Share of posts naming each affected party; the coding is
multi-select, so the bars do not sum to 100\%.
\textbf{(c)} Share of posts by vulnerability label; this coding is
single-select and does sum to 100\%.}
\label{fig:consequence_profile}
\end{figure}

\Cref{fig:consequence_profile} describes the consequence profile of the \dataset corpus. Severity covered the full 1--7 scale, with a median of 4.0 and an interquartile range of 3.0--5.0. Nineteen of the 135 dilemmas were rated 2 or below, indicating relatively minor matters of taste, etiquette, or annoyance, whereas 48 were rated 5 or above, indicating serious and lasting consequences. High-severity dilemmas concerned child welfare, family rejection, criminal reporting, and other lasting harms, whereas low-severity dilemmas concerned queues, seating, and minor etiquette disputes (\Cref{tab:consequence_exemplars}). The corpus therefore ranges from everyday friction to life-altering harm but is concentrated in the middle and upper-middle of the scale.

\begin{figure}[!htbp]
\centering
\includegraphics[width=\linewidth]{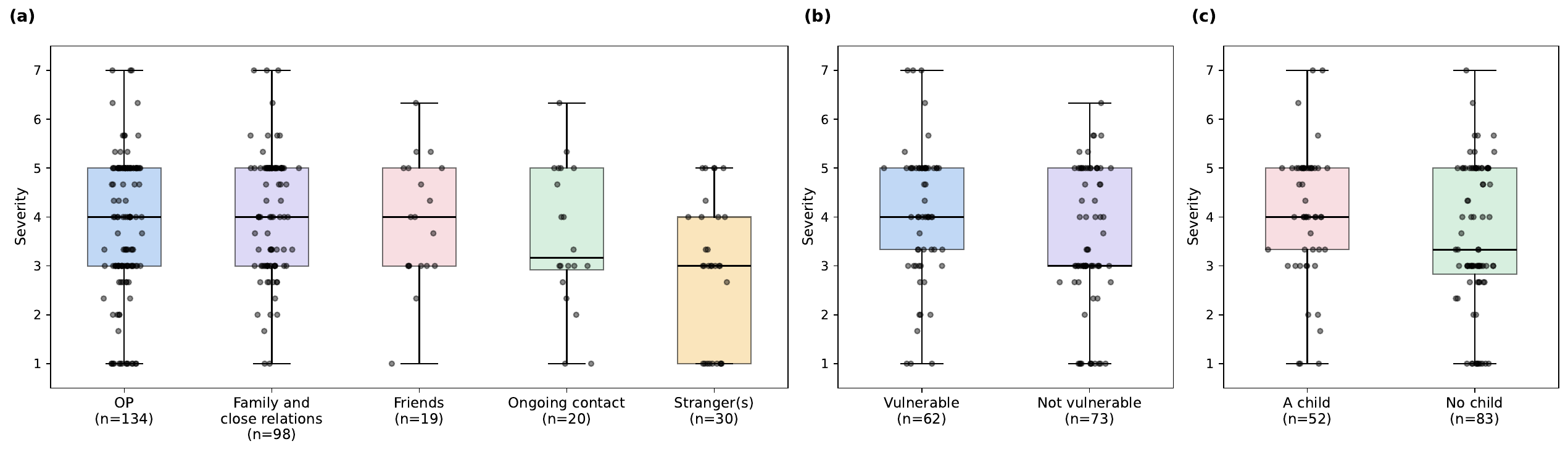}
\caption{\textbf{Severity by consequence feature.} Box plots with individual
dilemmas overlaid.
\textbf{(a)} By affected party. Because the coding is multi-select, a dilemma
contributes to every box it names, so the groups overlap.
\textbf{(b)} By whether any vulnerable person is affected.
\textbf{(c)} By whether a child specifically is affected.
Severity is higher when a vulnerable party is involved (median 4.0 vs 3.0) and
lower when only strangers are (median 3.0, never above 5.0).}
\label{fig:consequence_severity_by_group}
\end{figure}

For the affected-party profile, the author is implicated almost universally (99\%), which is unsurprising given that the author is narrating their own conflict. Beyond the author, the corpus divides across family (73\%), strangers (22\%), ongoing contacts (15\%), and friends (14\%).Just under half of the dilemmas (46\%) affect someone identifiably vulnerable, most often a child (39\%).
When intersecting severity with the affected-party profile, severity correlates only weakly with the presence of a vulnerable party ($r = 0.22$) and modestly and negatively with stranger involvement ($r = -0.31$).
Dilemmas affecting strangers never exceed 5.0 on the severity scale, whereas family- and child-involving dilemmas span its entire range (\Cref{fig:consequence_severity_by_group}).

\begin{table}[!htbp]
\centering
\caption{The ten highest- and ten lowest-severity posts in \dataset, with their annotations. Titles are given in full, as posted. Affected party is multi-select, so several posts name more than one group.}
\label{tab:consequence_exemplars}
\small
\begin{tabular}{@{}p{0.56\linewidth}ccp{0.15\linewidth}@{}}
\toprule
\textbf{Title} & \textbf{Severity} & \textbf{Vulnerable} & \textbf{Affected party} \\
\midrule
\multicolumn{4}{@{}l}{\textit{Ten highest severity}} \\
AITA for surrendering my sister's child to protective services when she forced me to babysit due to mental health? & 7.00 & child & OP, Family \\
AITA for telling my mom our illnesses aren't the same? & 7.00 & other & OP, Family \\
WIBTA If i don't adopt my sister's daughter & 7.00 & child & OP, Family \\
AITA for not wanting my son. & 6.33 & child & OP, Family \\
WIBTA if I snitched on a former friend to the feds for a 30k reward? & 6.33 & --- & OP, Friends, Ongoing \\
AITA for having my boyfriends dog put down without telling him? & 5.67 & --- & OP, Family \\
AITA-Don't Want Relationship w/Bio Child-reposted due to word-count & 5.67 & --- & OP, Family \\
AITA for making my father pay for a more expensive college out of spite? & 5.67 & --- & OP, Family \\
AITA for catfishing my underaged sister on Tinder and humiliating her in order to teach her a lesson? & 5.67 & child & OP, Family \\
AITA for leaving my drunk friend to fend for herself? & 5.33 & --- & OP, Friends \\
\midrule
\multicolumn{4}{@{}l}{\textit{Ten lowest severity}} \\
AITA - I saw a video of this guy video taping himself giving sandwiches to homeless people and I commented that he was vain & 1.00 & --- & OP, Stranger \\
AITA for calling my sister a fake vegan? & 1.00 & --- & OP, Family \\
AITA for refusing to allow someone to cut in front of me just because they have kids waiting? & 1.00 & child & OP, Stranger \\
WIBTA if I gift a globe to my flat earth friend? & 1.00 & --- & OP, Friends \\
AITA for trying to go to the front of the plane? & 1.00 & --- & OP, Stranger \\
AITA For comparing my wife to a cow? & 1.00 & --- & OP, Family \\
AITA for honking at another driver who cut in front of me at a drive thru each time she tries to order & 1.00 & --- & OP, Stranger \\
AITA for reaming out a woman in front of her kids? & 1.00 & child & OP, Stranger \\
AITA for refusing to move down from my spot at the bar? & 1.00 & --- & OP, Stranger \\
AITA for turning off the dryer on someone who removed my laundry before it was done? & 1.00 & --- & OP, Ongoing \\
\bottomrule
\end{tabular}
\end{table}

\paragraph{Results: consequences and judgment revision}\label{subsubsec:consequence_outcomes}

\begin{figure}[!htbp]
\centering
\includegraphics[width=\linewidth]{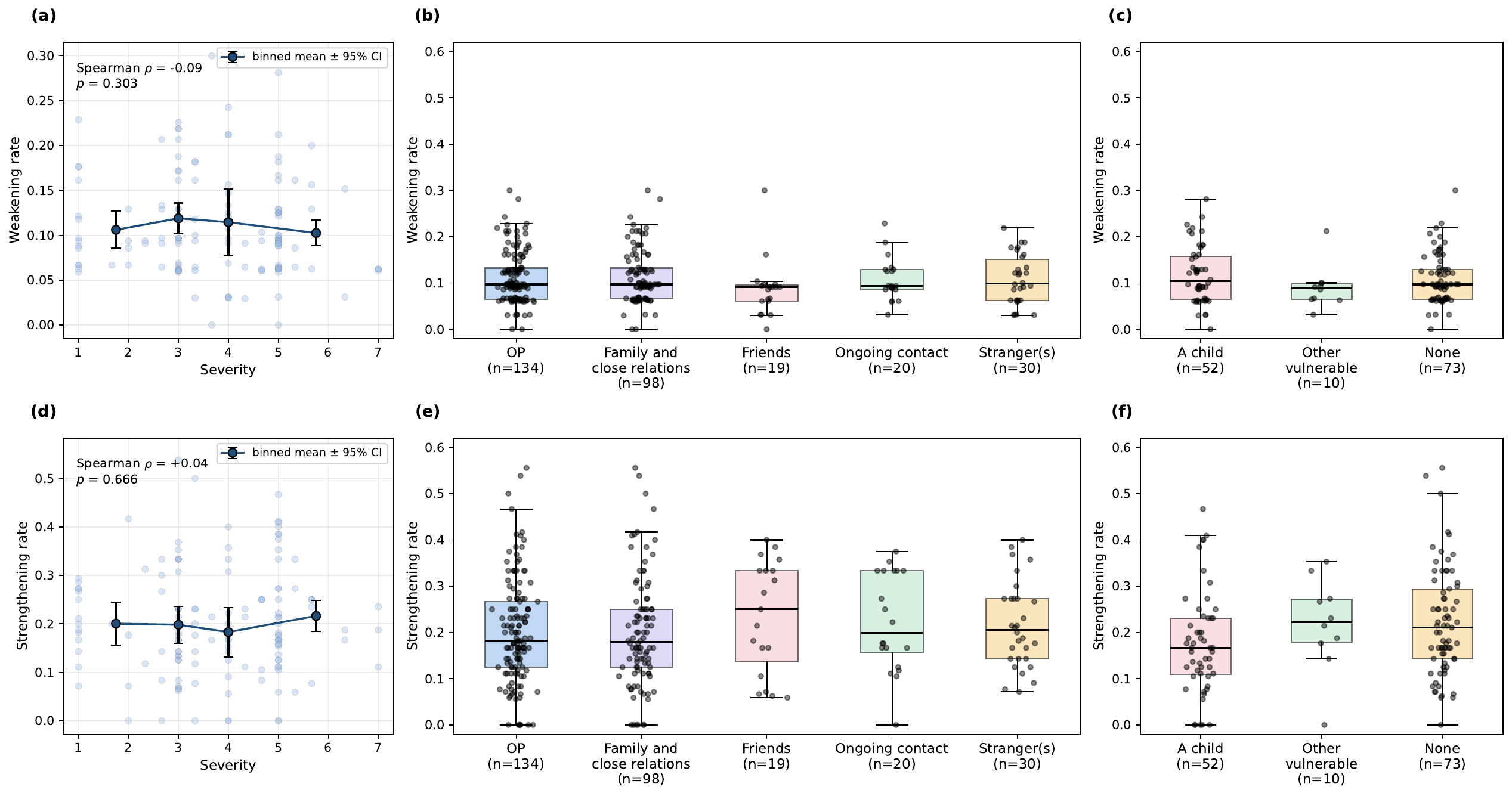}
\caption{\textbf{Dilemma-level revision rates against each stake feature.}
Each point is one dilemma (n = 135): its weakening rate over all responses to that
dilemma (top row) and its strengthening rate over responses with room to strengthen
(bottom row). Collapsing to the dilemma handles the nesting of responses within
dilemmas by construction, so no clustering correction is needed.
\textbf{(a, d)} Against severity, with Spearman $\rho$ inset; marker area is
proportional to the number of responses.
\textbf{(b, e)} By affected party. The coding is multi-select, so a dilemma appears in every box it names and the groups overlap.
\textbf{(c, f)} By vulnerability, which is single-select and does partition the
corpus.}
\label{fig:consequence_postlevel_rates}
\end{figure}

We next ask whether these features track how participants revised.
We report two rates.
The weakening rate is the share of all responses in a group that weakened.
The strengthening rate is the share that strengthened among responses with room to strengthen, that is, with an initial confidence below the ceiling.
Participants who begin at ceiling confidence cannot strengthen, so including them would measure how many started certain rather than how many were moved.
We then collapse each dilemma to a single pair of rates, so that every dilemma counts once regardless of how many responses it drew and the nesting of responses within dilemmas is handled by construction.
\Cref{fig:consequence_postlevel_rates} reports them, and \Cref{tab:consequence_postlevel_tests} gives the corresponding contrasts.
Because affected party is multi-select its categories overlap, so each pair is compared on disjoint subsets, and the author category is set aside as a near-constant, being named in 134 of the 135 dilemmas.

\Cref{fig:consequence_postlevel_rates} shows that severity and most affected-party and vulnerability categories had little association with dilemma-level weakening or strengthening rates. Two contrasts stood out, but we treat them as exploratory rather than as findings: dilemmas affecting family weakened more often than dilemmas affecting friends ($r = +0.61$, $p = .003$), and dilemmas involving a child strengthened less often than dilemmas involving nobody vulnerable ($r = -0.23$, $p = .031$). These contrasts require caution because the friends group contains only nine dilemmas, and each dilemma-level rate is based on only 6--10 responses per condition. Future work on how the consequences of moral dilemmas affect moral judgment change should include more dilemmas, balanced across affected-party and vulnerability categories, and more responses per dilemma to test whether the differences are reliable.

\begin{table}[!htbp]
\centering
\caption{Dilemma-level comparisons behind \Cref{fig:consequence_postlevel_rates}, panels (b), (c), (e) and (f). Each dilemma contributes one rate; contrasts are Mann-Whitney $U$ on those rates, with $r$ the rank-biserial correlation (positive means the first group is higher). A dilemma can name several affected parties, so the categories overlap. Each pair is therefore compared using only the dilemmas that tell them apart, those having affected party A but not B against those having affected party B but not A, with dilemmas having both affected parties set aside. The author category is omitted, being affected in 134 of 135 dilemmas and so having almost nothing to be contrasted against. Vulnerability is single-select.}
\label{tab:consequence_postlevel_tests}
\small
\begin{tabular}{llrrrrrr}
\toprule
\textbf{Outcome} & \textbf{Contrast (A vs B)} & $n_A$ & $n_B$ & med$_A$ & med$_B$ & $r$ & $p$ \\
\midrule
\multirow{9}{*}{Weakening}
 & \multicolumn{7}{l}{\textit{Affected party}} \\
 & Family vs Friends & 88 & 9 & 0.097 & 0.065 & +0.61 & 0.003$^{*}$ \\
 & Family vs Ongoing contact & 93 & 15 & 0.097 & 0.094 & +0.13 & 0.418 \\
 & Family vs Stranger & 89 & 21 & 0.097 & 0.118 & +0.02 & 0.909 \\
 & Friends vs Ongoing contact & 15 & 16 & 0.091 & 0.109 & -0.33 & 0.118 \\
 & Friends vs Stranger & 15 & 26 & 0.091 & 0.119 & -0.27 & 0.151 \\
 & Ongoing contact vs Stranger & 17 & 27 & 0.091 & 0.094 & -0.02 & 0.904 \\
 & \multicolumn{7}{l}{\textit{Vulnerability}} \\
 & a child vs other vulnerable & 52 & 10 & 0.104 & 0.088 & +0.23 & 0.250 \\
 & a child vs none & 52 & 73 & 0.104 & 0.097 & +0.07 & 0.524 \\
 & other vulnerable vs none & 10 & 73 & 0.088 & 0.097 & -0.18 & 0.348 \\
\midrule
\multirow{9}{*}{Strengthening}
 & \multicolumn{7}{l}{\textit{Affected party}} \\
 & Family vs Friends & 88 & 9 & 0.176 & 0.286 & -0.25 & 0.223 \\
 & Family vs Ongoing contact & 93 & 15 & 0.182 & 0.250 & -0.21 & 0.197 \\
 & Family vs Stranger & 89 & 21 & 0.182 & 0.211 & -0.16 & 0.255 \\
 & Friends vs Ongoing contact & 15 & 16 & 0.250 & 0.199 & +0.09 & 0.678 \\
 & Friends vs Stranger & 15 & 26 & 0.182 & 0.185 & +0.03 & 0.903 \\
 & Ongoing contact vs Stranger & 17 & 27 & 0.176 & 0.200 & +0.02 & 0.933 \\
 & \multicolumn{7}{l}{\textit{Vulnerability}} \\
 & a child vs other vulnerable & 52 & 10 & 0.167 & 0.223 & -0.31 & 0.123 \\
 & a child vs none & 52 & 73 & 0.167 & 0.211 & -0.23 & 0.031$^{*}$ \\
 & other vulnerable vs none & 10 & 73 & 0.223 & 0.211 & +0.07 & 0.711 \\
\bottomrule
\end{tabular}
\vspace{2pt}
\end{table}

\textbf{Summary.}
The \dataset dilemmas are diverse along three axes: how much is at stake, who bears the cost, and whether a vulnerable person is involved. The dilemmas cover a wide severity scale but are concentrated at medium to high stakes, with family or partners the most common affected party beyond the author and almost half involving a vulnerable person. Severity is higher when vulnerable people are involved and lower when strangers are involved, while dilemmas involving family or children cover the full severity range. Weakening and strengthening rates are otherwise largely similar across severity, affected-party, and vulnerability categories. Two exceptions stand out as suggestive patterns rather than confirmatory: dilemmas affecting family weaken more often than dilemmas affecting friends, and dilemmas involving a child strengthen less often than dilemmas involving nobody vulnerable.

\clearpage
\subsubsection{Participant feedback}\label{app:overall_feedback}
\begin{figure}[h]
    \centering
    \begin{subfigure}[t]{0.48\linewidth}
        \centering
        \includegraphics[width=\linewidth]{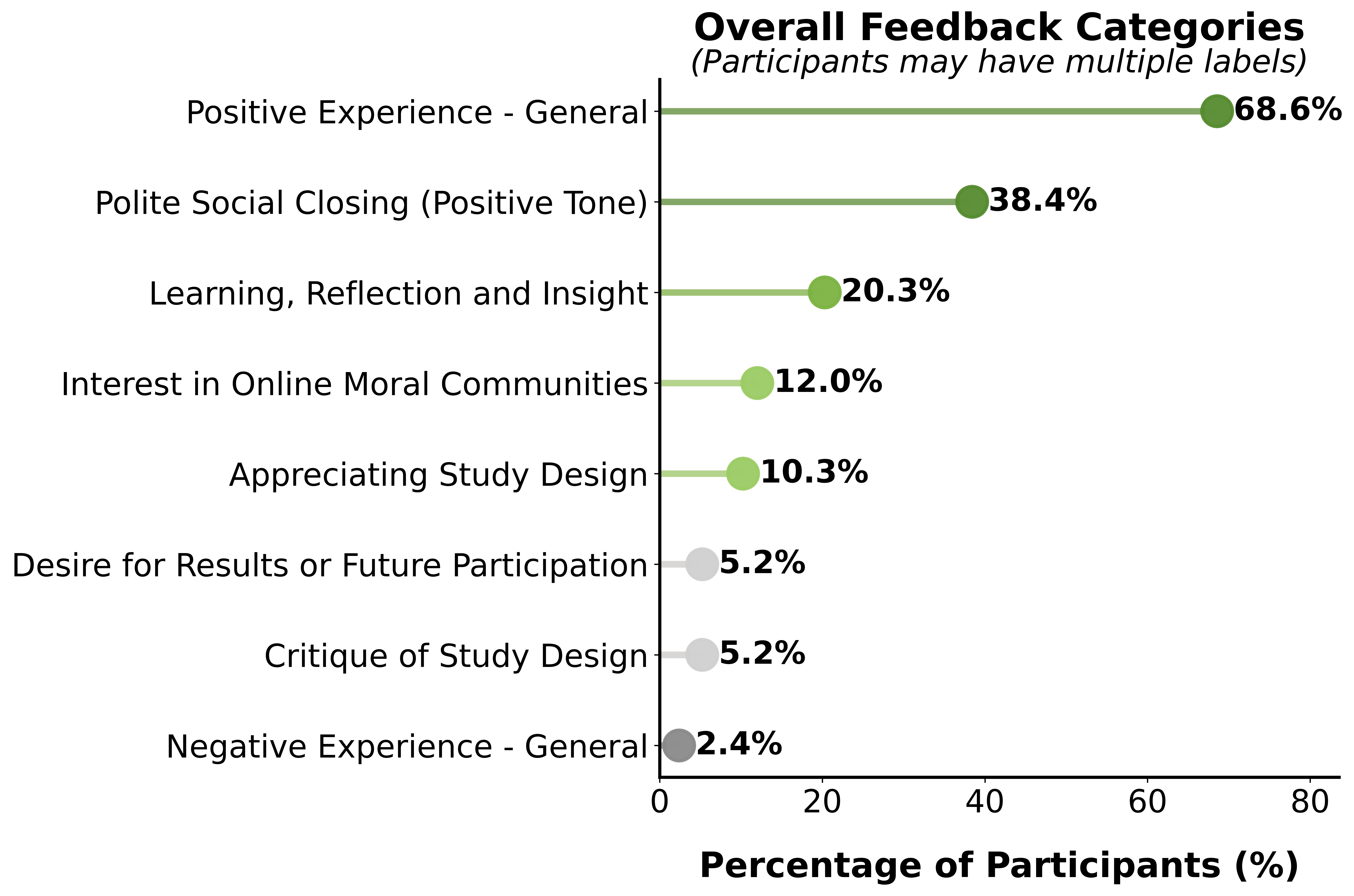}
        \caption{}
        \label{fig:overall_feedback_dist}
    \end{subfigure}
    \hfill
    \begin{subfigure}[t]{0.5\linewidth}
        \centering
        \includegraphics[width=\linewidth]{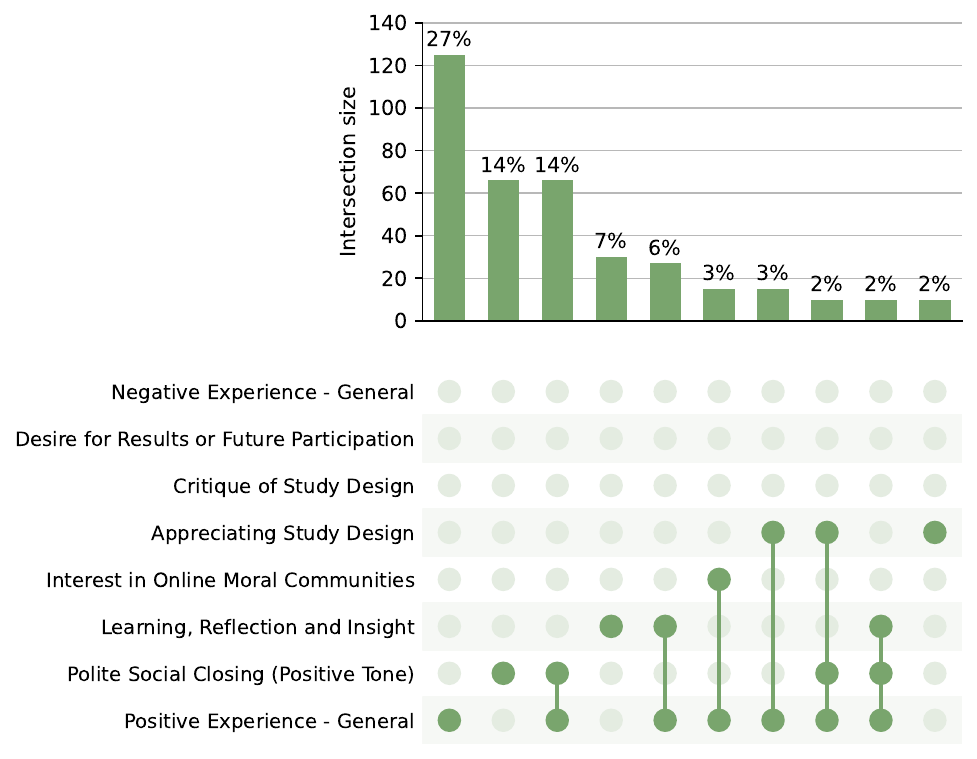}
        \caption{}
        \label{fig:overall_feedback_upset}
    \end{subfigure}
    \caption{\textbf{Participant feedback categories.}
    \textbf{(a)} Distribution of feedback categories among the 458 meaningful responses retained after quality filtering and the exclusion of neutral or empty entries. Bars show the percentage of responses assigned each category; a response may carry more than one category.
    \textbf{(b)} The ten most frequent category combinations, shown as an UpSet plot. Each column is one combination of co-occurring categories, ordered by frequency.}
    \label{fig:overall_feedback}
\end{figure}

After completing the full survey, participants could optionally leave overall feedback on their experience.
After applying the participant-level quality filter, 573 feedback entries remained. We excluded 115 entries coded as neutral or empty, leaving 458 meaningful responses (21.2\% of the full sample) for category analysis.
We find that the overall open responses fall into eight categories: Positive Experience (General), Polite Social Closing (Positive Tone), Learning, Reflection and Insight, Interest in Online Moral Communities, Appreciating Study Design, Critique of Study Design, Desire for Results or Future Participation, and Negative Experience (General).
A single response could carry more than one category.

Figure~\ref{fig:overall_feedback_dist} shows the share of feedback givers assigned to each category.
Most feedback was positive.
Positive Experience (General) was the most common category, at 68.6\%.
Polite Social Closing followed, at 38.4\%.
Learning, Reflection and Insight reached 20.3\%.
Interest in Online Moral Communities reached 12.0\%.
Appreciating Study Design reached 10.3\%.
The remaining categories were small.
Critique of Study Design and Desire for Results or Future Participation each appeared in 5.2\% of responses.
Negative Experience (General) appeared in 2.4\%.
Figure~\ref{fig:overall_feedback_upset} shows the ten most frequent category combinations.
Positive Experience on its own was the largest combination, at 27.3\%.
Polite Social Closing on its own and Positive Experience together with Polite Social Closing each accounted for 14.4\%.
Most of the remaining combinations also paired a positive category with one further category, such as Learning, Reflection and Insight or Appreciating Study Design.
This pattern shows that positive responses were common.
It also shows that participants often noted more than one positive feature of the task in the same response.

Table~\ref{tab:feedback-examples} lists representative quotes by category and treatment arm.
The quotes show the responses behind the category counts.
Several participants noted that the AITA material linked the task to a real online community.
Some described the task as a chance to reflect on their own judgment.
A small number raised design concerns.
One participant questioned the second judgment page, where no new information was added before the second response.
In sum, positive or reflective feedback was common among those who responded, and participants often valued more than one feature of the task.
Because feedback was optional, these patterns may not represent the full sample.

\definecolor{tickgreen}{RGB}{85,139,47}
\definecolor{tCtrl}{HTML}{C8C8C8}
\definecolor{tCntr}{HTML}{F4C3CB}
\definecolor{tCert}{HTML}{8FB8ED}
\definecolor{tBoth}{HTML}{C2BBF0}
\providecommand{\cmark}{\textcolor{tickgreen}{\checkmark}}
\providecommand{\tdot}[1]{\textcolor{#1}{\ensuremath{\blacksquare}}}

\begin{table}[!htbp]
\centering
\caption{Representative participant feedback by category. Quotes are drawn from the 458 meaningful responses retained after quality filtering and the exclusion of neutral or empty entries. A check mark indicates that the quote was coded into that category; quotes may carry more than one category.
Treatment is shown by the color of the dot:
gray~=~Control, pink~=~Controversy, blue~=~Certainty, purple~=~Controversy+Certainty.
Column headers map to: Positive Exp.\ = Positive Experience (General); Polite Closing = Polite Social Closing (Positive Tone); Learning = Learning, Reflection and Insight; Community Interest = Interest in Online Moral Communities; Praise (Design) = Appreciating Study Design; Critique (Design) = Critique of Study Design; Future Run = Desire for Results or Future Participation; Negative Exp.\ = Negative Experience (General).}
\label{tab:feedback-examples}
\footnotesize
\setlength{\tabcolsep}{3pt}
\renewcommand{\arraystretch}{1.25}
\begin{tabularx}{\linewidth}{c X *{8}{c}}
\toprule
\textbf{Tr.} & \textbf{Quote} & \makecell[l]{\textbf{Positive}\\\textbf{Exp.}} & \makecell[l]{\textbf{Polite}\\\textbf{Closing}} & \makecell[l]{\textbf{Learning,}\\\textbf{Reflection}} & \makecell[l]{\textbf{Community}\\\textbf{Interest}} & \makecell[l]{\textbf{Praise}\\\textbf{(Design)}} & \makecell[l]{\textbf{Critique}\\\textbf{(Design)}} & \makecell[l]{\textbf{Future}\\\textbf{Run}} & \makecell[l]{\textbf{Negative}\\\textbf{Exp.}} \\
\midrule
\mbox{\tdot{tCtrl}} & ``The study was easy to navigate and the two stories provided were very interesting.'' & \cmark &  &  &  & \cmark &  &  &  \\
\mbox{\tdot{tCtrl}} & ``I found the study very interesting. Thank you for allowing me to participate in your study. I look forward to more studies from you.'' & \cmark & \cmark &  &  &  &  & \cmark &  \\
\mbox{\tdot{tCntr}} & ``This was a fun study wouldn't mind doing more judgements.'' & \cmark &  &  &  &  &  & \cmark &  \\
\mbox{\tdot{tCntr}} & ``The study was fun and educational'' & \cmark &  & \cmark &  &  &  &  &  \\
\mbox{\tdot{tCntr}} & ``I enjoyed this study because it is connected to reality (subreddit r/AITA) :)'' & \cmark &  &  & \cmark &  &  &  &  \\
\mbox{\tdot{tCntr}} & ``AITA on Reddit are the best reads, truly an interesting study!'' & \cmark &  &  & \cmark &  &  &  &  \\
\mbox{\tdot{tCntr}} & ``This was one of the most interesting things I have ever done, I will look for reddit and read more of what people are saying.'' & \cmark &  &  & \cmark &  &  &  &  \\
\mbox{\tdot{tCntr}} & ``This survey was clear, great and engaging.'' & \cmark &  &  &  & \cmark &  &  &  \\
\mbox{\tdot{tCert}} & ``It was a nice survey and a great experience'' & \cmark &  &  &  &  &  &  &  \\
\mbox{\tdot{tCert}} & ``Eye opening and interesting.'' & \cmark &  & \cmark &  &  &  &  &  \\
\mbox{\tdot{tCert}} & ``Makes someone think outside the box'' & \cmark &  & \cmark &  &  &  &  &  \\
\mbox{\tdot{tCert}} & ``I love reddit and spend time on it. This was fun!'' & \cmark &  &  & \cmark &  &  &  &  \\
\mbox{\tdot{tCert}} & ``This was cool! Id never heatd of this before and im going to check out the reddit now'' & \cmark &  &  & \cmark &  &  &  &  \\
\mbox{\tdot{tCert}} & ``The survey was engaging and well structured.'' & \cmark &  &  &  & \cmark &  &  &  \\
\mbox{\tdot{tCert}} & ``The instructions were clear, I enjoyed taking part in this survey'' & \cmark &  &  &  & \cmark &  &  &  \\
\mbox{\tdot{tBoth}} & ``I enjoyed the survey! Thank you for inviting participants to engage in fun and thought-provoking research.'' & \cmark & \cmark &  &  &  &  &  &  \\
\mbox{\tdot{tBoth}} & ``It was interesting study and behavioral questions may include social life scenarios to identify real nature of decision making.'' & \cmark &  & \cmark &  &  &  &  &  \\
\mbox{\tdot{tBoth}} & ``I liked this study as it made me learn about my decision making too and proved (at least to me) that I think I judge a situation correctly and that I am not easily swayed by the opinions of others'' & \cmark &  & \cmark &  &  &  &  &  \\
\mbox{\tdot{tBoth}} & ``Fun study! I am part of that subreddit and always enjoy reading the discourse.'' & \cmark &  &  & \cmark &  &  &  &  \\
\mbox{\tdot{tBoth}} & ``Everything was good,just the font used for reddit posts were not the most convinent to read'' & \cmark &  &  &  &  & \cmark &  &  \\
\mbox{\tdot{tBoth}} & ``A very different and unique survey that was engaging and interesting to participate in. The AITA was fun and would have loved to do more for more money.'' & \cmark &  &  & \cmark &  &  & \cmark &  \\
\mbox{\tdot{tCntr}} & ``I think people should stop judging a situation that they have not been exposed to before.'' &  &  & \cmark &  &  &  &  &  \\
\mbox{\tdot{tCert}} & ``Reddit is a cesspool. It represents the worst people society has to offer. The opinions there dont represent society accurately.'' &  &  &  & \cmark &  &  &  & \cmark \\
\mbox{\tdot{tCtrl}} & ``I pointed out in the first box but I'm not sure what's the point of the 2nd page of each post where I'm asked if I changed my decision. No extra information is added so I can't really change my mind and will just press the same options I picked prior.'' &  &  &  &  &  & \cmark &  & \cmark \\
\bottomrule
\end{tabularx}
\end{table}

\clearpage

\renewcommand{\refname}{Appendix references}
\putbib[Reference]
\end{bibunit}


\begin{thebibliography}{}
\renewcommand{\doi}[1]{\url{https://doi.org/#1}}
\bibcommenthead

\bibitem [\protect \citeauthoryear {%
{AITA Moderators}%
}{%
{AITA Moderators}%
}{%
{\protect \APACyear {2026}}%
}]{%
reddit_aita_stats_2026}
\APACinsertmetastar {%
reddit_aita_stats_2026}%
\begin{APACrefauthors}%
{AITA Moderators}%
\end{APACrefauthors}%
\unskip\
\newblock
\APACrefYearMonthDay{2026}{}{}.
\newblock
\APACrefbtitle {r/{A}m{I}the{A}sshole Subreddit Community Statistics.} {r/{A}m{I}the{A}sshole subreddit community statistics.}
\newblock
\APAChowpublished {Reddit}.
\newblock
\begin{APACrefURL} {https://www.reddit.com/r/AmItheAsshole/} \end{APACrefURL}
\newblock
\APACrefnote{Accessed: 2026-08-13; Displays live metrics: 2.8M Weekly Visitors (labeled as ``Potential Assholes'') and 130K Weekly Contributions (labeled as ``Judging you right now'')}
\PrintBackRefs{\CurrentBib}

\bibitem [\protect \citeauthoryear {%
Awad%
\ \protect \BOthers {.}}{%
Awad%
\ \protect \BOthers {.}}{%
{\protect \APACyear {2018}}%
}]{%
Awad2018TheMM}
\APACinsertmetastar {%
Awad2018TheMM}%
\begin{APACrefauthors}%
Awad, E.%
, Dsouza, S.%
, Kim, R.%
, Schulz, J.F.%
, Henrich, J.%
, Shariff, A.F.%
\BDBL {}Rahwan, I.%
\end{APACrefauthors}%
\unskip\
\newblock
\APACrefYearMonthDay{2018}{}{}.
\newblock
{\BBOQ}\APACrefatitle {The Moral Machine experiment} {The moral machine experiment}.{\BBCQ}
\newblock
\APACjournalVolNumPages{Nature}{563}{7729}{59--64,}
\newblock
\begin{APACrefDOI} \doi{10.1038/s41586-018-0637-6} \end{APACrefDOI}
\newblock

\newblock

\PrintBackRefs{\CurrentBib}

\bibitem [\protect \citeauthoryear {%
Balietti%
, Getoor%
, Goldstein%
\BCBL {}\ \BBA {} Watts%
}{%
Balietti%
\ \protect \BOthers {.}}{%
{\protect \APACyear {2021}}%
}]{%
Balietti2021ReducingOP}
\APACinsertmetastar {%
Balietti2021ReducingOP}%
\begin{APACrefauthors}%
Balietti, S.%
, Getoor, L.%
, Goldstein, D.G.%
\BCBL {} Watts, D.J.%
\end{APACrefauthors}%
\unskip\
\newblock
\APACrefYearMonthDay{2021}{}{}.
\newblock
{\BBOQ}\APACrefatitle {Reducing opinion polarization: Effects of exposure to similar people with differing political views} {Reducing opinion polarization: Effects of exposure to similar people with differing political views}.{\BBCQ}
\newblock
\APACjournalVolNumPages{Proceedings of the National Academy of Sciences of the United States of America}{118}{52}{e2112552118,}
\newblock
\begin{APACrefDOI} \doi{10.1073/pnas.2112552118} \end{APACrefDOI}
\newblock

\newblock

\PrintBackRefs{\CurrentBib}

\bibitem [\protect \citeauthoryear {%
Bang%
\ \protect \BOthers {.}}{%
Bang%
\ \protect \BOthers {.}}{%
{\protect \APACyear {2017}}%
}]{%
bang2017confidence}
\APACinsertmetastar {%
bang2017confidence}%
\begin{APACrefauthors}%
Bang, D.%
, Aitchison, L.%
, Moran, R.%
, Herce~Casta{\~n}{\'o}n, S.%
, Rafiee, B.%
, Mahmoodi, A.%
\BDBL {}Summerfield, C.%
\end{APACrefauthors}%
\unskip\
\newblock
\APACrefYearMonthDay{2017}{}{}.
\newblock
{\BBOQ}\APACrefatitle {Confidence Matching in Group Decision-Making} {Confidence matching in group decision-making}.{\BBCQ}
\newblock
\APACjournalVolNumPages{Nature Human Behaviour}{1}{6}{0117,}
\newblock
\begin{APACrefDOI} \doi{10.1038/s41562-017-0117} \end{APACrefDOI}
\newblock

\newblock

\PrintBackRefs{\CurrentBib}

\bibitem [\protect \citeauthoryear {%
Barnard%
}{%
Barnard%
}{%
{\protect \APACyear {1945}}%
}]{%
barnard1945new}
\APACinsertmetastar {%
barnard1945new}%
\begin{APACrefauthors}%
Barnard, G.A.%
\end{APACrefauthors}%
\unskip\
\newblock
\APACrefYearMonthDay{1945}{}{}.
\newblock
{\BBOQ}\APACrefatitle {A new test for 2×2 tables} {A new test for 2×2 tables}.{\BBCQ}
\newblock
\APACjournalVolNumPages{Nature}{156}{}{177,}
\newblock
\begin{APACrefDOI} \doi{10.1038/156177a0} \end{APACrefDOI}
\newblock

\newblock

\PrintBackRefs{\CurrentBib}

\bibitem [\protect \citeauthoryear {%
Benjamini%
\ \BBA {} Hochberg%
}{%
Benjamini%
\ \BBA {} Hochberg%
}{%
{\protect \APACyear {1995}}%
}]{%
benjamini1995controlling}
\APACinsertmetastar {%
benjamini1995controlling}%
\begin{APACrefauthors}%
Benjamini, Y.%
\BCBT {}\ \BBA {} Hochberg, Y.%
\end{APACrefauthors}%
\unskip\
\newblock
\APACrefYearMonthDay{1995}{}{}.
\newblock
{\BBOQ}\APACrefatitle {Controlling the false discovery rate: a practical and powerful approach to multiple testing} {Controlling the false discovery rate: a practical and powerful approach to multiple testing}.{\BBCQ}
\newblock
\APACjournalVolNumPages{Journal of the Royal Statistical Society: Series B (Methodological)}{57}{1}{289--300,}
\newblock
\begin{APACrefDOI} \doi{10.1111/j.2517-6161.1995.tb02031.x} \end{APACrefDOI}
\newblock

\newblock

\PrintBackRefs{\CurrentBib}

\bibitem [\protect \citeauthoryear {%
Boot%
, Simons%
, Stothart%
\BCBL {}\ \BBA {} Stutts%
}{%
Boot%
\ \protect \BOthers {.}}{%
{\protect \APACyear {2013}}%
}]{%
boot2013pervasive}
\APACinsertmetastar {%
boot2013pervasive}%
\begin{APACrefauthors}%
Boot, W.R.%
, Simons, D.J.%
, Stothart, C.%
\BCBL {} Stutts, C.%
\end{APACrefauthors}%
\unskip\
\newblock
\APACrefYearMonthDay{2013}{}{}.
\newblock
{\BBOQ}\APACrefatitle {The Pervasive Problem With Placebos in Psychology: Why Active Control Groups Are Not Sufficient to Rule Out Placebo Effects} {The pervasive problem with placebos in psychology: Why active control groups are not sufficient to rule out placebo effects}.{\BBCQ}
\newblock
\APACjournalVolNumPages{Perspectives on Psychological Science}{8}{4}{445--454,}
\newblock
\begin{APACrefDOI} \doi{10.1177/1745691613491271} \end{APACrefDOI}
\newblock

\newblock

\PrintBackRefs{\CurrentBib}

\bibitem [\protect \citeauthoryear {%
Budescu%
, Por%
\BCBL {}\ \BBA {} Broomell%
}{%
Budescu%
\ \protect \BOthers {.}}{%
{\protect \APACyear {2012}}%
}]{%
Budescu2012EffectiveCO}
\APACinsertmetastar {%
Budescu2012EffectiveCO}%
\begin{APACrefauthors}%
Budescu, D.V.%
, Por, H\BHBI H.%
\BCBL {} Broomell, S.B.%
\end{APACrefauthors}%
\unskip\
\newblock
\APACrefYearMonthDay{2012}{}{}.
\newblock
{\BBOQ}\APACrefatitle {Effective communication of uncertainty in the IPCC reports} {Effective communication of uncertainty in the ipcc reports}.{\BBCQ}
\newblock
\APACjournalVolNumPages{Climatic Change}{113}{2}{181--200,}
\newblock
\begin{APACrefDOI} \doi{10.1007/s10584-011-0330-3} \end{APACrefDOI}
\newblock

\newblock

\PrintBackRefs{\CurrentBib}

\bibitem [\protect \citeauthoryear {%
Candia%
, Morales%
, Monti%
\BCBL {}\ \BBA {} Bonchi%
}{%
Candia%
\ \protect \BOthers {.}}{%
{\protect \APACyear {2022}}%
}]{%
Candia2022SocialNO}
\APACinsertmetastar {%
Candia2022SocialNO}%
\begin{APACrefauthors}%
Candia, S.D.%
, Morales, G.D.F.%
, Monti, C.%
\BCBL {} Bonchi, F.%
\end{APACrefauthors}%
\unskip\
\newblock
\APACrefYearMonthDay{2022}{}{}.
\newblock
{\BBOQ}\APACrefatitle {Social Norms on Reddit: A Demographic Analysis} {Social norms on reddit: A demographic analysis}.{\BBCQ}
\newblock
 \APACrefbtitle {Proceedings of the 14th ACM Web Science Conference 2022} {Proceedings of the 14th acm web science conference 2022}\ (\BPGS\ 139--147).
\PrintBackRefs{\CurrentBib}

\bibitem [\protect \citeauthoryear {%
Christensen%
}{%
Christensen%
}{%
{\protect \APACyear {2007}}%
}]{%
christensen2007epistemology}
\APACinsertmetastar {%
christensen2007epistemology}%
\begin{APACrefauthors}%
Christensen, D.%
\end{APACrefauthors}%
\unskip\
\newblock
\APACrefYearMonthDay{2007}{}{}.
\newblock
{\BBOQ}\APACrefatitle {Epistemology of disagreement: The good news} {Epistemology of disagreement: The good news}.{\BBCQ}
\newblock
\APACjournalVolNumPages{The Philosophical Review}{116}{2}{187--217,}
\newblock
\begin{APACrefDOI} \doi{10.1215/00318108-2006-035} \end{APACrefDOI}
\newblock

\newblock

\PrintBackRefs{\CurrentBib}

\bibitem [\protect \citeauthoryear {%
Clifford%
, Iyengar%
, Cabeza%
\BCBL {}\ \BBA {} Sinnott-Armstrong%
}{%
Clifford%
\ \protect \BOthers {.}}{%
{\protect \APACyear {2015}}%
}]{%
clifford2015moral}
\APACinsertmetastar {%
clifford2015moral}%
\begin{APACrefauthors}%
Clifford, S.%
, Iyengar, V.%
, Cabeza, R.%
\BCBL {} Sinnott-Armstrong, W.%
\end{APACrefauthors}%
\unskip\
\newblock
\APACrefYearMonthDay{2015}{}{}.
\newblock
{\BBOQ}\APACrefatitle {Moral foundations vignettes: A standardized stimulus database of scenarios based on moral foundations theory} {Moral foundations vignettes: A standardized stimulus database of scenarios based on moral foundations theory}.{\BBCQ}
\newblock
\APACjournalVolNumPages{Behavior Research Methods}{47}{4}{1178--1198,}
\newblock
\begin{APACrefDOI} \doi{10.3758/s13428-014-0551-2} \end{APACrefDOI}
\newblock

\newblock

\PrintBackRefs{\CurrentBib}

\bibitem [\protect \citeauthoryear {%
Cohen%
, Cohen%
, Aiken%
\BCBL {}\ \BBA {} West%
}{%
Cohen%
\ \protect \BOthers {.}}{%
{\protect \APACyear {1999}}%
}]{%
Cohen1999POMP}
\APACinsertmetastar {%
Cohen1999POMP}%
\begin{APACrefauthors}%
Cohen, P.%
, Cohen, J.%
, Aiken, L.S.%
\BCBL {} West, S.G.%
\end{APACrefauthors}%
\unskip\
\newblock
\APACrefYearMonthDay{1999}{}{}.
\newblock
{\BBOQ}\APACrefatitle {The Problem of Units and the Circumstance for {POMP}} {The problem of units and the circumstance for {POMP}}.{\BBCQ}
\newblock
\APACjournalVolNumPages{Multivariate Behavioral Research}{34}{3}{315--346,}
\newblock
\begin{APACrefDOI} \doi{10.1207/S15327906MBR3403_2} \end{APACrefDOI}
\newblock

\newblock

\PrintBackRefs{\CurrentBib}

\bibitem [\protect \citeauthoryear {%
Costello%
, Pennycook%
\BCBL {}\ \BBA {} Rand%
}{%
Costello%
\ \protect \BOthers {.}}{%
{\protect \APACyear {2024}}%
}]{%
costello2024durably}
\APACinsertmetastar {%
costello2024durably}%
\begin{APACrefauthors}%
Costello, T.H.%
, Pennycook, G.%
\BCBL {} Rand, D.G.%
\end{APACrefauthors}%
\unskip\
\newblock
\APACrefYearMonthDay{2024}{}{}.
\newblock
{\BBOQ}\APACrefatitle {Durably reducing conspiracy beliefs through dialogues with {AI}} {Durably reducing conspiracy beliefs through dialogues with {AI}}.{\BBCQ}
\newblock
\APACjournalVolNumPages{Science}{385}{6714}{eadq1814,}
\newblock
\begin{APACrefDOI} \doi{10.1126/science.adq1814} \end{APACrefDOI}
\newblock

\newblock

\PrintBackRefs{\CurrentBib}

\bibitem [\protect \citeauthoryear {%
FeldmanHall%
\ \BBA {} Shenhav%
}{%
FeldmanHall%
\ \BBA {} Shenhav%
}{%
{\protect \APACyear {2019}}%
}]{%
feldmanhall2019resolving}
\APACinsertmetastar {%
feldmanhall2019resolving}%
\begin{APACrefauthors}%
FeldmanHall, O.%
\BCBT {}\ \BBA {} Shenhav, A.%
\end{APACrefauthors}%
\unskip\
\newblock
\APACrefYearMonthDay{2019}{}{}.
\newblock
{\BBOQ}\APACrefatitle {Resolving uncertainty in a social world} {Resolving uncertainty in a social world}.{\BBCQ}
\newblock
\APACjournalVolNumPages{Nature Human Behaviour}{3}{5}{426--435,}
\newblock
\begin{APACrefDOI} \doi{10.1038/s41562-019-0590-x} \end{APACrefDOI}
\newblock

\newblock

\PrintBackRefs{\CurrentBib}

\bibitem [\protect \citeauthoryear {%
Gale{\v{s}}i{\'c}%
, Garcia-Retamero%
\BCBL {}\ \BBA {} Gigerenzer%
}{%
Gale{\v{s}}i{\'c}%
\ \protect \BOthers {.}}{%
{\protect \APACyear {2009}}%
}]{%
GalesicGarciaRetameroGigerenzer2009}
\APACinsertmetastar {%
GalesicGarciaRetameroGigerenzer2009}%
\begin{APACrefauthors}%
Gale{\v{s}}i{\'c}, M.%
, Garcia-Retamero, R.%
\BCBL {} Gigerenzer, G.%
\end{APACrefauthors}%
\unskip\
\newblock
\APACrefYearMonthDay{2009}{}{}.
\newblock
{\BBOQ}\APACrefatitle {Using Icon Arrays to Communicate Medical Risks: Overcoming Low Numeracy} {Using icon arrays to communicate medical risks: Overcoming low numeracy}.{\BBCQ}
\newblock
\APACjournalVolNumPages{Health Psychology}{28}{2}{210--216,}
\newblock
\begin{APACrefDOI} \doi{10.1037/a0014474} \end{APACrefDOI}
\newblock

\newblock

\PrintBackRefs{\CurrentBib}

\bibitem [\protect \citeauthoryear {%
Gneiting%
\ \BBA {} Raftery%
}{%
Gneiting%
\ \BBA {} Raftery%
}{%
{\protect \APACyear {2007}}%
}]{%
gneiting2007strictly}
\APACinsertmetastar {%
gneiting2007strictly}%
\begin{APACrefauthors}%
Gneiting, T.%
\BCBT {}\ \BBA {} Raftery, A.E.%
\end{APACrefauthors}%
\unskip\
\newblock
\APACrefYearMonthDay{2007}{}{}.
\newblock
{\BBOQ}\APACrefatitle {Strictly Proper Scoring Rules, Prediction, and Estimation} {Strictly proper scoring rules, prediction, and estimation}.{\BBCQ}
\newblock
\APACjournalVolNumPages{Journal of the American Statistical Association}{102}{477}{359--378,}
\newblock
\begin{APACrefDOI} \doi{10.1198/016214506000001437} \end{APACrefDOI}
\newblock

\newblock

\PrintBackRefs{\CurrentBib}

\bibitem [\protect \citeauthoryear {%
Gosling%
, Rentfrow%
\BCBL {}\ \BBA {} Swann%
}{%
Gosling%
\ \protect \BOthers {.}}{%
{\protect \APACyear {2003}}%
}]{%
Gosling2003AVB}
\APACinsertmetastar {%
Gosling2003AVB}%
\begin{APACrefauthors}%
Gosling, S.D.%
, Rentfrow, P.J.%
\BCBL {} Swann, W.B.%
\end{APACrefauthors}%
\unskip\
\newblock
\APACrefYearMonthDay{2003}{}{}.
\newblock
{\BBOQ}\APACrefatitle {A very brief measure of the Big-Five personality domains} {A very brief measure of the big-five personality domains}.{\BBCQ}
\newblock
\APACjournalVolNumPages{Journal of Research in Personality}{37}{6}{504--528,}
\newblock
\begin{APACrefDOI} \doi{10.1016/S0092-6566(03)00046-1} \end{APACrefDOI}
\newblock

\newblock

\PrintBackRefs{\CurrentBib}

\bibitem [\protect \citeauthoryear {%
He%
, Lien%
\BCBL {}\ \BBA {} Zheng%
}{%
He%
\ \protect \BOthers {.}}{%
{\protect \APACyear {2026}}%
}]{%
HeLienZheng2026}
\APACinsertmetastar {%
HeLienZheng2026}%
\begin{APACrefauthors}%
He, Y.%
, Lien, J.W.%
\BCBL {} Zheng, J.%
\end{APACrefauthors}%
\unskip\
\newblock
\APACrefYearMonthDay{2026}{}{}.
\newblock
{\BBOQ}\APACrefatitle {Majority views and confidence information promote informed decisions} {Majority views and confidence information promote informed decisions}.{\BBCQ}
\newblock
\APACjournalVolNumPages{Humanities and Social Sciences Communications}{13}{1}{363,}
\newblock
\begin{APACrefDOI} \doi{10.1057/s41599-026-06668-4} \end{APACrefDOI}
\newblock
\begin{APACrefURL} {https://doi.org/10.1057/s41599-026-06668-4} \end{APACrefURL}
\newblock

\newblock

\PrintBackRefs{\CurrentBib}

\bibitem [\protect \citeauthoryear {%
Hong%
\ \BBA {} Page%
}{%
Hong%
\ \BBA {} Page%
}{%
{\protect \APACyear {2004}}%
}]{%
HongPage2004}
\APACinsertmetastar {%
HongPage2004}%
\begin{APACrefauthors}%
Hong, L.%
\BCBT {}\ \BBA {} Page, S.E.%
\end{APACrefauthors}%
\unskip\
\newblock
\APACrefYearMonthDay{2004}{}{}.
\newblock
{\BBOQ}\APACrefatitle {Groups of Diverse Problem Solvers Can Outperform Groups of High-Ability Problem Solvers} {Groups of diverse problem solvers can outperform groups of high-ability problem solvers}.{\BBCQ}
\newblock
\APACjournalVolNumPages{Proceedings of the National Academy of Sciences}{101}{46}{16385--16389,}
\newblock
\begin{APACrefDOI} \doi{10.1073/pnas.0403723101} \end{APACrefDOI}
\newblock

\newblock

\PrintBackRefs{\CurrentBib}

\bibitem [\protect \citeauthoryear {%
Johnson%
\ \BBA {} Slovic%
}{%
Johnson%
\ \BBA {} Slovic%
}{%
{\protect \APACyear {1995}}%
}]{%
Johnson1995PresentingUI}
\APACinsertmetastar {%
Johnson1995PresentingUI}%
\begin{APACrefauthors}%
Johnson, B.B.%
\BCBT {}\ \BBA {} Slovic, P.%
\end{APACrefauthors}%
\unskip\
\newblock
\APACrefYearMonthDay{1995}{}{}.
\newblock
{\BBOQ}\APACrefatitle {Presenting uncertainty in health risk assessment: initial studies of its effects on risk perception and trust.} {Presenting uncertainty in health risk assessment: initial studies of its effects on risk perception and trust.}{\BBCQ}
\newblock
\APACjournalVolNumPages{Risk Analysis}{15}{4}{485--494,}
\newblock
\begin{APACrefDOI} \doi{10.1111/j.1539-6924.1995.tb00341.x} \end{APACrefDOI}
\newblock

\newblock

\PrintBackRefs{\CurrentBib}

\bibitem [\protect \citeauthoryear {%
Kahneman%
\ \BBA {} Tversky%
}{%
Kahneman%
\ \BBA {} Tversky%
}{%
{\protect \APACyear {1982}}%
}]{%
kahneman1982variants}
\APACinsertmetastar {%
kahneman1982variants}%
\begin{APACrefauthors}%
Kahneman, D.%
\BCBT {}\ \BBA {} Tversky, A.%
\end{APACrefauthors}%
\unskip\
\newblock
\APACrefYearMonthDay{1982}{}{}.
\newblock
{\BBOQ}\APACrefatitle {Variants of uncertainty} {Variants of uncertainty}.{\BBCQ}
\newblock
\APACjournalVolNumPages{Cognition}{11}{2}{143--157,}
\newblock
\begin{APACrefDOI} \doi{10.1016/0010-0277(82)90023-3} \end{APACrefDOI}
\newblock

\newblock

\PrintBackRefs{\CurrentBib}

\bibitem [\protect \citeauthoryear {%
Kappes%
, Harvey%
, Lohrenz%
, Montague%
\BCBL {}\ \BBA {} Sharot%
}{%
Kappes%
\ \protect \BOthers {.}}{%
{\protect \APACyear {2020}}%
}]{%
Kappes2020}
\APACinsertmetastar {%
Kappes2020}%
\begin{APACrefauthors}%
Kappes, A.%
, Harvey, A.H.%
, Lohrenz, T.%
, Montague, P.R.%
\BCBL {} Sharot, T.%
\end{APACrefauthors}%
\unskip\
\newblock
\APACrefYearMonthDay{2020}{}{}.
\newblock
{\BBOQ}\APACrefatitle {Confirmation bias in the utilization of others' opinion strength} {Confirmation bias in the utilization of others' opinion strength}.{\BBCQ}
\newblock
\APACjournalVolNumPages{Nature Neuroscience}{23}{1}{130--137,}
\newblock
\begin{APACrefDOI} \doi{10.1038/s41593-019-0549-2} \end{APACrefDOI}
\newblock

\newblock

\PrintBackRefs{\CurrentBib}

\bibitem [\protect \citeauthoryear {%
Leary%
\ \protect \BOthers {.}}{%
Leary%
\ \protect \BOthers {.}}{%
{\protect \APACyear {2017}}%
}]{%
leary2017IH}
\APACinsertmetastar {%
leary2017IH}%
\begin{APACrefauthors}%
Leary, M.R.%
, Diebels, K.J.%
, Davisson, E.J.%
, Jongman-Sereno, K.P.%
, Isherwood, J.C.%
, Raimi, K.T.%
\BDBL {}Hoyle, R.H.%
\end{APACrefauthors}%
\unskip\
\newblock
\APACrefYearMonthDay{2017}{}{}.
\newblock
{\BBOQ}\APACrefatitle {Cognitive and Interpersonal Features of Intellectual Humility} {Cognitive and interpersonal features of intellectual humility}.{\BBCQ}
\newblock
\APACjournalVolNumPages{Personality and Social Psychology Bulletin}{43}{6}{793--813,}
\newblock
\begin{APACrefDOI} \doi{10.1177/0146167217697695} \end{APACrefDOI}
\newblock
\APACrefnote{Original work published 2017}
\newblock

\newblock

\PrintBackRefs{\CurrentBib}

\bibitem [\protect \citeauthoryear {%
Lipkus%
}{%
Lipkus%
}{%
{\protect \APACyear {2007}}%
}]{%
Lipkus2007Numeric}
\APACinsertmetastar {%
Lipkus2007Numeric}%
\begin{APACrefauthors}%
Lipkus, I.M.%
\end{APACrefauthors}%
\unskip\
\newblock
\APACrefYearMonthDay{2007}{}{}.
\newblock
{\BBOQ}\APACrefatitle {Numeric, Verbal, and Visual Formats of Conveying Health Risks: Suggested Best Practices and Future Recommendations} {Numeric, verbal, and visual formats of conveying health risks: Suggested best practices and future recommendations}.{\BBCQ}
\newblock
\APACjournalVolNumPages{Medical Decision Making}{27}{5}{696--713,}
\newblock
\begin{APACrefDOI} \doi{10.1177/0272989X07307271} \end{APACrefDOI}
\newblock

\newblock

\PrintBackRefs{\CurrentBib}

\bibitem [\protect \citeauthoryear {%
Lourie%
, Bras%
\BCBL {}\ \BBA {} Choi%
}{%
Lourie%
\ \protect \BOthers {.}}{%
{\protect \APACyear {2021}}%
}]{%
Lourie2020ScruplesAC}
\APACinsertmetastar {%
Lourie2020ScruplesAC}%
\begin{APACrefauthors}%
Lourie, N.%
, Bras, R.L.%
\BCBL {} Choi, Y.%
\end{APACrefauthors}%
\unskip\
\newblock
\APACrefYearMonthDay{2021}{}{}.
\newblock
{\BBOQ}\APACrefatitle {Scruples: A Corpus of Community Ethical Judgments on 32,000 Real-Life Anecdotes} {Scruples: A corpus of community ethical judgments on 32,000 real-life anecdotes}.{\BBCQ}
\newblock
\APACjournalVolNumPages{Proceedings of the AAAI Conference on Artificial Intelligence}{35}{15}{13470--13479,}
\newblock
\begin{APACrefDOI} \doi{10.1609/aaai.v35i15.17589} \end{APACrefDOI}
\newblock

\newblock

\PrintBackRefs{\CurrentBib}

\bibitem [\protect \citeauthoryear {%
Moscovici%
, Lage%
\BCBL {}\ \BBA {} Naffrechoux%
}{%
Moscovici%
\ \protect \BOthers {.}}{%
{\protect \APACyear {1969}}%
}]{%
Moscovici1969Minority}
\APACinsertmetastar {%
Moscovici1969Minority}%
\begin{APACrefauthors}%
Moscovici, S.%
, Lage, E.%
\BCBL {} Naffrechoux, M.%
\end{APACrefauthors}%
\unskip\
\newblock
\APACrefYearMonthDay{1969}{}{}.
\newblock
{\BBOQ}\APACrefatitle {Influence of a Consistent Minority on the Responses of a Majority in a Color Perception Task} {Influence of a consistent minority on the responses of a majority in a color perception task}.{\BBCQ}
\newblock
\APACjournalVolNumPages{Sociometry}{32}{4}{365--380,}
\newblock
\begin{APACrefDOI} \doi{10.2307/2786541} \end{APACrefDOI}
\newblock

\newblock

\PrintBackRefs{\CurrentBib}

\bibitem [\protect \citeauthoryear {%
Moussa{\"i}d%
, K{\"a}mmer%
, Analytis%
\BCBL {}\ \BBA {} Neth%
}{%
Moussa{\"i}d%
\ \protect \BOthers {.}}{%
{\protect \APACyear {2013}}%
}]{%
Moussaid2013SocialInfluence}
\APACinsertmetastar {%
Moussaid2013SocialInfluence}%
\begin{APACrefauthors}%
Moussa{\"i}d, M.%
, K{\"a}mmer, J.E.%
, Analytis, P.P.%
\BCBL {} Neth, H.%
\end{APACrefauthors}%
\unskip\
\newblock
\APACrefYearMonthDay{2013}{}{}.
\newblock
{\BBOQ}\APACrefatitle {Social Influence and the Collective Dynamics of Opinion Formation} {Social influence and the collective dynamics of opinion formation}.{\BBCQ}
\newblock
\APACjournalVolNumPages{PLOS ONE}{8}{11}{e78433,}
\newblock
\begin{APACrefDOI} \doi{10.1371/journal.pone.0078433} \end{APACrefDOI}
\newblock
\begin{APACrefURL} {https://doi.org/10.1371/journal.pone.0078433} \end{APACrefURL}
\newblock

\newblock

\PrintBackRefs{\CurrentBib}

\bibitem [\protect \citeauthoryear {%
Nguyen%
\ \protect \BOthers {.}}{%
Nguyen%
\ \protect \BOthers {.}}{%
{\protect \APACyear {2022}}%
}]{%
Nguyen2022MappingTI}
\APACinsertmetastar {%
Nguyen2022MappingTI}%
\begin{APACrefauthors}%
Nguyen, T.D.%
, Lyall, G.%
, Tran, A.%
, Shin, M.%
, Carroll, N.G.%
, Klein, C.%
\BCBL {} Xie, L.%
\end{APACrefauthors}%
\unskip\
\newblock
\APACrefYearMonthDay{2022}{May}{}.
\newblock
{\BBOQ}\APACrefatitle {Mapping Topics in 100,000 Real-Life Moral Dilemmas} {Mapping topics in 100,000 real-life moral dilemmas}.{\BBCQ}
\newblock
\APACjournalVolNumPages{Proceedings of the International AAAI Conference on Web and Social Media}{16}{1}{699--710,}
\newblock
\begin{APACrefDOI} \doi{10.1609/icwsm.v16i1.19327} \end{APACrefDOI}
\newblock
\begin{APACrefURL} {https://ojs.aaai.org/index.php/ICWSM/article/view/19327} \end{APACrefURL}
\newblock

\newblock

\PrintBackRefs{\CurrentBib}

\bibitem [\protect \citeauthoryear {%
Oktar%
\ \BBA {} Lombrozo%
}{%
Oktar%
\ \BBA {} Lombrozo%
}{%
{\protect \APACyear {2025}}%
}]{%
oktar2025aggregated}
\APACinsertmetastar {%
oktar2025aggregated}%
\begin{APACrefauthors}%
Oktar, K.%
\BCBT {}\ \BBA {} Lombrozo, T.%
\end{APACrefauthors}%
\unskip\
\newblock
\APACrefYearMonthDay{2025}{}{}.
\newblock
{\BBOQ}\APACrefatitle {How aggregated opinions shape beliefs} {How aggregated opinions shape beliefs}.{\BBCQ}
\newblock
\APACjournalVolNumPages{Nature Reviews Psychology}{4}{2}{81--95,}
\newblock
\begin{APACrefDOI} \doi{10.1038/s44159-024-00398-7} \end{APACrefDOI}
\newblock
\begin{APACrefURL} {https://www.nature.com/articles/s44159-024-00398-7} \end{APACrefURL}
\newblock

\newblock

\PrintBackRefs{\CurrentBib}

\bibitem [\protect \citeauthoryear {%
Oktar%
, Lombrozo%
\BCBL {}\ \BBA {} Griffiths%
}{%
Oktar%
\ \protect \BOthers {.}}{%
{\protect \APACyear {2024}}%
}]{%
OktarLombrozoGriffiths2024}
\APACinsertmetastar {%
OktarLombrozoGriffiths2024}%
\begin{APACrefauthors}%
Oktar, K.%
, Lombrozo, T.%
\BCBL {} Griffiths, T.L.%
\end{APACrefauthors}%
\unskip\
\newblock
\APACrefYearMonthDay{2024}{}{}.
\newblock
{\BBOQ}\APACrefatitle {Learning from aggregated opinion} {Learning from aggregated opinion}.{\BBCQ}
\newblock
\APACjournalVolNumPages{Psychological Science}{35}{9}{1010--1024,}
\newblock
\begin{APACrefDOI} \doi{10.1177/09567976241251741} \end{APACrefDOI}
\newblock

\newblock

\PrintBackRefs{\CurrentBib}

\bibitem [\protect \citeauthoryear {%
Paivio%
}{%
Paivio%
}{%
{\protect \APACyear {1986}}%
}]{%
Paivio1986Mental}
\APACinsertmetastar {%
Paivio1986Mental}%
\begin{APACrefauthors}%
Paivio, A.%
\end{APACrefauthors}%
\unskip\
\newblock
\APACrefYear{1986}.
\newblock
\APACrefbtitle {Mental Representations: A Dual Coding Approach} {Mental representations: A dual coding approach}.
\newblock
\APACaddressPublisher{New York}{Oxford University Press}.
\PrintBackRefs{\CurrentBib}

\bibitem [\protect \citeauthoryear {%
Pryor%
, Perfors%
\BCBL {}\ \BBA {} Howe%
}{%
Pryor%
\ \protect \BOthers {.}}{%
{\protect \APACyear {2019}}%
}]{%
pryor2019even}
\APACinsertmetastar {%
pryor2019even}%
\begin{APACrefauthors}%
Pryor, C.%
, Perfors, A.%
\BCBL {} Howe, P.D.L.%
\end{APACrefauthors}%
\unskip\
\newblock
\APACrefYearMonthDay{2019}{}{}.
\newblock
{\BBOQ}\APACrefatitle {Even arbitrary norms influence moral decision-making} {Even arbitrary norms influence moral decision-making}.{\BBCQ}
\newblock
\APACjournalVolNumPages{Nature Human Behaviour}{3}{1}{57--62,}
\newblock
\begin{APACrefDOI} \doi{10.1038/s41562-018-0489-y} \end{APACrefDOI}
\newblock

\newblock

\PrintBackRefs{\CurrentBib}

\bibitem [\protect \citeauthoryear {%
Sachdeva%
\ \BBA {} van Nuenen%
}{%
Sachdeva%
\ \BBA {} van Nuenen%
}{%
{\protect \APACyear {2025}}%
}]{%
sachdeva2025normative}
\APACinsertmetastar {%
sachdeva2025normative}%
\begin{APACrefauthors}%
Sachdeva, P.S.%
\BCBT {}\ \BBA {} van Nuenen, T.%
\end{APACrefauthors}%
\unskip\
\newblock
\APACrefYearMonthDay{2025}{}{}.
\newblock
{\BBOQ}\APACrefatitle {Normative Evaluation of Large Language Models with Everyday Moral Dilemmas} {Normative evaluation of large language models with everyday moral dilemmas}.{\BBCQ}
\newblock
 \APACrefbtitle {Proceedings of the 2025 ACM Conference on Fairness, Accountability, and Transparency (FAccT ’25)} {Proceedings of the 2025 acm conference on fairness, accountability, and transparency (facct ’25)}\ (\BPGS\ 690--709).
\newblock
\APACaddressPublisher{}{ACM}.
\newblock
\begin{APACrefURL} {https://doi.org/10.1145/3715275.3732044} \end{APACrefURL}
\PrintBackRefs{\CurrentBib}

\bibitem [\protect \citeauthoryear {%
Salvi%
, Horta~Ribeiro%
, Gallotti%
\BCBL {}\ \BBA {} West%
}{%
Salvi%
\ \protect \BOthers {.}}{%
{\protect \APACyear {2025}}%
}]{%
salvi2025conversational}
\APACinsertmetastar {%
salvi2025conversational}%
\begin{APACrefauthors}%
Salvi, F.%
, Horta~Ribeiro, M.%
, Gallotti, R.%
\BCBL {} West, R.%
\end{APACrefauthors}%
\unskip\
\newblock
\APACrefYearMonthDay{2025}{}{}.
\newblock
{\BBOQ}\APACrefatitle {On the conversational persuasiveness of {GPT-4}} {On the conversational persuasiveness of {GPT-4}}.{\BBCQ}
\newblock
\APACjournalVolNumPages{Nature Human Behaviour}{9}{8}{1645--1653,}
\newblock
\begin{APACrefDOI} \doi{10.1038/s41562-025-02194-6} \end{APACrefDOI}
\newblock

\newblock

\PrintBackRefs{\CurrentBib}

\bibitem [\protect \citeauthoryear {%
Schuster%
\ \BBA {} Kilov%
}{%
Schuster%
\ \BBA {} Kilov%
}{%
{\protect \APACyear {2025}}%
}]{%
SchusterKilov2025MoralDisagreement}
\APACinsertmetastar {%
SchusterKilov2025MoralDisagreement}%
\begin{APACrefauthors}%
Schuster, N.%
\BCBT {}\ \BBA {} Kilov, D.%
\end{APACrefauthors}%
\unskip\
\newblock
\APACrefYearMonthDay{2025}{}{}.
\newblock
{\BBOQ}\APACrefatitle {Moral disagreement and the limits of {AI} value alignment: a dual challenge of epistemic justification and political legitimacy} {Moral disagreement and the limits of {AI} value alignment: a dual challenge of epistemic justification and political legitimacy}.{\BBCQ}
\newblock
\APACjournalVolNumPages{AI \& Society}{40}{8}{6073--6087,}
\newblock
\begin{APACrefDOI} \doi{10.1007/s00146-025-02427-2} \end{APACrefDOI}
\newblock

\newblock

\PrintBackRefs{\CurrentBib}

\bibitem [\protect \citeauthoryear {%
Scott%
\ \BBA {} Bruce%
}{%
Scott%
\ \BBA {} Bruce%
}{%
{\protect \APACyear {1995}}%
}]{%
scott1995decision}
\APACinsertmetastar {%
scott1995decision}%
\begin{APACrefauthors}%
Scott, S.G.%
\BCBT {}\ \BBA {} Bruce, R.A.%
\end{APACrefauthors}%
\unskip\
\newblock
\APACrefYearMonthDay{1995}{}{}.
\newblock
{\BBOQ}\APACrefatitle {Decision-making style: The development and assessment of a new measure} {Decision-making style: The development and assessment of a new measure}.{\BBCQ}
\newblock
\APACjournalVolNumPages{Educational and Psychological Measurement}{55}{5}{818--831,}
\newblock
\begin{APACrefDOI} \doi{10.1177/0013164495055005017} \end{APACrefDOI}
\newblock

\newblock

\PrintBackRefs{\CurrentBib}

\bibitem [\protect \citeauthoryear {%
Surowiecki%
}{%
Surowiecki%
}{%
{\protect \APACyear {2004}}%
}]{%
Surowiecki2004}
\APACinsertmetastar {%
Surowiecki2004}%
\begin{APACrefauthors}%
Surowiecki, J.%
\end{APACrefauthors}%
\unskip\
\newblock
\APACrefYear{2004}.
\newblock
\APACrefbtitle {The Wisdom of Crowds} {The wisdom of crowds}.
\newblock
\APACaddressPublisher{New York}{Doubleday}.
\PrintBackRefs{\CurrentBib}

\bibitem [\protect \citeauthoryear {%
Tessler%
\ \protect \BOthers {.}}{%
Tessler%
\ \protect \BOthers {.}}{%
{\protect \APACyear {2024}}%
}]{%
tessler2024ai}
\APACinsertmetastar {%
tessler2024ai}%
\begin{APACrefauthors}%
Tessler, M.H.%
, Bakker, M.A.%
, Jarrett, D.%
, Sheahan, H.%
, Chadwick, M.J.%
, Koster, R.%
\BDBL {}Summerfield, C.%
\end{APACrefauthors}%
\unskip\
\newblock
\APACrefYearMonthDay{2024}{}{}.
\newblock
{\BBOQ}\APACrefatitle {{AI} can help humans find common ground in democratic deliberation} {{AI} can help humans find common ground in democratic deliberation}.{\BBCQ}
\newblock
\APACjournalVolNumPages{Science}{386}{6719}{eadq2852,}
\newblock
\begin{APACrefDOI} \doi{10.1126/science.adq2852} \end{APACrefDOI}
\newblock

\newblock

\PrintBackRefs{\CurrentBib}

\bibitem [\protect \citeauthoryear {%
van~der Bles%
, van~der Linden%
, Freeman%
\BCBL {}\ \BBA {} Spiegelhalter%
}{%
van~der Bles%
\ \protect \BOthers {.}}{%
{\protect \APACyear {2020}}%
}]{%
vanderbles2020effects}
\APACinsertmetastar {%
vanderbles2020effects}%
\begin{APACrefauthors}%
van~der Bles, A.M.%
, van~der Linden, S.%
, Freeman, A.L.J.%
\BCBL {} Spiegelhalter, D.J.%
\end{APACrefauthors}%
\unskip\
\newblock
\APACrefYearMonthDay{2020}{}{}.
\newblock
{\BBOQ}\APACrefatitle {The Effects of Communicating Uncertainty on Public Trust in Facts and Numbers} {The effects of communicating uncertainty on public trust in facts and numbers}.{\BBCQ}
\newblock
\APACjournalVolNumPages{Proceedings of the National Academy of Sciences}{117}{14}{7672--7683,}
\newblock
\begin{APACrefDOI} \doi{10.1073/pnas.1913678117} \end{APACrefDOI}
\newblock

\newblock

\PrintBackRefs{\CurrentBib}

\bibitem [\protect \citeauthoryear {%
Vavova%
}{%
Vavova%
}{%
{\protect \APACyear {2014}}%
}]{%
vavova2014confidence}
\APACinsertmetastar {%
vavova2014confidence}%
\begin{APACrefauthors}%
Vavova, K.%
\end{APACrefauthors}%
\unskip\
\newblock
\APACrefYearMonthDay{2014}{}{}.
\newblock
{\BBOQ}\APACrefatitle {Confidence, evidence, and disagreement} {Confidence, evidence, and disagreement}.{\BBCQ}
\newblock
\APACjournalVolNumPages{Erkenntnis}{79}{Suppl 1}{173--183,}
\newblock
\begin{APACrefDOI} \doi{10.1007/s10670-013-9451-6} \end{APACrefDOI}
\newblock

\newblock

\PrintBackRefs{\CurrentBib}

\bibitem [\protect \citeauthoryear {%
Wood%
, Lundgren%
, Ouellette%
, Busceme%
\BCBL {}\ \BBA {} Blackstone%
}{%
Wood%
\ \protect \BOthers {.}}{%
{\protect \APACyear {1994}}%
}]{%
Wood1994}
\APACinsertmetastar {%
Wood1994}%
\begin{APACrefauthors}%
Wood, W.%
, Lundgren, S.%
, Ouellette, J.A.%
, Busceme, S.%
\BCBL {} Blackstone, T.%
\end{APACrefauthors}%
\unskip\
\newblock
\APACrefYearMonthDay{1994}{}{}.
\newblock
{\BBOQ}\APACrefatitle {Minority influence: A meta-analytic review of social influence processes} {Minority influence: A meta-analytic review of social influence processes}.{\BBCQ}
\newblock
\APACjournalVolNumPages{Psychological Bulletin}{115}{3}{323--345,}
\newblock
\begin{APACrefDOI} \doi{10.1037/0033-2909.115.3.323} \end{APACrefDOI}
\newblock

\newblock

\PrintBackRefs{\CurrentBib}

\bibitem [\protect \citeauthoryear {%
Yarkoni%
\ \BBA {} Westfall%
}{%
Yarkoni%
\ \BBA {} Westfall%
}{%
{\protect \APACyear {2017}}%
}]{%
yarkoni2017choosing}
\APACinsertmetastar {%
yarkoni2017choosing}%
\begin{APACrefauthors}%
Yarkoni, T.%
\BCBT {}\ \BBA {} Westfall, J.%
\end{APACrefauthors}%
\unskip\
\newblock
\APACrefYearMonthDay{2017}{}{}.
\newblock
{\BBOQ}\APACrefatitle {Choosing Prediction Over Explanation in Psychology: {Lessons} from Machine Learning} {Choosing prediction over explanation in psychology: {Lessons} from machine learning}.{\BBCQ}
\newblock
\APACjournalVolNumPages{Perspectives on Psychological Science}{12}{6}{1100--1122,}
\newblock
\begin{APACrefDOI} \doi{10.1177/1745691617693393} \end{APACrefDOI}
\newblock

\newblock

\PrintBackRefs{\CurrentBib}

\bibitem [\protect \citeauthoryear {%
Zikmund-Fisher%
\ \protect \BOthers {.}}{%
Zikmund-Fisher%
\ \protect \BOthers {.}}{%
{\protect \APACyear {2014}}%
}]{%
ZikmundFisher2014Blocks}
\APACinsertmetastar {%
ZikmundFisher2014Blocks}%
\begin{APACrefauthors}%
Zikmund-Fisher, B.J.%
, Witteman, H.O.%
, Dickson, M.%
, Fuhrel-Forbis, A.%
, Kahn, V.C.%
, Exe, N.L.%
\BDBL {}Fagerlin, A.%
\end{APACrefauthors}%
\unskip\
\newblock
\APACrefYearMonthDay{2014}{}{}.
\newblock
{\BBOQ}\APACrefatitle {Blocks, Ovals, or People? Icon Type Affects Risk Perceptions and Recall of Pictographs} {Blocks, ovals, or people? icon type affects risk perceptions and recall of pictographs}.{\BBCQ}
\newblock
\APACjournalVolNumPages{Medical Decision Making}{34}{4}{443--453,}
\newblock
\begin{APACrefDOI} \doi{10.1177/0272989X13511706} \end{APACrefDOI}
\newblock

\newblock

\PrintBackRefs{\CurrentBib}

\end{thebibliography}


\begin{thebibliography}{}
\renewcommand{\doi}[1]{\url{https://doi.org/#1}}
\bibcommenthead

\bibitem [\protect \citeauthoryear {%
Adam%
\ \protect \BOthers {.}}{%
Adam%
\ \protect \BOthers {.}}{%
{\protect \APACyear {2026}}%
}]{%
2026gptzero}
\APACinsertmetastar {%
2026gptzero}%
\begin{APACrefauthors}%
Adam, G.A.%
, Cui, A.%
, Thomas, E.%
, Napier, E.%
, Shmatko, N.%
, Schnell, J.%
\BDBL {}Lee, D.%
\end{APACrefauthors}%
\unskip\
\newblock
\APACrefYearMonthDay{2026}{}{}.
\newblock
\APACrefbtitle {GPTZero: Robust Detection of LLM-Generated Texts.} {Gptzero: Robust detection of llm-generated texts.}
\newblock
\begin{APACrefURL} {https://arxiv.org/abs/2602.13042} \end{APACrefURL}
\PrintBackRefs{\CurrentBib}

\bibitem [\protect \citeauthoryear {%
Aramovich%
, Lytle%
\BCBL {}\ \BBA {} Skitka%
}{%
Aramovich%
\ \protect \BOthers {.}}{%
{\protect \APACyear {2012}}%
}]{%
aramovich2012opposing}
\APACinsertmetastar {%
aramovich2012opposing}%
\begin{APACrefauthors}%
Aramovich, N.P.%
, Lytle, B.L.%
\BCBL {} Skitka, L.J.%
\end{APACrefauthors}%
\unskip\
\newblock
\APACrefYearMonthDay{2012}{}{}.
\newblock
{\BBOQ}\APACrefatitle {Opposing torture: Moral conviction and resistance to majority influence} {Opposing torture: Moral conviction and resistance to majority influence}.{\BBCQ}
\newblock
\APACjournalVolNumPages{Social Influence}{7}{1}{21--34,}
\newblock
\begin{APACrefDOI} \doi{10.1080/15534510.2011.640199} \end{APACrefDOI}
\newblock

\newblock

\PrintBackRefs{\CurrentBib}

\bibitem [\protect \citeauthoryear {%
Asch%
}{%
Asch%
}{%
{\protect \APACyear {1955}}%
}]{%
asch1955opinions}
\APACinsertmetastar {%
asch1955opinions}%
\begin{APACrefauthors}%
Asch, S.E.%
\end{APACrefauthors}%
\unskip\
\newblock
\APACrefYearMonthDay{1955}{}{}.
\newblock
{\BBOQ}\APACrefatitle {Opinions and Social Pressure} {Opinions and social pressure}.{\BBCQ}
\newblock
\APACjournalVolNumPages{Scientific American}{193}{5}{31--35,}
\newblock
\begin{APACrefDOI} \doi{10.1038/scientificamerican1155-31} \end{APACrefDOI}
\newblock

\newblock

\PrintBackRefs{\CurrentBib}

\bibitem [\protect \citeauthoryear {%
Awad%
\ \protect \BOthers {.}}{%
Awad%
\ \protect \BOthers {.}}{%
{\protect \APACyear {2018}}%
}]{%
Awad2018TheMM}
\APACinsertmetastar {%
Awad2018TheMM}%
\begin{APACrefauthors}%
Awad, E.%
, Dsouza, S.%
, Kim, R.%
, Schulz, J.F.%
, Henrich, J.%
, Shariff, A.F.%
\BDBL {}Rahwan, I.%
\end{APACrefauthors}%
\unskip\
\newblock
\APACrefYearMonthDay{2018}{}{}.
\newblock
{\BBOQ}\APACrefatitle {The Moral Machine experiment} {The moral machine experiment}.{\BBCQ}
\newblock
\APACjournalVolNumPages{Nature}{563}{7729}{59--64,}
\newblock
\begin{APACrefDOI} \doi{10.1038/s41586-018-0637-6} \end{APACrefDOI}
\newblock

\newblock

\PrintBackRefs{\CurrentBib}

\bibitem [\protect \citeauthoryear {%
Axsom%
, Yates%
\BCBL {}\ \BBA {} Chaiken%
}{%
Axsom%
\ \protect \BOthers {.}}{%
{\protect \APACyear {1987}}%
}]{%
Axsom1987Audience}
\APACinsertmetastar {%
Axsom1987Audience}%
\begin{APACrefauthors}%
Axsom, D.%
, Yates, S.%
\BCBL {} Chaiken, S.%
\end{APACrefauthors}%
\unskip\
\newblock
\APACrefYearMonthDay{1987}{}{}.
\newblock
{\BBOQ}\APACrefatitle {Audience response as a heuristic cue in persuasion.} {Audience response as a heuristic cue in persuasion.}{\BBCQ}
\newblock
\APACjournalVolNumPages{Journal of Personality and Social Psychology}{53}{1}{30--40,}
\newblock
\begin{APACrefDOI} \doi{10.1037/0022-3514.53.1.30} \end{APACrefDOI}
\newblock

\newblock

\PrintBackRefs{\CurrentBib}

\bibitem [\protect \citeauthoryear {%
Bahrami%
\ \protect \BOthers {.}}{%
Bahrami%
\ \protect \BOthers {.}}{%
{\protect \APACyear {2012}}%
}]{%
bahrami2012whatfailure}
\APACinsertmetastar {%
bahrami2012whatfailure}%
\begin{APACrefauthors}%
Bahrami, B.%
, Olsen, K.%
, Bang, D.%
, Roepstorff, A.%
, Rees, G.%
\BCBL {} Frith, C.%
\end{APACrefauthors}%
\unskip\
\newblock
\APACrefYearMonthDay{2012}{}{}.
\newblock
{\BBOQ}\APACrefatitle {What Failure in Collective Decision-Making Tells Us About Metacognition} {What failure in collective decision-making tells us about metacognition}.{\BBCQ}
\newblock
\APACjournalVolNumPages{Philosophical Transactions of the Royal Society B: Biological Sciences}{367}{1594}{1350--1365,}
\newblock
\begin{APACrefDOI} \doi{10.1098/rstb.2011.0420} \end{APACrefDOI}
\newblock

\newblock

\PrintBackRefs{\CurrentBib}

\bibitem [\protect \citeauthoryear {%
Bail%
\ \protect \BOthers {.}}{%
Bail%
\ \protect \BOthers {.}}{%
{\protect \APACyear {2018}}%
}]{%
Bail2018Exposure}
\APACinsertmetastar {%
Bail2018Exposure}%
\begin{APACrefauthors}%
Bail, C.A.%
, Argyle, L.P.%
, Brown, T.W.%
, Bumpus, J.P.%
, Chen, H.%
, Hunzaker, M.B.F.%
\BDBL {}Volfovsky, A.%
\end{APACrefauthors}%
\unskip\
\newblock
\APACrefYearMonthDay{2018}{}{}.
\newblock
{\BBOQ}\APACrefatitle {Exposure to Opposing Views on Social Media Can Increase Political Polarization} {Exposure to opposing views on social media can increase political polarization}.{\BBCQ}
\newblock
\APACjournalVolNumPages{Proceedings of the National Academy of Sciences}{115}{37}{9216--9221,}
\newblock
\begin{APACrefDOI} \doi{10.1073/pnas.1804840115} \end{APACrefDOI}
\newblock

\newblock

\PrintBackRefs{\CurrentBib}

\bibitem [\protect \citeauthoryear {%
Balietti%
, Getoor%
, Goldstein%
\BCBL {}\ \BBA {} Watts%
}{%
Balietti%
\ \protect \BOthers {.}}{%
{\protect \APACyear {2021}}%
}]{%
Balietti2021ReducingOP}
\APACinsertmetastar {%
Balietti2021ReducingOP}%
\begin{APACrefauthors}%
Balietti, S.%
, Getoor, L.%
, Goldstein, D.G.%
\BCBL {} Watts, D.J.%
\end{APACrefauthors}%
\unskip\
\newblock
\APACrefYearMonthDay{2021}{}{}.
\newblock
{\BBOQ}\APACrefatitle {Reducing opinion polarization: Effects of exposure to similar people with differing political views} {Reducing opinion polarization: Effects of exposure to similar people with differing political views}.{\BBCQ}
\newblock
\APACjournalVolNumPages{Proceedings of the National Academy of Sciences of the United States of America}{118}{52}{e2112552118,}
\newblock
\begin{APACrefDOI} \doi{10.1073/pnas.2112552118} \end{APACrefDOI}
\newblock

\newblock

\PrintBackRefs{\CurrentBib}

\bibitem [\protect \citeauthoryear {%
Bang%
\ \protect \BOthers {.}}{%
Bang%
\ \protect \BOthers {.}}{%
{\protect \APACyear {2017}}%
}]{%
bang2017confidence}
\APACinsertmetastar {%
bang2017confidence}%
\begin{APACrefauthors}%
Bang, D.%
, Aitchison, L.%
, Moran, R.%
, Herce~Casta{\~n}{\'o}n, S.%
, Rafiee, B.%
, Mahmoodi, A.%
\BDBL {}Summerfield, C.%
\end{APACrefauthors}%
\unskip\
\newblock
\APACrefYearMonthDay{2017}{}{}.
\newblock
{\BBOQ}\APACrefatitle {Confidence Matching in Group Decision-Making} {Confidence matching in group decision-making}.{\BBCQ}
\newblock
\APACjournalVolNumPages{Nature Human Behaviour}{1}{6}{0117,}
\newblock
\begin{APACrefDOI} \doi{10.1038/s41562-017-0117} \end{APACrefDOI}
\newblock

\newblock

\PrintBackRefs{\CurrentBib}

\bibitem [\protect \citeauthoryear {%
Barnard%
}{%
Barnard%
}{%
{\protect \APACyear {1945}}%
}]{%
barnard1945new}
\APACinsertmetastar {%
barnard1945new}%
\begin{APACrefauthors}%
Barnard, G.A.%
\end{APACrefauthors}%
\unskip\
\newblock
\APACrefYearMonthDay{1945}{}{}.
\newblock
{\BBOQ}\APACrefatitle {A new test for 2×2 tables} {A new test for 2×2 tables}.{\BBCQ}
\newblock
\APACjournalVolNumPages{Nature}{156}{}{177,}
\newblock
\begin{APACrefDOI} \doi{10.1038/156177a0} \end{APACrefDOI}
\newblock

\newblock

\PrintBackRefs{\CurrentBib}

\bibitem [\protect \citeauthoryear {%
Beel%
, Xiang%
, Soni%
\BCBL {}\ \BBA {} Yang%
}{%
Beel%
\ \protect \BOthers {.}}{%
{\protect \APACyear {2022}}%
}]{%
Beel2022LinguisticCharacterization}
\APACinsertmetastar {%
Beel2022LinguisticCharacterization}%
\begin{APACrefauthors}%
Beel, J.%
, Xiang, T.%
, Soni, S.%
\BCBL {} Yang, D.%
\end{APACrefauthors}%
\unskip\
\newblock
\APACrefYearMonthDay{2022}{May}{}.
\newblock
{\BBOQ}\APACrefatitle {Linguistic Characterization of Divisive Topics Online: Case Studies on Contentiousness in Abortion, Climate Change, and Gun Control} {Linguistic characterization of divisive topics online: Case studies on contentiousness in abortion, climate change, and gun control}.{\BBCQ}
\newblock
\APACjournalVolNumPages{Proceedings of the International AAAI Conference on Web and Social Media}{16}{1}{32--42,}
\newblock
\begin{APACrefDOI} \doi{10.1609/icwsm.v16i1.19270} \end{APACrefDOI}
\newblock
\begin{APACrefURL} {https://ojs.aaai.org/index.php/ICWSM/article/view/19270} \end{APACrefURL}
\newblock

\newblock

\PrintBackRefs{\CurrentBib}

\bibitem [\protect \citeauthoryear {%
Benjamini%
\ \BBA {} Hochberg%
}{%
Benjamini%
\ \BBA {} Hochberg%
}{%
{\protect \APACyear {1995}}%
}]{%
benjamini1995controlling}
\APACinsertmetastar {%
benjamini1995controlling}%
\begin{APACrefauthors}%
Benjamini, Y.%
\BCBT {}\ \BBA {} Hochberg, Y.%
\end{APACrefauthors}%
\unskip\
\newblock
\APACrefYearMonthDay{1995}{}{}.
\newblock
{\BBOQ}\APACrefatitle {Controlling the false discovery rate: a practical and powerful approach to multiple testing} {Controlling the false discovery rate: a practical and powerful approach to multiple testing}.{\BBCQ}
\newblock
\APACjournalVolNumPages{Journal of the Royal Statistical Society: Series B (Methodological)}{57}{1}{289--300,}
\newblock
\begin{APACrefDOI} \doi{10.1111/j.2517-6161.1995.tb02031.x} \end{APACrefDOI}
\newblock

\newblock

\PrintBackRefs{\CurrentBib}

\bibitem [\protect \citeauthoryear {%
Brady%
, Crockett%
\BCBL {}\ \BBA {} Van~Bavel%
}{%
Brady%
\ \protect \BOthers {.}}{%
{\protect \APACyear {2020}}%
}]{%
brady2020mad}
\APACinsertmetastar {%
brady2020mad}%
\begin{APACrefauthors}%
Brady, W.J.%
, Crockett, M.J.%
\BCBL {} Van~Bavel, J.J.%
\end{APACrefauthors}%
\unskip\
\newblock
\APACrefYearMonthDay{2020}{}{}.
\newblock
{\BBOQ}\APACrefatitle {The MAD model of moral contagion: The role of motivation, attention, and design in the spread of moralized content online} {The mad model of moral contagion: The role of motivation, attention, and design in the spread of moralized content online}.{\BBCQ}
\newblock
\APACjournalVolNumPages{Perspectives on Psychological Science}{15}{4}{978--1010,}
\newblock
\begin{APACrefDOI} \doi{10.1177/1745691620917336} \end{APACrefDOI}
\newblock

\newblock

\PrintBackRefs{\CurrentBib}

\bibitem [\protect \citeauthoryear {%
Brennan%
, Eriksson%
, Goodin%
\BCBL {}\ \BBA {} Southwood%
}{%
Brennan%
\ \protect \BOthers {.}}{%
{\protect \APACyear {2013}}%
}]{%
Brennanetal2013}
\APACinsertmetastar {%
Brennanetal2013}%
\begin{APACrefauthors}%
Brennan, G.%
, Eriksson, L.%
, Goodin, R.E.%
\BCBL {} Southwood, N.%
\end{APACrefauthors}%
\unskip\
\newblock
\APACrefYear{2013}.
\newblock
\APACrefbtitle {Explaining Norms} {Explaining norms}.
\newblock
\APACaddressPublisher{}{Oxford University Press}.
\PrintBackRefs{\CurrentBib}

\bibitem [\protect \citeauthoryear {%
Candia%
, Morales%
, Monti%
\BCBL {}\ \BBA {} Bonchi%
}{%
Candia%
\ \protect \BOthers {.}}{%
{\protect \APACyear {2022}}%
}]{%
Candia2022SocialNO}
\APACinsertmetastar {%
Candia2022SocialNO}%
\begin{APACrefauthors}%
Candia, S.D.%
, Morales, G.D.F.%
, Monti, C.%
\BCBL {} Bonchi, F.%
\end{APACrefauthors}%
\unskip\
\newblock
\APACrefYearMonthDay{2022}{}{}.
\newblock
{\BBOQ}\APACrefatitle {Social Norms on Reddit: A Demographic Analysis} {Social norms on reddit: A demographic analysis}.{\BBCQ}
\newblock
 \APACrefbtitle {Proceedings of the 14th ACM Web Science Conference 2022} {Proceedings of the 14th acm web science conference 2022}\ (\BPGS\ 139--147).
\PrintBackRefs{\CurrentBib}

\bibitem [\protect \citeauthoryear {%
Card%
\ \protect \BOthers {.}}{%
Card%
\ \protect \BOthers {.}}{%
{\protect \APACyear {2022}}%
}]{%
Card2022ComputationalAO}
\APACinsertmetastar {%
Card2022ComputationalAO}%
\begin{APACrefauthors}%
Card, D.%
, Chang, S.%
, Becker, C.%
, Mendelsohn, J.%
, Voigt, R.%
, Boustan, L.P.%
\BDBL {}Jurafsky, D.%
\end{APACrefauthors}%
\unskip\
\newblock
\APACrefYearMonthDay{2022}{}{}.
\newblock
{\BBOQ}\APACrefatitle {Computational analysis of 140 years of US political speeches reveals more positive but increasingly polarized framing of immigration} {Computational analysis of 140 years of us political speeches reveals more positive but increasingly polarized framing of immigration}.{\BBCQ}
\newblock
\APACjournalVolNumPages{Proceedings of the National Academy of Sciences of the United States of America}{119}{31}{e2120510119,}
\newblock
\begin{APACrefDOI} \doi{10.1073/pnas.2120510119} \end{APACrefDOI}
\newblock

\newblock

\PrintBackRefs{\CurrentBib}

\bibitem [\protect \citeauthoryear {%
Christensen%
}{%
Christensen%
}{%
{\protect \APACyear {2007}}%
}]{%
christensen2007epistemology}
\APACinsertmetastar {%
christensen2007epistemology}%
\begin{APACrefauthors}%
Christensen, D.%
\end{APACrefauthors}%
\unskip\
\newblock
\APACrefYearMonthDay{2007}{}{}.
\newblock
{\BBOQ}\APACrefatitle {Epistemology of disagreement: The good news} {Epistemology of disagreement: The good news}.{\BBCQ}
\newblock
\APACjournalVolNumPages{The Philosophical Review}{116}{2}{187--217,}
\newblock
\begin{APACrefDOI} \doi{10.1215/00318108-2006-035} \end{APACrefDOI}
\newblock

\newblock

\PrintBackRefs{\CurrentBib}

\bibitem [\protect \citeauthoryear {%
Cialdini%
, Wosińska%
, Barrett%
, Butner%
\BCBL {}\ \BBA {} Górnik-Durose%
}{%
Cialdini%
\ \protect \BOthers {.}}{%
{\protect \APACyear {1999}}%
}]{%
Cialdini1999Compliance}
\APACinsertmetastar {%
Cialdini1999Compliance}%
\begin{APACrefauthors}%
Cialdini, R.%
, Wosińska, W.%
, Barrett, D.W.%
, Butner, J.%
\BCBL {} Górnik-Durose, M.%
\end{APACrefauthors}%
\unskip\
\newblock
\APACrefYearMonthDay{1999}{}{}.
\newblock
{\BBOQ}\APACrefatitle {Compliance with a Request in Two Cultures: The Differential Influence of Social Proof and Commitment/Consistency on Collectivists and Individualists} {Compliance with a request in two cultures: The differential influence of social proof and commitment/consistency on collectivists and individualists}.{\BBCQ}
\newblock
\APACjournalVolNumPages{Personality and Social Psychology Bulletin}{25}{10}{1242--1253,}
\newblock
\begin{APACrefDOI} \doi{10.1177/0146167299258006} \end{APACrefDOI}
\newblock

\newblock

\PrintBackRefs{\CurrentBib}

\bibitem [\protect \citeauthoryear {%
Clifford%
, Iyengar%
, Cabeza%
\BCBL {}\ \BBA {} Sinnott-Armstrong%
}{%
Clifford%
\ \protect \BOthers {.}}{%
{\protect \APACyear {2015}}%
}]{%
clifford2015moral}
\APACinsertmetastar {%
clifford2015moral}%
\begin{APACrefauthors}%
Clifford, S.%
, Iyengar, V.%
, Cabeza, R.%
\BCBL {} Sinnott-Armstrong, W.%
\end{APACrefauthors}%
\unskip\
\newblock
\APACrefYearMonthDay{2015}{}{}.
\newblock
{\BBOQ}\APACrefatitle {Moral foundations vignettes: A standardized stimulus database of scenarios based on moral foundations theory} {Moral foundations vignettes: A standardized stimulus database of scenarios based on moral foundations theory}.{\BBCQ}
\newblock
\APACjournalVolNumPages{Behavior Research Methods}{47}{4}{1178--1198,}
\newblock
\begin{APACrefDOI} \doi{10.3758/s13428-014-0551-2} \end{APACrefDOI}
\newblock

\newblock

\PrintBackRefs{\CurrentBib}

\bibitem [\protect \citeauthoryear {%
Combs%
\ \protect \BOthers {.}}{%
Combs%
\ \protect \BOthers {.}}{%
{\protect \APACyear {2023}}%
}]{%
Combs2023ReducingPP}
\APACinsertmetastar {%
Combs2023ReducingPP}%
\begin{APACrefauthors}%
Combs, A.%
, Tierney, G.%
, Guay, B.M.%
, Merhout, F.%
, Bail, C.A.%
, Hillygus, D.S.%
\BCBL {} Volfovsky, A.%
\end{APACrefauthors}%
\unskip\
\newblock
\APACrefYearMonthDay{2023}{}{}.
\newblock
{\BBOQ}\APACrefatitle {Reducing political polarization in the United States with a mobile chat platform} {Reducing political polarization in the united states with a mobile chat platform}.{\BBCQ}
\newblock
\APACjournalVolNumPages{Nature Human Behaviour}{7}{9}{1454--1461,}
\newblock
\begin{APACrefDOI} \doi{10.1038/s41562-023-01655-0} \end{APACrefDOI}
\newblock

\newblock

\PrintBackRefs{\CurrentBib}

\bibitem [\protect \citeauthoryear {%
Costello%
, Pennycook%
\BCBL {}\ \BBA {} Rand%
}{%
Costello%
\ \protect \BOthers {.}}{%
{\protect \APACyear {2024}}%
}]{%
costello2024durably}
\APACinsertmetastar {%
costello2024durably}%
\begin{APACrefauthors}%
Costello, T.H.%
, Pennycook, G.%
\BCBL {} Rand, D.G.%
\end{APACrefauthors}%
\unskip\
\newblock
\APACrefYearMonthDay{2024}{}{}.
\newblock
{\BBOQ}\APACrefatitle {Durably reducing conspiracy beliefs through dialogues with {AI}} {Durably reducing conspiracy beliefs through dialogues with {AI}}.{\BBCQ}
\newblock
\APACjournalVolNumPages{Science}{385}{6714}{eadq1814,}
\newblock
\begin{APACrefDOI} \doi{10.1126/science.adq1814} \end{APACrefDOI}
\newblock

\newblock

\PrintBackRefs{\CurrentBib}

\bibitem [\protect \citeauthoryear {%
Emelin%
, Le~Bras%
, Hwang%
, Forbes%
\BCBL {}\ \BBA {} Choi%
}{%
Emelin%
\ \protect \BOthers {.}}{%
{\protect \APACyear {2021}}%
}]{%
emelin-etal-2021-moral}
\APACinsertmetastar {%
emelin-etal-2021-moral}%
\begin{APACrefauthors}%
Emelin, D.%
, Le~Bras, R.%
, Hwang, J.D.%
, Forbes, M.%
\BCBL {} Choi, Y.%
\end{APACrefauthors}%
\unskip\
\newblock
\APACrefYearMonthDay{2021}{{\APACmonth{11}}}{}.
\newblock
{\BBOQ}\APACrefatitle {Moral Stories: Situated Reasoning about Norms, Intents, Actions, and their Consequences} {Moral stories: Situated reasoning about norms, intents, actions, and their consequences}.{\BBCQ}
\newblock
 M\BHBI F.~Moens, X.~Huang, L.~Specia\BCBL {}\ \BBA {} S.W\BHBI t.~Yih\ (\BEDS), \APACrefbtitle {Proceedings of the 2021 Conference on Empirical Methods in Natural Language Processing} {Proceedings of the 2021 conference on empirical methods in natural language processing}\ (\BPGS\ 698--718).
\newblock
\APACaddressPublisher{Online and Punta Cana, Dominican Republic}{Association for Computational Linguistics}.
\newblock
\begin{APACrefURL} {https://aclanthology.org/2021.emnlp-main.54} \end{APACrefURL}
\PrintBackRefs{\CurrentBib}

\bibitem [\protect \citeauthoryear {%
Farkas%
, Vincze%
, M{\'o}ra%
, Csirik%
\BCBL {}\ \BBA {} Szarvas%
}{%
Farkas%
\ \protect \BOthers {.}}{%
{\protect \APACyear {2010}}%
}]{%
farkas2010conll}
\APACinsertmetastar {%
farkas2010conll}%
\begin{APACrefauthors}%
Farkas, R.%
, Vincze, V.%
, M{\'o}ra, G.%
, Csirik, J.%
\BCBL {} Szarvas, G.%
\end{APACrefauthors}%
\unskip\
\newblock
\APACrefYearMonthDay{2010}{}{}.
\newblock
{\BBOQ}\APACrefatitle {The CoNLL-2010 Shared Task: Learning to Detect Hedges and their Scope in Natural Language Text} {The conll-2010 shared task: Learning to detect hedges and their scope in natural language text}.{\BBCQ}
\newblock
 \APACrefbtitle {Proceedings of the Fourteenth Conference on Computational Natural Language Learning -- Shared Task} {Proceedings of the fourteenth conference on computational natural language learning -- shared task}\ (\BPGS\ 1--12).
\newblock
\APACaddressPublisher{Uppsala, Sweden}{Association for Computational Linguistics}.
\newblock
\begin{APACrefURL} {https://aclanthology.org/W10-3001/} \end{APACrefURL}
\PrintBackRefs{\CurrentBib}

\bibitem [\protect \citeauthoryear {%
Farquhar%
, Kossen%
, Kuhn%
\BCBL {}\ \BBA {} Gal%
}{%
Farquhar%
\ \protect \BOthers {.}}{%
{\protect \APACyear {2024}}%
}]{%
farquhar2024detecting}
\APACinsertmetastar {%
farquhar2024detecting}%
\begin{APACrefauthors}%
Farquhar, S.%
, Kossen, J.%
, Kuhn, L.%
\BCBL {} Gal, Y.%
\end{APACrefauthors}%
\unskip\
\newblock
\APACrefYearMonthDay{2024}{}{}.
\newblock
{\BBOQ}\APACrefatitle {Detecting Hallucinations in Large Language Models Using Semantic Entropy} {Detecting hallucinations in large language models using semantic entropy}.{\BBCQ}
\newblock
\APACjournalVolNumPages{Nature}{630}{8017}{625--630,}
\newblock
\begin{APACrefDOI} \doi{10.1038/s41586-024-07421-0} \end{APACrefDOI}
\newblock

\newblock

\PrintBackRefs{\CurrentBib}

\bibitem [\protect \citeauthoryear {%
Feinberg%
, Kovacheff%
, Teper%
\BCBL {}\ \BBA {} Inbar%
}{%
Feinberg%
\ \protect \BOthers {.}}{%
{\protect \APACyear {2019}}%
}]{%
feinberg2019understanding}
\APACinsertmetastar {%
feinberg2019understanding}%
\begin{APACrefauthors}%
Feinberg, M.%
, Kovacheff, C.%
, Teper, R.%
\BCBL {} Inbar, Y.%
\end{APACrefauthors}%
\unskip\
\newblock
\APACrefYearMonthDay{2019}{}{}.
\newblock
{\BBOQ}\APACrefatitle {Understanding the process of moralization: How eating meat becomes a moral issue.} {Understanding the process of moralization: How eating meat becomes a moral issue.}{\BBCQ}
\newblock
\APACjournalVolNumPages{Journal of personality and social psychology}{117}{1}{50--72,}
\newblock
\begin{APACrefDOI} \doi{10.1037/pspa0000149} \end{APACrefDOI}
\newblock

\newblock

\PrintBackRefs{\CurrentBib}

\bibitem [\protect \citeauthoryear {%
Fleming%
}{%
Fleming%
}{%
{\protect \APACyear {2024}}%
}]{%
fleming2024metacognition}
\APACinsertmetastar {%
fleming2024metacognition}%
\begin{APACrefauthors}%
Fleming, S.M.%
\end{APACrefauthors}%
\unskip\
\newblock
\APACrefYearMonthDay{2024}{}{}.
\newblock
{\BBOQ}\APACrefatitle {Metacognition and Confidence: A Review and Synthesis} {Metacognition and confidence: A review and synthesis}.{\BBCQ}
\newblock
\APACjournalVolNumPages{Annual Review of Psychology}{75}{}{241--268,}
\newblock
\begin{APACrefDOI} \doi{10.1146/annurev-psych-022423-032425} \end{APACrefDOI}
\newblock

\newblock

\PrintBackRefs{\CurrentBib}

\bibitem [\protect \citeauthoryear {%
Forbes%
, Hwang%
, Shwartz%
, Sap%
\BCBL {}\ \BBA {} Choi%
}{%
Forbes%
\ \protect \BOthers {.}}{%
{\protect \APACyear {2020}}%
}]{%
Forbes2020SocialC1}
\APACinsertmetastar {%
Forbes2020SocialC1}%
\begin{APACrefauthors}%
Forbes, M.%
, Hwang, J.D.%
, Shwartz, V.%
, Sap, M.%
\BCBL {} Choi, Y.%
\end{APACrefauthors}%
\unskip\
\newblock
\APACrefYearMonthDay{2020}{{\APACmonth{11}}}{}.
\newblock
{\BBOQ}\APACrefatitle {Social Chemistry 101: Learning to Reason about Social and Moral Norms} {Social chemistry 101: Learning to reason about social and moral norms}.{\BBCQ}
\newblock
 B.~Webber, T.~Cohn, Y.~He\BCBL {}\ \BBA {} Y.~Liu\ (\BEDS), \APACrefbtitle {Proceedings of the 2020 Conference on Empirical Methods in Natural Language Processing (EMNLP)} {Proceedings of the 2020 conference on empirical methods in natural language processing (emnlp)}\ (\BPGS\ 653--670).
\newblock
\APACaddressPublisher{Online}{Association for Computational Linguistics}.
\newblock
\begin{APACrefURL} {https://aclanthology.org/2020.emnlp-main.48} \end{APACrefURL}
\PrintBackRefs{\CurrentBib}

\bibitem [\protect \citeauthoryear {%
Giorgi%
, Zhao%
, Feng%
\BCBL {}\ \BBA {} Martin%
}{%
Giorgi%
\ \protect \BOthers {.}}{%
{\protect \APACyear {2023}}%
}]{%
Giorgi2023AuthorAC}
\APACinsertmetastar {%
Giorgi2023AuthorAC}%
\begin{APACrefauthors}%
Giorgi, S.%
, Zhao, K.%
, Feng, A.H.%
\BCBL {} Martin, L.J.%
\end{APACrefauthors}%
\unskip\
\newblock
\APACrefYearMonthDay{2023}{}{}.
\newblock
{\BBOQ}\APACrefatitle {Author as Character and Narrator: Deconstructing Personal Narratives from the r/AmITheAsshole Reddit Community} {Author as character and narrator: Deconstructing personal narratives from the r/amitheasshole reddit community}.{\BBCQ}
\newblock
\APACjournalVolNumPages{Proceedings of the International AAAI Conference on Web and Social Media}{17}{1}{233--244,}
\newblock
\begin{APACrefDOI} \doi{10.1609/icwsm.v17i1.22141} \end{APACrefDOI}
\newblock

\newblock

\PrintBackRefs{\CurrentBib}

\bibitem [\protect \citeauthoryear {%
Gneiting%
\ \BBA {} Raftery%
}{%
Gneiting%
\ \BBA {} Raftery%
}{%
{\protect \APACyear {2007}}%
}]{%
gneiting2007strictly}
\APACinsertmetastar {%
gneiting2007strictly}%
\begin{APACrefauthors}%
Gneiting, T.%
\BCBT {}\ \BBA {} Raftery, A.E.%
\end{APACrefauthors}%
\unskip\
\newblock
\APACrefYearMonthDay{2007}{}{}.
\newblock
{\BBOQ}\APACrefatitle {Strictly Proper Scoring Rules, Prediction, and Estimation} {Strictly proper scoring rules, prediction, and estimation}.{\BBCQ}
\newblock
\APACjournalVolNumPages{Journal of the American Statistical Association}{102}{477}{359--378,}
\newblock
\begin{APACrefDOI} \doi{10.1198/016214506000001437} \end{APACrefDOI}
\newblock

\newblock

\PrintBackRefs{\CurrentBib}

\bibitem [\protect \citeauthoryear {%
Gosling%
, Rentfrow%
\BCBL {}\ \BBA {} Swann%
}{%
Gosling%
\ \protect \BOthers {.}}{%
{\protect \APACyear {2003}}%
}]{%
Gosling2003AVB}
\APACinsertmetastar {%
Gosling2003AVB}%
\begin{APACrefauthors}%
Gosling, S.D.%
, Rentfrow, P.J.%
\BCBL {} Swann, W.B.%
\end{APACrefauthors}%
\unskip\
\newblock
\APACrefYearMonthDay{2003}{}{}.
\newblock
{\BBOQ}\APACrefatitle {A very brief measure of the Big-Five personality domains} {A very brief measure of the big-five personality domains}.{\BBCQ}
\newblock
\APACjournalVolNumPages{Journal of Research in Personality}{37}{6}{504--528,}
\newblock
\begin{APACrefDOI} \doi{10.1016/S0092-6566(03)00046-1} \end{APACrefDOI}
\newblock

\newblock

\PrintBackRefs{\CurrentBib}

\bibitem [\protect \citeauthoryear {%
Graham%
, Haidt%
\BCBL {}\ \BBA {} Nosek%
}{%
Graham%
\ \protect \BOthers {.}}{%
{\protect \APACyear {2009}}%
}]{%
Graham2009Liberals}
\APACinsertmetastar {%
Graham2009Liberals}%
\begin{APACrefauthors}%
Graham, J.%
, Haidt, J.%
\BCBL {} Nosek, B.A.%
\end{APACrefauthors}%
\unskip\
\newblock
\APACrefYearMonthDay{2009}{}{}.
\newblock
{\BBOQ}\APACrefatitle {Liberals and conservatives rely on different sets of moral foundations.} {Liberals and conservatives rely on different sets of moral foundations.}{\BBCQ}
\newblock
\APACjournalVolNumPages{Journal of Personality and Social Psychology}{96}{5}{1029--1046,}
\newblock
\begin{APACrefDOI} \doi{10.1037/a0015141} \end{APACrefDOI}
\newblock

\newblock

\PrintBackRefs{\CurrentBib}

\bibitem [\protect \citeauthoryear {%
Guo%
, Pleiss%
, Sun%
\BCBL {}\ \BBA {} Weinberger%
}{%
Guo%
\ \protect \BOthers {.}}{%
{\protect \APACyear {2017}}%
}]{%
guo2017calibration}
\APACinsertmetastar {%
guo2017calibration}%
\begin{APACrefauthors}%
Guo, C.%
, Pleiss, G.%
, Sun, Y.%
\BCBL {} Weinberger, K.Q.%
\end{APACrefauthors}%
\unskip\
\newblock
\APACrefYearMonthDay{2017}{}{}.
\newblock
{\BBOQ}\APACrefatitle {On Calibration of Modern Neural Networks} {On calibration of modern neural networks}.{\BBCQ}
\newblock
 \APACrefbtitle {Proceedings of the 34th International Conference on Machine Learning (ICML)} {Proceedings of the 34th international conference on machine learning (icml)}\ (\BVOL~70, \BPGS\ 1321--1330).
\newblock
\APACaddressPublisher{}{PMLR}.
\newblock
\begin{APACrefURL} {https://proceedings.mlr.press/v70/guo17a.html} \end{APACrefURL}
\PrintBackRefs{\CurrentBib}

\bibitem [\protect \citeauthoryear {%
Haworth%
\ \protect \BOthers {.}}{%
Haworth%
\ \protect \BOthers {.}}{%
{\protect \APACyear {2021}}%
}]{%
Haworth2021ClassifyingRI}
\APACinsertmetastar {%
Haworth2021ClassifyingRI}%
\begin{APACrefauthors}%
Haworth, E.%
, Grover, T.%
, Langston, J.%
, Patel, A.%
, West, J.%
\BCBL {} Williams, A.C.%
\end{APACrefauthors}%
\unskip\
\newblock
\APACrefYearMonthDay{2021}{}{}.
\newblock
{\BBOQ}\APACrefatitle {Classifying Reasonability in Retellings of Personal Events Shared on Social Media: A Preliminary Case Study with /r/AmITheAsshole} {Classifying reasonability in retellings of personal events shared on social media: A preliminary case study with /r/amitheasshole}.{\BBCQ}
\newblock
\APACjournalVolNumPages{Proceedings of the International AAAI Conference on Web and Social Media}{15}{1}{1075--1079,}
\newblock
\begin{APACrefDOI} \doi{10.1609/icwsm.v15i1.18133} \end{APACrefDOI}
\newblock

\newblock

\PrintBackRefs{\CurrentBib}

\bibitem [\protect \citeauthoryear {%
He%
, Lien%
\BCBL {}\ \BBA {} Zheng%
}{%
He%
\ \protect \BOthers {.}}{%
{\protect \APACyear {2026}}%
}]{%
HeLienZheng2026}
\APACinsertmetastar {%
HeLienZheng2026}%
\begin{APACrefauthors}%
He, Y.%
, Lien, J.W.%
\BCBL {} Zheng, J.%
\end{APACrefauthors}%
\unskip\
\newblock
\APACrefYearMonthDay{2026}{}{}.
\newblock
{\BBOQ}\APACrefatitle {Majority views and confidence information promote informed decisions} {Majority views and confidence information promote informed decisions}.{\BBCQ}
\newblock
\APACjournalVolNumPages{Humanities and Social Sciences Communications}{13}{1}{363,}
\newblock
\begin{APACrefDOI} \doi{10.1057/s41599-026-06668-4} \end{APACrefDOI}
\newblock
\begin{APACrefURL} {https://doi.org/10.1057/s41599-026-06668-4} \end{APACrefURL}
\newblock

\newblock

\PrintBackRefs{\CurrentBib}

\bibitem [\protect \citeauthoryear {%
Horberg%
, Oveis%
, Keltner%
\BCBL {}\ \BBA {} Cohen%
}{%
Horberg%
\ \protect \BOthers {.}}{%
{\protect \APACyear {2009}}%
}]{%
Horberg2009Disgust}
\APACinsertmetastar {%
Horberg2009Disgust}%
\begin{APACrefauthors}%
Horberg, E.%
, Oveis, C.%
, Keltner, D.%
\BCBL {} Cohen, A.%
\end{APACrefauthors}%
\unskip\
\newblock
\APACrefYearMonthDay{2009}{}{}.
\newblock
{\BBOQ}\APACrefatitle {Disgust and the moralization of purity.} {Disgust and the moralization of purity.}{\BBCQ}
\newblock
\APACjournalVolNumPages{Journal of Personality and Social Psychology}{97}{6}{963--976,}
\newblock
\begin{APACrefDOI} \doi{10.1037/a0017423} \end{APACrefDOI}
\newblock

\newblock

\PrintBackRefs{\CurrentBib}

\bibitem [\protect \citeauthoryear {%
H{\"u}llermeier%
\ \BBA {} Waegeman%
}{%
H{\"u}llermeier%
\ \BBA {} Waegeman%
}{%
{\protect \APACyear {2021}}%
}]{%
hullermeier2021aleatoric}
\APACinsertmetastar {%
hullermeier2021aleatoric}%
\begin{APACrefauthors}%
H{\"u}llermeier, E.%
\BCBT {}\ \BBA {} Waegeman, W.%
\end{APACrefauthors}%
\unskip\
\newblock
\APACrefYearMonthDay{2021}{}{}.
\newblock
{\BBOQ}\APACrefatitle {Aleatoric and Epistemic Uncertainty in Machine Learning: An Introduction to Concepts and Methods} {Aleatoric and epistemic uncertainty in machine learning: An introduction to concepts and methods}.{\BBCQ}
\newblock
\APACjournalVolNumPages{Machine Learning}{110}{3}{457--506,}
\newblock
\begin{APACrefDOI} \doi{10.1007/s10994-021-05946-3} \end{APACrefDOI}
\newblock

\newblock

\PrintBackRefs{\CurrentBib}

\bibitem [\protect \citeauthoryear {%
Jenke%
}{%
Jenke%
}{%
{\protect \APACyear {2017}}%
}]{%
Jenke2017Vote}
\APACinsertmetastar {%
Jenke2017Vote}%
\begin{APACrefauthors}%
Jenke, L.%
\end{APACrefauthors}%
\unskip\
\newblock
\APACrefYearMonthDay{2017}{}{}.
\newblock
{\BBOQ}\APACrefatitle {The Unique Utility Function of Morally Convicted Voters} {The unique utility function of morally convicted voters}.{\BBCQ}
\newblock
\APACjournalVolNumPages{SSRN Electronic Journal}{}{}{,}
\newblock
\begin{APACrefDOI} \doi{10.2139/SSRN.2919036} \end{APACrefDOI}
\newblock

\newblock

\PrintBackRefs{\CurrentBib}

\bibitem [\protect \citeauthoryear {%
L.~Jiang%
\ \protect \BOthers {.}}{%
L.~Jiang%
\ \protect \BOthers {.}}{%
{\protect \APACyear {2021}}%
}]{%
Jiang2021DelphiTM}
\APACinsertmetastar {%
Jiang2021DelphiTM}%
\begin{APACrefauthors}%
Jiang, L.%
, Hwang, J.D.%
, Bhagavatula, C.%
, Bras, R.L.%
, Forbes, M.%
, Borchardt, J.%
\BDBL {}Choi, Y.%
\end{APACrefauthors}%
\unskip\
\newblock
\APACrefYearMonthDay{2021}{}{}.
\newblock
{\BBOQ}\APACrefatitle {Delphi: Towards Machine Ethics and Norms} {Delphi: Towards machine ethics and norms}.{\BBCQ}
\newblock
\APACjournalVolNumPages{arXiv preprint arXiv:2110.07574}{}{}{,}
\newblock
\begin{APACrefDOI} \doi{10.48550/arXiv.2110.07574} \end{APACrefDOI}
\newblock
\begin{APACrefURL} {https://arxiv.org/abs/2110.07574} \end{APACrefURL}
\newblock
{\href{https://arxiv.org/abs/2110.07574}{{arXiv:2110.07574}}}
\newblock

\PrintBackRefs{\CurrentBib}

\bibitem [\protect \citeauthoryear {%
Y.~Jiang%
, Marcowski%
, Ryazanov%
\BCBL {}\ \BBA {} Winkielman%
}{%
Y.~Jiang%
\ \protect \BOthers {.}}{%
{\protect \APACyear {2023}}%
}]{%
Jiang2023PeopleCT}
\APACinsertmetastar {%
Jiang2023PeopleCT}%
\begin{APACrefauthors}%
Jiang, Y.%
, Marcowski, P.%
, Ryazanov, A.%
\BCBL {} Winkielman, P.%
\end{APACrefauthors}%
\unskip\
\newblock
\APACrefYearMonthDay{2023}{}{}.
\newblock
{\BBOQ}\APACrefatitle {People conform to social norms when gambling with lives or money} {People conform to social norms when gambling with lives or money}.{\BBCQ}
\newblock
\APACjournalVolNumPages{Scientific Reports}{13}{1}{853,}
\newblock
\begin{APACrefDOI} \doi{10.1038/s41598-023-27462-1} \end{APACrefDOI}
\newblock
\begin{APACrefURL} {https://doi.org/10.1038/s41598-023-27462-1} \end{APACrefURL}
\newblock

\newblock

\PrintBackRefs{\CurrentBib}

\bibitem [\protect \citeauthoryear {%
Kappes%
, Harvey%
, Lohrenz%
, Montague%
\BCBL {}\ \BBA {} Sharot%
}{%
Kappes%
\ \protect \BOthers {.}}{%
{\protect \APACyear {2020}}%
}]{%
Kappes2020}
\APACinsertmetastar {%
Kappes2020}%
\begin{APACrefauthors}%
Kappes, A.%
, Harvey, A.H.%
, Lohrenz, T.%
, Montague, P.R.%
\BCBL {} Sharot, T.%
\end{APACrefauthors}%
\unskip\
\newblock
\APACrefYearMonthDay{2020}{}{}.
\newblock
{\BBOQ}\APACrefatitle {Confirmation bias in the utilization of others' opinion strength} {Confirmation bias in the utilization of others' opinion strength}.{\BBCQ}
\newblock
\APACjournalVolNumPages{Nature Neuroscience}{23}{1}{130--137,}
\newblock
\begin{APACrefDOI} \doi{10.1038/s41593-019-0549-2} \end{APACrefDOI}
\newblock

\newblock

\PrintBackRefs{\CurrentBib}

\bibitem [\protect \citeauthoryear {%
Kendall%
\ \BBA {} Gal%
}{%
Kendall%
\ \BBA {} Gal%
}{%
{\protect \APACyear {2017}}%
}]{%
kendallgal2017uncertainties}
\APACinsertmetastar {%
kendallgal2017uncertainties}%
\begin{APACrefauthors}%
Kendall, A.%
\BCBT {}\ \BBA {} Gal, Y.%
\end{APACrefauthors}%
\unskip\
\newblock
\APACrefYearMonthDay{2017}{}{}.
\newblock
{\BBOQ}\APACrefatitle {What Uncertainties Do We Need in Bayesian Deep Learning for Computer Vision?} {What uncertainties do we need in bayesian deep learning for computer vision?}{\BBCQ}
\newblock
 \APACrefbtitle {Advances in Neural Information Processing Systems (NeurIPS)} {Advances in neural information processing systems (neurips)}\ (\BVOL~30, \BPGS\ 5574--5584).
\newblock
\begin{APACrefURL} {https://proceedings.neurips.cc/paper\_files/paper/2017/hash/2650d6089a6d640c5e85b2b88265dc2b-Abstract.html} \end{APACrefURL}
\PrintBackRefs{\CurrentBib}

\bibitem [\protect \citeauthoryear {%
Kubin%
, Puryear%
, Schein%
\BCBL {}\ \BBA {} Gray%
}{%
Kubin%
\ \protect \BOthers {.}}{%
{\protect \APACyear {2021}}%
}]{%
kubin2021personal}
\APACinsertmetastar {%
kubin2021personal}%
\begin{APACrefauthors}%
Kubin, E.%
, Puryear, C.%
, Schein, C.%
\BCBL {} Gray, K.%
\end{APACrefauthors}%
\unskip\
\newblock
\APACrefYearMonthDay{2021}{}{}.
\newblock
{\BBOQ}\APACrefatitle {Personal experiences bridge moral and political divides better than facts} {Personal experiences bridge moral and political divides better than facts}.{\BBCQ}
\newblock
\APACjournalVolNumPages{Proceedings of the National Academy of Sciences}{118}{6}{e2008389118,}
\newblock
\begin{APACrefDOI} \doi{10.1073/pnas.2008389118} \end{APACrefDOI}
\newblock

\newblock

\PrintBackRefs{\CurrentBib}

\bibitem [\protect \citeauthoryear {%
Leary%
\ \protect \BOthers {.}}{%
Leary%
\ \protect \BOthers {.}}{%
{\protect \APACyear {2017}}%
}]{%
leary2017IH}
\APACinsertmetastar {%
leary2017IH}%
\begin{APACrefauthors}%
Leary, M.R.%
, Diebels, K.J.%
, Davisson, E.J.%
, Jongman-Sereno, K.P.%
, Isherwood, J.C.%
, Raimi, K.T.%
\BDBL {}Hoyle, R.H.%
\end{APACrefauthors}%
\unskip\
\newblock
\APACrefYearMonthDay{2017}{}{}.
\newblock
{\BBOQ}\APACrefatitle {Cognitive and Interpersonal Features of Intellectual Humility} {Cognitive and interpersonal features of intellectual humility}.{\BBCQ}
\newblock
\APACjournalVolNumPages{Personality and Social Psychology Bulletin}{43}{6}{793--813,}
\newblock
\begin{APACrefDOI} \doi{10.1177/0146167217697695} \end{APACrefDOI}
\newblock
\APACrefnote{Original work published 2017}
\newblock

\newblock

\PrintBackRefs{\CurrentBib}

\bibitem [\protect \citeauthoryear {%
Lourie%
, Bras%
\BCBL {}\ \BBA {} Choi%
}{%
Lourie%
\ \protect \BOthers {.}}{%
{\protect \APACyear {2021}}%
}]{%
Lourie2020ScruplesAC}
\APACinsertmetastar {%
Lourie2020ScruplesAC}%
\begin{APACrefauthors}%
Lourie, N.%
, Bras, R.L.%
\BCBL {} Choi, Y.%
\end{APACrefauthors}%
\unskip\
\newblock
\APACrefYearMonthDay{2021}{}{}.
\newblock
{\BBOQ}\APACrefatitle {Scruples: A Corpus of Community Ethical Judgments on 32,000 Real-Life Anecdotes} {Scruples: A corpus of community ethical judgments on 32,000 real-life anecdotes}.{\BBCQ}
\newblock
\APACjournalVolNumPages{Proceedings of the AAAI Conference on Artificial Intelligence}{35}{15}{13470--13479,}
\newblock
\begin{APACrefDOI} \doi{10.1609/aaai.v35i15.17589} \end{APACrefDOI}
\newblock

\newblock

\PrintBackRefs{\CurrentBib}

\bibitem [\protect \citeauthoryear {%
MacAskill%
, Bykvist%
\BCBL {}\ \BBA {} Ord%
}{%
MacAskill%
\ \protect \BOthers {.}}{%
{\protect \APACyear {2020}}%
}]{%
macaskill2020moral}
\APACinsertmetastar {%
macaskill2020moral}%
\begin{APACrefauthors}%
MacAskill, W.%
, Bykvist, K.%
\BCBL {} Ord, T.%
\end{APACrefauthors}%
\unskip\
\newblock
\APACrefYear{2020}.
\newblock
\APACrefbtitle {Moral Uncertainty} {Moral uncertainty}.
\newblock
\APACaddressPublisher{Oxford}{Oxford University Press}.
\PrintBackRefs{\CurrentBib}

\bibitem [\protect \citeauthoryear {%
Milgram%
}{%
Milgram%
}{%
{\protect \APACyear {1963}}%
}]{%
milgram1963behavioral}
\APACinsertmetastar {%
milgram1963behavioral}%
\begin{APACrefauthors}%
Milgram, S.%
\end{APACrefauthors}%
\unskip\
\newblock
\APACrefYearMonthDay{1963}{}{}.
\newblock
{\BBOQ}\APACrefatitle {Behavioral study of obedience.} {Behavioral study of obedience.}{\BBCQ}
\newblock
\APACjournalVolNumPages{The Journal of abnormal and social psychology}{67}{4}{371--378,}
\newblock
\begin{APACrefDOI} \doi{10.1037/h0040525} \end{APACrefDOI}
\newblock

\newblock

\PrintBackRefs{\CurrentBib}

\bibitem [\protect \citeauthoryear {%
Mitra%
\ \BBA {} Gilbert%
}{%
Mitra%
\ \BBA {} Gilbert%
}{%
{\protect \APACyear {2014}}%
}]{%
Mitra2014The}
\APACinsertmetastar {%
Mitra2014The}%
\begin{APACrefauthors}%
Mitra, T.%
\BCBT {}\ \BBA {} Gilbert, E.%
\end{APACrefauthors}%
\unskip\
\newblock
\APACrefYearMonthDay{2014}{}{}.
\newblock
{\BBOQ}\APACrefatitle {The language that gets people to give: phrases that predict success on {Kickstarter}} {The language that gets people to give: phrases that predict success on {Kickstarter}}.{\BBCQ}
\newblock
 \APACrefbtitle {Proceedings of the 17th ACM Conference on Computer Supported Cooperative Work \& Social Computing} {Proceedings of the 17th acm conference on computer supported cooperative work \& social computing}\ (\BPGS\ 49--61).
\PrintBackRefs{\CurrentBib}

\bibitem [\protect \citeauthoryear {%
Moscovici%
, Lage%
\BCBL {}\ \BBA {} Naffrechoux%
}{%
Moscovici%
\ \protect \BOthers {.}}{%
{\protect \APACyear {1969}}%
}]{%
Moscovici1969Minority}
\APACinsertmetastar {%
Moscovici1969Minority}%
\begin{APACrefauthors}%
Moscovici, S.%
, Lage, E.%
\BCBL {} Naffrechoux, M.%
\end{APACrefauthors}%
\unskip\
\newblock
\APACrefYearMonthDay{1969}{}{}.
\newblock
{\BBOQ}\APACrefatitle {Influence of a Consistent Minority on the Responses of a Majority in a Color Perception Task} {Influence of a consistent minority on the responses of a majority in a color perception task}.{\BBCQ}
\newblock
\APACjournalVolNumPages{Sociometry}{32}{4}{365--380,}
\newblock
\begin{APACrefDOI} \doi{10.2307/2786541} \end{APACrefDOI}
\newblock

\newblock

\PrintBackRefs{\CurrentBib}

\bibitem [\protect \citeauthoryear {%
Moussa{\"i}d%
, K{\"a}mmer%
, Analytis%
\BCBL {}\ \BBA {} Neth%
}{%
Moussa{\"i}d%
\ \protect \BOthers {.}}{%
{\protect \APACyear {2013}}%
}]{%
Moussaid2013SocialInfluence}
\APACinsertmetastar {%
Moussaid2013SocialInfluence}%
\begin{APACrefauthors}%
Moussa{\"i}d, M.%
, K{\"a}mmer, J.E.%
, Analytis, P.P.%
\BCBL {} Neth, H.%
\end{APACrefauthors}%
\unskip\
\newblock
\APACrefYearMonthDay{2013}{}{}.
\newblock
{\BBOQ}\APACrefatitle {Social Influence and the Collective Dynamics of Opinion Formation} {Social influence and the collective dynamics of opinion formation}.{\BBCQ}
\newblock
\APACjournalVolNumPages{PLOS ONE}{8}{11}{e78433,}
\newblock
\begin{APACrefDOI} \doi{10.1371/journal.pone.0078433} \end{APACrefDOI}
\newblock
\begin{APACrefURL} {https://doi.org/10.1371/journal.pone.0078433} \end{APACrefURL}
\newblock

\newblock

\PrintBackRefs{\CurrentBib}

\bibitem [\protect \citeauthoryear {%
Myers%
\ \BBA {} Lamm%
}{%
Myers%
\ \BBA {} Lamm%
}{%
{\protect \APACyear {1976}}%
}]{%
Myers1976The}
\APACinsertmetastar {%
Myers1976The}%
\begin{APACrefauthors}%
Myers, D.%
\BCBT {}\ \BBA {} Lamm, H.%
\end{APACrefauthors}%
\unskip\
\newblock
\APACrefYearMonthDay{1976}{}{}.
\newblock
{\BBOQ}\APACrefatitle {The group polarization phenomenon.} {The group polarization phenomenon.}{\BBCQ}
\newblock
\APACjournalVolNumPages{Psychological Bulletin}{83}{4}{602--627,}
\newblock
\begin{APACrefDOI} \doi{10.1037/0033-2909.83.4.602} \end{APACrefDOI}
\newblock

\newblock

\PrintBackRefs{\CurrentBib}

\bibitem [\protect \citeauthoryear {%
Ng%
, Luke%
\BCBL {}\ \BBA {} Gawronski%
}{%
Ng%
\ \protect \BOthers {.}}{%
{\protect \APACyear {2023}}%
}]{%
Ng2023MoralJU}
\APACinsertmetastar {%
Ng2023MoralJU}%
\begin{APACrefauthors}%
Ng, N.L.%
, Luke, D.M.%
\BCBL {} Gawronski, B.%
\end{APACrefauthors}%
\unskip\
\newblock
\APACrefYearMonthDay{2023}{}{}.
\newblock
{\BBOQ}\APACrefatitle {Moral judgment under uncertainty: A {CNI} model analysis} {Moral judgment under uncertainty: A {CNI} model analysis}.{\BBCQ}
\newblock
\APACjournalVolNumPages{European Journal of Social Psychology}{53}{6}{1055--1077,}
\newblock
\begin{APACrefDOI} \doi{10.1002/ejsp.2952} \end{APACrefDOI}
\newblock

\newblock

\PrintBackRefs{\CurrentBib}

\bibitem [\protect \citeauthoryear {%
Nguyen%
\ \protect \BOthers {.}}{%
Nguyen%
\ \protect \BOthers {.}}{%
{\protect \APACyear {2024}}%
}]{%
Nguyen2023MeasuringMD}
\APACinsertmetastar {%
Nguyen2023MeasuringMD}%
\begin{APACrefauthors}%
Nguyen, T.D.%
, Chen, Z.%
, Carroll, N.G.%
, Tran, A.%
, Klein, C.%
\BCBL {} Xie, L.%
\end{APACrefauthors}%
\unskip\
\newblock
\APACrefYearMonthDay{2024}{}{}.
\newblock
{\BBOQ}\APACrefatitle {Measuring Moral Dimensions in Social Media with Mformer} {Measuring moral dimensions in social media with mformer}.{\BBCQ}
\newblock
\APACjournalVolNumPages{Proceedings of the International AAAI Conference on Web and Social Media}{18}{1}{1134--1147,}
\newblock
\begin{APACrefDOI} \doi{10.1609/icwsm.v18i1.31378} \end{APACrefDOI}
\newblock

\newblock

\PrintBackRefs{\CurrentBib}

\bibitem [\protect \citeauthoryear {%
Nguyen%
\ \protect \BOthers {.}}{%
Nguyen%
\ \protect \BOthers {.}}{%
{\protect \APACyear {2022}}%
}]{%
Nguyen2022MappingTI}
\APACinsertmetastar {%
Nguyen2022MappingTI}%
\begin{APACrefauthors}%
Nguyen, T.D.%
, Lyall, G.%
, Tran, A.%
, Shin, M.%
, Carroll, N.G.%
, Klein, C.%
\BCBL {} Xie, L.%
\end{APACrefauthors}%
\unskip\
\newblock
\APACrefYearMonthDay{2022}{May}{}.
\newblock
{\BBOQ}\APACrefatitle {Mapping Topics in 100,000 Real-Life Moral Dilemmas} {Mapping topics in 100,000 real-life moral dilemmas}.{\BBCQ}
\newblock
\APACjournalVolNumPages{Proceedings of the International AAAI Conference on Web and Social Media}{16}{1}{699--710,}
\newblock
\begin{APACrefDOI} \doi{10.1609/icwsm.v16i1.19327} \end{APACrefDOI}
\newblock
\begin{APACrefURL} {https://ojs.aaai.org/index.php/ICWSM/article/view/19327} \end{APACrefURL}
\newblock

\newblock

\PrintBackRefs{\CurrentBib}

\bibitem [\protect \citeauthoryear {%
Nickerson%
}{%
Nickerson%
}{%
{\protect \APACyear {1998}}%
}]{%
Nickerson1998ConfirmationBA}
\APACinsertmetastar {%
Nickerson1998ConfirmationBA}%
\begin{APACrefauthors}%
Nickerson, R.S.%
\end{APACrefauthors}%
\unskip\
\newblock
\APACrefYearMonthDay{1998}{}{}.
\newblock
{\BBOQ}\APACrefatitle {Confirmation Bias: A Ubiquitous Phenomenon in Many Guises} {Confirmation bias: A ubiquitous phenomenon in many guises}.{\BBCQ}
\newblock
\APACjournalVolNumPages{Review of General Psychology}{2}{2}{175--220,}
\newblock
\begin{APACrefDOI} \doi{10.1037/1089-2680.2.2.175} \end{APACrefDOI}
\newblock

\newblock

\PrintBackRefs{\CurrentBib}

\bibitem [\protect \citeauthoryear {%
Nyhan%
\ \protect \BOthers {.}}{%
Nyhan%
\ \protect \BOthers {.}}{%
{\protect \APACyear {2023}}%
}]{%
Nyhan2023LikemindedSO}
\APACinsertmetastar {%
Nyhan2023LikemindedSO}%
\begin{APACrefauthors}%
Nyhan, B.%
, Settle, J.E.%
, Thorson, E.A.%
, Wojcieszak, M.E.%
, Barber{\'a}, P.%
, Chen, A.Y.%
\BDBL {}Tucker, J.A.%
\end{APACrefauthors}%
\unskip\
\newblock
\APACrefYearMonthDay{2023}{}{}.
\newblock
{\BBOQ}\APACrefatitle {Like-minded sources on Facebook are prevalent but not polarizing} {Like-minded sources on facebook are prevalent but not polarizing}.{\BBCQ}
\newblock
\APACjournalVolNumPages{Nature}{620}{7972}{137--144,}
\newblock
\begin{APACrefDOI} \doi{10.1038/s41586-023-06297-w} \end{APACrefDOI}
\newblock

\newblock

\PrintBackRefs{\CurrentBib}

\bibitem [\protect \citeauthoryear {%
Ojea~Quintana%
, Reimann%
, Cheong%
, Alfano%
\BCBL {}\ \BBA {} Klein%
}{%
Ojea~Quintana%
\ \protect \BOthers {.}}{%
{\protect \APACyear {2022}}%
}]{%
Ojea-QuintanaPolarization22}
\APACinsertmetastar {%
Ojea-QuintanaPolarization22}%
\begin{APACrefauthors}%
Ojea~Quintana, I.%
, Reimann, R.%
, Cheong, M.%
, Alfano, M.%
\BCBL {} Klein, C.%
\end{APACrefauthors}%
\unskip\
\newblock
\APACrefYearMonthDay{2022}{12}{}.
\newblock
{\BBOQ}\APACrefatitle {Polarization and trust in the evolution of vaccine discourse on Twitter during COVID-19} {Polarization and trust in the evolution of vaccine discourse on twitter during covid-19}.{\BBCQ}
\newblock
\APACjournalVolNumPages{PLOS ONE}{17}{12}{e0277292,}
\newblock
\begin{APACrefDOI} \doi{10.1371/journal.pone.0277292} \end{APACrefDOI}
\newblock

\newblock

\PrintBackRefs{\CurrentBib}

\bibitem [\protect \citeauthoryear {%
Oktar%
\ \BBA {} Lombrozo%
}{%
Oktar%
\ \BBA {} Lombrozo%
}{%
{\protect \APACyear {2025}}%
}]{%
oktar2025aggregated}
\APACinsertmetastar {%
oktar2025aggregated}%
\begin{APACrefauthors}%
Oktar, K.%
\BCBT {}\ \BBA {} Lombrozo, T.%
\end{APACrefauthors}%
\unskip\
\newblock
\APACrefYearMonthDay{2025}{}{}.
\newblock
{\BBOQ}\APACrefatitle {How aggregated opinions shape beliefs} {How aggregated opinions shape beliefs}.{\BBCQ}
\newblock
\APACjournalVolNumPages{Nature Reviews Psychology}{4}{2}{81--95,}
\newblock
\begin{APACrefDOI} \doi{10.1038/s44159-024-00398-7} \end{APACrefDOI}
\newblock
\begin{APACrefURL} {https://www.nature.com/articles/s44159-024-00398-7} \end{APACrefURL}
\newblock

\newblock

\PrintBackRefs{\CurrentBib}

\bibitem [\protect \citeauthoryear {%
Oktar%
, Lombrozo%
\BCBL {}\ \BBA {} Griffiths%
}{%
Oktar%
\ \protect \BOthers {.}}{%
{\protect \APACyear {2024}}%
}]{%
OktarLombrozoGriffiths2024}
\APACinsertmetastar {%
OktarLombrozoGriffiths2024}%
\begin{APACrefauthors}%
Oktar, K.%
, Lombrozo, T.%
\BCBL {} Griffiths, T.L.%
\end{APACrefauthors}%
\unskip\
\newblock
\APACrefYearMonthDay{2024}{}{}.
\newblock
{\BBOQ}\APACrefatitle {Learning from aggregated opinion} {Learning from aggregated opinion}.{\BBCQ}
\newblock
\APACjournalVolNumPages{Psychological Science}{35}{9}{1010--1024,}
\newblock
\begin{APACrefDOI} \doi{10.1177/09567976241251741} \end{APACrefDOI}
\newblock

\newblock

\PrintBackRefs{\CurrentBib}

\bibitem [\protect \citeauthoryear {%
Pei%
\ \BBA {} Jurgens%
}{%
Pei%
\ \BBA {} Jurgens%
}{%
{\protect \APACyear {2021}}%
}]{%
pei2021measuring}
\APACinsertmetastar {%
pei2021measuring}%
\begin{APACrefauthors}%
Pei, J.%
\BCBT {}\ \BBA {} Jurgens, D.%
\end{APACrefauthors}%
\unskip\
\newblock
\APACrefYearMonthDay{2021}{}{}.
\newblock
{\BBOQ}\APACrefatitle {Measuring Sentence-Level and Aspect-Level (Un)certainty in Science Communications} {Measuring sentence-level and aspect-level (un)certainty in science communications}.{\BBCQ}
\newblock
 \APACrefbtitle {Proceedings of the 2021 Conference on Empirical Methods in Natural Language Processing} {Proceedings of the 2021 conference on empirical methods in natural language processing}\ (\BPGS\ 9959--10011).
\newblock
\APACaddressPublisher{}{Association for Computational Linguistics}.
\newblock
\begin{APACrefURL} {https://aclanthology.org/2021.emnlp-main.784} \end{APACrefURL}
\PrintBackRefs{\CurrentBib}

\bibitem [\protect \citeauthoryear {%
Petrocelli%
, Tormala%
\BCBL {}\ \BBA {} Rucker%
}{%
Petrocelli%
\ \protect \BOthers {.}}{%
{\protect \APACyear {2007}}%
}]{%
petrocelli2007unpacking}
\APACinsertmetastar {%
petrocelli2007unpacking}%
\begin{APACrefauthors}%
Petrocelli, J.V.%
, Tormala, Z.L.%
\BCBL {} Rucker, D.D.%
\end{APACrefauthors}%
\unskip\
\newblock
\APACrefYearMonthDay{2007}{}{}.
\newblock
{\BBOQ}\APACrefatitle {Unpacking Attitude Certainty: Attitude Clarity and Attitude Correctness} {Unpacking attitude certainty: Attitude clarity and attitude correctness}.{\BBCQ}
\newblock
\APACjournalVolNumPages{Journal of Personality and Social Psychology}{92}{1}{30--41,}
\newblock
\begin{APACrefDOI} \doi{10.1037/0022-3514.92.1.30} \end{APACrefDOI}
\newblock

\newblock

\PrintBackRefs{\CurrentBib}

\bibitem [\protect \citeauthoryear {%
R.~Petty%
, Wegener%
\BCBL {}\ \BBA {} Fabrigar%
}{%
R.~Petty%
\ \protect \BOthers {.}}{%
{\protect \APACyear {1997}}%
}]{%
Petty1997Attitudes}
\APACinsertmetastar {%
Petty1997Attitudes}%
\begin{APACrefauthors}%
Petty, R.%
, Wegener, D.%
\BCBL {} Fabrigar, L.%
\end{APACrefauthors}%
\unskip\
\newblock
\APACrefYearMonthDay{1997}{}{}.
\newblock
{\BBOQ}\APACrefatitle {Attitudes and attitude change.} {Attitudes and attitude change.}{\BBCQ}
\newblock
\APACjournalVolNumPages{Annual Review of Psychology}{48}{}{609--647,}
\newblock
\begin{APACrefDOI} \doi{10.1146/ANNUREV.PSYCH.48.1.609} \end{APACrefDOI}
\newblock

\newblock

\PrintBackRefs{\CurrentBib}

\bibitem [\protect \citeauthoryear {%
R.E.~Petty%
\ \BBA {} Cacioppo%
}{%
R.E.~Petty%
\ \BBA {} Cacioppo%
}{%
{\protect \APACyear {1986}}%
}]{%
Petty1986TheEL}
\APACinsertmetastar {%
Petty1986TheEL}%
\begin{APACrefauthors}%
Petty, R.E.%
\BCBT {}\ \BBA {} Cacioppo, J.T.%
\end{APACrefauthors}%
\unskip\
\newblock
\APACrefYearMonthDay{1986}{}{}.
\newblock
{\BBOQ}\APACrefatitle {The Elaboration Likelihood Model of Persuasion} {The elaboration likelihood model of persuasion}.{\BBCQ}
\newblock
 \APACrefbtitle {Advances in Experimental Social Psychology} {Advances in experimental social psychology}\ (\BVOL~19, \BPGS\ 123--205).
\newblock
\APACaddressPublisher{}{Academic Press}.
\PrintBackRefs{\CurrentBib}

\bibitem [\protect \citeauthoryear {%
Pomerantz%
, Chaiken%
\BCBL {}\ \BBA {} Tordesillas%
}{%
Pomerantz%
\ \protect \BOthers {.}}{%
{\protect \APACyear {1995}}%
}]{%
Pomerantz1995Attitude}
\APACinsertmetastar {%
Pomerantz1995Attitude}%
\begin{APACrefauthors}%
Pomerantz, E.%
, Chaiken, S.%
\BCBL {} Tordesillas, R.S.%
\end{APACrefauthors}%
\unskip\
\newblock
\APACrefYearMonthDay{1995}{}{}.
\newblock
{\BBOQ}\APACrefatitle {Attitude strength and resistance processes.} {Attitude strength and resistance processes.}{\BBCQ}
\newblock
\APACjournalVolNumPages{Journal of Personality and Social Psychology}{69}{3}{408--419,}
\newblock
\begin{APACrefDOI} \doi{10.1037/0022-3514.69.3.408} \end{APACrefDOI}
\newblock

\newblock

\PrintBackRefs{\CurrentBib}

\bibitem [\protect \citeauthoryear {%
Pryor%
, Perfors%
\BCBL {}\ \BBA {} Howe%
}{%
Pryor%
\ \protect \BOthers {.}}{%
{\protect \APACyear {2019}}%
}]{%
pryor2019even}
\APACinsertmetastar {%
pryor2019even}%
\begin{APACrefauthors}%
Pryor, C.%
, Perfors, A.%
\BCBL {} Howe, P.D.L.%
\end{APACrefauthors}%
\unskip\
\newblock
\APACrefYearMonthDay{2019}{}{}.
\newblock
{\BBOQ}\APACrefatitle {Even arbitrary norms influence moral decision-making} {Even arbitrary norms influence moral decision-making}.{\BBCQ}
\newblock
\APACjournalVolNumPages{Nature Human Behaviour}{3}{1}{57--62,}
\newblock
\begin{APACrefDOI} \doi{10.1038/s41562-018-0489-y} \end{APACrefDOI}
\newblock

\newblock

\PrintBackRefs{\CurrentBib}

\bibitem [\protect \citeauthoryear {%
Pulford%
, Colman%
, Buabang%
\BCBL {}\ \BBA {} Krockow%
}{%
Pulford%
\ \protect \BOthers {.}}{%
{\protect \APACyear {2018}}%
}]{%
pulford2018persuasive}
\APACinsertmetastar {%
pulford2018persuasive}%
\begin{APACrefauthors}%
Pulford, B.D.%
, Colman, A.M.%
, Buabang, E.K.%
\BCBL {} Krockow, E.M.%
\end{APACrefauthors}%
\unskip\
\newblock
\APACrefYearMonthDay{2018}{}{}.
\newblock
{\BBOQ}\APACrefatitle {The Persuasive Power of Knowledge: Testing the Confidence Heuristic} {The persuasive power of knowledge: Testing the confidence heuristic}.{\BBCQ}
\newblock
\APACjournalVolNumPages{Journal of Experimental Psychology: General}{147}{10}{1431--1444,}
\newblock
\begin{APACrefDOI} \doi{10.1037/xge0000471} \end{APACrefDOI}
\newblock

\newblock

\PrintBackRefs{\CurrentBib}

\bibitem [\protect \citeauthoryear {%
Ruff%
\ \BBA {} Fehr%
}{%
Ruff%
\ \BBA {} Fehr%
}{%
{\protect \APACyear {2014}}%
}]{%
Ruff2014TheNO}
\APACinsertmetastar {%
Ruff2014TheNO}%
\begin{APACrefauthors}%
Ruff, C.C.%
\BCBT {}\ \BBA {} Fehr, E.%
\end{APACrefauthors}%
\unskip\
\newblock
\APACrefYearMonthDay{2014}{}{}.
\newblock
{\BBOQ}\APACrefatitle {The neurobiology of rewards and values in social decision making} {The neurobiology of rewards and values in social decision making}.{\BBCQ}
\newblock
\APACjournalVolNumPages{Nature Reviews Neuroscience}{15}{8}{549--562,}
\newblock
\begin{APACrefDOI} \doi{10.1038/nrn3776} \end{APACrefDOI}
\newblock

\newblock

\PrintBackRefs{\CurrentBib}

\bibitem [\protect \citeauthoryear {%
Salvi%
, Horta~Ribeiro%
, Gallotti%
\BCBL {}\ \BBA {} West%
}{%
Salvi%
\ \protect \BOthers {.}}{%
{\protect \APACyear {2025}}%
}]{%
salvi2025conversational}
\APACinsertmetastar {%
salvi2025conversational}%
\begin{APACrefauthors}%
Salvi, F.%
, Horta~Ribeiro, M.%
, Gallotti, R.%
\BCBL {} West, R.%
\end{APACrefauthors}%
\unskip\
\newblock
\APACrefYearMonthDay{2025}{}{}.
\newblock
{\BBOQ}\APACrefatitle {On the conversational persuasiveness of {GPT-4}} {On the conversational persuasiveness of {GPT-4}}.{\BBCQ}
\newblock
\APACjournalVolNumPages{Nature Human Behaviour}{9}{8}{1645--1653,}
\newblock
\begin{APACrefDOI} \doi{10.1038/s41562-025-02194-6} \end{APACrefDOI}
\newblock

\newblock

\PrintBackRefs{\CurrentBib}

\bibitem [\protect \citeauthoryear {%
Schuster%
\ \BBA {} Kilov%
}{%
Schuster%
\ \BBA {} Kilov%
}{%
{\protect \APACyear {2025}}%
}]{%
SchusterKilov2025MoralDisagreement}
\APACinsertmetastar {%
SchusterKilov2025MoralDisagreement}%
\begin{APACrefauthors}%
Schuster, N.%
\BCBT {}\ \BBA {} Kilov, D.%
\end{APACrefauthors}%
\unskip\
\newblock
\APACrefYearMonthDay{2025}{}{}.
\newblock
{\BBOQ}\APACrefatitle {Moral disagreement and the limits of {AI} value alignment: a dual challenge of epistemic justification and political legitimacy} {Moral disagreement and the limits of {AI} value alignment: a dual challenge of epistemic justification and political legitimacy}.{\BBCQ}
\newblock
\APACjournalVolNumPages{AI \& Society}{40}{8}{6073--6087,}
\newblock
\begin{APACrefDOI} \doi{10.1007/s00146-025-02427-2} \end{APACrefDOI}
\newblock

\newblock

\PrintBackRefs{\CurrentBib}

\bibitem [\protect \citeauthoryear {%
Scott%
\ \BBA {} Bruce%
}{%
Scott%
\ \BBA {} Bruce%
}{%
{\protect \APACyear {1995}}%
}]{%
scott1995decision}
\APACinsertmetastar {%
scott1995decision}%
\begin{APACrefauthors}%
Scott, S.G.%
\BCBT {}\ \BBA {} Bruce, R.A.%
\end{APACrefauthors}%
\unskip\
\newblock
\APACrefYearMonthDay{1995}{}{}.
\newblock
{\BBOQ}\APACrefatitle {Decision-making style: The development and assessment of a new measure} {Decision-making style: The development and assessment of a new measure}.{\BBCQ}
\newblock
\APACjournalVolNumPages{Educational and Psychological Measurement}{55}{5}{818--831,}
\newblock
\begin{APACrefDOI} \doi{10.1177/0013164495055005017} \end{APACrefDOI}
\newblock

\newblock

\PrintBackRefs{\CurrentBib}

\bibitem [\protect \citeauthoryear {%
Skitka%
\ \BBA {} Bauman%
}{%
Skitka%
\ \BBA {} Bauman%
}{%
{\protect \APACyear {2008}}%
}]{%
skitka2008moral}
\APACinsertmetastar {%
skitka2008moral}%
\begin{APACrefauthors}%
Skitka, L.J.%
\BCBT {}\ \BBA {} Bauman, C.W.%
\end{APACrefauthors}%
\unskip\
\newblock
\APACrefYearMonthDay{2008}{}{}.
\newblock
{\BBOQ}\APACrefatitle {Moral conviction and political engagement} {Moral conviction and political engagement}.{\BBCQ}
\newblock
\APACjournalVolNumPages{Political Psychology}{29}{1}{29--54,}
\newblock
\begin{APACrefDOI} \doi{10.1111/j.1467-9221.2007.00611.x} \end{APACrefDOI}
\newblock

\newblock

\PrintBackRefs{\CurrentBib}

\bibitem [\protect \citeauthoryear {%
Skitka%
, Hanson%
, Morgan%
\BCBL {}\ \BBA {} Wisneski%
}{%
Skitka%
\ \protect \BOthers {.}}{%
{\protect \APACyear {2021}}%
}]{%
skitka2021psychology}
\APACinsertmetastar {%
skitka2021psychology}%
\begin{APACrefauthors}%
Skitka, L.J.%
, Hanson, B.E.%
, Morgan, G.S.%
\BCBL {} Wisneski, D.C.%
\end{APACrefauthors}%
\unskip\
\newblock
\APACrefYearMonthDay{2021}{}{}.
\newblock
{\BBOQ}\APACrefatitle {The psychology of moral conviction} {The psychology of moral conviction}.{\BBCQ}
\newblock
\APACjournalVolNumPages{Annual Review of Psychology}{72}{}{347--366,}
\newblock
\begin{APACrefDOI} \doi{10.1146/annurev-psych-063020-030612} \end{APACrefDOI}
\newblock

\newblock

\PrintBackRefs{\CurrentBib}

\bibitem [\protect \citeauthoryear {%
Skitka%
\ \BBA {} Morgan%
}{%
Skitka%
\ \BBA {} Morgan%
}{%
{\protect \APACyear {2014}}%
}]{%
skitka2014social}
\APACinsertmetastar {%
skitka2014social}%
\begin{APACrefauthors}%
Skitka, L.J.%
\BCBT {}\ \BBA {} Morgan, G.S.%
\end{APACrefauthors}%
\unskip\
\newblock
\APACrefYearMonthDay{2014}{}{}.
\newblock
{\BBOQ}\APACrefatitle {The social and political implications of moral conviction} {The social and political implications of moral conviction}.{\BBCQ}
\newblock
\APACjournalVolNumPages{Political psychology}{35}{S1}{95--110,}
\newblock
\begin{APACrefDOI} \doi{10.1111/pops.12166} \end{APACrefDOI}
\newblock

\newblock

\PrintBackRefs{\CurrentBib}

\bibitem [\protect \citeauthoryear {%
Smith%
, Hogg%
, Martin%
\BCBL {}\ \BBA {} Terry%
}{%
Smith%
\ \protect \BOthers {.}}{%
{\protect \APACyear {2007}}%
}]{%
Smith2007UncertaintyAT}
\APACinsertmetastar {%
Smith2007UncertaintyAT}%
\begin{APACrefauthors}%
Smith, J.R.%
, Hogg, M.A.%
, Martin, R.%
\BCBL {} Terry, D.J.%
\end{APACrefauthors}%
\unskip\
\newblock
\APACrefYearMonthDay{2007}{}{}.
\newblock
{\BBOQ}\APACrefatitle {Uncertainty and the influence of group norms in the attitude-behaviour relationship.} {Uncertainty and the influence of group norms in the attitude-behaviour relationship.}{\BBCQ}
\newblock
\APACjournalVolNumPages{British Journal of Social Psychology}{46}{4}{769--792,}
\newblock
\begin{APACrefDOI} \doi{10.1348/014466606X164439} \end{APACrefDOI}
\newblock

\newblock

\PrintBackRefs{\CurrentBib}

\bibitem [\protect \citeauthoryear {%
Strimling%
, Vartanova%
, Jansson%
\BCBL {}\ \BBA {} Eriksson%
}{%
Strimling%
\ \protect \BOthers {.}}{%
{\protect \APACyear {2019}}%
}]{%
Strimling2018TheCB}
\APACinsertmetastar {%
Strimling2018TheCB}%
\begin{APACrefauthors}%
Strimling, P.%
, Vartanova, I.%
, Jansson, F.%
\BCBL {} Eriksson, K.%
\end{APACrefauthors}%
\unskip\
\newblock
\APACrefYearMonthDay{2019}{}{}.
\newblock
{\BBOQ}\APACrefatitle {The connection between moral positions and moral arguments drives opinion change} {The connection between moral positions and moral arguments drives opinion change}.{\BBCQ}
\newblock
\APACjournalVolNumPages{Nature Human Behaviour}{3}{9}{922--930,}
\newblock
\begin{APACrefDOI} \doi{10.1038/s41562-019-0647-x} \end{APACrefDOI}
\newblock

\newblock

\PrintBackRefs{\CurrentBib}

\bibitem [\protect \citeauthoryear {%
Tan%
, Niculae%
, Danescu-Niculescu-Mizil%
\BCBL {}\ \BBA {} Lee%
}{%
Tan%
\ \protect \BOthers {.}}{%
{\protect \APACyear {2016}}%
}]{%
Tan2016WinningAI}
\APACinsertmetastar {%
Tan2016WinningAI}%
\begin{APACrefauthors}%
Tan, C.%
, Niculae, V.%
, Danescu-Niculescu-Mizil, C.%
\BCBL {} Lee, L.%
\end{APACrefauthors}%
\unskip\
\newblock
\APACrefYearMonthDay{2016}{}{}.
\newblock
{\BBOQ}\APACrefatitle {Winning Arguments: Interaction Dynamics and Persuasion Strategies in Good-faith Online Discussions} {Winning arguments: Interaction dynamics and persuasion strategies in good-faith online discussions}.{\BBCQ}
\newblock
 \APACrefbtitle {Proceedings of the 25th International Conference on World Wide Web} {Proceedings of the 25th international conference on world wide web}\ (\BPGS\ 613--624).
\PrintBackRefs{\CurrentBib}

\bibitem [\protect \citeauthoryear {%
Tarsney%
, Thomas%
\BCBL {}\ \BBA {} MacAskill%
}{%
Tarsney%
\ \protect \BOthers {.}}{%
{\protect \APACyear {2024}}%
}]{%
tarsney2024moral}
\APACinsertmetastar {%
tarsney2024moral}%
\begin{APACrefauthors}%
Tarsney, C.%
, Thomas, T.%
\BCBL {} MacAskill, W.%
\end{APACrefauthors}%
\unskip\
\newblock
\APACrefYearMonthDay{2024}{}{}.
\newblock
{\BBOQ}\APACrefatitle {Moral Decision-Making Under Uncertainty} {Moral decision-making under uncertainty}.{\BBCQ}
\newblock
 E.N.~Zalta\ \BBA {} U.~Nodelman\ (\BEDS), \APACrefbtitle {The Stanford Encyclopedia of Philosophy} {The stanford encyclopedia of philosophy}\ (\PrintOrdinal{Spring 2024}\ \BEd).
\newblock
\APACaddressPublisher{}{Metaphysics Research Lab, Stanford University}.
\newblock
\begin{APACrefURL} {https://plato.stanford.edu/archives/spr2024/entries/moral-decision-uncertainty/} \end{APACrefURL}
\PrintBackRefs{\CurrentBib}

\bibitem [\protect \citeauthoryear {%
Tessler%
\ \protect \BOthers {.}}{%
Tessler%
\ \protect \BOthers {.}}{%
{\protect \APACyear {2024}}%
}]{%
tessler2024ai}
\APACinsertmetastar {%
tessler2024ai}%
\begin{APACrefauthors}%
Tessler, M.H.%
, Bakker, M.A.%
, Jarrett, D.%
, Sheahan, H.%
, Chadwick, M.J.%
, Koster, R.%
\BDBL {}Summerfield, C.%
\end{APACrefauthors}%
\unskip\
\newblock
\APACrefYearMonthDay{2024}{}{}.
\newblock
{\BBOQ}\APACrefatitle {{AI} can help humans find common ground in democratic deliberation} {{AI} can help humans find common ground in democratic deliberation}.{\BBCQ}
\newblock
\APACjournalVolNumPages{Science}{386}{6719}{eadq2852,}
\newblock
\begin{APACrefDOI} \doi{10.1126/science.adq2852} \end{APACrefDOI}
\newblock

\newblock

\PrintBackRefs{\CurrentBib}

\bibitem [\protect \citeauthoryear {%
Tetlock%
}{%
Tetlock%
}{%
{\protect \APACyear {1983}}%
}]{%
tetlock1983accountability}
\APACinsertmetastar {%
tetlock1983accountability}%
\begin{APACrefauthors}%
Tetlock, P.E.%
\end{APACrefauthors}%
\unskip\
\newblock
\APACrefYearMonthDay{1983}{}{}.
\newblock
{\BBOQ}\APACrefatitle {Accountability and complexity of thought.} {Accountability and complexity of thought.}{\BBCQ}
\newblock
\APACjournalVolNumPages{Journal of personality and social psychology}{45}{1}{74--83,}
\newblock
\begin{APACrefDOI} \doi{10.1037/0022-3514.45.1.74} \end{APACrefDOI}
\newblock

\newblock

\PrintBackRefs{\CurrentBib}

\bibitem [\protect \citeauthoryear {%
Tetlock%
\ \BBA {} Boettger%
}{%
Tetlock%
\ \BBA {} Boettger%
}{%
{\protect \APACyear {1989}}%
}]{%
Tetlock1989AccountabilityAS}
\APACinsertmetastar {%
Tetlock1989AccountabilityAS}%
\begin{APACrefauthors}%
Tetlock, P.E.%
\BCBT {}\ \BBA {} Boettger, R.%
\end{APACrefauthors}%
\unskip\
\newblock
\APACrefYearMonthDay{1989}{}{}.
\newblock
{\BBOQ}\APACrefatitle {Accountability: a social magnifier of the dilution effect.} {Accountability: a social magnifier of the dilution effect.}{\BBCQ}
\newblock
\APACjournalVolNumPages{Journal of Personality and Social Psychology}{57}{3}{388--398,}
\newblock
\begin{APACrefDOI} \doi{10.1037/0022-3514.57.3.388} \end{APACrefDOI}
\newblock

\newblock

\PrintBackRefs{\CurrentBib}

\bibitem [\protect \citeauthoryear {%
Tetlock%
, Skitka%
\BCBL {}\ \BBA {} Boettger%
}{%
Tetlock%
\ \protect \BOthers {.}}{%
{\protect \APACyear {1989}}%
}]{%
tetlock1989social}
\APACinsertmetastar {%
tetlock1989social}%
\begin{APACrefauthors}%
Tetlock, P.E.%
, Skitka, L.%
\BCBL {} Boettger, R.%
\end{APACrefauthors}%
\unskip\
\newblock
\APACrefYearMonthDay{1989}{}{}.
\newblock
{\BBOQ}\APACrefatitle {Social and cognitive strategies for coping with accountability: conformity, complexity, and bolstering.} {Social and cognitive strategies for coping with accountability: conformity, complexity, and bolstering.}{\BBCQ}
\newblock
\APACjournalVolNumPages{Journal of personality and social psychology}{57}{4}{632--640,}
\newblock
\begin{APACrefDOI} \doi{10.1037/0022-3514.57.4.632} \end{APACrefDOI}
\newblock

\newblock

\PrintBackRefs{\CurrentBib}

\bibitem [\protect \citeauthoryear {%
Toelch%
\ \BBA {} Dolan%
}{%
Toelch%
\ \BBA {} Dolan%
}{%
{\protect \APACyear {2015}}%
}]{%
Toelch2015InformationalAN}
\APACinsertmetastar {%
Toelch2015InformationalAN}%
\begin{APACrefauthors}%
Toelch, U.%
\BCBT {}\ \BBA {} Dolan, R.J.%
\end{APACrefauthors}%
\unskip\
\newblock
\APACrefYearMonthDay{2015}{}{}.
\newblock
{\BBOQ}\APACrefatitle {Informational and Normative Influences in Conformity from a Neurocomputational Perspective} {Informational and normative influences in conformity from a neurocomputational perspective}.{\BBCQ}
\newblock
\APACjournalVolNumPages{Trends in Cognitive Sciences}{19}{10}{579--589,}
\newblock
\begin{APACrefDOI} \doi{10.1016/j.tics.2015.07.007} \end{APACrefDOI}
\newblock

\newblock

\PrintBackRefs{\CurrentBib}

\bibitem [\protect \citeauthoryear {%
Tormala%
\ \BBA {} Petty%
}{%
Tormala%
\ \BBA {} Petty%
}{%
{\protect \APACyear {2002}}%
}]{%
Tormala2002What}
\APACinsertmetastar {%
Tormala2002What}%
\begin{APACrefauthors}%
Tormala, Z.L.%
\BCBT {}\ \BBA {} Petty, R.%
\end{APACrefauthors}%
\unskip\
\newblock
\APACrefYearMonthDay{2002}{}{}.
\newblock
{\BBOQ}\APACrefatitle {What doesn't kill me makes me stronger: the effects of resisting persuasion on attitude certainty.} {What doesn't kill me makes me stronger: the effects of resisting persuasion on attitude certainty.}{\BBCQ}
\newblock
\APACjournalVolNumPages{Journal of Personality and Social Psychology}{83}{6}{1298--1313,}
\newblock
\begin{APACrefDOI} \doi{10.1037/0022-3514.83.6.1298} \end{APACrefDOI}
\newblock

\newblock

\PrintBackRefs{\CurrentBib}

\bibitem [\protect \citeauthoryear {%
Tormala%
\ \BBA {} Rucker%
}{%
Tormala%
\ \BBA {} Rucker%
}{%
{\protect \APACyear {2018}}%
}]{%
tormala2018attitude}
\APACinsertmetastar {%
tormala2018attitude}%
\begin{APACrefauthors}%
Tormala, Z.L.%
\BCBT {}\ \BBA {} Rucker, D.D.%
\end{APACrefauthors}%
\unskip\
\newblock
\APACrefYearMonthDay{2018}{}{}.
\newblock
{\BBOQ}\APACrefatitle {Attitude certainty: Antecedents, consequences, and new directions} {Attitude certainty: Antecedents, consequences, and new directions}.{\BBCQ}
\newblock
\APACjournalVolNumPages{Consumer Psychology Review}{1}{1}{72--89,}
\newblock
\begin{APACrefDOI} \doi{10.1002/arcp.1004} \end{APACrefDOI}
\newblock

\newblock

\PrintBackRefs{\CurrentBib}

\bibitem [\protect \citeauthoryear {%
van Baar%
\ \BBA {} FeldmanHall%
}{%
van Baar%
\ \BBA {} FeldmanHall%
}{%
{\protect \APACyear {2022}}%
}]{%
vanBaar2021ThePM}
\APACinsertmetastar {%
vanBaar2021ThePM}%
\begin{APACrefauthors}%
van Baar, J.M.%
\BCBT {}\ \BBA {} FeldmanHall, O.%
\end{APACrefauthors}%
\unskip\
\newblock
\APACrefYearMonthDay{2022}{}{}.
\newblock
{\BBOQ}\APACrefatitle {The polarized mind in context: Interdisciplinary approaches to the psychology of political polarization.} {The polarized mind in context: Interdisciplinary approaches to the psychology of political polarization.}{\BBCQ}
\newblock
\APACjournalVolNumPages{American Psychologist}{77}{3}{394--408,}
\newblock
\begin{APACrefDOI} \doi{10.1037/amp0000814} \end{APACrefDOI}
\newblock

\newblock

\PrintBackRefs{\CurrentBib}

\bibitem [\protect \citeauthoryear {%
Van~Bavel%
, Reinero%
, Spring%
, Harris%
\BCBL {}\ \BBA {} Duke%
}{%
Van~Bavel%
\ \protect \BOthers {.}}{%
{\protect \APACyear {2021}}%
}]{%
van2021speaking}
\APACinsertmetastar {%
van2021speaking}%
\begin{APACrefauthors}%
Van~Bavel, J.J.%
, Reinero, D.A.%
, Spring, V.%
, Harris, E.A.%
\BCBL {} Duke, A.%
\end{APACrefauthors}%
\unskip\
\newblock
\APACrefYearMonthDay{2021}{}{}.
\newblock
{\BBOQ}\APACrefatitle {Speaking my truth: Why personal experiences can bridge divides but mislead} {Speaking my truth: Why personal experiences can bridge divides but mislead}.{\BBCQ}
\newblock
\APACjournalVolNumPages{Proceedings of the National Academy of Sciences}{118}{8}{e2100280118,}
\newblock
\begin{APACrefDOI} \doi{10.1073/pnas.2100280118} \end{APACrefDOI}
\newblock

\newblock

\PrintBackRefs{\CurrentBib}

\bibitem [\protect \citeauthoryear {%
van~der Bles%
\ \protect \BOthers {.}}{%
van~der Bles%
\ \protect \BOthers {.}}{%
{\protect \APACyear {2019}}%
}]{%
vanderbles2019communicating}
\APACinsertmetastar {%
vanderbles2019communicating}%
\begin{APACrefauthors}%
van~der Bles, A.M.%
, van~der Linden, S.%
, Freeman, A.L.J.%
, Mitchell, J.%
, Galvao, A.B.%
, Zaval, L.%
\BCBL {} Spiegelhalter, D.J.%
\end{APACrefauthors}%
\unskip\
\newblock
\APACrefYearMonthDay{2019}{}{}.
\newblock
{\BBOQ}\APACrefatitle {Communicating Uncertainty About Facts, Numbers and Science} {Communicating uncertainty about facts, numbers and science}.{\BBCQ}
\newblock
\APACjournalVolNumPages{Royal Society Open Science}{6}{5}{181870,}
\newblock
\begin{APACrefDOI} \doi{10.1098/rsos.181870} \end{APACrefDOI}
\newblock

\newblock

\PrintBackRefs{\CurrentBib}

\bibitem [\protect \citeauthoryear {%
van~der Bles%
, van~der Linden%
, Freeman%
\BCBL {}\ \BBA {} Spiegelhalter%
}{%
van~der Bles%
\ \protect \BOthers {.}}{%
{\protect \APACyear {2020}}%
}]{%
vanderbles2020effects}
\APACinsertmetastar {%
vanderbles2020effects}%
\begin{APACrefauthors}%
van~der Bles, A.M.%
, van~der Linden, S.%
, Freeman, A.L.J.%
\BCBL {} Spiegelhalter, D.J.%
\end{APACrefauthors}%
\unskip\
\newblock
\APACrefYearMonthDay{2020}{}{}.
\newblock
{\BBOQ}\APACrefatitle {The Effects of Communicating Uncertainty on Public Trust in Facts and Numbers} {The effects of communicating uncertainty on public trust in facts and numbers}.{\BBCQ}
\newblock
\APACjournalVolNumPages{Proceedings of the National Academy of Sciences}{117}{14}{7672--7683,}
\newblock
\begin{APACrefDOI} \doi{10.1073/pnas.1913678117} \end{APACrefDOI}
\newblock

\newblock

\PrintBackRefs{\CurrentBib}

\bibitem [\protect \citeauthoryear {%
Vavova%
}{%
Vavova%
}{%
{\protect \APACyear {2014}}%
}]{%
vavova2014confidence}
\APACinsertmetastar {%
vavova2014confidence}%
\begin{APACrefauthors}%
Vavova, K.%
\end{APACrefauthors}%
\unskip\
\newblock
\APACrefYearMonthDay{2014}{}{}.
\newblock
{\BBOQ}\APACrefatitle {Confidence, evidence, and disagreement} {Confidence, evidence, and disagreement}.{\BBCQ}
\newblock
\APACjournalVolNumPages{Erkenntnis}{79}{Suppl 1}{173--183,}
\newblock
\begin{APACrefDOI} \doi{10.1007/s10670-013-9451-6} \end{APACrefDOI}
\newblock

\newblock

\PrintBackRefs{\CurrentBib}

\bibitem [\protect \citeauthoryear {%
Wei%
\ \protect \BOthers {.}}{%
Wei%
\ \protect \BOthers {.}}{%
{\protect \APACyear {2022}}%
}]{%
wei2022chain}
\APACinsertmetastar {%
wei2022chain}%
\begin{APACrefauthors}%
Wei, J.%
, Wang, X.%
, Schuurmans, D.%
, Bosma, M.%
, Ichter, B.%
, Xia, F.%
\BDBL {}Zhou, D.%
\end{APACrefauthors}%
\unskip\
\newblock
\APACrefYearMonthDay{2022}{}{}.
\newblock
{\BBOQ}\APACrefatitle {Chain-of-Thought Prompting Elicits Reasoning in Large Language Models} {Chain-of-thought prompting elicits reasoning in large language models}.{\BBCQ}
\newblock
 \APACrefbtitle {Advances in Neural Information Processing Systems} {Advances in neural information processing systems}\ (\BVOL~35, \BPGS\ 24824--24837).
\newblock
\begin{APACrefURL} {https://proceedings.neurips.cc/paper\_files/paper/2022/hash/9d5609613524ecf4f15af0f7b31abca4-Abstract.html} \end{APACrefURL}
\PrintBackRefs{\CurrentBib}

\bibitem [\protect \citeauthoryear {%
Wood%
, Lundgren%
, Ouellette%
, Busceme%
\BCBL {}\ \BBA {} Blackstone%
}{%
Wood%
\ \protect \BOthers {.}}{%
{\protect \APACyear {1994}}%
}]{%
Wood1994}
\APACinsertmetastar {%
Wood1994}%
\begin{APACrefauthors}%
Wood, W.%
, Lundgren, S.%
, Ouellette, J.A.%
, Busceme, S.%
\BCBL {} Blackstone, T.%
\end{APACrefauthors}%
\unskip\
\newblock
\APACrefYearMonthDay{1994}{}{}.
\newblock
{\BBOQ}\APACrefatitle {Minority influence: A meta-analytic review of social influence processes} {Minority influence: A meta-analytic review of social influence processes}.{\BBCQ}
\newblock
\APACjournalVolNumPages{Psychological Bulletin}{115}{3}{323--345,}
\newblock
\begin{APACrefDOI} \doi{10.1037/0033-2909.115.3.323} \end{APACrefDOI}
\newblock

\newblock

\PrintBackRefs{\CurrentBib}

\bibitem [\protect \citeauthoryear {%
Xiong%
\ \protect \BOthers {.}}{%
Xiong%
\ \protect \BOthers {.}}{%
{\protect \APACyear {2024}}%
}]{%
xiong2024llms}
\APACinsertmetastar {%
xiong2024llms}%
\begin{APACrefauthors}%
Xiong, M.%
, Hu, Z.%
, Lu, X.%
, Li, Y.%
, Fu, J.%
, He, J.%
\BCBL {} Hooi, B.%
\end{APACrefauthors}%
\unskip\
\newblock
\APACrefYearMonthDay{2024}{}{}.
\newblock
{\BBOQ}\APACrefatitle {Can {LLMs} Express Their Uncertainty? An Empirical Evaluation of Confidence Elicitation in {LLMs}} {Can {LLMs} express their uncertainty? an empirical evaluation of confidence elicitation in {LLMs}}.{\BBCQ}
\newblock
 \APACrefbtitle {The Twelfth International Conference on Learning Representations (ICLR).} {The twelfth international conference on learning representations (iclr).}
\newblock
\begin{APACrefURL} {https://openreview.net/forum?id=gjeQKFxFpZ} \end{APACrefURL}
\PrintBackRefs{\CurrentBib}

\bibitem [\protect \citeauthoryear {%
Yarkoni%
\ \BBA {} Westfall%
}{%
Yarkoni%
\ \BBA {} Westfall%
}{%
{\protect \APACyear {2017}}%
}]{%
yarkoni2017choosing}
\APACinsertmetastar {%
yarkoni2017choosing}%
\begin{APACrefauthors}%
Yarkoni, T.%
\BCBT {}\ \BBA {} Westfall, J.%
\end{APACrefauthors}%
\unskip\
\newblock
\APACrefYearMonthDay{2017}{}{}.
\newblock
{\BBOQ}\APACrefatitle {Choosing Prediction Over Explanation in Psychology: {Lessons} from Machine Learning} {Choosing prediction over explanation in psychology: {Lessons} from machine learning}.{\BBCQ}
\newblock
\APACjournalVolNumPages{Perspectives on Psychological Science}{12}{6}{1100--1122,}
\newblock
\begin{APACrefDOI} \doi{10.1177/1745691617693393} \end{APACrefDOI}
\newblock

\newblock

\PrintBackRefs{\CurrentBib}

\bibitem [\protect \citeauthoryear {%
Ziems%
\ \protect \BOthers {.}}{%
Ziems%
\ \protect \BOthers {.}}{%
{\protect \APACyear {2024}}%
}]{%
ziems2024can}
\APACinsertmetastar {%
ziems2024can}%
\begin{APACrefauthors}%
Ziems, C.%
, Held, W.%
, Shaikh, O.%
, Chen, J.%
, Zhang, Z.%
\BCBL {} Yang, D.%
\end{APACrefauthors}%
\unskip\
\newblock
\APACrefYearMonthDay{2024}{}{}.
\newblock
{\BBOQ}\APACrefatitle {Can Large Language Models Transform Computational Social Science?} {Can large language models transform computational social science?}{\BBCQ}
\newblock
\APACjournalVolNumPages{Computational Linguistics}{50}{1}{237--291,}
\newblock
\begin{APACrefDOI} \doi{10.1162/coli_a_00502} \end{APACrefDOI}
\newblock

\newblock

\PrintBackRefs{\CurrentBib}

\bibitem [\protect \citeauthoryear {%
Ziems%
, Yu%
, Wang%
, Halevy%
\BCBL {}\ \BBA {} Yang%
}{%
Ziems%
\ \protect \BOthers {.}}{%
{\protect \APACyear {2022}}%
}]{%
Ziems2022TheMI}
\APACinsertmetastar {%
Ziems2022TheMI}%
\begin{APACrefauthors}%
Ziems, C.%
, Yu, J.%
, Wang, Y\BHBI C.%
, Halevy, A.%
\BCBL {} Yang, D.%
\end{APACrefauthors}%
\unskip\
\newblock
\APACrefYearMonthDay{2022}{{\APACmonth{05}}}{}.
\newblock
{\BBOQ}\APACrefatitle {The Moral Integrity Corpus: A Benchmark for Ethical Dialogue Systems} {The moral integrity corpus: A benchmark for ethical dialogue systems}.{\BBCQ}
\newblock
 S.~Muresan, P.~Nakov\BCBL {}\ \BBA {} A.~Villavicencio\ (\BEDS), \APACrefbtitle {Proceedings of the 60th Annual Meeting of the Association for Computational Linguistics (Volume 1: Long Papers)} {Proceedings of the 60th annual meeting of the association for computational linguistics (volume 1: Long papers)}\ (\BPGS\ 3755--3773).
\newblock
\APACaddressPublisher{Dublin, Ireland}{Association for Computational Linguistics}.
\newblock
\begin{APACrefURL} {https://aclanthology.org/2022.acl-long.261} \end{APACrefURL}
\PrintBackRefs{\CurrentBib}

\bibitem [\protect \citeauthoryear {%
Zuwerink%
\ \BBA {} Devine%
}{%
Zuwerink%
\ \BBA {} Devine%
}{%
{\protect \APACyear {1996}}%
}]{%
Zuwerink1996Attitude}
\APACinsertmetastar {%
Zuwerink1996Attitude}%
\begin{APACrefauthors}%
Zuwerink, J.R.%
\BCBT {}\ \BBA {} Devine, P.%
\end{APACrefauthors}%
\unskip\
\newblock
\APACrefYearMonthDay{1996}{}{}.
\newblock
{\BBOQ}\APACrefatitle {Attitude importance and resistance to persuasion: It's not just the thought that counts} {Attitude importance and resistance to persuasion: It's not just the thought that counts}.{\BBCQ}
\newblock
\APACjournalVolNumPages{Journal of Personality and Social Psychology}{70}{5}{931--944,}
\newblock
\begin{APACrefDOI} \doi{10.1037/0022-3514.70.5.931} \end{APACrefDOI}
\newblock

\newblock

\PrintBackRefs{\CurrentBib}

\end{thebibliography}
\end{document}